%% file: AArXiv20261030.tex
\documentclass[12pt]{article}
\usepackage[nodisplayskipstretch]{setspace}
\usepackage[margin=1in]{geometry}
\usepackage{caption} 
\usepackage{amsmath, amssymb}
\usepackage{longtable}
\renewcommand{\arraystretch}{1.2} 
\allowdisplaybreaks  
\usepackage{amsthm}
\usepackage{graphicx}
\usepackage{dcolumn}
\newcolumntype{d}[1]{D{.}{.}{#1}}
\usepackage{pifont}
\usepackage{enumitem} 
\usepackage{enumerate}
\usepackage{natbib}
\usepackage{makecell}   
\usepackage{algorithm}
\usepackage{algorithmic}

\usepackage{titlesec}
\titlespacing*{\section}{0pt}{*1}{*0} 
\titlespacing*{\subsection}{0pt}{*1}{*0} 
\titlespacing*{\subparagraph}{0pt}{\baselineskip}{0pt}
\usepackage{bm} 
\usepackage{bbm}
\usepackage{booktabs}
\usepackage{multirow}
\usepackage{xcolor}
\usepackage{pstricks}
\usepackage{array}
\usepackage{mathtools}
\usepackage{siunitx}
\usepackage{tikz}
\usepackage{collcell}
\usepackage{float}
\usepackage{placeins}
\usepackage{comment}
\usepackage{subcaption}
\usepackage{dcolumn}
\usepackage{url} 
\usepackage{amsthm}
\usepackage{array}
\usepackage{makecell}

\newcolumntype{C}[1]{>{\centering\arraybackslash}m{#1}} 
\renewcommand\arraystretch{1.1} 

\newcommand{\change}[1]{#1}
\newenvironment{changeblock}{}{}
\newenvironment{changetable}[1][htbp]{\begin{table}[#1]}{\end{table}}
\newcommand{\thesisversion}{1}   
\input{macros}

\begin{document}

\begin{center}
  {\LARGE\bfseries Exact, Nonparametric Sensitivity Analysis for Observational Studies of Contingency Tables}
  \par
  \vspace{1.2em}
  Elaine K. Chiu$^{*}$ and Hyunseung Kang \\[0.6em]
  \textit{Department of Statistics, University of Wisconsin--Madison, Madison, WI 53706, U.S.A.} \\[0.4em]
  $^{*}$Corresponding author. Email: chiuk2@gene.com \hspace{2em} hyunseung@stat.wisc.edu
\end{center}

\begin{abstract}
In observational studies, contingency tables are commonly used to examine associations between categorical variables. However, any test of association in contingency tables may be biased by unmeasured confounding, and existing sensitivity analyses typically assume a binary treatment or impose strong parametric assumptions on non-binary treatments. We develop an exact (non-asymptotic) and nonparametric sensitivity analysis for unmeasured confounding in $I \times J$ and $I\times J\times K$ contingency tables, accommodating both non-binary treatments and outcomes. Extending Rosenbaum’s generic bias sensitivity model, we derive a general method to compute the exact worst-case null distribution for any count-based test, including chi-squared and likelihood-based tests of association. We further provide specialized results for subfamilies of count-based tests that enable more efficient computation of the worst-case null distribution. Finally, we investigate test power in sensitivity analyses and show that tests exploiting all treatment and outcome levels can achieve higher power than tests that dichotomize the categorical variables. We illustrate the proposed methods with a re-analysis of the effect of pre-kindergarten care on math achievement using data from the Early Childhood Longitudinal Study. 
The \texttt{R} package \texttt{sensitivityIxJ} that implements the method is on CRAN.
\end{abstract}

\noindent%
\newpage
\section{Introduction}

\subsection{Motivation: Contingency Tables in Observational Studies and Bias from Unmeasured Confounding}  
In observational studies concerning the effect of a categorical treatment on a categorical outcome, the observed data are often structured as an \(I \times J\) contingency table, where the rows represent treatment levels, the columns represent outcome levels, and each cell records the number of subjects with a particular combination of treatment and outcome. In some cases, the data are structured as an \(I \times J \times K\) contingency table, where \(K\) indexes strata defined by pre-treatment, measured confounders. Once framed as a contingency table, several test statistics can be used to test the null hypothesis of no treatment effect—or equivalently, no association between the rows and the columns of the table. Common choices include the chi-squared test, the \(G^2\) test (i.e., a likelihood-based test), and ordinal tests (i.e., score-based tests); see Chapter 3 of \citet{agresti2012categorical} for an overview of testing in contingency tables.

Unfortunately, unmeasured confounding is unavoidable in observational studies and a statistically significant \(p\)-value from any of the aforementioned tests may reflect bias from unmeasured confounding. As a concrete example, consider an observational study on the effect of pre-kindergarten care (i.e., treatment) on math achievement (i.e., outcome) among Black children from low-income families—an important topic in early childhood education; see \citet{lee2022effects} for a recent review. In our re-analysis of the data in Section~\ref{sec:data_analysis}, there are three types of pre-kindergarten care (no nonparental care, relative care, and center-based care) and three types of math achievement (mastery in number and shape, relative size, and ordinality and sequences), all of which are categorical variables. We stratified the data by gender (boys and girls), so the data can be organized as a \(3 \times 3 \times 2\) contingency table; see Table~\ref{tb:black_girl_boy_low_income_single_table}. Under the assumption of no unmeasured confounding, the null hypothesis of no effect of pre-kindergarten care on math achievement is rejected at the $0.05$ significance level, with \(p\)-values of 0.006 and 0.013 for girls and boys, respectively. 
However, in the presence of unmeasured confounders such as parenting style or home learning environment, 
these \(p\)-values may provide misleading evidence about the effect of pre-kindergarten care.

Sensitivity analysis for unmeasured confounding assesses how a study's conclusion under no unmeasured confounding changes in the presence of one. Briefly, to test the null of no treatment effect, we compute the \emph{worst-case \(p\)-value}: the largest \(p\)-value that could arise after adjusting for the \emph{worst-case unmeasured confounder}, whose effect on the treatment or outcome is parametrized by a sensitivity parameter $\gamma$, with larger $\gamma$ corresponding to a larger effect. If the worst-case \(p\)-value remains below the significance level even for large $\gamma$, the study's conclusion is said to be robust, or insensitive, to such a confounder. For more details, see Section~\ref{sec:def_worst_case}.

Sensitivity analysis for unmeasured confounding has been extensively developed for binary treatment, i.e., when $I=2$ in a contingency table (e.g., \citet{rosenbaum1987sensitivity,tan2006distributional,ding2016sensitivity,fogarty2017studentized,huang2024variance, dorn2024sharp,wu2025sensitivity}). However, the literature on sensitivity analysis for non-binary, categorical treatments (i.e., when $I > 2$) is limited. Among the works on a non-binary treatment variable in a sensitivity analysis, most rely on a linear outcome model where the outcome is assumed to be a linear function of both the treatment and unmeasured confounder (e.g., \citet{frank2000impact,imbens_sensitivity,cinelli2020making}). 
\citet{rosenbaum1989sensitivity},  \citet{gastwirth1998dual}, and \citet{zhang2024sensitivity} proposed a sensitivity analysis for a non-binary treatment variable without parametric assumptions, specifically under Rosenbaum's sensitivity model \citep{rosenbaum1987sensitivity}. However, they assumed a matched study design 
and \citet{zhang2024sensitivity} further assumed a binary outcome (i.e., $J=2$ in a contingency table). \citet{Jesson2022Scalable} proposed a sensitivity analysis for continuous treatment without parametric assumptions under the marginal sensitivity model of \citet{tan2006distributional}, but they do not have inferential guarantees. \citet{chernozhukov2022long} and \citet{bonvini2022sensitivity} proposed a general framework for sensitivity analysis with binary and non-binary treatments. However, their inferential guarantees are asymptotic (i.e., non-exact). 

In practice, a simple, yet naive way to conduct sensitivity analysis with a non-binary treatment variable is to dichotomize the treatment variable and apply existing methods for sensitivity analysis with a binary treatment; see \citet{zhai2013estimating} and \citet{shen2025calibrated} for examples using Rosenbaum's sensitivity model and \citet{tan2006distributional}'s model, respectively. Unfortunately, such procedures based on dichotomization lose valuable information about treatment dose and have poor or no statistical power \citep{zhang2024sensitivity}. As we show in Section~\ref{sec:power_main}, using all levels of the treatment variable can deliver higher statistical power compared to a dichotomized version of the treatment variable.

\subsection{Our Contributions: Exact, Nonparametric Sensitivity Analysis of Unmeasured Confounding for Categorical Data }
\begin{figure}[ht]
  \centering
  \includegraphics[width=1.00\textwidth]{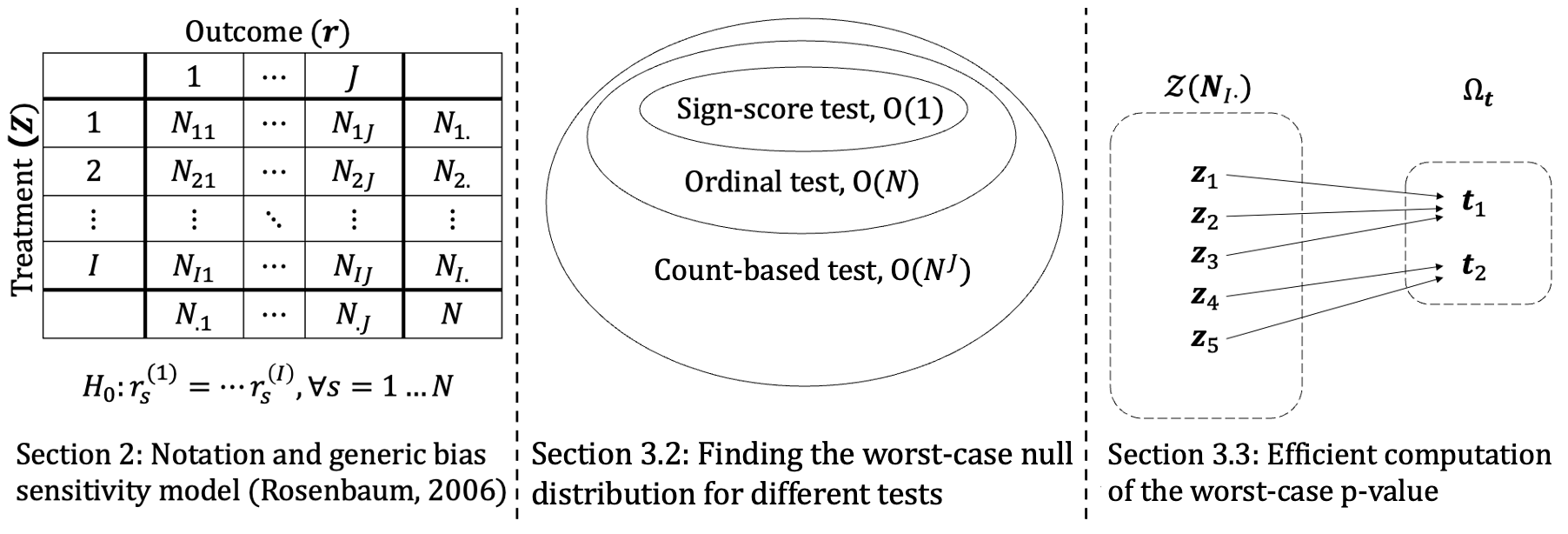} 
  \caption{A visual summary of our exact, nonparametric sensitivity analysis for contingency tables. The rows of the contingency table represent $I$ treatment levels while the columns represent $J$ outcome levels. The term $r_{s}^{(i)}$ represents the potential outcome of subject $s$ under treatment level $i \in \{1,\ldots,I\}$. The $O(\cdot)$ notation represents the size of the candidate set to find the worst-case null distribution. This size is a function of the sample size $N$ and $J$. The third column illustrates the computation in Theorem~\ref{thm:recursive_probability} by showing the mapping between the space of biased treatment assignments $\mathcal{Z}(\mathbf{N}_{I\cdot})$ to a smaller space of ``biased'' contingency tables $\Omega_{\mathbf{t}}$ with the counts in Theorem~\ref{thm:recursive_probability}.}
  \label{fig:contribution_figure}
\end{figure}

The main contribution of this paper is a nonparametric framework to conduct an exact sensitivity analysis for an $I \times J$ or an $I \times J \times K$ contingency table; see Figure \ref{fig:contribution_figure} for an overview. We use the qualifier “exact” because our sensitivity analysis does not rely on asymptotic approximations, thereby ensuring valid statistical inference, such as Type I error rate control, for any sample size and for any contingency table, such as a sparse table where many entries are zero and traditional asymptotics can fail to achieve Type I error control \citep[Section 1]{agresti2001exact}. 
Also, we use the qualifier ``nonparametric'' in the sense that we do not make parametric modeling assumptions about the joint relationship between the treatment, the outcome, and the unmeasured confounder.

At a high level, our sensitivity analysis extends Rosenbaum’s sensitivity model for generic bias   \citep{rosenbaum2006differential} to a non-binary treatment variable.  
Under the extended sensitivity model, we propose a general framework to compute the worst-case p-value for an $I \times J$ table or an $I \times J \times K$ table. The framework is broken into two parts: (a) a reduction of the original non-convex global optimization problem to iterating over at most polynomially (in $N$) many candidate worst-case null distributions and
(b) numerically evaluating the objective function in the reduced problem, leading to an efficient computation of the worst-case p-value. The results from part (a) depend on the complexity of the test statistic, with the sign-score test statistic being the ``easiest'' among the test statistics considered; see Figure \ref{fig:contribution_figure}. The results from part (b) utilize a novel transformation from the space of biased treatment assignments to the space of biased contingency tables. This transformation accelerates the computation of the exact worst-case p-value by roughly \change{750,000} times compared to the original formulation; see Section \ref{sec:exact_worst_null} for details.

A major conceptual difference between a sensitivity analysis for binary treatment and the proposed sensitivity analysis for non-binary, categorical treatment is that the unmeasured confounder can have non-constant effects on each treatment level, where the unmeasured confounder may have a larger effect on some treatment levels than others. As we show below, our extension of the generic bias model in \citet{rosenbaum2006differential} to categorical treatment captures these differential effects while simultaneously enabling exact and nonparametric inference; see Sections~\ref{sec:Model_for_sensitivity_analysis} and~\ref{sec:Maximizer_Characterization_and_Bounds}. We also illustrate the differences between sensitivity analyses for binary and non-binary treatment by examining the power of tests in a sensitivity analysis \citep[Chapter 14]{rosenbaum2010design}, characterizing settings in which tests that use all treatment levels can improve power relative to tests dichotomizing treatment levels.

Finally, we summarize our secondary results. First, when the outcome is binary, we show that the exact, worst-case null distribution of the sign-score statistic is a linear transformation of a multivariate extended hypergeometric distribution for any nondecreasing sensitivity weights $\phi(i)$, including those in \citet{zhang2024sensitivity}'s sensitivity model; see Corollary~\ref{cor:unique_corner_multivariate_hyper} in Section~\ref{sec:exact_worst_null} and Section~\sref{app:C4} of the Supplementary Material. Second, we show that the worst-case null distribution across several candidate vectors $\bm{\delta}^{1},\ldots,\bm{\delta}^{M}$ is the maximum of the $M$ worst-case null distributions; see Section~\ref{sec:integrate_delta}. Third, we establish sharp bounds for a class of stratified sum tests that aggregate evidence across $K$ strata, extending \citet[equation~(2)]{rosenbaum2017adaptive} from binary treatments to non-binary treatments and non-paired designs; see Section~\ref{sec:ijk_contingency_table}. Fourth, we derive the exact first two moments of the cell counts to facilitate a normal approximation, extending \citet{rosenbaum1990sensitivity}'s results under binary treatment; see Sections~\sref{app:D} and \sref{app:F8} of the Supplementary Material.

\section{Setup}
\subsection{Notation}
\label{sec:notations_and_review}
There are \(N\) subjects indexed by \(s = 1, \dots, N\), \(I\) treatment levels indexed by \(i = 1, \dots, I\), and \(J\) outcome levels indexed by \(j = 1, \dots, J\). The indices \(i\) and \(j\) do not necessarily imply that either the treatment or the outcome is ordinal. For each subject \(s\), we write \(Z_s = i\) if the subject was assigned treatment level \(i\), and \(r_s = j\) if the subject's observed outcome was level \(j\). Let $\mathbf{Z}=(Z_1,\ldots,Z_{N})$ and $\rr = (r_1,\ldots,r_N)$ denote the treatments and outcomes of all $N$ subjects.

We define the following statistics of $\mathbf{Z}$ and $\rr$. First, let $N_{ij} =  \sum_{s=1}^{N} \mathbbm{1}\{Z_s = i, r_s = j\}$ denote the number of subjects who received treatment level \(i\) and exhibited outcome level \(j\), where \(\mathbbm{1}\{\cdot\}\) is the indicator function. Second, let $ N_{i\cdot} = \sum_{j=1}^{J} N_{ij}$
denote the number of subjects who received treatment level $i$, and let $N_{\cdot j} = \sum_{i=1}^{I} N_{ij}$ denote the number of subjects who had outcome level $j$. Let \(\mathbf{N}_{I\cdot} = (N_{1\cdot}, \dots, N_{I\cdot})^\intercal\) and \(\mathbf{N}_{\cdot J} = (N_{\cdot 1}, \dots, N_{\cdot J})^\intercal\) be the vectors defined by $N_{i\cdot}$ and $N_{\cdot j}$, respectively. Third, let $\mathbf{N} \in \mathbb{R}^{I \times J}$ be the matrix whose \((i,j)\)-th entry is \(N_{ij}\).

Note that the \(I \times J\) contingency table is the matrix $\mathbf{N}$, with \(N_{i\cdot}\) the $i$th row sum and \(N_{\cdot j}\) the $j$th column sum. Also, the contingency table $\mathbf{N}$ is a matrix-valued statistic of the observed data $\mathbf{Z}$ and $\rr$. Whenever appropriate, we use either $\mathbf{N}$ or $\mathbf{N}(\mathbf{Z},\rr)$ to denote the contingency table where the latter notation emphasizes the connection of the contingency table to the observed data vectors $\mathbf{Z}$ and $\rr$.

\subsection{Review: Potential Outcomes, Randomized Experiments, and Testing the Sharp Null of No Effect} \label{review:setup}
We briefly review the potential outcomes framework, randomized experiments, and testing the sharp null of no effect in an $I$ by $J$ contingency table. For each subject $s$, let \( r_s^{(i)} \in \{1,\ldots,J\} \) denote the potential outcome of subject \( s \) under treatment \( i \) and the observed outcome for each subject is a realization of one of the potential outcomes:
\begin{equation}\label{eq:SUTVA}
r_{s} = r_{s}^{(Z_s)}.
\end{equation}
We work under a design-based paradigm where the potential outcomes are fixed and only the treatment variable $Z_s$ is random. 
For each subject $s$, let $x_s$ denote the subject's observed confounder and $u_s$ denote the subject's unobserved confounder. Let $\bm{x} = (x_1,\ldots,x_N)^\intercal $, and $\mathbf{u} = (u_1,\ldots, u_N)^\intercal$ denote the observed and unobserved confounders for all $N$ subjects. We denote the collection of all potential outcomes, measured confounders, and unmeasured confounders as $\mathcal{F} = \{r_s^{(i)}, x_s, u_s \;\mid\; s=1,\ldots,N;\, i=1,\ldots,I\}$. 

Consider a completely randomized experiment where exactly $N_{i\cdot} > 0$ subjects are assigned to treatment level $i \in \{1,\ldots,I\}$ and let $\mathcal{Z}(\mathbf{N}_{I\cdot})$ denote the set of all possible treatment assignments of $N$ subjects, i.e., $\mathcal{Z}(\mathbf{N}_{I\cdot
}) = \{\mathbf{z} \in \{1,\ldots,I\}^N \mid \sum_{s=1}^{N}\mathbbm{1}\{z_s = i\} = N_{i\cdot} > 0, \forall i = 1,\ldots,I\}.$  Then, the distribution of $\mathbf{Z}$ can be written as
\begin{equation}\label{eq:strong_ig}  
\mathbb{P}(\mathbf{Z} = \mathbf{z}  
  \mid \mathcal{F},\mathcal{Z}(\mathbf{N}_{I\cdot}))  
\;= \frac{1}{|\mathcal{Z}(\mathbf{N}_{I\cdot})|} = \left( \frac{N!}{\prod_{i=1}^{I} N_{i\cdot}!} \right)^{-1}.
\end{equation}  
The term $|\mathcal{Z}(\mathbf{N}_{I\cdot})|$ denotes the size of the set $\mathcal{Z}(\mathbf{N}_{I\cdot})$. Importantly, the right-hand side of the first equality in equation \eqref{eq:strong_ig} does not depend on the potential outcomes or unmeasured confounders and can be explicitly computed. 

Fisher's sharp null hypothesis of no treatment effect posits that the potential outcomes are identical across all treatment levels for all $N$ subjects:    
\begin{equation}\label{eq:null_sharp}
H_0: \; r_s^{(1)} = \cdots = r_s^{(I)} 
\quad \text{for all } s = 1,\ldots,N.
\end{equation}
Conditional on $\mathcal{F}$, the sharp null and equation \eqref{eq:SUTVA} imply that $r_s = r_{s}^{(1)} = \cdots = r_{s}^{(I)}$ and the column totals of the \(I \times J\) contingency table (i.e., \(\mathbf{N}_{\cdot J}\)) are fixed. Also, under equations ~\eqref{eq:SUTVA} and~\eqref{eq:strong_ig}, the exact null distribution of a test statistic $T(\mathbf{Z}, \mathbf{r}) \in \mathbb{R}$ is given by
\begin{equation}
\mathbb{P}_{H_0}\!\left(
T(\mathbf{Z}, \mathbf{r}) \ge \criticalt 
\;\middle|\; \mathcal{F}, \mathcal{Z}(\mathbf{N}_{I\cdot})
\right)
=
\sum_{\mathbf{z} \in \mathcal{Z}(\mathbf{N}_{I\cdot})}
\mathbbm{1}\!\left\{
T(\mathbf{z}, \mathbf{r}) \ge \criticalt
\right\}
\left(
\frac{N!}{\prod_{i=1}^I N_{i\cdot}!}
\right)^{-1}.
\label{eq:p_value_strong_ig}
\end{equation}
The one-sided exact $p$-value is obtained by replacing $\criticalt$ in~\eqref{eq:p_value_strong_ig} with the observed value $T(\mathbf{Z}, \mathbf{r})$.

Throughout the paper, we consider test statistics of the contingency table $\mathbf{N}$, which we call count-based test statistics.

\begin{defn}[Count-Based Test Statistic]
\label{def:count_based}
A test statistic $T(\mathbf{Z},\rr)$ is said to be count-based if it is a function of the contingency table $\mathbf{N}$, which is itself a function of $\mathbf{Z}$ and $\rr$. For brevity, we write a count-based test statistic as $T(\mathbf{N})$ or $T(\mathbf{N}(\mathbf{Z},\rr))$.
\end{defn}
\noindent Count-based test statistics include the chi-squared statistic, $T(\mathbf{N}) = \sum_{i=1}^{I}\sum_{j=1}^{J}\frac{(N_{ij} - N_{i\cdot}N_{\cdot j}/N)^2}{N_{i\cdot}N_{\cdot j}/N}$, and the $G^2$ statistic, $T(\mathbf{N}) = 2\sum_{i=1}^{I}\sum_{j=1}^{J}N_{ij} \log\left(\frac{N_{ij}}{N_{i\cdot}N_{\cdot j}/N}\right)$. Count-based test statistics are a general family of test statistics because, under multinomial sampling for contingency tables, \citet[Sections 2.1.5 and 3.5.1]{agresti2012categorical} showed that $\mathbf{N}$ is a sufficient statistic. We also introduce two sub-families of count-based test statistics: ordinal test statistics and sign-score test statistics.
\begin{defn}[Ordinal Test Statistic]
\label{eq:ordinal_test}
A count-based test statistic $T(\mathbf{N})$ is an ordinal test statistic if it can be written as
\[
T(\mathbf{N}) = \sum_{i=1}^{I} \sum_{j=1}^{J} w_i v_j N_{ij}
\]
for some fixed weights $w_1 \leq \cdots \leq w_I$ and $v_1 \leq \cdots \leq v_J$.
\end{defn}
\begin{defn}[Sign-Score Test Statistic]
\label{eq:signed_score_test}
Suppose the outcome is binary (i.e., $J = 2$). A count-based test statistic $T(\mathbf{N})$ is a sign-score test statistic if it can be written as
\[
T(\mathbf{N}) = \sum_{i=1}^{I} w_i N_{i2}
\]
for some fixed weights $w_1 \leq \cdots \leq w_I$.
\end{defn}
\noindent An ordinal test is designed to detect a monotonic trend in a contingency table, and the weights $w_1,\ldots,w_I$ and $v_1,\ldots,v_J$ are pre-specified by the investigator based on the hypothesized trend; see Section \ref{sec:data_analysis} and \citet[Section 3.4]{agresti2012categorical} for example weights. Note that sign-score test statistics are a special case of ordinal test statistics, which are in turn a special case of count-based test statistics. 

We briefly remark that any count-based test statistic is a permutation-invariant test statistic, in that simultaneously permuting any two indices of $\mathbf{Z}$ and $\rr$ does not change the value of the test statistic; see \citet[Section~3]{rosenbaum1990sensitivity} for discussions on permutation-invariant tests in a sensitivity analysis. Furthermore, Sections~\sref{app:B2}--\sref{app:B4} of the Supplementary Material show that permutation-invariance is central to construct the equivalence classes underlying Theorems~\ref{thm:set_size_of_U} and \ref{thm:score_test_maximizer_nonbinary}.

\subsection{Model for Sensitivity Analysis}
\label{sec:Model_for_sensitivity_analysis}
\subsubsection{Sensitivity Model for Binary Treatment}
Suppose the treatment assignment is no longer randomized and does not follow equation \eqref{eq:strong_ig}. Sensitivity analysis provides a framework to quantify departures from this assumption due to an unmeasured confounder. Before we introduce our main sensitivity analysis for contingency tables, we briefly review sensitivity analysis for binary treatment (i.e., $I=2$). 

For binary treatment, a popular model that describes departure from random treatment assignment is Rosenbaum's sensitivity model \citep{rosenbaum1987sensitivity,rosenbaum1990sensitivity,rosenbaum2002observational}:
\begin{equation} \label{eq:binary_sens_analysis_model_1}
\mathbb{P}(Z_s = 1 \mid \mathcal{F})
= \frac{\exp (\xi_1(x_s) + \gamma u_s)}
{\sum_{b \in \{0,1\}} \exp(\xi_b(x_s) + \gamma b u_s)}, \quad  \gamma \geq 0, u_s \in [0,1].
\end{equation}
In words, Rosenbaum's sensitivity model asserts that subject $s$'s treatment probability depends on measured confounders $x_s$ via an unknown nonparametric function $\xi(\cdot)$ and an unmeasured confounder $u_s$ via the parametric term $\gamma u_s$. In particular, the magnitude of the parameter $\gamma$, which is referred to as the sensitivity parameter, quantifies the departure from random assignment by increasing the effect of the unmeasured confounder on the treatment assignment probability. Note that setting $\gamma = 0$ corresponds to random treatment assignment.

The magnitude of the sensitivity parameter $\gamma$ can also be interpreted on the odds ratio scale. Specifically, exponentiation of $\gamma$ serves as the upper and lower bounds of the odds ratio of the treatment assignment probabilities of any two subjects $s$ and $s'$ with identical $x_s = x_{s'}$, but with potentially different values of $u_s$ and $u_{s'}$:
\begin{equation*}
    \frac{1}{\Gamma} \leq \frac{\mathbb{P}(Z_s = 1\mid \mathcal{F})}{\mathbb{P}(Z_s = 0\mid \mathcal{F})} \bigg/
    \frac{\mathbb{P}(Z_{s'} = 1\mid \mathcal{F})}{\mathbb{P}(Z_{s'} = 0\mid \mathcal{F})} \leq \Gamma, \quad{} \Gamma = \exp(\gamma) \in [1,\infty).
\end{equation*}
Like before, setting $\Gamma = 1$, which is equivalent to setting $\gamma =0$, corresponds to random treatment assignment. Increasing $\Gamma$ away from $1$ allows two subjects' odds of being assigned treatment to differ up to a factor of $\Gamma$ due to differences in their unmeasured confounders.

\subsubsection{Sensitivity Model for $I$ by $J$ Contingency Tables} \label{sec:sens_model_ibyj}
This section presents a sensitivity model for a categorical treatment variable in an $I$ by $J$ contingency table. Similar to the previous section, suppose the treatment assignment is no longer randomized and does not follow equation \eqref{eq:strong_ig}. Instead, the probability of getting assigned treatment level $i \in \{1,\ldots,I\}$ depends on an unmeasured confounder $u_s \in [0,1]$:
\begin{equation} \label{eq:generic_bias_sensitivity}
\mathbb{P}(Z_s = i \mid \mathcal{F}) = \frac{\exp(\xi_{i}(x_s) + \gamma \delta_{i} u_s)}
{\sum_{b \in \{1,\ldots,I\}} \exp(\xi_b(x_s) + \gamma \delta_{b} u_s)}, \quad{} \gamma \geq 0, \ \delta_i \in \{0,1\}, u_s \in [0,1].
\end{equation}
In words, subject $s$'s probability of being assigned treatment level $i \in \{1,\ldots,I\}$ depends on an unknown nonparametric function $\xi(\cdot)$, which is a function of measured covariates $x_s$, and a parametric term $\gamma \delta_i u_s$. Notably, unlike the sensitivity model for binary treatment in \eqref{eq:binary_sens_analysis_model_1}, the departure from random assignment with non-binary treatment is now measured by two sensitivity parameters: a scalar number $\gamma$ and the binary vector $\bm{\delta} = (\delta_1,\ldots,\delta_I)$. 

Similar to the sensitivity model for binary treatment, setting $\gamma = 0$ in the $I \times J$ sensitivity model \eqref{eq:generic_bias_sensitivity} corresponds to random treatment assignment.  
However, when $\gamma \neq 0$, model \eqref{eq:generic_bias_sensitivity} implies that subjects with larger unmeasured confounders $u_s$ are more likely to be assigned to treatment levels $i$ where $\delta_i = 1$ than treatment levels where $\delta_i = 0$. Note that if two treatment levels $i$ and $i'$ satisfy $\delta_i = \delta_{i'}$, the effect of $u_s$ on the treatment assignment probability to either treatment level $i$ or $i'$ is identical. In other words, even though each subject's unmeasured confounder is a scalar value $u_s$, $\bm{\delta}$ enables some treatment levels to have different effects by $u_s$.

To better understand the role of $\bm{\delta}$ and $\gamma$, we present an alternative interpretation of $\delta_i$ and $\gamma$ on the odds ratio scale. Specifically, the odds ratio of getting assigned treatment level $i$ versus $i'$ for any two subjects $s$ and $s'$ with identical $x_s = x_{s'}$, but potentially different $u_s, u_{s'}$ is bounded as follows.
\begin{equation*}
\frac{1}{\Gamma^{\lvert\delta_i - \delta_{i'}\rvert}} \leq 
\frac{\mathbb{P}(Z_s = i \mid \mathcal{F})}{\mathbb{P}(Z_s = i' \mid \mathcal{F})} \bigg/
\frac{\mathbb{P}(Z_{s'} = i \mid \mathcal{F})}{\mathbb{P}(Z_{s'} = i' \mid \mathcal{F})} \leq \Gamma^{\lvert\delta_i - \delta_{i'}\rvert}, \quad{} \Gamma = \exp(\gamma) \in [1,\infty).
\end{equation*}
If $\delta_i = \delta_{i'}$, the odds of getting assigned treatment level $i$ versus $i'$ are the same between the two subjects.  If $\delta_i \neq \delta_{i'}$, the odds can differ between the two subjects by a factor of $\Gamma$. Notably, the odds ratio formulation reveals that if all treatment levels satisfy $\delta_1 = \cdots = \delta_I$, then the bound equals $1$ for any pair of treatment levels and the model reduces to random treatment assignment. As such, we exclude these homogeneous cases ($\delta_1 = \cdots = \delta_I = 0$ and $\delta_1 = \cdots = \delta_I = 1$) from further consideration.

\subsubsection{Interpreting the sensitivity parameter $\bm{\delta}$}
To help interpret the sensitivity parameter $\bm{\delta}$, we provide  a comparison between the sensitivity model for contingency tables in \eqref{eq:generic_bias_sensitivity} and other sensitivity models for unmeasured confounding. First, under binary treatment, our model \eqref{eq:generic_bias_sensitivity} reduces to the sensitivity model for binary treatment in \eqref{eq:binary_sens_analysis_model_1} if $\delta_1 = 0$ and $\delta_2 = 1$; in fact, in the binary case there is no need to introduce the parameter $\bm{\delta}$, leaving $\gamma$ as the sole sensitivity parameter. Second, under continuous treatment, the sensitivity parameter $\delta_i$ is commonly replaced with $i$ (i.e., $\gamma \delta_i u_s$ in \eqref{eq:generic_bias_sensitivity} becomes $\gamma i u_s$) to reflect that the unmeasured confounder $u_s$ has a monotonic effect on the treatment assignment probability with increasing levels of treatment $i$ \citep{rosenbaum1989sensitivity,gastwirth1998dual,zhang2024sensitivity,zhang2024bridging}. Notably, the term $\gamma \delta_i u_s$ represents a more discrete relationship between the unmeasured confounder and treatment assignment probabilities compared to the term $\gamma i u_s$, which represents a more continuous, smooth relationship. 

In general, under non-binary treatment, it is common to specify multiple sensitivity parameters to reflect the heterogeneous effect of the unmeasured confounder across different treatment levels. The proposed sensitivity model in \eqref{eq:generic_bias_sensitivity} does this through the $I$-dimensional binary vector $\bm{\delta}$ where each component of the vector represents one of the $I$ treatment levels. The aforementioned works on continuous treatment do this by specifying a smooth, dose-response function $\phi(i) = i$. For more discussions on sensitivity models for non-binary treatment and their implications for exact inference, see Section \ref{sec:discussion}.


Finally, our sensitivity model \eqref{eq:generic_bias_sensitivity} for $I \times J$ contingency tables builds on work by \citet{rosenbaum2006differential}, who proposed a sensitivity model for generic and differential unobserved biases in 2 x 2 factorial designs. Briefly, generic unobserved bias is a phenomenon where an unmeasured confounder influences treatment assignment probabilities in the same direction and magnitude. In contrast, differential unobserved bias is a phenomenon where an unmeasured confounder influences treatment assignment probabilities in a different direction with a different magnitude. In our setting, comparing treatment assignment probabilities among treatment levels with identical $\delta_i$s constitutes generic unobserved biases whereas comparing treatment assignment probabilities among treatment levels with different $\delta_i$s constitutes differential unobserved biases. 

The connection to generic and differential unobserved biases provides another explanation of the sensitivity parameter $\bm{\delta}$ and we illustrate it by re-interpreting the two examples from \citet{rosenbaum2006differential}, but in the context of an $I$ by $J$ contingency table. First, consider an observational study of illicit drug use, specifically crack cocaine, heroin, or methamphetamine, versus none.  Using any one of these illicit drugs may be driven by a shared unmeasured factor such as an individual's risk tolerance and this phenomenon can be captured in our sensitivity analysis by setting $\delta_i = 1$ for all drug categories and $\delta_i = 0$ for ``none'' in \eqref{eq:generic_bias_sensitivity}. 
Second, in an observational study of non-steroidal, anti-inflammatory drug use, an individual's pain tolerance (i.e., an unmeasured confounder)  may affect ibuprofen and acetaminophen use and if so, we can represent this in \eqref{eq:generic_bias_sensitivity} by setting $\delta_i = 1$ for both medications and $\delta_i = 0$ for non-use. In both examples, a comparison between treatment levels with identical $\delta_i$ is not biased by the unmeasured confounder because the unmeasured confounder's effect on the treatment assignment probabilities for these specific treatment levels is identical.

\section{Sensitivity Analysis for $I$ by $J$ Table} \label{sec:sens_ibyj}

\subsection{Worst-Case Unmeasured Confounder and P-value} \label{sec:def_worst_case}

\par Suppose all subjects share the same observed covariates in an $I \times J$ contingency table, i.e., $x_{s} = x_{s'}$ for all $s, s' \in \{1,\ldots,N\}$. Then, for any $\bm{\delta}$, $\mathbf{u}$, and $\gamma \geq 0$, the exact conditional distribution of $\mathbf{Z}$ under the sensitivity model in equation \eqref{eq:generic_bias_sensitivity} is
\begin{equation*}
\mathbb{P}\left(\mathbf{Z} = \mathbf{z} 
     \mid \mathcal{F},\mathcal{Z}(\mathbf{N}_{I\cdot})\right)
= 
\frac{\exp \left\{ \gamma \sum_{s=1}^{N} \sum_{i =1}^{I} \delta_i \mathbbm{1}(z_s = i) u_s \right\}}{\sum_{\mathbf{b} \in \mathcal{Z}(\mathbf{N}_{I\cdot})} \exp \left\{ \gamma \sum_{s=1}^{N} \sum_{i=1}^{I} \delta_i \mathbbm{1}(b_s = i) u_s \right\}}, \quad{} u_{s} \in [0,1].
\label{eq:conditional_distribution_treatment_with_sensitivity}
\end{equation*}
Also, the exact null distribution of the test statistic $T(\mathbf{N}(\mathbf{Z},\rr))$ under the sensitivity model is 
\begin{align}
\mathbb{P}_{H_0}\left(T(\mathbf{N}(\mathbf{Z}, \rr)) \geq \criticalt \mid \mathcal{F},\mathcal{Z}(\mathbf{N}_{I\cdot})\right) &= 
\frac{\sum_{\mathbf{z} \in \mathcal{Z}(\mathbf{N}_{I\cdot})} \mathbbm{1}\{T(\mathbf{N}(\mathbf{z}, \rr)) \geq c\} \exp \left\{ \gamma \sum_{s=1}^{N} \sum_{i=1}^I \delta_i \mathbbm{1}\{z_s = i\} u_s \right\}}{\sum_{\mathbf{b} \in \mathcal{Z}(\mathbf{N}_{I\cdot})} \exp \left\{ \gamma \sum_{s=1}^{N} \sum_{i =1}^I \delta_i \mathbbm{1}\{b_s = i\} u_s \right\}}. \label{eq:raw_sens_test}
\end{align}
\noindent If the unmeasured confounders $\mathbf{u}$ were known, the exact null distribution of the test statistic $T$ can be obtained by using \eqref{eq:raw_sens_test}. However, because $\mathbf{u}$ is unobserved, it is common to study the exact, \emph{worst-case null distribution} of $T$:
\begin{equation} 
\label{eq:worst_case_null}
\max_{\mathbf{u} \in [0,1]^N} \alpha(T,\rr,\mathbf{u}), \text{ where } 
\alpha(T,\rr,\mathbf{u}) = \mathbb{P}_{H_0}\left(T(\mathbf{N}(\mathbf{Z}, \rr)) \geq \criticalt \mid \mathcal{F},\mathcal{Z}(\mathbf{N}_{I\cdot})\right).
\end{equation}
A maximizer of~\eqref{eq:worst_case_null} is referred to as a \emph{worst-case unmeasured confounder} and is denoted as $\mathbf{u}^+$:
\begin{equation*} 
    \mathbf{u}^+ = \operatorname{argmax}_{\mathbf{u} \in [0,1]^N} \alpha(T,\rr, \mathbf{u}).
\end{equation*} 
Given $\mathbf{u}^+$, the exact, \emph{worst-case, one-sided p-value}, denoted as $\overline{p}_{\gamma, \bm{\delta}}(T_{\text{obs}})$, is given by \eqref{eq:raw_sens_test} where we replace $\mathbf{u}$ with $\mathbf{u}^+$ and $c$ with the observed value of the test statistic $T_{\rm obs}$:
\[
\overline{p}_{\gamma, \bm{\delta}}(T_{\text{obs}}) 
= \max_{\mathbf{u}\in[0,1]^N} 
\mathbb{P}_{H_0}\!\left(T(\mathbf{N}(\mathbf{Z},\rr))\geq T_{\text{obs}} \,\middle|\, 
\mathcal{F}, \mathcal{Z}(\mathbf{N}_{I\cdot})\right).
\]
The main technical difficulty is solving the optimization problem in \eqref{eq:worst_case_null}. Without any constraints on the test statistic $T$, it is a ratio of convex functions in $\mathbf{u}$, which is neither quasiconvex nor quasiconcave, and is a global optimization problem \citep{benson_2006_maximizing}. Also, from \eqref{eq:raw_sens_test}, evaluating the objective function $\alpha(T,\mathbf{r},\mathbf{u})$ requires enumerating the set $\mathcal{Z}(\mathbf{N}_{I\cdot})$, whose size is exponential in sample size. The next two sections show how the structures defining the three classes of test statistics in Section~\ref{review:setup} render \eqref{eq:worst_case_null} computationally feasible.

\subsection{Finding the Worst-Case Null Distribution}
\label{sec:Maximizer_Characterization_and_Bounds}

Before we present the main results of this section, we review some notation and concepts from discrete mathematics; see Section 0.2 of \citet{Balakrishnan1991} for a textbook introduction. 
For a fixed $\rr$ and $\mathbf{u}\in\{0,1\}^N$, let $\overline{\mathbf{u}}$ be the vector whose $j$th component is $\overline{u}_j = \sum_{s=1}^{N}\mathbbm{1}\{r_s = j, u_s=1\}$. For any elements $\mathbf{u}, \mathbf{u}' \in \{0,1\}^N$, we define an equivalence relation $\mathbf{u} \sim \mathbf{u}'$  if and only if $\overline{\mathbf{u}} = \overline{\mathbf{u}}'$. Also, for each $\mathbf{u} \in \{0,1\}^N$, we define an equivalence class $[\mathbf{u}] = \{ \mathbf{u}' \in \{0,1\}^N \mid \mathbf{u}' \sim \mathbf{u}\}$.
Finally, let $\mathcal{U}_{\rm C} = \{0,1\}^N/\!\sim$ be the set of all equivalence classes of the set $\{0,1\}^N$. Note that the set of equivalence classes forms a partition of the set $\{0,1\}^N$. For example, suppose $N=3$, $J=2$, and $\rr = (1,1,2)$. If $\mathbf{u} = (1,0,1)$, we have $\overline{\mathbf{u}} = (1,1)$
and the equivalence class of $\mathbf{u}$ is $[\mathbf{u}] = [(1,0,1)] = \{(1,0,1), (0,1,1)\}$. The set of equivalence classes is $\mathcal{U}_{C} = \{[(0,0,0)],[(0,0,1)],[(0,1,0)],[(1,1,0)],[(1,0,1)],[(1,1,1)]\}$. 

Theorem~\ref{thm:set_size_of_U} states important equivalence relationships for any count-based test statistic. 
\begin{thm}[Worst-Case Null Distribution for Count-Based Test Statistics]
\label{thm:set_size_of_U} For any count-based test statistic $T$ and $\rr$, the following holds.  
\begin{itemize}
\item [(i)] The worst-case null distribution satisfies 
$\alpha(T,\rr,\mathbf{u}^+) = \max_{\mathbf{u} \in \{0,1\}^N} \alpha(T,\rr,\mathbf{u})$.
\item [(ii)] If $\mathbf{u},\mathbf{u}'\in\{0,1\}^N$ satisfy $\mathbf{u}\sim\mathbf{u}'$, then $\alpha(T,\rr,\mathbf{u})=\alpha(T,\rr,\mathbf{u}').$  
\item[(iii)] 
Let $\tilde{\alpha}(T,\rr,[\mathbf{u}]) := \alpha(T,\rr,\mathbf{u}')$ for any $\mathbf{u}' \in [\mathbf{u}]$.
 Then $\alpha(T,\rr,\mathbf{u}^+) =
\max_{[\mathbf{u}]\in\mathcal{U}_{\rm C}}\tilde{\alpha}(T,\rr,[\mathbf{u}]).$
\item [(iv)] The size of the set of equivalence classes, which we refer to as the candidate set, satisfies 
$\lvert\mathcal{U}_{\rm C}\rvert \le N^J$. 
\end{itemize}
\end{thm}
See Sections \sref{app:B2}--\sref{app:B4} of the Supplementary Material for the proof. At a high level, Theorem~\ref{thm:set_size_of_U} produces a useful reduction of the original optimization problem of finding the worst-case null distribution in \eqref{eq:worst_case_null}.  
Specifically, part (i) states that the worst-case null distribution can be found by maximizing over a set of binary vectors $\{0,1\}^N$, thereby reducing the feasible space from the entire hypercube (i.e., $[0,1]^N$) to the corners of the hypercube (i.e., $\{0,1\}^N$); the result also implies that a binary vector is one of the worst-case unmeasured confounders $\mathbf{u}^+$. Part (ii) states that if two binary vectors $\mathbf{u}$ and $\mathbf{u}'$ form an equivalence relation, the two vectors lead to the same null distribution. Part (iii) states that the worst-case null distribution can be found by optimizing over the set of equivalence classes $\mathcal{U}_{\rm C}$. Finally, part (iv) shows the size of the candidate set is at most polynomial in the sample size $N$.

An immediate implication of Theorem~\ref{thm:set_size_of_U} is that we can find the worst-case null distribution by iterating over the candidate set $\mathcal{U}_{\rm C}$ and the number of iterations is at most $N^J$. Interestingly, the number of treatment levels $I$ or the sensitivity parameters $\gamma$ and $\bm{\delta}$ do not affect the maximum number of iterations. \change{The number of candidates is one of two factors in the total cost of the worst-case p-value; the other is the cost of evaluating $\alpha(T,\rr,\mathbf{u})$ at each candidate, which Section~\ref{sec:exact_worst_null} addresses and Section~\sref{app:G} of the Supplementary Material measures.}

Next, Theorem~\ref{thm:score_test_maximizer_nonbinary} shows that, if we restrict the test statistics to the class of ordinal test statistics, the size of the candidate set shrinks to $N+1$.
\change{\begin{thm}[Worst-Case Null Distribution for Ordinal Test Statistics] \label{thm:score_test_maximizer_nonbinary}
Suppose $r_1 \le r_2 \le \cdots \le r_N$, and suppose the treatment scores $w_1 \le w_2 \le \cdots \le w_I$ and sensitivity parameters $\delta_1 \le \delta_2 \le \cdots \le \delta_I$ are simultaneously nondecreasing. For any $\gamma \ge 0$, the worst-case null distribution of an ordinal test $T$ satisfies $\alpha(T,\mathbf{r},\mathbf{u}^+) = \max_{\mathbf{u} \in \mathcal{U}_{\rm O}} \alpha(T,\mathbf{r}, \mathbf{u})$, where
\[
\mathcal{U}_{\rm O} = \left\{ \mathbf{u} \in \{0,1\}^N \mid u_1 \le u_2 \le \cdots \le u_N \right\} = \{(0,\ldots,0), (0,\ldots,0,1),\ldots,(1,\ldots,1)\},
\]
and the size of the candidate set $\mathcal{U}_{\rm O}$ is $N + 1$, i.e., $|\mathcal{U}_{\rm O}| = N+1$.
\end{thm}
}
Similar to Theorem~\ref{thm:set_size_of_U}, the candidate set $\mathcal{U}_{\rm O}$ is independent of $\gamma$ and $\bm{\delta}$. Unlike Theorem~\ref{thm:set_size_of_U}, however, its size does not depend on the dimensions of the contingency table: for any $I$ by $J$ table and any ordinal test statistic, we iterate through the same $N+1$ candidates and take the maximizer of $\alpha(T,\rr,\mathbf{u})$. Theorem~\ref{thm:score_test_maximizer_nonbinary} extends Proposition 2 of \citet{rosenbaum1990sensitivity} to non-binary treatments. 

Technically speaking, the size of the candidate set for ordinal test statistics is generally smaller than the size of the candidate set for count-based test statistics because ordinal test statistics are arrangement-increasing with respect to $Z_{s}$ and $r_s$; see \citet{hollander_proschan_sethuraman_1977} for definition of arrangement-increasing quantities. The arrangement-increasing nature of ordinal test statistics leads to an ordering relationship between a pair of equivalence classes, i.e., $\tilde{\alpha}(T,\mathbf{r},[\mathbf{u}]) \leq \tilde{\alpha}(T,\mathbf{r},[\mathbf{u}'])$. In contrast, count-based test statistics may not be arrangement-increasing (e.g., the chi-squared test), which implies that we need to search through the entire set of equivalence classes. 

Finally, when the outcome is binary so that the dimension of the contingency table is $I \times 2$, a worst-case unmeasured confounder can be characterized exactly.
\change{\begin{thm}[Worst-Case Unmeasured Confounder for Sign-Score Test Statistics] \label{thm:score_test_maximizer_binary}
Suppose $r_1 \le r_2 \le \cdots \le r_N$, and suppose the treatment scores $w_1 \le w_2 \le \cdots \le w_I$ and sensitivity parameters $\delta_1 \le \delta_2 \le \cdots \le \delta_I$ are simultaneously nondecreasing. For any $\gamma \ge 0$, a worst-case unmeasured confounder $\mathbf{u}^+$ for a sign-score test is equal to an $N$-dimensional binary vector, where the first $N_{\cdot 1}$ elements are zeros and the rest $N_{\cdot 2} = N - N_{\cdot 1}$ are ones, i.e., $\mathbf{u}^+ = (\underbrace{0, \dots, 0}_{N_{\cdot 1}}, \underbrace{1, \dots, 1}_{N_{\cdot 2}})$.
\end{thm}}

We briefly remark that Theorem~\ref{thm:score_test_maximizer_binary} also holds if the sensitivity parameter $\bm{\delta}$ is replaced by a vector $\bm{\phi} = (\phi(1),\ldots,\phi(I))^\intercal \in \mathbb{R}^I$ with $\phi(1) \leq \phi(2)\leq \cdots \leq \phi(I)$, which is the sensitivity model discussed in \citet{zhang2024sensitivity}.
\change{Theorems~\ref{thm:score_test_maximizer_nonbinary} and~\ref{thm:score_test_maximizer_binary} require treatment scores and $\bm{\delta}$ to be simultaneously nondecreasing; for nonmonotone patterns (e.g., $\bm{\delta} = (0, 1, 0)$), Theorem~\ref{thm:set_size_of_U} applies directly via candidate set $\mathcal{U}_{\rm C}$ (see Section~\sref{app:B_nonmonotone} of the Supplementary Material).}

\subsection{Efficient Computation of the Exact Worst-Case P-Value}
\label{sec:exact_worst_null}
This section develops an efficient computational procedure for evaluating the objective function $\alpha(T,\mathbf{r},\mathbf{u})$
in \eqref{eq:worst_case_null}, which leads to an efficient computation of the worst-case p-value, i.e., $\alpha(T, \rr, \mathbf{u}^+)$ where $c$ is replaced with the observed value of the test statistic. In short, the proposed procedure avoids direct enumeration of every element of the exponentially large set $\mathcal{Z}(\mathbf{N}_{I\cdot})$ in the definition of $\alpha(T, \mathbf{r}, \mathbf{u})$ and the procedure is applicable to any binary unmeasured confounder $\mathbf{u} \in \{0,1\}^N$, which we showed in Theorems~\ref{thm:set_size_of_U}--\ref{thm:score_test_maximizer_binary} is sufficient for finding the exact, worst-case null distribution. Importantly, the procedure is applicable to \emph{any} count-based test statistic (i.e., all test statistics considered in the paper)\change{, and it uses that the sensitivity parameters $\bm{\delta}$ in \eqref{eq:generic_bias_sensitivity} are binary}.

\change{\begin{thm}[Efficient Computation of $\alpha(T,\rr,\mathbf{u})$] \label{thm:recursive_probability}
Let $\bm{\delta} \in \{0,1\}^I$ as in \eqref{eq:generic_bias_sensitivity}, let $\mathcal{I}_1 = \{i : \delta_i = 1\}$, and let $\overline{N} = \sum_{i \in \mathcal{I}_1} N_{i\cdot}$. For a given $\mathbf{u} \in \{0,1\}^N$, let $\overline{u}_j = \sum_{s=1}^{N}\mathbbm{1}\{r_s = j, u_s = 1\}$ as in Section~\ref{sec:Maximizer_Characterization_and_Bounds} and $\overline{u} = \sum_{j=1}^{J}\overline{u}_j$; for a table $\mathbf{t} \in \mathbb{Z}_{\ge 0}^{I \times J}$, let $\overline{t}_j = \sum_{i \in \mathcal{I}_1} t_{ij}$ and $\overline{\mathbf{t}} = (\overline{t}_1, \ldots, \overline{t}_J)$. The summation variables below have the following interpretation: for an assignment $\mathbf{z}$, $\overline{n}_j = \sum_{s=1}^{N}\mathbbm{1}\{z_s \in \mathcal{I}_1,\ r_s = j,\ u_s = 1\}$ is the number of subjects with outcome $j$ and $u_s = 1$ assigned to a treatment level in $\mathcal{I}_1$, $\overline{\mathbf{n}} = (\overline{n}_1, \ldots, \overline{n}_J)$, and $\overline{q} = \sum_{j=1}^{J} \overline{n}_j$. The exact null distribution of any count-based test statistic $T(\mathbf{N})$ can be rewritten as
\begin{align}
\alpha(T,\rr,\mathbf{u}) 
&= 
\dfrac{
\sum_{\mathbf{t} \in \Omega_{\mathbf{t}}} 
\sum_{\overline{\mathbf{n}} \in \Omega_{\overline{\mathbf{n}}}} 
\exp\big(\gamma \textstyle\sum_{j=1}^{J} \overline{n}_j\big) 
\, \kappa(\mathbf{t}) \prod_{j=1}^{J} \binom{\overline{u}_j}{\overline{n}_j}\binom{N_{\cdot j} - \overline{u}_j}{\overline{t}_j - \overline{n}_j}
}{
\sum_{\overline{q} \in \Omega_{\overline{q}}} 
\exp(\gamma \overline{q}) 
\, \kappa_0 \binom{\overline{u}}{\overline{q}}\binom{N - \overline{u}}{\overline{N} - \overline{q}}
},
\label{eq:exact_distribution_kernel_form}
\end{align}
where
\begin{align}
\Omega_{\overline{q}} 
&= 
\left\{
\overline{q} \in \mathbb{Z}_{\ge 0} \,\middle|\,
\max(0, \overline{u} + \overline{N} - N) \le \overline{q} \le \min(\overline{u}, \overline{N})
\right\}, \nonumber \\[0.4em]
\Omega_{\overline{\mathbf{n}}} 
&= 
\left\{
\overline{\mathbf{n}} \in \mathbb{Z}_{\ge 0}^{J} \,\middle|\,
\max(0, \overline{u}_j + \overline{t}_j - N_{\cdot j}) \le \overline{n}_j \le \min(\overline{u}_j, \overline{t}_j), \;
\forall j
\right\}, \nonumber \\[0.4em]
\Omega_{\mathbf{t}} 
&= 
\left\{
\mathbf{t} \in \mathbb{Z}_{\ge 0}^{I \times J} \,\middle|\,
T(\mathbf{t}) \ge \criticalt, \;
\sum_{i=1}^{I} t_{ij} = N_{\cdot j}, \;
\sum_{j=1}^{J} t_{ij} = N_{i\cdot}, \;
\forall i,j
\right\}. \nonumber
\end{align}
The multinomial constants $\kappa(\mathbf{t})$ and $\kappa_0$ do not involve $\mathbf{u}$ or $\gamma$ and are given in Equation~\seqref{eq:kappa_def} of the Supplementary Material.
\end{thm}}

\change{Theorem~\ref{thm:recursive_probability} reformulates the null distribution $\alpha(T,\rr,\mathbf{u})$ from a distribution over treatment assignments $\mathcal{Z}(\mathbf{N}_{I\cdot})$ to a distribution over pairs $(\mathbf{t}, \overline{\mathbf{n}})$, where $\mathbf{t}$ is an $I \times J$ table of cell counts with the observed margins (i.e., $\Omega_{\mathbf{t}}$) and $\overline{\mathbf{n}} = (\overline{n}_1, \ldots, \overline{n}_J)$ is a $J$-dimensional vector whose $j$-th component is the number of subjects with outcome $j$ and $u_s = 1$ who are assigned to a treatment level in $\mathcal{I}_1$ (i.e., $\Omega_{\overline{\mathbf{n}}}$). The result exploits the fact that multiple treatment vectors in $\mathcal{Z}(\mathbf{N}_{I\cdot})$ may induce the same table in $\Omega_{\mathbf{t}}$ and the same vector in $\Omega_{\overline{\mathbf{n}}}$, so the number of pairs $(\mathbf{t}, \overline{\mathbf{n}})$ is substantially smaller than $|\mathcal{Z}(\mathbf{N}_{I\cdot})|$. For example, for a $3\times 3$ table with $\rr=(1,1,2,2,3,3)$, $\mathbf{u}=(1,1,0,0,0,0)$, and $\bm{\delta}=(0,1,1)$, the three treatment vectors $\mathbf{z}_1=(1,2,1,2,3,3)$, $\mathbf{z}_2=(2,1,2,1,3,3)$, $\mathbf{z}_3=(1,2,2,1,3,3)$ produce the same $3 \times 3$ table $\mathbf{t} \in \Omega_{\mathbf{t}}$ (i.e., $N_{11}=N_{12}=N_{21}=N_{22}=1$, $N_{33}=2$, and the remaining cells zero) and the same vector $\overline{\mathbf{n}} \in \Omega_{\overline{\mathbf{n}}}$ (i.e., $\overline{\mathbf{n}}=(1,0,0)$). \citet{pagano1981algorithm} considered a similar transformation for $I \times J$ tables in the absence of unmeasured confounding.}
We illustrate the computational implications of Theorem \ref{thm:recursive_probability} in Table~\ref{tb:main_text_exact_computation_time_permutation_and_kernel} by comparing the computation time for the exact $p$-value between two formulations of $\alpha(T,\mathbf{r},\mathbf{u})$. For a $3$ by $3$ contingency table with $N = 18$, the original formulation in \eqref{eq:raw_sens_test} (i.e., formulation based on $\mathcal{Z}(\mathbf{N}_{I\cdot})$) takes \change{451.0} seconds to evaluate whereas the formulation in Theorem \ref{thm:recursive_probability} \change{(i.e., formulation based on $\Omega_{\mathbf{t}}$, $\Omega_{\overline{q}}$, and $\Omega_{\overline{\mathbf{n}}}$)} takes \change{0.0006} seconds. Section \sref{app:F1} of the Supplementary Material provides additional numerical results and computational details.

\begin{table}[ht]
\centering
\small
\begin{tabular}{@{}C{22mm} c d{2.2} c c@{}}
\toprule
\thead{Observed\\Table} & $\mathbf{u}$ 
& \multicolumn{1}{c}{\(p\)-value}
& \multicolumn{2}{c}{Computation Time (in seconds)} \\ 
\cmidrule(lr){4-5}
& & & \multicolumn{1}{c}{Theorem~\ref{thm:recursive_probability}} & \multicolumn{1}{c}{\change{Equation \eqref{eq:raw_sens_test}}} \\
\midrule
\multirow{3}{*}{{%
  $\begin{bmatrix}
    3 & 2 & 1 \\
    0 & 2 & 4 \\
    0 & 1 & 5
  \end{bmatrix}$%
}} & \multicolumn{1}{c}{$(\underbrace{0, \dots, 0}_{13}, \underbrace{1, \dots, 1}_{5})$} & 0.01 
& \change{$0.00060 \pm 0.00005$} & \change{$450.974 \pm 3.373$} \\  
  & \multicolumn{1}{c}{$(\underbrace{0, \dots, 0}_{8}, \underbrace{1, \dots, 1}_{10})$} & 0.04 
& \change{$0.00059 \pm 0.00003$} & \change{$456.565 \pm 7.369$} \\
  & \multicolumn{1}{c}{$(\underbrace{0, \dots, 0}_{4}, \underbrace{1, \dots, 1}_{14})$} & 0.03 
& \change{$0.00060 \pm 0.00003$} & \change{$460.657 \pm 4.571$} \\
\bottomrule
\end{tabular}
\caption{A computational comparison between two approaches of evaluating the exact p-value, one based on Theorem~\ref{thm:recursive_probability} and another based on the original formulation in \eqref{eq:raw_sens_test}. 
Each row corresponds to a particular $\mathbf{u}$ 
and for each~$\mathbf{u}$, we report the one-sided p-value based on an ordinal test with weights $(w_1, w_2, w_3) = (0, 1, 2)$ and $(v_1, v_2, v_3) = (0, 1, 2)$. The sensitivity parameters are set to $\gamma = 1$ and $\bm{\delta} = (0, 1, 1)$. The mean\,\(\pm\)\,standard deviation of the computation time, measured in seconds, is evaluated over 10 runs on Posit Cloud with 8 CPUs and 4~GB RAM.}
\label{tb:main_text_exact_computation_time_permutation_and_kernel}
\end{table}

Theorem~\ref{thm:recursive_probability} can also be combined with fast table-sampling algorithms \citep{ChenEtAl2005, Eisinger2017} that stochastically explore $\Omega_{\mathbf{t}}$ to approximate the exact p-value. In Section \sref{app:F2} of the Supplementary Material, we show that sampling over $\Omega_{\mathbf{t}}$ converges faster to the exact p-value than sampling over $\mathcal{Z}(\mathbf{N}_{I\cdot})$.

We conclude by showing that in the special setting where the outcome is binary, Theorem~\ref{thm:recursive_probability} reduces to a known family of distributions.
\begin{corollary}[Binary Outcome and Multivariate Extended Hypergeometric Distribution] \label{cor:unique_corner_multivariate_hyper} Consider the setting in Theorem \ref{thm:score_test_maximizer_binary}. 
 The exact, worst-case null distribution of the cell counts $(N_{12},\ldots,N_{I2})$ is a multivariate extended hypergeometric distribution with margins $\mathbf{N}_{I\cdot} = (N_{1\cdot}, N_{2\cdot}, \dots, N_{I\cdot})$, $N_{\cdot 2}$, and odds $(\exp(\gamma \delta_1),\ldots, \exp(\gamma \delta_I))$; see equations (7)-(9) in \citet{Fog2008} for the definition of the multivariate extended hypergeometric distribution.
\end{corollary}
Corollary~\ref{cor:unique_corner_multivariate_hyper} also holds when the sensitivity parameter \( \bm{\delta} \) is replaced by a monotone-increasing function of treatment \( \phi(i) \), as in \citet{zhang2024sensitivity}. The odds then become \( (\exp(\gamma \phi(1)), \dots, \exp(\gamma \phi(I))) \) and our results become an extension of their result by deriving a closed-form distribution for the sign-score statistic under their sensitivity model; see Section \sref{app:C4} of the Supplementary Material for details.

\section{Extensions}
\subsection{Integrating Multiple $\bm{\delta}$ Vectors} \label{sec:integrate_delta}
Suppose the investigator believes that $M$ vectors of sensitivity parameters
$\bm{\delta}^{1},\ldots, \boldsymbol{\delta}^{M}$ could plausibly describe the effects
of the unmeasured confounder
and wishes to integrate all $M$ vectors in a sensitivity analysis. 
The worst-case $p$-value across $\bm{\delta}^{1},\ldots, \boldsymbol{\delta}^{M}$ can be
formalized as
\begin{equation} \label{eq:integrate_delta}
\max_{\mathbf{u} \in [0,1]^N,\; \boldsymbol{\delta} \in
\{\boldsymbol{\delta}^{1}, \dots, \boldsymbol{\delta}^{M}\}}
\mathbb{P}_{H_0}\!\left(T(\mathbf{N}(\mathbf{Z}, \rr)) \ge T_{\rm obs}
\mid \mathcal{F},\mathcal{Z}(\mathbf{N}_{I\cdot})\right),
\end{equation}
\noindent Corollary \ref{thm:several_sensitivity_models_main} shows that the solution to the
optimization problem in \eqref{eq:integrate_delta} is the maximum of the $M$ worst-case $p$-values.

\begin{corollary}[Worst-Case $P$-Value Across Multiple Vectors of $\bm{\delta}$]
\label{thm:several_sensitivity_models_main}
For each $\bm{\delta}^m$, let $\mathbf{u}_m$ be the worst-case unmeasured confounder under
$\bm{\delta}^m$, which can be solved by the techniques in
Section \ref{sec:Maximizer_Characterization_and_Bounds}, and let
$\overline{p}_{\gamma,\bm{\delta}^m}(T_{\rm obs})$ be the worst-case $p$-value under $\bm{\delta}^m$, which is
defined in Section \ref{sec:def_worst_case}. Then, the solution to equation
\eqref{eq:integrate_delta} is
\begin{equation} \label{eq:integrate_delta_result}
\max_{\mathbf{u} \in [0,1]^N,\; \boldsymbol{\delta} \in
\{\boldsymbol{\delta}^{1}, \dots, \boldsymbol{\delta}^{M}\}}
\mathbb{P}_{H_0}\!\left(T(\mathbf{N}(\mathbf{Z}, \rr)) \ge T_{\rm obs}
\mid \mathcal{F},\mathcal{Z}(\mathbf{N}_{I\cdot})\right)
= \max\big(\overline{p}_{\gamma,\bm{\delta}^1}(T_{\rm obs}),\dots,\overline{p}_{\gamma,\bm{\delta}^M}(T_{\rm obs})\big).
\end{equation}
\end{corollary}
\noindent The proof is in Section~\sref{app:B8} of the Supplementary Material.

\subsection{{$I \times J \times K$} Contingency Tables}
\label{sec:ijk_contingency_table}
In observational studies, subjects are often divided into $K$ strata defined by observed covariates, forming an  $I\times J \times K$ contingency table. This section describes how the exact methods from Section \ref{sec:sens_ibyj} can be extended to such contingency tables.

For each stratum $k$, suppose there are $N_{k}$ subjects. Let $Z_{ks} \in \{1,\ldots,I\}$ and $r_{ks} \in \{1,\ldots,J\}$ denote the treatment assignment and the observed outcome, respectively, of subject $s = 1,\ldots,N_{k}$. Let $x_{ks}$ and $u_{ks}$ denote the observed and unobserved confounders, respectively, of subject $s$ in stratum $k$. Within each stratum $k$, all subjects have the same $x_{ks}$, i.e., $x_{ks} = x_{ks'}$ for all $1 \leq s < s' \leq N_k$, but they may have different values of $u_{ks}$, i.e., $u_{ks} \neq u_{ks'}$ for some $s \neq s'$. Finally, let $r_{ks}^{(i)}$ denote the potential outcome of subject $s$ in stratum $k$ if the subject, contrary to fact, was assigned to treatment level $i \in \{1,\ldots,I\}$. As before, we assume $r_{ks} = r_{ks}^{(Z_{ks})}$. We also let $\mathcal{F}_{k} = \{ (r_{ks}^{(i)}, x_{ks}, u_{ks}): s = 1,\dots,N_k;\ i = 1,\dots,I \}$ denote the collection of all potential outcomes, observed confounders, and unmeasured confounders for every subject in stratum $k$ and let $\mathbf{u}_{k} = (u_{k1},\ldots,u_{kN_k})$.

We define the following statistics for each stratum $k$. Let $N_{kij} = \sum_{s=1}^{N_k}\mathbbm{1}(Z_{ks}=i, r_{ks}=j)$ denote the number of subjects in stratum $k$ who received treatment level $i \in \{1,\ldots,I\}$ and exhibited outcome level $j \in \{1,\ldots,J\}$. Let $N_{ki\cdot} = \sum_{j=1}^J N_{kij}$ denote the number of subjects in stratum $k$ who received treatment level $i$ and let $N_{k\cdot j} = \sum_{i=1}^{I} N_{kij}$ denote the number of subjects in stratum $k$ with outcome level $j$. Finally, let $\mathbf{N}_{kI\cdot} = (N_{k1\cdot},\ldots,N_{kI\cdot})$ and $\mathbf{N}_{k \cdot J} = (N_{k\cdot 1},\ldots,N_{k \cdot J})$.

For each stratum $k$, the Fisher's sharp null hypothesis of no treatment effect asserts $H_{k0}: r_{ks}^{(1)} = r_{ks}^{(2)} = \cdots = r_{ks}^{(I)}$ for all subjects $s=1,\ldots,N_k$. The joint sharp null hypothesis asserts that all stratum-specific sharp null hypotheses hold, i.e., $H_{\text{joint}} = \bigcap_{k=1}^{K}H_{k0}.$ The goal is to test both $H_{k0}$ and $H_{\text{joint}}$ when the treatment assignment is no longer random within each stratum. We will first discuss testing $H_{\text{joint}}$ and then testing $H_{k0}$ for each $k=1,\ldots,K$. 

Let $\mathbf{Z}_{k} = (Z_{k1},\ldots,Z_{kN_{k}})$ denote the vector of treatment assignments for each stratum $k$ and suppose the treatment assignments are independent across strata, i.e.,
\begin{equation*}
\begin{aligned}
\mathbb{P}\big(&\mathbf{Z}_{1} = \mathbf{z}_{1},\, \dots,\, \mathbf{Z}_{K} = \mathbf{z}_{K} | 
    \mathcal{F}_{1} \dots, \mathcal{F}_{K}, \mathcal{Z}(\mathbf{N}_{1 I\cdot})\dots,\, \mathcal{Z}(\mathbf{N}_{K I\cdot}) 
\big) = \prod_{k=1}^K \mathbb{P}\big(\mathbf{Z}_{k} = \mathbf{z}_{k} \mid \mathcal{F}_{k},\, \mathcal{Z}(\mathbf{N}_{kI\cdot})\big).
\end{aligned}
\end{equation*}
Within each stratum $k$, we assume the sensitivity model \eqref{eq:generic_bias_sensitivity} holds whereby 
the probability of getting assigned treatment level $i$ for subject $s$ depends on both measured and unmeasured confounders: 
\begin{equation} \label{eq:stratum_sensitivity_model}
    \mathbb{P}(Z_{ks} = i \mid \mathcal{F}_k) = \frac{\exp\{\xi_{ki}(x_{ks}) + \gamma \delta_i u_{ks}\}}{\sum_{i'=1}^I \exp\{\xi_{ki'}(x_{ks}) + \gamma \delta_{i'} u_{ks}\}}, \quad \gamma \geq 0, \delta_i \in \{0,1\}, u_{ks} \in [0,1]. 
\end{equation}
Similar to \eqref{eq:generic_bias_sensitivity}, the sensitivity model in  \eqref{eq:stratum_sensitivity_model} has two sensitivity parameters $\gamma$ and $\bm{\delta}=(\delta_1,\ldots,\delta_I)$. 

Let $T_k$ denote the test statistic using data from stratum $k$ and let
$\bm{T} = (T_1,\dots,T_K)$. Because of the independence of treatment assignment across strata, the worst-case p-values computed from each stratum will be independent of each other and we can combine these p-values using various methods for combining independent p-values in multiple testing problems. Specifically, let $\overline{p}_{\gamma, \bm{\delta}}(T_{1}),\ldots,\overline{p}_{\gamma, \bm{\delta}}(T_{K})$ denote these $K$ independent p-values. One approach to combine the p-values is the truncated product method from \citet{zaykin2002truncated} for some truncation point $\tau \in (0,1)$, i.e., $W = \prod_{k=1}^{K} \overline{p}_{\gamma,\bm{\delta}}(T_k)^{\mathbbm{1}\{\overline{p}_{\gamma,\bm{\delta}}(T_k) \le \tau\}}$. The null distribution of $W$ under $H_{\rm joint}$ can be found in \citet[Equation (1)]{zaykin2002truncated}. \citet{HsuSmallRosenbaum2013} found that smaller values of $\tau$ increased the power of a sensitivity analysis. Importantly, the test based on $W$ is valid for any test statistic $T_k$, including the count-based test statistic, the ordinal test statistic, and the sign-score test statistic.

If the outcomes are binary so that the contingency table becomes an $I \times 2 \times K$ table, Corollary \ref{cor:i2k} shows that we can obtain a closed-form expression of the exact, worst-case null distribution for any monotone-increasing function of sign-score tests, also referred to as a stratified sum test.  
\begin{corollary}[Exact, Worst-Case Null Distribution of the Stratified Sum Test on an $I\times2\times K$ Table]
\label{cor:i2k}
Suppose we have an $I \times 2 \times K$ contingency table with binary outcomes. 
Let $g : \mathbb{R}^K \to \mathbb{R}$ be a component-wise monotone-increasing function, i.e.,
$g(\mathbf{T}) \le g(\mathbf{T}')$ whenever $T_k \le T_k'$ for all $k = 1, \ldots, K$. 
Let $\mathbf{T} = (T_1,\ldots,T_K)$ be the vector of sign-score statistics across $K$ strata, 
where $T_{k} = \sum_{i=1}^{I} w_{ki} N_{ki2}$ and, for every $k$, the treatment scores $w_{k1} \le \cdots \le w_{kI}$ and sensitivity parameters $\delta_1 \le \cdots \le \delta_I$ are simultaneously nondecreasing.  
Under the joint null hypothesis $H_{\mathrm{joint}}$ and the sensitivity model in 
\eqref{eq:stratum_sensitivity_model}, 
the exact worst-case null distribution of $g(\mathbf{T})$ for all $c \in \mathbb{R}$ is
\begin{equation}
\max_{\substack{\mathbf{u}_1 \in [0,1]^{N_1},\,\ldots,\,\mathbf{u}_K \in [0,1]^{N_K}}}
\mathbb{P}_{H_{\mathrm{joint}}}\!\big(g(\mathbf{T}) \ge \criticalt 
\mid \mathcal{F}_{1},\ldots,\mathcal{F}_{K},
\mathcal{Z}(\mathbf{N}_{1I\cdot}),\ldots,\mathcal{Z}(\mathbf{N}_{KI\cdot})\big)
=
\mathbb{P}_{H_{\mathrm{joint}}}\!\big(g(\overline{\mathbf{T}}) \ge \criticalt\big).
\label{eq:creating_valid_test_binary}
\end{equation}
The random vector $\overline{\mathbf{T}} = (\overline{T}_1,\ldots,\overline{T}_K)$ consists of $\overline{T}_{k} = \sum_{i=1}^I w_{ki} M_{ki}$, and  
$\mathbf{M}_{k} = (M_{k1},\allowbreak \ldots,M_{kI})$ follows a multivariate extended hypergeometric 
distribution with parameters $\mathbf{N}_{kI\cdot} = (N_{k1\cdot},\ldots,N_{kI\cdot})$, $N_{k\cdot 2}$, and 
odds $(\exp(\gamma \delta_{1}), \ldots, \exp(\gamma \delta_{I}))$.
\end{corollary}
The proof of Corollary \ref{cor:i2k} follows from Theorem \ref{thm:score_test_maximizer_binary}, Corollary~\ref{cor:unique_corner_multivariate_hyper}, and classic results on stochastic ordering; see Chapter 17 of \citet{marshall_olkin_arnold_2011} and
Section \sref{app:B9} of the Supplementary Material. Also, Corollary \ref{cor:i2k} holds for any monotone-increasing function $g$ and one popular example of $g$ is the (unweighted) linear sum of sign-score test statistics, i.e., $g(\mathbf{T}) = \sum_{k=1}^K T_k$. 

The investigator may also wish to test the stratum-specific $H_{k0}$ while controlling the family-wise Type I error rate. A popular approach under Rosenbaum's sensitivity model is the closed-testing procedure of \citet{marcus1976closed} and for more details, see \citet{HsuSmallRosenbaum2013} and \citet{lee2018discovering}.
\section{Power of Tests in Sensitivity Analysis} \label{sec:power_main}
\subsection{Motivation and Setup}

\label{sec:power_analysis_size_control}
In a sensitivity analysis for unmeasured confounding, \citet{rosenbaum2004design} proposed to compare tests by studying their power under the so-called favorable situation where there is a non-zero treatment effect and there is no unmeasured confounding. Several works have studied the power of a sensitivity analysis under binary treatment (e.g., \citet{rosenbaum2012design,fogarty2017studentized,huang2025design}). Under non-binary treatment, we are aware only of \citet{zhang2024sensitivity}, who studied power for continuous treatment rather than for contingency tables.

To this end, we conduct a simulation study to study the power of tests in a sensitivity analysis for $I$ by $J$ contingency tables. We focus on  $3 \times 3$ contingency tables 
and follow \citet{agresti1987empirical}'s analysis of power in contingency tables. Specifically, we consider a log-linear model to specify an alternative hypothesis where there is a linear-by-linear association between the rows and the columns of the $I$ by $J$ table:
\begin{equation}
\label{eq:log_linear_model_DGP}
\log(\mathbb{E}[N_{ij}])
\;=\;
\lambda
\;+\;
\lambda_i^{Z}
\;+\;
\lambda_j^{r}
\;+\;
\beta w^*_i \, v^*_j,
\quad
w^*_1 \leq w^*_2 \leq \cdots\leq w^*_I, \quad
v^*_1 \leq v^*_2  \leq \cdots\leq v^*_J.
\end{equation}
The parameter $\lambda$ represents the overall mean of the expected counts in the table, $\lambda_i^{Z}$ is the ``row effect,'' $\lambda_j^{r}$ is the ``column effect,'' and $\beta w^*_i v^*_j$ is the interaction effect between the rows and the columns. Setting $\beta \neq 0$ indicates a treatment effect, while $w_i^*$ and $v_j^*$ control the strength of row-column association. For additional details on the log-linear models and power analysis in contingency tables, see Section 2 of \citet{agresti1987empirical}, Chapter 3.4 of \citet{agresti2012categorical}, and Section \sref{app:F4} of the Supplementary Material.
Throughout the simulation study, we fix $\lambda = 0$,  $\beta = 1$, treatment margins $\mathbf{N}_{3\cdot}=(20,20,20)$ so that $N = 60$. We consider four cases of $(w_i^*, v_j^*)$; see Figure \ref{fig:power_of_collapsed_test_60} for the exact specifications under each case. We also set the row and column effects to be  \( (\lambda_1^Z, \lambda_2^Z, \lambda_3^Z) = (1, 0, 0) \), and \( (\lambda_1^r, \lambda_2^r, \lambda_3^r) = (1, 0.2, 0) \) for cases I and II, and \( (\lambda_1^Z, \lambda_2^Z, \lambda_3^Z) =  (\lambda_1^r, \lambda_2^r, \lambda_3^r) = (0,0,0) \) for cases III and IV. In summary, case II posits well-separated treatment and outcome levels where every adjacent difference, $w_i^{*}-w_{i-1}^{*}$ or $v_j^{*}-v_{j-1}^{*}$, is substantive. Case I posits well-separated treatment levels but similar higher outcome levels, i.e., $v_3^{*}-v_2^{*}$ is small. Cases III and IV represent situations in which the lower treatment and outcome levels are similar, i.e., $w_2^{*}-w_1^{*}$ and $v_2^{*}-v_1^{*}$ are small.

\begin{figure}[ht]
    \centering
\includegraphics[width=1\textwidth]{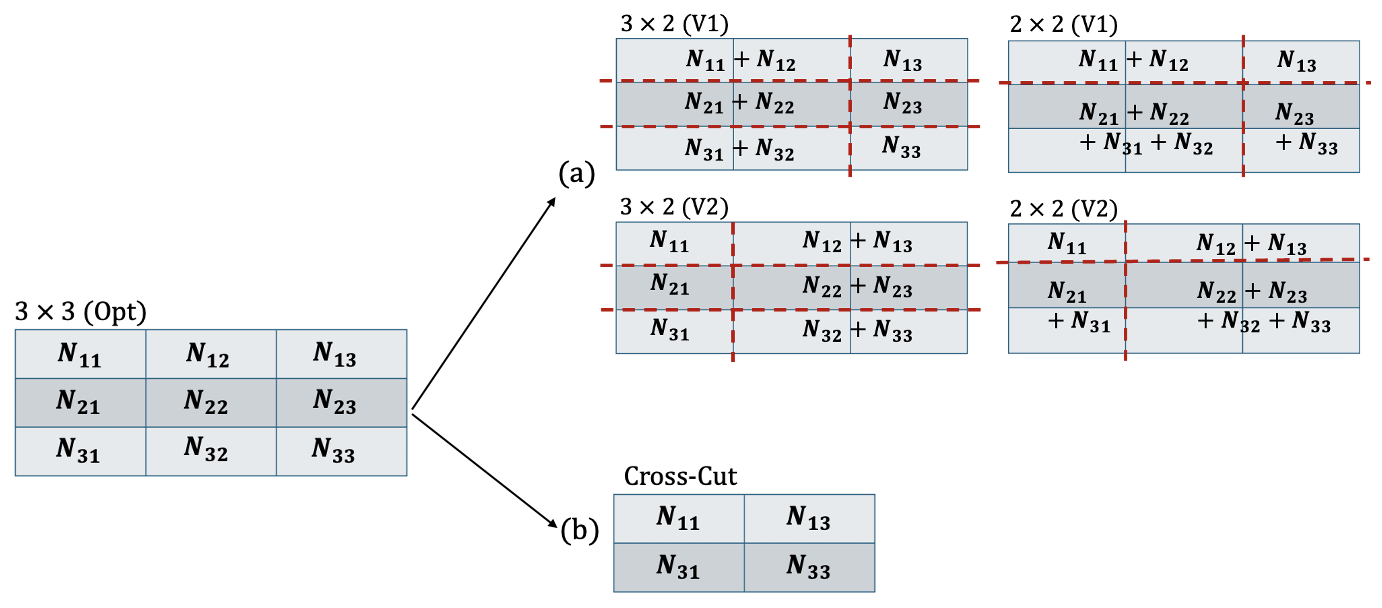}
    \caption{Construction of smaller tables by either (a) collapsing or (b) cutting the rows or columns. Collapsed tables are formed by summing the counts within the red boundaries. 
    The cross-cut table  \citep{rosenbaum2016crosscut} keeps only the ``extreme'' cells of the 3 by 3 table.} 
\label{fig:collapsing_tables}
\end{figure}

For test statistics, we follow \citet{agresti2012categorical} and consider the family of ordinal tests in Definition~\ref{eq:ordinal_test}, i.e., $T(\mathbf{N}) = \sum_{i=1}^{3} \sum_{j=1}^{3} w_i v_j N_{ij}$. Under no unmeasured confounding and the linear-by-linear association model, ordinal tests are known to be more powerful than the chi-squared or the $G^2$ test. The weights $w_i$ and $v_j$ are specified by the investigator based on beliefs about the alternative hypothesis or by data-driven methods such as midrank scores \citep{agresti2012categorical} or profile-likelihood scores (see Section~\sref{app:F3} of the Supplementary Material). We consider three sets of weights: (a) ``$3 \times 3$ (Opt),'' the ``optimal'' weights $w_i^*$ and $v_j^*$ that match the data generating model; (b) ``$3 \times 2$ (V1),'' the optimal $w_i^*$s with $v_{1} = v_{2} = 0$ and $v_{3} = 1$; and (c) ``$3 \times 2$ (V2),'' the optimal $w_i^*$s with $v_{1}=0$ and $v_2 = v_3 =1$. Tests (b) and (c) collapse the columns of the $3 \times 3$ table into a $3 \times 2$ table; see Figure~\ref{fig:collapsing_tables}. All three tests use $\bm{\delta} = (0,1,1)$ and their p-values are computed under the sensitivity model in \eqref{eq:generic_bias_sensitivity}.

We also compare the ordinal tests to two Fisher's exact tests computed by collapsing the $3 \times 3$ table into a $2 \times 2$ table by merging treatment levels 2 and 3 (so that $\bm{\delta} = (0,1)$); see Figure \ref{fig:collapsing_tables}. The two Fisher's exact tests correspond to an investigator who, given the limitation of existing sensitivity analysis to binary treatments, may first collapse the contingency tables and use the exact methods available for conducting sensitivity analysis with $2 \times 2$ tables \citep{rosenbaum1990sensitivity}. The two Fisher's exact tests are denoted as ``$2 \times 2$ (V1)'' and ``$2 \times 2$ (V2).'' 

Finally, we include the cross-cut test proposed by \citet{rosenbaum2016crosscut}, which evaluates only the extreme cells of the \(3\times3\) contingency table (see Figure~\ref{fig:collapsing_tables}); we denote this test as ``Cross-Cut.'' 
This test differs from the ones above, and its performance has not been previously examined under contingency tables where both the outcome and the treatment are discrete. The sensitivity model in~\eqref{eq:generic_bias_sensitivity} implies the corresponding sensitivity models for the Fisher's exact tests mentioned above or for the cross-cut test in \citet{rosenbaum1990sensitivity}; see Section \sref{app:B10} of the Supplementary Material for details.

\subsection{Results}
\begin{figure}[ht]
    \centering
    \includegraphics[width=1.0\textwidth]{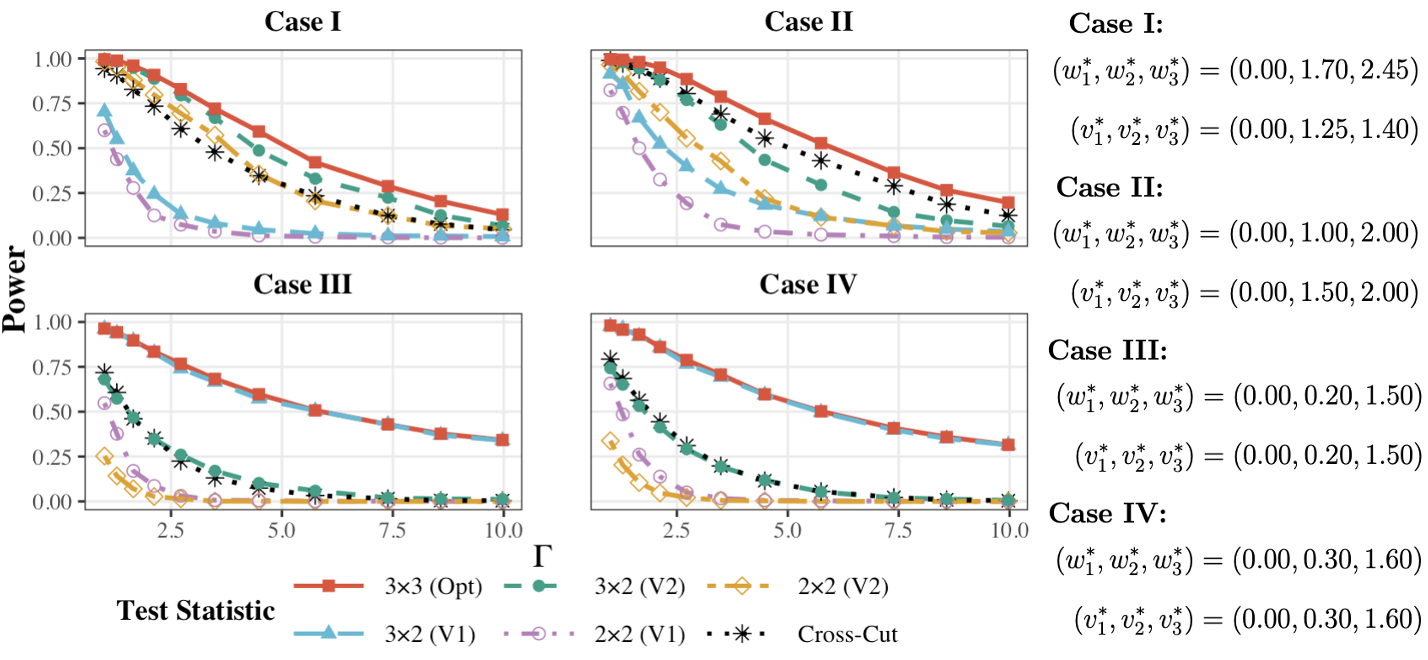}
    \caption{Power of tests under different $\Gamma = \exp(\gamma)$ in a 3 by 3 contingency table with $N=60$ subjects. We consider four data generating processes and six different test statistics.}
\label{fig:power_of_collapsed_test_60}
\end{figure}

Figure \ref{fig:power_of_collapsed_test_60} presents the results. First, the $3\times 3$ (Opt) test achieves the highest or essentially tied-highest power across all four cases. This is not surprising as the test statistic uses the weights that correspond to the data generating model in \eqref{eq:log_linear_model_DGP}. Second, the two tests $3 \times 2$ (V1) and $3 \times 2$ (V2) can come close to the power of the $3\times 3$ (Opt) test when the column collapse combines outcome categories with similar interaction weights $v_j^*$. For example, in cases III and IV the $3\times 2$ (V1) test performs almost as well as the $3 \times 3$ (Opt) test since $3\times 2$ (V1) combines the first two outcome levels and the simulation model assigns similar interaction effects to the first two levels of the outcome (i.e., $(v^{*}_1, v^{*}_2, v^{*}_3) = (0.00, 0.20, 1.50)$ under case III and $(v^{*}_1, v^{*}_2, v^{*}_3) = (0.00, 0.30, 1.60)$ under case IV); likewise, in case I the $3 \times 2$ (V2) test, which merges the two similar higher outcome levels, tracks the $3 \times 3$ (Opt) test closely. Third, the $2\times 2$ (V1) test, $2\times 2$ (V2) test, and the cross-cut test generally have lower power than $3\times 3$ (Opt) test.

We conduct additional simulation studies in Sections \sref{app:F5} to \sref{app:F7} of the Supplementary Material where we include sparse contingency tables and the results are broadly similar: the $3\times 3$ (Opt) test has the highest or near-highest power in most settings. More broadly, our simulations highlight the potential benefit of incorporating all treatment and outcome levels in a sensitivity analysis.

\section{Data Analysis} \label{sec:data_analysis_main}

\subsection{Background}
\label{sec:data_analysis}
We illustrate our method using the Early Childhood Longitudinal Study, Kindergarten Class of 1998--99 (ECLS-K) \citep{NCES2009ECLSK}, examining how pre-kindergarten care affects kindergarten math achievement under potential unmeasured confounding.
Following \citet{gormley_gayer_2005,magnuson_ruhm_waldfogel_2007,lee2022effects}, we focus on two subgroups: Black girls and Black boys raised in low-income
(weekly family income below the median), single-parent families. The outcome is a three-level ordinal variable for the highest math
proficiency mastered by Fall of kindergarten --- (1) number and shape, (2) relative size, or (3) ordinality and sequence. Treatment has three
levels: (1) no non-parental care, (2) relative-based care, or (3) center-based care (daycare, preschool, public pre-K). Table~\ref{tb:black_girl_boy_low_income_single_table} displays the resulting
$3 \times 3 \times 2$ contingency table; see data dictionaries
\texttt{C1R4MPF} and \texttt{P1PRIMPK} for further detail.

\begin{table}[ht]
\centering
\setlength{\tabcolsep}{5pt}
\begin{tabular}{ll|ccc}
\toprule
\multicolumn{2}{c|}{} & \multicolumn{3}{c}{\textbf{Highest Math Proficiency}} \\
\cmidrule(lr){3-5}
\textbf{Black} & \textbf{Pre-K Care} & Number \& & Relative & Ordinality \& \\ 
& & shape & size & sequence \\ 
\midrule
\multirow{3}{*}{Girls} 
    & No care           & 12 & 3  & 0 \\ 
    & Relative care     & 18 & 12 & 3 \\ 
    & Center-based care & 17 & 25 & 4 \\ 
\midrule
\multirow{3}{*}{Boys} 
    & No care           & 10 & 8  & 1 \\ 
    & Relative care     & 29 & 11 & 3 \\ 
    & Center-based care & 20 & 24 & 6 \\ 
\bottomrule
\end{tabular}
\caption{A $3 \times 3 \times 2$ contingency table displaying pre-kindergarten care type (rows) and highest math proficiency (columns) among Black girls and boys in low-income, single-parent families. The data comes from the ECLS-K (1998-99) cohort.}
\label{tb:black_girl_boy_low_income_single_table}
\end{table}

\change{For the sensitivity analysis we assume generic bias and set $\bm{\delta} = (0, 1, 1)$, with $\delta_1 = 0$ for no non-parental care and $\delta_2 = \delta_3 = 1$ for relative and center-based care. This $\bm{\delta}$ implies that the unmeasured confounder systematically increases the likelihood of receiving any form of care. This choice follows \citet{meyers2006choice} and \citet{harknett_schneider_luhr_2022}, who argue that, after race and income are accounted for, no systematic confounder influences the choice between center-based and relative-based care; instead, parental work schedule regularity acts as the unmeasured confounder that increases care receipt while influencing academic performance. We consider two ordinal tests with treatment weights either $(w_1, w_2, w_3) = (0.00, 0.25, 1.50)$ or $(0.00, 0.20, 1.60)$ and outcome weights either $(v_1, v_2, v_3) = (0.00, 1.00, 1.50)$ or $(0.00, 0.67, 0.70)$. The ordinal score choices reflect a positive treatment-outcome association, with treatment gaps encoding that center-based care provides a richer learning environment than relative or no non-parental care, and outcome gaps reflecting that Relative Size and Ordinality \& Sequence represent closely related, advanced concepts compared to basic Number \& Shape. For further discussion on score choices and additional test results, see Section~\sref{app:F3} of the Supplementary Material. We use the procedure described in Section~\ref{sec:ijk_contingency_table} to conduct sensitivity analysis across the two subgroups.}

\subsection{Results}
The results of the data analysis are in
Table~\ref{tb:data_analysis_result_table}. At $\alpha = 0.05$, the joint null and the subgroup nulls for Black girls and boys are all rejected
under no unmeasured confounding ($\Gamma = \exp(\gamma) = 1$). Under the prior scores, the null is rejected for all $\Gamma \leq 2.5$ for Black girls but only for $\Gamma \leq 1.5$ for Black boys, suggesting that the evidence of a benefit from pre-K care is more robust to unmeasured confounding for girls than for boys.

\setlength{\tabcolsep}{12pt} 
\begin{table}[ht]
\centering
\begin{tabular}{
  l
  l
  d{1.3}@{}l@{\hspace{4pt}}
  d{1.3}@{}l@{\hspace{4pt}}
  d{1.3}@{}l@{\hspace{4pt}}
  d{1.3}@{}l@{\hspace{4pt}}
  d{1.3}@{}l@{\hspace{4pt}}
  d{1.3}@{}l@{\hspace{4pt}}
  d{1.3}@{}l@{\hspace{4pt}}
  d{1.3}@{}l
}
\toprule
\multicolumn{1}{l}{Score} &
\multicolumn{1}{l}{$\Gamma$} 
  & \multicolumn{2}{c}{$1.0$}
  & \multicolumn{2}{c}{$1.5$}
  & \multicolumn{2}{c}{$2.0$}
  & \multicolumn{2}{c}{$2.5$}
  & \multicolumn{2}{c}{$3.0$}
  & \multicolumn{2}{c}{$3.5$}
  & \multicolumn{2}{c}{$4.0$}
  & \multicolumn{2}{c}{$4.5$} \\
\midrule
\multirow{3}{*}{%
   Prior score
  } & Girls       
  & 0.006 & $^*$
  & 0.015 & $^*$
  & 0.028 & $^*$
  & 0.041 & $^*$
  & 0.054 & 
  & 0.067 & 
  & 0.080 & 
  & 0.091 & \\
& Boys        
  & 0.013 & $^*$
  & 0.032 & $^*$
  & 0.056 & 
  & 0.082 & 
  & 0.106 & 
  & 0.130 & 
  & 0.152 & 
  & 0.172 &  \\
& Joint
  & 0.001 & {}
  & 0.003 & {}
  & 0.009 & {}
  & 0.017 & {}
  & 0.026 & {}
  & 0.036 & {}
  & 0.046 & {}
  & 0.055 & {} \\
\midrule
\multirow{3}{*}{%
   Profile score
  } & Girls       
  & 0.004 & $^*$
  & 0.011 & $^*$
  & 0.021 & $^*$
  & 0.033 & $^*$
  & 0.046 & $^*$
  & 0.059 & {}
  & 0.071 & {}
  & 0.083 & {} \\
& Boys        
  & 0.012 & $^*$
  & 0.028 & $^*$
  & 0.045 & $^*$
  & 0.064 & {}
  & 0.082 & {}
  & 0.099 & {}
  & 0.117 & {}
  & 0.133 & {} \\
& Joint
  & 0.001 & {}
  & 0.002 & {}
  & 0.006 & {}
  & 0.012 & {}
  & 0.019 & {}
  & 0.027 & {}
  & 0.035 & {}
  & 0.043 & {} \\
\bottomrule
\end{tabular}
\caption{%
  Results of the sensitivity analysis for our data. Prior score sets the weights of the ordinal test to be $(w_1,w_2,w_3) = (0.00,0.25,1.50)$, $(v_1,v_2,v_3)=(0.00,1.00,1.50)$ and the profile score sets them to be $(w_1,w_2,w_3) = (0.00,0.20,1.60)$, $(v_1,v_2,v_3)=(0.00,0.67,0.70)$.
  Each entry is the exact, worst-case p-value for a given $\Gamma$. 
  The row labeled ``Joint'' reports the p-value for the joint null hypothesis of no effect for boys and girls, obtained by the truncated product method of Section~\ref{sec:ijk_contingency_table} with $\tau = 0.2$. An asterisk ($^*$) indicates rejection of the subgroup null hypothesis at the 0.05 family-wise error rate using the closed-testing procedure in Section \ref{sec:ijk_contingency_table}.}
\label{tb:data_analysis_result_table}
\end{table}
\setlength{\tabcolsep}{6pt}

The result of the sensitivity analysis parallels some findings from \citet{joo2010long}, who reported significant improvements in reading and math scores among low-income girls who attended Head Start, but not among boys. 
\citet{Ewing2009} theorized that these gender differences may be explained by girls' advantages in temperament and attention regulation, which foster better relationships with instructors and support early learning. 

\section{Discussion and Summary} \label{sec:discussion}
This paper develops an exact, nonparametric sensitivity analysis for contingency tables. Extending Rosenbaum’s generic bias model to non-binary treatments, we propose a framework to compute the worst-case null distribution for a broad class of test statistics. Practically, our approach allows sensitivity analysis without dichotomizing the treatment variable and our simulation study showed that using all treatment and outcome levels can improve power.

As discussed immediately after Theorem~\ref{thm:score_test_maximizer_binary} and Corollary~\ref{cor:unique_corner_multivariate_hyper}, some of our results extend to a larger class of sensitivity models, notably the monotone dose sensitivity model in \citet{zhang2024sensitivity}. Beyond this, our results may not generalize and we conjecture that this is due to a fundamental limitation of sensitivity analysis with non-binary treatments rather than due to the specific setup in the paper. Specifically, in Appendices~\sref{app:E1} and \sref{app:E2}, we provide a numerical counterexample where a key intermediate result in our theory no longer holds under a sensitivity model that is more general than the generic bias sensitivity model and discuss the implications of the counterexample for exact inference.

\change{Finally, on the computational aspect of our framework, Section~\sref{app:G} of the Supplementary Material reports end-to-end running times and memory of the exact worst-case p-value on a machine with 8 CPUs and 4~GB of RAM. For ordinal and sign-score tests, the worst-case p-value is obtained within about 100 seconds for every table dimension and $N \le 200$ we examined. For a general count-based test such as the chi-square test, the computation takes seconds for $J = 2$ and $N \le 200$, about two hours for $J = 3$ and $N = 200$, and about two hours already at $J = 4$ and $N = 100$. If the investigator's data is larger than these dimensions and she is not using an ordinal or sign-score test statistic,} we outline how to obtain a stochastic approximation of the exact p-value through Monte Carlo in Section \ref{sec:exact_worst_null} and Section \sref{app:F2} of the Supplementary Material.

\section*{Acknowledgments}
The authors gratefully acknowledge BLANK, BLANK, BLANK, BLANK, and BLANK for their feedback. BLANK was supported in part by NIH Grant BLANK. 

\section*{Data Availability Statement}
The data that support the findings of this study are openly available in Kindergarten Class of 1998-99 (ECLS-K) at \url{https://nces.ed.gov/ecls/kindergarten.asp}.

\section*{Disclosure Statement}
No potential competing interest was reported by the author(s).

\clearpage
\setcounter{section}{0}
\renewcommand{\thesection}{\Alph{section}}
\renewcommand{\thesubsection}{\thesection.\arabic{subsection}}
\titleformat{\section}[block]{\normalfont\Large\bfseries}{}{0pt}{Appendix~\thesection. }
\titleformat{\subsection}[block]{\normalfont\large\bfseries}{}{0pt}{\thesubsection\ }
\setcounter{equation}{0}\renewcommand{\theequation}{S\arabic{equation}}
\setcounter{table}{0}\renewcommand{\thetable}{S\arabic{table}}
\setcounter{figure}{0}\renewcommand{\thefigure}{S\arabic{figure}}
\setcounter{defn}{0}\renewcommand{\thedefn}{S\arabic{defn}}
\setcounter{corollary}{0}\renewcommand{\thecorollary}{S\arabic{corollary}}
\begin{center}
  {\LARGE\bfseries Supplementary Material}
\end{center}

\section{Mathematical Equivalence and Interpretations of the Sensitivity Model} \label{app:A}
This section establishes the mathematical equivalence among several formulations of the sensitivity model introduced in the main text. In addition, we present further properties and interpretations of the model, providing insight into its practical implications.
\par For model equivalence, the first expression states that:
\begin{equation}
\label{eq:generic_bias_sensitivity_supp}
\mathbb{P}(Z_s = i \mid \mathcal{F}) 
= 
\frac{\exp\left(\xi_i(x_s) + \gamma\delta_i\,u_s\right)}
{\sum_{i'=1}^{I} \exp\left(\xi_{i'}(x_s) + \gamma\,\delta_{i'}\,u_s\right)},
\;\delta_i \in \{0,1\},
\;
\gamma \geq 0,
\;
u_s \in [0,1].
\end{equation}

\bigskip

\par \noindent The second expression states that, for any pair \(i,i' \in \{1,\dots,I\}\):
\begin{equation}
\label{eq:log-odds-ratio_supp}
\log (
    \frac{\mathbb{P}(Z_s = i' \mid \mathcal{F})}{\mathbb{P}(Z_s = i \mid \mathcal{F})})=
\xi_{i'}(x_s)-\xi_i(x_s) +\gamma\left(\delta_{i'} - \delta_i\right)u_s,\; u_s \in [0,1].
\end{equation}

\bigskip

\par \noindent Finally, Proposition~\ref{prop:equivalence_odds_inequality} establishes that \eqref{eq:log-odds-ratio_supp} is equivalent to the probability ratio constraints \eqref{eq:odds_constraint1} and \eqref{eq:odds_constraint2}. That is, these constraints are both necessary and sufficient for the sensitivity model.

\begin{proposition}[Equivalence of Sensitivity Models]
\label{prop:equivalence_odds_inequality}  
\noindent Let \(\gamma \geq 0\) be a sensitivity parameter, and define \(\Gamma = \exp(\gamma)\). 
The sensitivity model in \eqref{eq:log-odds-ratio_supp} holds if and only if the following constraints hold for any pair of subjects \(s, s'\) with identical observed covariates, i.e., \(x_s = x_{s'}\):
\begin{itemize}
    \item The probability ratio of receiving a treatment with \(\delta_i = 1\) versus one with \(\delta_i = 0\) may differ by at most a factor of \(\Gamma\):
    \begin{align}
        \frac{1}{\Gamma} 
        \leq
        \frac{\mathbb{P}(Z_s = i' \mid \mathcal{F})\, \mathbb{P}(Z_{s'} = i \mid \mathcal{F})}
             {\mathbb{P}(Z_s = i \mid \mathcal{F})\,\mathbb{P}(Z_{s'} = i' \mid \mathcal{F})} 
        \leq \Gamma, 
        \quad \text{if } |\delta_{i'} - \delta_i| = 1.
        \label{eq:odds_constraint1}
    \end{align}
 \item If \(\delta_{i'} = \delta_i\), the probability ratio of receiving treatment \(i'\) versus \(i\) is the same:
    \begin{align}
        \frac{\mathbb{P}(Z_s = i' \mid \mathcal{F})\,\mathbb{P}(Z_{s'} = i \mid \mathcal{F})}
             {\mathbb{P}(Z_s = i \mid \mathcal{F})\,\mathbb{P}(Z_{s'} = i' \mid \mathcal{F})} 
        = 1.
        \label{eq:odds_constraint2}
    \end{align}
\end{itemize}
\end{proposition}

\subsection{\eqref{eq:generic_bias_sensitivity_supp} $\Leftrightarrow$ \eqref{eq:log-odds-ratio_supp}} \label{app:A1}
\begin{proof}
\noindent
\eqref{eq:generic_bias_sensitivity_supp} $\Rightarrow$ \eqref{eq:log-odds-ratio_supp}:  
Taking the ratio of the probabilities of receiving treatments 
\( i' \) and \( i \) on both sides of \eqref{eq:generic_bias_sensitivity_supp}, and then applying the logarithm, we obtain:
\[
\log\left(\frac{\mathbb{P}(Z_s = i' \mid \mathcal{F})}{\mathbb{P}(Z_s = i \mid \mathcal{F})}\right) 
= \xi_{i'}(x_s) - \xi_i(x_s) + \gamma(\delta_{i'}-\delta_i)\,u_s, 
\]
which matches the form in \eqref{eq:log-odds-ratio_supp}.  

\medskip
\par \noindent
\eqref{eq:log-odds-ratio_supp} $\Rightarrow$ \eqref{eq:generic_bias_sensitivity_supp}:  
Setting \( i=1 \) in \eqref{eq:log-odds-ratio_supp}, we obtain:
\begin{align}
\mathbb{P}(Z_s = i' \mid \mathcal{F}) 
= \mathbb{P}(Z_s = 1 \mid \mathcal{F}) 
  \exp(\xi_{i'}(x_s) - \xi_1(x_s) + \gamma(\delta_{i'} - \delta_1)u_s).
  \label{eq:exponential_proportion}
\end{align}
Since a subject must receive exactly one of the \( I \) possible treatment levels, we have: 
\[
\sum_{i'=1}^{I} \mathbb{P}(Z_s = i' \mid \mathcal{F}) = 1.
\]
Substituting the expression for \( \mathbb{P}(Z_s = i' \mid \mathcal{F}) \) from \eqref{eq:exponential_proportion}, we obtain
\[
1 = \mathbb{P}(Z_s = 1 \mid \mathcal{F}) 
  \sum_{i'=1}^{I} \exp\bigl(\xi_{i'}(x_s) + \gamma\delta_{i'} u_s - \xi_1(x_s) - \gamma\delta_1 u_s\bigr).
\]
Solving for \( \mathbb{P}(Z_s = 1 \mid \mathcal{F}) \), we obtain:
\[
\mathbb{P}(Z_s=1 \mid \mathcal{F})
= \frac{\exp(\xi_1(x_s) + \gamma\delta_1 u_s)}
       {\sum_{i'=1}^I \exp(\xi_{i'}(x_s) + \gamma\,\delta_{i'}\,u_s)}.
\]
For \( i \neq 1 \), applying the relationship in \eqref{eq:exponential_proportion}, we obtain:
\[
\mathbb{P}(Z_s = i \mid \mathcal{F}) 
= \frac{\exp(\xi_i(x_s) + \gamma\,\delta_i\,u_s)}
       {\sum_{i'=1}^I \exp(\xi_{i'}(x_s) + \gamma\,\delta_{i'}\,u_s)}.
\]
Thus, we recover the form of \eqref{eq:generic_bias_sensitivity_supp}.
\end{proof}
\subsection{\eqref{eq:log-odds-ratio_supp} $\Leftrightarrow$ \eqref{eq:odds_constraint1} \text{and} \eqref{eq:odds_constraint2}} \label{app:A2}
\begin{proof}
\eqref{eq:log-odds-ratio_supp} $\Rightarrow$ \eqref{eq:odds_constraint1} and \eqref{eq:odds_constraint2}:  
Suppose \eqref{eq:log-odds-ratio_supp} holds, then we have:
\begin{align}
    \frac{\mathbb{P}(Z_s = i' \mid \mathcal{F})\;\mathbb{P}(Z_{s'} = i \mid \mathcal{F})}
    {\mathbb{P}(Z_s = i \mid \mathcal{F})\;\mathbb{P}(Z_{s'} = i' \mid \mathcal{F})} 
    = \exp\big(\gamma(\delta_{i'} - \delta_{i})(u_s - u_{s'})\big).
    \label{eq:odds_ratio_two_subjects}
\end{align}
For \( |\delta_{i'} - \delta_i| = 1 \), since \( u_s, u_{s'} \in [0,1] \), we obtain:
\begin{align*}
    \frac{1}{\Gamma} = \exp(-\gamma) \leq \exp\big(\gamma(\delta_{i'} - \delta_{i})(u_s - u_{s'})\big) \leq \exp(\gamma) = \Gamma,
\end{align*}
implying \eqref{eq:odds_constraint1}. If instead \( \delta_{i'} = \delta_i \), then \eqref{eq:odds_ratio_two_subjects} reduces to \( 1 \), implying \eqref{eq:odds_constraint2}.

\medskip  
\par \noindent \eqref{eq:odds_constraint1} and \eqref{eq:odds_constraint2} $\Rightarrow$ \eqref{eq:log-odds-ratio_supp}:  
Assume that \eqref{eq:odds_constraint1} and \eqref{eq:odds_constraint2} hold. For each covariate value \(x\), let \(s'\) denote the subject with the smallest probability ratio of receiving a treatment with \(\delta_{i'} = 1\) versus one with \(\delta_i = 0\):
\begin{align}
s' = \operatorname{argmin}_{s : x_s = x}
 \frac{\mathbb{P}(Z_s = i' \mid \mathcal{F})}
        {\mathbb{P}(Z_s = i \mid \mathcal{F})}. 
\label{eq:definition_of_the_us_zero_person}
\end{align}
Without loss of generality, we may choose any \( i' \) such that $\delta_{i'}=1$ and any \( i \) such that $\delta_{i}=0$ to start this construction, as the subject minimizing  
\[
\frac{\mathbb{P}(Z_s = i' \mid \mathcal{F})}{\mathbb{P}(Z_s = i \mid \mathcal{F})}
\] 
is also the subject minimizing  
\[
\frac{\mathbb{P}(Z_s = i'_2 \mid \mathcal{F})}{\mathbb{P}(Z_s = i_2 \mid \mathcal{F})}, \quad \delta_{i'_2} = 1,\; \delta_{i_2} = 0, \quad i'_2 \neq i', i_2 \neq i,
\]
under \eqref{eq:odds_constraint2}.  
To see this, let \( s_1 \) and \( s_2 \) be two subjects with the same observed covariate value. If \( i' \) and \( i_2' \) are both $1$ and \( i \) and \( i_2 \) are both $0$, then by \eqref{eq:odds_constraint2}, we have:  
\[
\frac{\mathbb{P}(Z_{s_1} = i \mid \mathcal{F})}{\mathbb{P}(Z_{s_1} = i_2 \mid \mathcal{F})} = \frac{\mathbb{P}(Z_{s_2} = i \mid \mathcal{F})}{\mathbb{P}(Z_{s_2} = i_2 \mid \mathcal{F})}, \quad
\frac{\mathbb{P}(Z_{s_1} = i_2' \mid \mathcal{F})}{\mathbb{P}(Z_{s_1} = i' \mid \mathcal{F})} = \frac{\mathbb{P}(Z_{s_2} = i_2' \mid \mathcal{F})}{\mathbb{P}(Z_{s_2} = i' \mid \mathcal{F})}.
\]
Thus, it follows that:  
\begin{align}
\frac{\mathbb{P}(Z_{s_1} = i \mid \mathcal{F})\mathbb{P}(Z_{s_1} = i_2' \mid \mathcal{F})}{\mathbb{P}(Z_{s_1} = i' \mid \mathcal{F})\mathbb{P}(Z_{s_1} = i_2 \mid \mathcal{F})} = \frac{\mathbb{P}(Z_{s_2} = i \mid \mathcal{F})\mathbb{P}(Z_{s_2} = i_2' \mid \mathcal{F})}{\mathbb{P}(Z_{s_2} = i_2 \mid \mathcal{F})\mathbb{P}(Z_{s_2} = i' \mid \mathcal{F})}.
\label{eq:within_the_same_group_odds_ratio}
\end{align}
Thus, if  
\[
\frac{\mathbb{P}(Z_{s_1} = i' \mid \mathcal{F})}{\mathbb{P}(Z_{s_1} = i \mid \mathcal{F})} \geq \frac{\mathbb{P}(Z_{s_2} = i' \mid \mathcal{F})}{\mathbb{P}(Z_{s_2} = i \mid \mathcal{F})},
\]
then by multiplying the left-hand side (right-hand side) of the inequality by the left-hand side (right-hand side) of \eqref{eq:within_the_same_group_odds_ratio}, we obtain:  
\[
\frac{\mathbb{P}(Z_{s_1} = i_2' \mid \mathcal{F})}{\mathbb{P}(Z_{s_1} = i_2 \mid \mathcal{F})} \geq \frac{\mathbb{P}(Z_{s_2} = i_2' \mid \mathcal{F})}{\mathbb{P}(Z_{s_2} = i_2 \mid \mathcal{F})}.
\]
Set \(u_{s'} = 0\) and let:
\begin{align*}
\mathbb{P}(Z_{s'}=i \mid \mathcal{F}) = \frac{\exp(\xi_{i}(x))}{\sum_{i'=1}^{I}\exp(\xi_{i'}(x))}, \;\forall i =1,\dots,I.
\end{align*}
\noindent If \(\gamma > 0\) and there is another subject $s$ with $x_s = x$, set:
\begin{align}
u_s & = \frac{1}{\gamma}(\log(\frac{\mathbb{P}(Z_s=i' \mid \mathcal{F})}{\mathbb{P}(Z_s=i \mid \mathcal{F})})-(\xi_{i'}(x)-\xi_i(x)))\label{eq:set_us_logit_form} \\
& = \frac{1}{\gamma}(\log(\frac{\mathbb{P}(Z_s=i'
 \mid \mathcal{F})\mathbb{P}(Z_{s'}=i \mid \mathcal{F})}{\mathbb{P}(Z_s = i \mid \mathcal{F})\mathbb{P}(Z_{s'}=i' \mid \mathcal{F})})) \label{eq:us_and_log_ratio} 
\end{align}
Note that \eqref{eq:set_us_logit_form} is the same as \eqref{eq:log-odds-ratio_supp}.
Using \eqref{eq:odds_constraint2} again, \eqref{eq:set_us_logit_form} also implies:
\begin{align*}
u_s & = \frac{1}{\gamma}(\log(\frac{\mathbb{P}(Z_s=i'_2
 \mid \mathcal{F})\mathbb{P}(Z_{s'}=i_2 \mid \mathcal{F})}{\mathbb{P}(Z_s = i_2 \mid \mathcal{F})\mathbb{P}(Z_{s'}=i_2' \mid \mathcal{F})})) \\
 & = \frac{1}{\gamma}(\log(\frac{\mathbb{P}(Z_s=i'_2\mid \mathcal{F})}{\mathbb{P}(Z_s = i_2 \mid \mathcal{F})})-(\xi_{i'_2}(x)-\xi_{i_2}(x))),
 \end{align*}
for any $i'_2$ such that $\delta_{i'_2}=1$ and $i_2$ such that $\delta_{i_2}=0$. Thus, this recovers the logit model for all pairs of treatments. 
Using \eqref{eq:odds_constraint1} and \eqref{eq:us_and_log_ratio}, $u_s \leq 1$. 
Furthermore, since 
\[
\frac{\mathbb{P}(Z_{s'}=i' \mid \mathcal{F})}{\mathbb{P}(Z_{s'} = i \mid \mathcal{F})} \leq \frac{\mathbb{P}(Z_{s}=i' \mid \mathcal{F})}{\mathbb{P}(Z_{s}=i \mid \mathcal{F})}
\]
by construction, it follows from \eqref{eq:us_and_log_ratio} that $u_{s} \geq 0$. Therefore, the constraint on $u_s$ in \eqref{eq:log-odds-ratio_supp} holds.
For \(\gamma = 0\), we set \(u_s = 0\) for all \(s\). 
\end{proof}

\medskip
\noindent
The following results, A.3 and A.4, provide further insights on the implications of the sensitivity model.

\subsection{The probability of receiving a treatment $i$ with $\delta_i = 1$ increases with \(u_s\) when \(\gamma > 0\).} \label{app:A3}

\begin{proof}
From the probability expression in \eqref{eq:generic_bias_sensitivity_supp}, if \(\delta_i = 1\):
\[
\mathbb{P}(Z_s = i \mid \mathcal{F}) 
= \frac{\exp\bigl(\xi_i(x_s) + \gamma\,u_s\bigr)}
       {\sum_{i'=1}^I \exp\bigl(\xi_{i'}(x_s) + \gamma\delta_{i'}\,u_s\bigr)}.
\]
Differentiating \(\mathbb{P}(Z_s = i \mid \mathcal{F})\) with respect to \(u_s\) gives
\begin{align*}
\frac{\gamma\,\exp(\xi_i(x_s) + \gamma u_s)
         \Bigl(\sum_{i'=1}^I \exp(\xi_{i'}(x_s) + \gamma\delta_{i'}u_s)(1-\delta_{i'})\Bigr)}
        {\Bigl(\sum_{i'=1}^I \exp(\xi_{i'}(x_s) + \gamma\,\delta_{i'}u_s)\Bigr)^2 }.
\end{align*}
Given $\gamma>0$, the sign depends on 
\[
\sum_{i'=1}^{I}\exp(\xi_{i'}(x_s)+\gamma \delta_{i'}u_s)(1-\delta_{i'}).
\]
Since \(\delta_{i'}\in\{0,1\}\), this expression is nonnegative and strictly positive if at least one treatment level \(i'\) has $\delta_{i'} = 0$. Thus, \(\mathbb{P}(Z_s = i \mid \mathcal{F})\) increases with \(u_s\).
\end{proof}

\subsection{The conditional probability of $\mathbf{{Z}}$ given fixed treatment margins} \label{app:A4}
This proof obtains the conditional probability in the main text under model \eqref{eq:generic_bias_sensitivity_supp} assuming $x_s = x$ for all $s=1,\dots,N$.  
\begin{proof}
We assume that the treatment assignment of each subject is independent before conditioning on $\mathcal{Z}(\mathbf{N}_{I\cdot})$. Therefore, the joint probability can be written as:
\[
\mathbb{P}({\mathbf{Z}} = {\mathbf{z}} \mid \mathcal{F}) = \prod_{s=1}^N \frac{\sum_{i=1}^I \mathbbm{1}\{z_s = i\} \exp(\xi_i(x) + \gamma \delta_i u_s)}{\sum_{i=1}^I \exp(\xi_i(x) + \gamma \delta_i u_s)}.
\]

\noindent Conditioning on the marginal constraints \(\mathbf{N}_{I.}\), we have:
\begin{align*}
\mathbb{P}({\mathbf{Z}} = {\mathbf{z}}\mid {\mathcal{Z}}(\mathbf{N}_{I.}), \mathcal{F}) 
&= \frac{\Pi_{i=1}^I \exp(\xi_i(x))^{N_{i\cdot}}\Bigl(\prod_{s=1}^N \sum_{i=1}^I \mathbbm{1}\{z_s = i\}\exp(\gamma \delta_i u_s)\Bigr)}
{\Pi_{i=1}^I \exp(\xi_i(x))^{N_{i\cdot}}\Bigl(\sum_{{\mathbf{b}} \in {\mathcal{Z}}(\mathbf{N}_{I.})}\prod_{s=1}^N \sum_{i=1}^I \mathbbm{1}\{b_s = i\}\exp(\gamma \delta_i u_s)\Bigr)}\\
&= \frac{\prod_{s=1}^N \sum_{i=1}^I \mathbbm{1}\{z_s = i\}\exp(\gamma \delta_i u_s)}
{\sum_{{\mathbf{b}} \in {\mathcal{Z}}(\mathbf{N}_{I.})}\prod_{s=1}^N \sum_{i=1}^I \mathbbm{1}\{b_s = i\}\exp(\gamma \delta_i u_s)}.
\end{align*}
The last line simplifies to the conditional probability expression in the main text.
\end{proof}

\section{Derivations for Finding the Worst-Case Unmeasured Confounder $\mathbf{u}^+$} \label{app:B}
\label{sec:Appendix_Bound_on_inference}
This appendix presents key technical results concerning the maximizer of the p-value under the sensitivity model. We begin by introducing the binary equivalent representation of treatment and outcome. Sections~\ref{app:B2}--\ref{app:B4} prove Theorem~\mref{thm:set_size_of_U}, which characterizes the set of maximizer candidates for count-based test statistics. Section~\ref{app:B5} establishes Theorem~\mref{thm:score_test_maximizer_nonbinary} for ordinal tests, and Section~\ref{app:B6} proves Theorem~\mref{thm:score_test_maximizer_binary} for the sign-score tests. The appendix concludes with a discussion of implications for multi-table tests and multiple sensitivity models.

\subsection{Binary Representation of Treatment and Outcome} \label{app:B1}

For certain derivations involving permutations, it is more convenient to represent treatment and outcome using binary matrices. Define
\(\widetilde{\mathbf{Z}} \in \{0,1\}^{N \times I}\), where each entry \(\widetilde{Z}_{si}\) indicates whether subject \(s\) received treatment level \(i\). Each row of \(\widetilde{\mathbf{Z}}\) corresponds to the treatment assignment of a subject. Similarly, define \(\trr \in \{0,1\}^{N \times J}\), where each entry \(\tilde{r}_{sj}\) indicates whether subject \(s\) exhibited outcome level \(j\). Given a treatment vector \(\mathbf{Z} = (Z_1, \ldots, Z_N)\) and an outcome vector \(\rr = (r_1, \ldots, r_N)\), the following one-to-one mapping holds:
\begin{equation}
\begin{aligned}
    \widetilde{Z}_{si} &= \mathbbm{1}(Z_s = i),  
    &\quad Z_s &= \sum_{i=1}^I i \cdot \widetilde{Z}_{si}, \\
    \tilde{r}_{sj} &= \mathbbm{1}(r_s = j),  
    &\quad r_s &= \sum_{j=1}^J j \cdot \tilde{r}_{sj}.
\end{aligned}
\label{eq:integer_dummy_mapping}
\end{equation}

By construction in \eqref{eq:integer_dummy_mapping}, each row of \(\widetilde{\mathbf{Z}}\) and \(\tilde{\rr}\) sums to 1. The column sums of \(\widetilde{\mathbf{Z}}\) match the row totals (treatment margins) of the contingency table, while the column sums of $\trr$ match the column totals (outcome margins) of the contingency table.

Recall that under the integer representation, the set of all treatment assignments with fixed margins \(\mathbf{N}_{I\cdot} = (N_{1\cdot}, \ldots, N_{I\cdot})^\intercal\) is defined as:
\begin{equation}
\mathcal{Z}(\mathbf{N}_{I\cdot}) = 
\left\{
\mathbf{z} \in \{1,\ldots,I\}^N 
\;\middle|\; 
\sum_{s=1}^{N} \mathbbm{1}\{z_s = i\} = N_{i\cdot},\; \forall i = 1,\ldots,I 
\right\}.
\label{eq:integer_permutation_treatment_set}
\end{equation}
Under the mapping in \eqref{eq:integer_dummy_mapping}, each \(\mathbf{z} \in \mathcal{Z}(\mathbf{N}_{I\cdot})\) corresponds uniquely to a binary matrix \(\tilde{\mathbf{z}} \in \widetilde{\mathcal{Z}}(\mathbf{N}_{I\cdot})\), where:
\begin{equation}
\widetilde{\mathcal{Z}}(\mathbf{N}_{I\cdot}) = 
\left\{
\tilde{\mathbf{z}} \in \{0,1\}^{N\times I} 
\;\middle|\; 
\tilde{\mathbf{z}}^\intercal \mathbf{1}_{N} = (N_{1\cdot}, \ldots, N_{I\cdot})^\intercal,\;
\sum_{i=1}^{I} \tilde{z}_{si} = 1,\; \forall s = 1,\ldots,N
\right\}.
\label{eq:binary_matrix_permutation_treatment_set}
\end{equation}

\begin{figure}[ht]
    \centering
\includegraphics[width=0.6\textwidth]{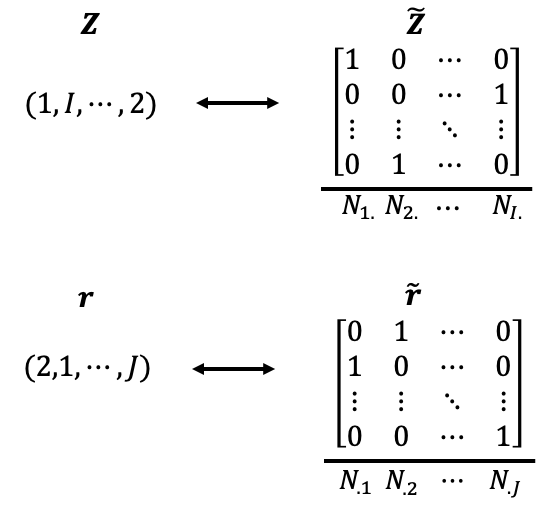}
    \caption{A visualization of the mapping defined in \eqref{eq:integer_dummy_mapping}
    }
\label{fig:integer_dummy_mapping}
\end{figure}

Using these binary matrices, the contingency table \(\mathbf{N}\) can be written as:
\[
\mathbf{N} = \trZ^\intercal \trr,
\quad \text{so that} \quad
N_{ij} = \sum_{s=1}^N \widetilde{Z}_{si}\, \tilde{r}_{sj}.
\]
The row and column margins are:
\[
\mathbf{N}_{I\cdot} = \widetilde{\mathbf{Z}}^\intercal \mathbf{1}_{N},
\qquad
\mathbf{N}_{\cdot J} = {\trr}^\intercal \mathbf{1}_{N},
\]
where \(\mathbf{1}_{N}\) denotes an \(N\)-dimensional column vector of ones. Finally, we will use $\trZ_{i}$ and $\trr_{j}$ to denote the $i$-th column of $\widetilde{\mathbf{Z}}$ and the $j$-th column of $\trr$, respectively. Since $\trr$ is in one-to-one correspondence with $\rr$, we write $\alpha(T,\rr,\mathbf{u})$ as its equivalent expression $\alpha(T, \trr,\mathbf{u})$ whenever the derivation is facilitated by the binary matrix representation. Figure \ref{fig:integer_dummy_mapping} gives a visual example of the one-to-one correspondence between the integer vector and the binary matrix representation.

Before establishing Theorem \mref{thm:set_size_of_U}, we first establish the following two results.

\subsection{Corner Solution} \label{app:B2}
We first show that a worst-case $\mathbf{u}^+$ can always be located at a corner (vertex) of the unit cube, that is, $\mathbf{u}^+ \in \{0,1\}^N$ can be assumed.

\begin{lem}\mbox{}
\label{lem:corner_solution}
\noindent For each \(s\), \(\alpha(T, \trr, \mathbf{u} + \Delta \mathbf{e}_s)\) is monotone in \(\Delta\in \mathbb{R}\), where $\mathbf{e}_s$ is an $N$-dimensional vector with $1$ in the $s$-th entry and $0$ elsewhere.
As a result, a maximizer can be chosen with
\[
\mathbf{u}^+ \in \{0,1\}^N \subset [0,1]^N.
\]
\end{lem}
Lemma~\ref{lem:corner_solution} establishes that the search for the worst-case unmeasured confounder $\mathbf{u}^+$ can be restricted to the $2^N$ corners of the unit cube, i.e., $\mathbf{u} \in \{0,1\}^N$. We note, however, that this property does not hold for every sensitivity model. A counterexample and further discussion of this limitation are provided in Appendix~\ref{app:E}.
\begin{proof}
We can rewrite $\alpha(T,\rr,\mathbf{u})$ into its equivalent form based on B.1:
\[
\alpha(T,\trr,\mathbf{u}) 
\;=\;
\frac{\sum_{\tilde{\mathbf{z}}\in \widetilde{\mathcal{Z}}(\mathbf{N}_{I.})}\mathbbm{1}\bigl\{T(\tilde{\mathbf{z}}, \trr)\geq \criticalt \bigr\}\exp\bigl(\gamma (\tilde{\mathbf{z}}\bm{\delta})^\intercal\mathbf{u}\bigr)}
{\sum_{\tilde{\mathbf{b}} \in \widetilde{\mathcal{Z}}(\mathbf{N}_{I.})}\exp\bigl(\gamma (\tilde{\mathbf{b}}\bm{\delta})^\intercal\mathbf{u}\bigr)}.
\]

Fix a specific index $s$, and write $\widetilde{{\mathcal{Z}}}(\mathbf{N}_{I.}) = \Omega_{1} \cup \Omega_{0}$, where
\[
\Omega_1 = \{\tilde{\mathbf{z}}\in \widetilde{\mathcal{Z}}(\mathbf{N}_{I.}) \mid (\tilde{\mathbf{z}}\bm{\delta})
_{s} = 1\}, 
\]
\[
\Omega_0 = \{\tilde{\mathbf{z}}\in \widetilde{\mathcal{Z}}(\mathbf{N}_{I.}) \mid (\tilde{\mathbf{z}}\bm{\delta})
_{s} = 0\}, 
\]
and $(\tilde{\mathbf{z}}\bm{\delta})_s$ denotes the $s$-th entry of the N-dimensional vector $\tilde{\mathbf{z}} \bm{\delta}$.
Also define
\[
\omega_{\criticalt} =
\{\tilde{\mathbf{z}}\in \widetilde{\mathcal{Z}}(\mathbf{N}_{I.}) 
  \mid T(\tilde{\mathbf{z}}, \trr) \geq \criticalt \}.
\]
Let
\begin{alignat*}{2}
A_0 &\;=\;& 
  \sum_{\tilde{\mathbf{z}} \in \Omega_0 \cap \omega_{\criticalt}} 
    \exp\bigl(\gamma (\tilde{\mathbf{z}}\bm{\delta})^\intercal \mathbf{u}\bigr), 
&\quad
A_1 \;=\; 
  \sum_{\tilde{\mathbf{z}} \in \Omega_1 \cap \omega_{\criticalt}}
    \exp\bigl(\gamma (\tilde{\mathbf{z}}\bm{\delta})^\intercal \mathbf{u}\bigr),
\\[6pt]
D_1 &\;=\;& 
  \sum_{\tilde{\mathbf{z}} \in \Omega_1} 
    \exp\bigl(\gamma (\tilde{\mathbf{z}}\bm{\delta})^\intercal \mathbf{u}\bigr), 
&\quad
D_0 \;=\; 
  \sum_{\tilde{\mathbf{z}} \in \Omega_0}
    \exp\bigl(\gamma (\tilde{\mathbf{z}}\bm{\delta})^\intercal \mathbf{u}\bigr).
\end{alignat*}
Then,
\[
\alpha(T,\trr, \mathbf{u}) 
\;=\; 
\frac{A_0 + A_1}{D_0 + D_1},
\]
which does not involve $\Delta$. Furthermore, 
\[
\alpha(T,\trr,\mathbf{u}+\Delta \mathbf{e}_s) 
\;=\; 
\frac{A_0 + A_1 \exp({\gamma \Delta})}{D_0 + D_1\exp({\gamma \Delta})}.
\]
Taking the first-order derivative with respect to $\Delta$ yields
\begin{align}
\frac{\partial \alpha\bigl(T,\trr, \mathbf{u}+\Delta \mathbf{e}_s\bigr)}{\partial \Delta}
&= 
\frac{\gamma \exp(\gamma \Delta)}{\bigl(D_0 + D_1 \exp(\gamma \Delta)\bigr)^2}
  \bigl(A_1 \,D_0 \;-\; A_0 \,D_1\bigr).
\label{eq:corner_solution_first_order_derivative}
\end{align}
Although the sign of~\eqref{eq:corner_solution_first_order_derivative} cannot be determined because $A_1 D_0 - A_0 D_1$ may be positive or negative, it is independent of $\Delta$. Consequently, $\alpha(T,\trr,\mathbf{u})$ is monotone in~$\Delta$.
\end{proof}

\subsection{Some Corners Yield the Same Null Distribution} \label{app:B3}

Lemma~\ref{lem:corner_solution} restricts the search to the corners of $[0,1]^N$. Lemma~\ref{lem:conditions_corners_giving_the_same_sig} further reduces the search space by showing that any two $\mathbf{u}, \mathbf{u}' \in \{0,1\}^N$ with identical joint counts $\sum_{s=1}^{N} \mathbbm{1}(r_s = j, u_s = 1)$ for all $j$ yield the same null distribution. Therefore, to compute the upper bound on the p-value, it suffices to search $\mathcal{U}_{\rm C}$, which contains the $\mathbf{u}$'s whose vectors $(\sum_{s=1}^{N}\mathbbm{1}\{r_s = 1, u_s=1\}, \ldots, \sum_{s=1}^{N}\mathbbm{1}\{r_s = J, u_s=1\})$ are unique.

\begin{lem}\mbox{}
\label{lem:conditions_corners_giving_the_same_sig}
\noindent For \(\mathbf{u}, \mathbf{u'} \in \{0,1\}^N\), if for all $j = 1,\ldots,J$, 
\[
\sum_{s=1}^{N}\mathbbm{1}\{r_s=j,u_s=1\} = \sum_{s=1}^{N} \mathbbm{1}\{r_s = j, u'_s=1\}, 
\]
then, 
\[\alpha(T, \rr, \mathbf{u}) = \alpha(T, \rr, \mathbf{u'}).
\]
\end{lem}

\subsubsection*{Permutation-Invariance in the Binary Matrix Representation}
Since the proof of Lemma~\ref{lem:conditions_corners_giving_the_same_sig} is clearer when expressed in terms of $\widetilde{\mathbf{Z}}$ and $\trr$, we first extend the notion of permutation invariance to binary matrices, thereby generalizing the definition of permutation-invariant tests.  

Recall that $\mathbf{Z} \in \mathcal{Z}(\mathbf{N}_{I\cdot})$ is bijectively mapped to $\widetilde{\mathbf{Z}} \in \widetilde{\mathcal{Z}}(\mathbf{N}_{I\cdot})$ via \eqref{eq:integer_dummy_mapping}. Consequently, exchanging the entries at positions $a, b$ in $\mathbf{Z}$, where $a < b$,  
\[
\mathbf{Z} = (Z_1,\dots,Z_a,\dots,Z_b,\dots,Z_N) \quad \longrightarrow \quad 
\prescript{}{ab}{\mathbf{Z}} = (Z_1,\dots,Z_b, \dots,Z_a,\dots,Z_N),
\]
is equivalent to exchanging rows $a$ and $b$ in $\widetilde{\mathbf{Z}}$: 
\[
\widetilde{\mathbf{Z}} =
\begin{bmatrix}
\widetilde{\mathbf{Z}}_{1\cdot} \\
\vdots\\
\widetilde{\mathbf{Z}}_{a\cdot} \\
\vdots \\
\widetilde{\mathbf{Z}}_{b\cdot}\\
\vdots\\
\widetilde{\mathbf{Z}}_{N\cdot}
\end{bmatrix}
\quad \longrightarrow \quad 
\prescript{}{ab}{\widetilde{\mathbf{Z}}} =
\begin{bmatrix}
\widetilde{\mathbf{Z}}_{1\cdot} \\
\vdots\\
\widetilde{\mathbf{Z}}_{b\cdot} \\
\vdots \\
\widetilde{\mathbf{Z}}_{a\cdot}\\
\vdots\\
\widetilde{\mathbf{Z}}_{N\cdot}
\end{bmatrix}
\]
Here, $\widetilde{\mathbf{Z}}_{s \cdot}$ denotes the $s$-th row of $\widetilde{\mathbf{Z}}$. 

A similar equivalence holds for $\rr$ and $\trr$. With this, we provide an alternative definition of permutation invariance.
 
\begin{defn}
\label{def:matrix_permutation_invariance}
A function $f$ of two binary matrices $\widetilde{\mathbf{Z}}$ and $\trr$ is \textit{permutation-invariant} if  
\[
f(\widetilde{\mathbf{Z}},\trr) = f(\prescript{}{ab}{\widetilde{\mathbf{Z}}}, \prescript{}{ab}{\trr}), \quad \forall a, b.
\]
Here, $\prescript{}{ab}{\widetilde{\mathbf{Z}}}$ denotes the matrix obtained by exchanging rows $a$ and $b$ in $\widetilde{\mathbf{Z}}$. Each row of $\widetilde{\mathbf{Z}}$ and $\trr$ contains exactly one entry equal to 1.
\label{def:permutation_invariance_binary_matrix}
\end{defn}

For any test statistic $T$ that depends only on the cell counts of a contingency table,
\begin{align*}
T(\widetilde{\mathbf{Z}}^\intercal\trr) &= T(\widetilde{\mathbf{Z}}_1^\intercal\trr_1, \widetilde{\mathbf{Z}}_1^\intercal\trr_2, \ldots, \widetilde{\mathbf{Z}}_I^\intercal\trr_J) \\
&= T\big(\prescript{}{ab}{\widetilde{\mathbf{Z}}}_1^\intercal\, {}_{ab}\trr_1, \ldots, \prescript{}{ab}{\widetilde{\mathbf{Z}}}_I^\intercal\, {}_{ab}\trr_J\big) \\
&= T\big(\prescript{}{ab}{\widetilde{\mathbf{Z}}}^\intercal\, {}_{ab}\trr\big).
\end{align*}
The second equality follows because each $\widetilde{\mathbf{Z}}_i^\intercal\trr_j$ is permutation-invariant under the vector-based definition. Thus, $T$ is permutation-invariant in $\widetilde{\mathbf{Z}}$ and $\trr$ in the sense of Definition \ref{def:permutation_invariance_binary_matrix}. Similarly, $\widetilde{\mathbf{Z}}^\intercal \mathbf{u} = {(\widetilde{\mathbf{Z}}_1^\intercal \mathbf{u}, \ldots, \widetilde{\mathbf{Z}}_I^\intercal \mathbf{u})}^\intercal$ is permutation-invariant in $\widetilde{\mathbf{Z}}$ and $\mathbf{u}$.

\subsubsection*{The Existence of the Permutation Matrix $\mathbf{P}$}
Fix a matrix $\trr \in \{0,1\}^{N\times J}$ in which each row contains exactly one entry equal to $1$, and denote its $(s,j)$-th entry by $\tilde{r}_{sj}$. We now show that for any $\mathbf{u}, \mathbf{u}' \in \{0,1\}^N$ with $\mathbf{u} \neq \mathbf{u}'$, if $\trr^\intercal \mathbf{u} = \trr^\intercal \mathbf{u}'$, then there exists a permutation matrix $\mathbf{P} \in \{0,1\}^{N \times N}$ such that $\mathbf{P}\trr = \trr$ and $\mathbf{P}\mathbf{u} = \mathbf{u}'$. The proof proceeds by directly constructing $\mathbf{P}$ through the following claims.
\vspace{-0.2\baselineskip} 
\paragraph{Claim 1. The sets $\mathcal{S}_{j,\mathbf{u}>\mathbf{u}'}$ and $\mathcal{S}_{j,\mathbf{u}<\mathbf{u}'}$ have equal size.} 
\leavevmode
\par \noindent We define the following sets:  
\begin{align}
\text{for every } j, \quad 
\mathcal{S}_{j, \mathbf{u} > \mathbf{u'}} = \bigg\{s \in \{1, \ldots, N\} \;\bigg|\; \tilde{r}_{sj} (u_s - u_s') > 0\bigg\}, 
\label{def:s_{j,u>u'}}
\end{align}
\begin{align}\text{for every } j, \quad 
\mathcal{S}_{j, \mathbf{u} < \mathbf{u'}} = \bigg\{s \in \{1, \ldots, N\} \;\bigg|\; \tilde{r}_{sj} (u_s - u_s') < 0\bigg\}, 
\label{def:s_{j,u<u'}}
\end{align}
\begin{align}
\mathcal{S}_{\mathbf{u} = \mathbf{u'}} = \bigg\{s \in \{1, \ldots, N\} \;\bigg|\; u_s  = u'_s \bigg\},
\label{def:s_{u=u'}}
\end{align}
where $u_s$ and $u'_s$ denote the $s$-th entry of $\mathbf{u}$ and $\mathbf{u}'$, respectively. In words, $\mathcal{S}_{j,\mathbf{u}> \mathbf{u}'}$ is the set of indices $s$ such that the subject has outcome $j$ and $u_s=1$ and $u'_s=0$, while $\mathcal{S}_{\mathbf{u} = \mathbf{u}'}$ is the set of indices $s$ such that $u_s = u'_s$.

\noindent To show that, for all \(j\), the sets \(\mathcal{S}_{j, \mathbf{u} > \mathbf{u'}}\) and \(\mathcal{S}_{j, \mathbf{u} < \mathbf{u'}}\) have the same size, we recall that, by assumption,  
\[
\sum_{s=1}^N \mathbbm{1}\{r_s = j, u_s = 1\} = \trr_j^\intercal \mathbf{u} = \trr_j^\intercal \mathbf{u}' = \sum_{s=1}^N \mathbbm{1}\{r_s = j, u_s' = 1\}.
\]
\noindent This implies:   
\begin{align*}
0 &= \sum_{s=1}^N \mathbbm{1}\{r_s = j, u_s = 1, u_s' = 1\} 
+ \sum_{s=1}^N \mathbbm{1}\{r_s = j, u_s = 1, u_s' = 0\} \\
&\quad - \sum_{s=1}^N \mathbbm{1}\{r_s = j, u_s = 1, u_s' = 1\} 
- \sum_{s=1}^N \mathbbm{1}\{r_s = j, u_s = 0, u_s' = 1\}\\
\quad \Rightarrow 
0 & = \sum_{s=1}^N \mathbbm{1}\{r_s = j, u_s = 1, u_s' = 0\} 
- \sum_{s=1}^N \mathbbm{1}\{r_s = j, u_s = 0, u_s' = 1\}.
\end{align*}
Here, the first sum equals $|\mathcal{S}_{j,\mathbf{u}>\mathbf{u'}}|$, while the second equals $|\mathcal{S}_{j,\mathbf{u}<\mathbf{u'}}|$, so the two sets have the same size.  

\vspace{-0.2\baselineskip} 
\par \noindent \paragraph{Claim 2. $\mathcal{S}_{j, \mathbf{u} > \mathbf{u'}}, \mathcal{S}_{j, \mathbf{u} < \mathbf{u'}}, \; j = 1, \dots, J, \text{ and } \mathcal{S}_{\mathbf{u} = \mathbf{u'}}$ partition $\{1,\dots,N\}$.}
\leavevmode
\par \noindent We first verify the set identity
\begin{align}
\bigcup_{j=1}^{J} (\mathcal{S}_{j, \mathbf{u} > \mathbf{u'}} \cup \mathcal{S}_{j, \mathbf{u} < \mathbf{u'}}) \cup \mathcal{S}_{\mathbf{u} = \mathbf{u'}} = \{1, 2, \dots, N\}.
\label{eq:S_set_cover}
\end{align}

\noindent An index \(s\) belongs to \(\mathcal{S}_{j, \mathbf{u} > \mathbf{u'}}\) if and only if \(\tilde{r}_{sj} = 1\) and \(u_s > u_s'\). Since for each \( s \) there exists exactly one \( j \in \{1, \dots, J\} \) such that \(\tilde{r}_{sj} = 1\), we obtain:
\[
\bigcup_{j=1}^J \mathcal{S}_{j, \mathbf{u} > \mathbf{u'}} = \mathcal{S}_{\mathbf{u} > \mathbf{u'}}, \quad \text{where } \mathcal{S}_{\mathbf{u} > \mathbf{u'}} = \{s \in \{1,\dots,N\}\mid u_s > u_s'\}.
\]
Similarly, for \(\mathcal{S}_{j, \mathbf{u} < \mathbf{u'}}\):
\[
\bigcup_{j=1}^J \mathcal{S}_{j, \mathbf{u} < \mathbf{u'}} = \mathcal{S}_{\mathbf{u} < \mathbf{u'}}, \quad \text{where } \mathcal{S}_{\mathbf{u} < \mathbf{u'}} = \{s \in \{1,\dots,N\}\mid u_s < u_s'\}.
\]

\noindent Each \(s \in \{1, 2, \dots, N\}\) must satisfy exactly one of the conditions \(u_s > u_s'\), \(u_s = u_s'\), or \(u_s < u_s'\), implying:
\[
\mathcal{S}_{\mathbf{u} = \mathbf{u'}} \cup \mathcal{S}_{\mathbf{u} > \mathbf{u'}} \cup \mathcal{S}_{\mathbf{u} < \mathbf{u'}} = \{1, 2, \dots, N\}.
\]

\noindent Next we show disjointness. Since the conditions $u_s > u'_s, u_s = u'_s, u_s < u'_s$ are mutually exclusive, we obtain $\mathcal{S}_{j, \mathbf{u} > \mathbf{u}'} \cap \mathcal{S}_{j', \mathbf{u} < \mathbf{u}'}  = \emptyset$, $\mathcal{S}_{j, \mathbf{u} > \mathbf{u}'} \cap \mathcal{S}_{\mathbf{u} = \mathbf{u}'}  = \emptyset$, and $\mathcal{S}_{j, \mathbf{u} < \mathbf{u}'} \cap \mathcal{S}_{ \mathbf{u} = \mathbf{u}'}  = \emptyset$, for any $j,j' \in \{1,\dots,J\}$.  Moreover, for distinct \( j, j'' \in \{1, \dots, J\}, j \neq j''\), the condition \(\sum_{j=1}^J \tilde{r}_{sj} = 1\) ensures that no \( s \) can simultaneously satisfy \(\tilde{r}_{sj} = 1\) and \(\tilde{r}_{sj''} = 1\). By the definition of \(\mathcal{S}_{j, \mathbf{u} > \mathbf{u}'}\) and \(\mathcal{S}_{j'', \mathbf{u} > \mathbf{u}'}\), it follows that  
\[
\mathcal{S}_{j, \mathbf{u} > \mathbf{u}'} \cap \mathcal{S}_{j'', \mathbf{u} > \mathbf{u}'} = \emptyset.
\]
\noindent Similarly, 
    \[
    \mathcal{S}_{j, \mathbf{u} < \mathbf{u}'} \cap \mathcal{S}_{j'', \mathbf{u} < \mathbf{u}'} = \emptyset,
    \]
\noindent Therefore, the sets \(\mathcal{S}_{j,\mathbf{u}>\mathbf{u}'}\), \(\mathcal{S}_{j,\mathbf{u} <\mathbf{u}'}\) for all \( j \in \{1,\dots,J\} \), together with \(\mathcal{S}_{\mathbf{u}=\mathbf{u}'}\), are pairwise disjoint. Combined with \eqref{eq:S_set_cover}, these sets form a partition of \(\{1, 2, \dots, N\}\).

\noindent \paragraph{Claim 3. The construction of the permutation matrix}\mbox{} 
\par \noindent We construct \(\mathbf{P}\) as follows:
\begin{itemize}
  \item For each element \(s \in \mathcal{S}_{\mathbf{u} = \mathbf{u}'}\), set the diagonal element \(P_{ss} = 1\). 
    \item For each \(j \in \{1, 2, \dots, J\}\), pair elements from \(\mathcal{S}_{j, \mathbf{u} > \mathbf{u}'}\) with those of \(\mathcal{S}_{j, \mathbf{u} < \mathbf{u}'}\) iteratively:
    \begin{enumerate}
        \item While both sets \(\mathcal{S}_{j, \mathbf{u} > \mathbf{u}'}\) and \(\mathcal{S}_{j, \mathbf{u} < \mathbf{u}'}\) are non-empty:
        \begin{enumerate}
            \item Select \(s\) from \(\mathcal{S}_{j, \mathbf{u} > \mathbf{u}'}\) and \(s'\) from \(\mathcal{S}_{j, \mathbf{u} < \mathbf{u}'}\).
            \item Set \(P_{ss'} = P_{s's} = 1\).
            \item Remove \(s\) and \(s'\) from their respective sets.
        \end{enumerate}
        \item Repeat until both sets \(\mathcal{S}_{j, \mathbf{u} > \mathbf{u}'}\) and \(\mathcal{S}_{j, \mathbf{u} < \mathbf{u}'}\) are empty. By Claim 1, \(|\mathcal{S}_{j,\mathbf{u} > \mathbf{u}'}| = |\mathcal{S}_{j,\mathbf{u} < \mathbf{u}'}|\) for all \(j\), so the process terminates simultaneously for both sets.
    \end{enumerate}
     \item Set all remaining entries of $\mathbf{P}$ to $0$.
\end{itemize}
By Claim 2, the sets $\mathcal{S}_{j,\mathbf{u}>\mathbf{u}'}$, $\mathcal{S}_{j,\mathbf{u}<\mathbf{u}'}$ for all $j$, and $\mathcal{S}_{\mathbf{u} = \mathbf{u}'}$ partition $\{1,\dots,N\}$. Hence, under this construction, every row and column of $\mathbf{P}$ contains exactly one nonzero entry: 
\[
\sum_{s'=1}^{N}P_{ss'} = 1,  \quad \sum_{s'=1}^{N}P_{s's} = 1, \quad \forall s \in \{1,\ldots,N\},
\]
so $\mathbf{P}$ is a permutation matrix. 
\noindent \paragraph{Claim 4. $\mathbf{P}\mathbf{u} = \mathbf{u}'$}\mbox{}   
\par \noindent To show this, we prove that for all $s \in \{1,2,\dots,N\}$, 
\[
\sum_{s'=1}^{N} P_{ss'} u_{s'} = u'_s
\]
holds.
Since the sets $\mathcal{S}_{j,\mathbf{u}>\mathbf{u}'}$, $ \mathcal{S}_{j,\mathbf{u}<\mathbf{u}'}$ for all j and $\mathcal{S}_{\mathbf{u} = \mathbf{u}'}$ partition $\{1,2,\dots,N\}$, it suffices to check the following three cases.
\par \noindent \textbf{Case 1. $s \in \mathcal{S}_{\mathbf{u}=\mathbf{u}'}$}

By construction, $P_{ss}=1$ and all other entries in row $s$ are zero. Hence
\[
\sum_{s'=1}^N P_{ss'} u_{s'} = P_{ss} u_s = u_s = u'_s, 
\]
where the third equality follows from the definition of $\mathcal{S}_{\mathbf{u} = \mathbf{u}'}$.
\par \noindent \textbf{Case 2. $s \in \mathcal{S}_{j,\mathbf{u}>\mathbf{u}'}$ for some $j$} 

By construction of $\mathbf{P}$, there exists a \( s'' \in \mathcal{S}_{j, \mathbf{u} < \mathbf{u}'} \) such that \( P_{ss''} = P_{s''s} = 1 \) and \( P_{ss'} = 0 \) for \( s' \neq s'' \). Hence
\[
\sum_{s'=1}^N P_{ss'} u_{s'} = P_{ss''} u_{s''} = u_{s''} = 0 = u'_s,
\]
where the equalities $u_{s''} = 0$ and $u'_{s}$ follow respectively from the definitions of $\mathcal{S}_{j,\mathbf{u}< \mathbf{u}'}$ and $\mathcal{S}_{j,\mathbf{u}> \mathbf{u}'}$.
\par \noindent \textbf{Case 3. $s \in \mathcal{S}_{j,\mathbf{u}<\mathbf{u}'}$ for some $j$}

Similar to Case 2, there exists \( s'' \in \mathcal{S}_{j, \mathbf{u} > \mathbf{u}'} \) such that \( P_{ss''} = P_{s''s} = 1 \) and $P_{ss'} = 0$ for all $s' \neq s''$. Hence:
\[
\sum_{s'=1}^N P_{ss'} u_{s'} = P_{ss''} u_{s''} = u_{s''} = 1 = u'_s,
\]
where \( u_{s''} = 1 \) follows from the definition of \(\mathcal{S}_{j, \mathbf{u} > \mathbf{u}'} \), and \( u'_s = 1 \) follows from the definition of \(\mathcal{S}_{j, \mathbf{u} < \mathbf{u'}} \).

\par \noindent \paragraph{Claim 5. $\mathbf{P}\trr =\trr$}\mbox{}  
\par \noindent We claim that for all $s \in \{1, 2, \dots, N\}$ and for any $j \in \{1, 2, \dots, J\}$, the following holds:
\begin{align}
\sum_{s'=1}^{N}P_{ss'} \tilde{r}_{s'j} = \tilde{r}_{sj}.
\label{eq:PR=R}
\end{align}
\par \noindent Without loss of generality, we prove \eqref{eq:PR=R} for a fixed \( j \).
\par \noindent \textbf{Case 1. $s \in \mathcal{S}_{\mathbf{u} = \mathbf{u}'}$}
\[
\sum_{s'=1}^N P_{ss'} \tilde{r}_{s'j} = P_{ss} \tilde{r}_{sj} = \tilde{r}_{sj}.
\]
The first equality follows from the construction of $\mathbf{P}$, where $P_{ss} = 1$ and $P_{ss'} = 0$ for all $s' \neq s$ whenever $s \in \mathcal{S}_{\mathbf{u} = \mathbf{u}'}$.

\par \noindent If $s \notin \mathcal{S}_{ \mathbf{u} =  \mathbf{u}'}$, then by construction there exists $s'' \neq s$ such that $P_{ss''} = 1$, and $P_{ss'} = 0$ for all $s' \neq s''$. In this case,
\begin{align}
\sum_{s'=1}^N P_{ss'} \tilde{r}_{s'j} = P_{ss''} \tilde{r}_{s''j} = \tilde{r}_{s''j}. 
\label{eq:PR_simplified}
\end{align}
\eqref{eq:PR_simplified} shows that to establish \eqref{eq:PR=R} for $s \notin \mathcal{S}_{\mathbf{u}=\mathbf{u}'}$, it suffices to prove that $\tilde{r}_{s''j} = \tilde{r}_{sj}$. We now examine this condition in the remaining cases.
\par \noindent \textbf{Case 2. $s\in \mathcal{S}_{j,\mathbf{u}>\mathbf{u}'}$}  

By the construction of $\mathbf{P}$, there exists a unique index $s'' \in \mathcal{S}_{j,\mathbf{u}<\mathbf{u}'}$ such that $P_{ss''}=1$. By the definitions of $\mathcal{S}_{j,\mathbf{u}>\mathbf{u}'}$ and $\mathcal{S}_{j,\mathbf{u}<\mathbf{u}'}$, $\tilde{r}_{sj} = 1$ and $\tilde{r}_{s''j} = 1$. Thus, $\tilde{r}_{sj}=\tilde{r}_{s''j}$ as desired.
\par \noindent \textbf{Case 3. $s \in \mathcal{S}_{j,\mathbf{u}<\mathbf{u'}}$}

Similar to Case 2, by the construction of $\mathbf{P}$, there exists a unique index $s'' \in \mathcal{S}_{j, \mathbf{u}>\mathbf{u}'}$ such that $P_{ss''}=1$. By the definitions of $\mathcal{S}_{j,\mathbf{u}>\mathbf{u}'}$ and $\mathcal{S}_{j,\mathbf{u}<\mathbf{u}'}$, $\tilde{r}_{s''j} = 1 = \tilde{r}_{sj}$. Thus, $\tilde{r}_{sj}=\tilde{r}_{s''j}$ as desired. 

\par \noindent \textbf{Case 4. $s \in \mathcal{S}_{j', \mathbf{u}>\mathbf{u'}}, \; j' \neq j$}  

By the construction of $\mathbf{P}$, there exists a unique index $s'' \in \mathcal{S}_{j',\mathbf{u}<\mathbf{u'}}$ such that $P_{ss''}=1$. By the definitions in \eqref{def:s_{j,u<u'}}, \eqref{def:s_{j,u>u'}}, we have $\tilde{r}_{sj'} = 1$ and $\tilde{r}_{s''j'} = 1$. Since $\sum_{j=1}^{J} \tilde{r}_{sj} = 1$ for every $s$, it follows that $\tilde{r}_{sj} = 0$ and $\tilde{r}_{s''j} = 0$. Thus, $\tilde{r}_{sj} = \tilde{r}_{s''j}$, as desired. 
\par \noindent \textbf{Case 5. $s\in \mathcal{S}_{j',\mathbf{u}<\mathbf{u'}},\; j \neq j'$}

Similar to Case 4, we have $\tilde{r}_{sj} = 0 = \tilde{r}_{s''j}$. 

\medskip

\par Thus far, we have extended permutation-invariance to binary matrices and showed that both \( T(\trZ, \trr) \) and \( \trZ^\intercal \mathbf{u} \) are permutation-invariant. We have also constructed a permutation matrix \( \mathbf{P} \) such that \( \mathbf{P}\mathbf{u} = \mathbf{u}' \) and \( \mathbf{P}\trr = \trr \), assuming \( \trr^\intercal \mathbf{u} = \trr^\intercal \mathbf{u}' \). We now show these conditions ensure $\alpha(T,\trr,\mathbf{u}) = \alpha(T,\trr,\mathbf{u}')$. 

\par \noindent \paragraph{Claim 6. For every $\tilde{\mathbf{z}} \in \widetilde{\mathcal{Z}}(\mathbf{N}_{I\cdot})$, there exists a unique $\tilde{\mathbf{z}}' \in \widetilde{\mathcal{Z}}(\mathbf{N}_{I\cdot})$ such that $\mathbf{P} \tilde{\mathbf{z}} = \tilde{\mathbf{z}}'$, where $\mathbf{P}$ is constructed in Claim~3.}\mbox{}
\par \noindent First, since a permutation matrix $\mathbf{P}$ is invertible, the mapping $\tilde{\mathbf{z}} \mapsto \mathbf{P} \tilde{\mathbf{z}}$ is one-to-one, ensuring the uniqueness of $\tilde{\mathbf{z}}'$ satisfying $\mathbf{P} \tilde{\mathbf{z}} = \tilde{\mathbf{z}}'$. Second, permuting the entries of an all-ones vector leaves it unchanged:
\[
\mathbf{P}\mathbf{1}_{N} = \bm{1}_{N} \Rightarrow (\mathbf{P}\bm{1}_{N}
)^\intercal = \bm{1}_{N}^\intercal \Rightarrow\bm{1}_{N}^\intercal \mathbf{P}^\intercal = \bm{1}_{N}^\intercal. 
\]
Since a permutation matrix satisfies
${\mathbf{P}}^\intercal =  \mathbf{P}^{-1}$, it follows that
\begin{align}
\bm{1}_{N}^\intercal \mathbf{P}^\intercal = \bm{1}_{N}^\intercal \Rightarrow \bm{1}_{N}^\intercal \mathbf{P}^{-1} = \bm{1}_{N}^\intercal \ \Rightarrow \bm{1}_{N}^\intercal = \bm{1}_{N}^\intercal \mathbf{P}.
\label{eq:1_NT = 1_NP}
\end{align}
Now let $\mathbf{P} \tilde{\mathbf{z}} = \tilde{\mathbf{z}}'$. The $i$-th column sum of $\tilde{\mathbf{z}}'$ satisfies 
\begin{align}
\bm{1}_{N}^\intercal \tilde{\mathbf{z}}_i' = \bm{1}_{N}^\intercal (\mathbf{P}\tilde{\mathbf{z}}_i) = (\bm{1}_N^\intercal \mathbf{P})\tilde{\mathbf{z}}_i = \bm{1}_{N}^\intercal \tilde{\mathbf{z}}_i,
\label{eq:column_sum_the_same_after_permutation}
\end{align}  
where the last equality follows from \eqref{eq:1_NT = 1_NP}. Equation~\eqref{eq:column_sum_the_same_after_permutation} shows that column sums are preserved under multiplication by $\mathbf{P}$. Hence, if $\tilde{\mathbf{z}} \in \widetilde{\mathcal{Z}}(\mathbf{N}_{I\cdot})$, then $\tilde{\mathbf{z}}' = \mathbf{P}\tilde{\mathbf{z}} \in \widetilde{\mathcal{Z}}(\mathbf{N}_{I\cdot})$.

\par \noindent \paragraph{Claim 7. $\alpha(T,\trr,\mathbf{u}) = \alpha(T,\trr,\mathbf{u}')$, given $\trr^\intercal \mathbf{u} = \trr^\intercal \mathbf{u}'$}\mbox{}  
\par \noindent Take the $\mathbf{P}$ from Claim 3 and compute:
\begin{equation}
\begin{aligned}
\alpha(T,\trr, \mathbf{u}) 
&= \frac{\sum_{\mathbf{\tilde{z}} \in \widetilde{\mathcal{Z}}(\mathbf{N}_{I.})} \mathbbm{1}\big\{T(\tilde{\mathbf{z}}, \trr) \geq \criticalt\big\}\exp\big\{\gamma \bm{\delta}^\intercal \tilde{\mathbf{z}}^\intercal \mathbf{u}\big\}}
{\sum_{\tilde{\mathbf{b}} \in \widetilde{\mathcal{Z}}(\mathbf{N}_{I.})} \exp\big\{\gamma \bm{\delta}^\intercal \tilde{\mathbf{b}}^\intercal \mathbf{u}\big\}} \\[6pt]
&= \frac{\sum_{\tilde{\mathbf{z}} \in \widetilde{\mathcal{Z}}(\mathbf{N}_{I.})} \mathbbm{1}\big\{T(\mathbf{P}\tilde{\mathbf{z}},\mathbf{P}\trr) \geq \criticalt\big\}\exp\big\{\gamma \bm{\delta}^\intercal (\mathbf{P}\tilde{\mathbf{z}})^\intercal (\mathbf{P}\mathbf{u})\big\}}
{\sum_{\tilde{\mathbf{b}} \in \widetilde{\mathcal{Z}}(\mathbf{N}_{I.})} \exp\big\{\gamma \bm{\delta}^\intercal (\mathbf{P}\tilde{\mathbf{b}})^\intercal (\mathbf{P}\mathbf{u})\big\}} \\[6pt]
& = \frac{\sum_{\tilde{\mathbf{z}} \in\widetilde{\mathcal{Z}}(\mathbf{N}_{I.})} \mathbbm{1}\big\{T(\mathbf{P}\tilde{\mathbf{z}},\trr) \geq \criticalt\big\}\exp\big\{\gamma \bm{\delta}^\intercal ( \mathbf{P}\tilde{\mathbf{z}})^\intercal \mathbf{u}'\big\}}
{\sum_{\tilde{\mathbf{b}} \in \widetilde{\mathcal{Z}}(\mathbf{N}_{I.})} \exp\big\{\gamma \bm{\delta}^\intercal (\mathbf{P}\tilde{\mathbf{b}})^\intercal \mathbf{u}'\big\}}.
\label{eq:significance_level_permute_u}
\end{aligned}
\end{equation}

\par \noindent The second equality uses the permutation-invariance of $T(\tilde{\mathbf{z}}, \trr)$ and $\tilde{\mathbf{z}}^\intercal \mathbf{u}$. The third equality follows from $\mathbf{P} \mathbf{u} = \mathbf{u}'$ and $\mathbf{P} \trr = \trr$.  
By Claim 6, $\mathbf{P}$ defines a bijection on $\widetilde{\mathcal{Z}}(\mathbf{N}_{I\cdot})$, so summing over terms evaluated at $\mathbf{P} \tilde{\mathbf{z}}$ for all $\tilde{\mathbf{z}} \in \widetilde{\mathcal{Z}}(\mathbf{N}_{I\cdot})$ is equivalent to summing over terms evaluated at $\tilde{\mathbf{z}}$.  
Thus, equation~\eqref{eq:significance_level_permute_u} simplifies to:  
\[
\frac{\sum_{\tilde{\mathbf{z}} \in \widetilde{\mathcal{Z}}(\mathbf{N}_{I.})} \mathbbm{1}\big\{T(\tilde{\mathbf{z}}, \trr) \geq \criticalt\big\}\exp\big\{\gamma \bm{\delta}^\intercal \tilde{\mathbf{z}}^\intercal \mathbf{u}'\big\}}{\sum_{\tilde{\mathbf{b}} \in \widetilde{\mathcal{Z}}(\mathbf{N}_{I.})} \exp\big\{\gamma \bm{\delta}^\intercal \tilde{\mathbf{b}}^\intercal \mathbf{u'}\big\}}
= \alpha(T,\trr, \mathbf{u}'),
\]
which establishes the claim, since $\trr^\intercal\mathbf{u} = \trr^\intercal \mathbf{u}'$ is equivalent to 
\[
\sum_{s=1}^{N} \mathbbm{1}\{r_s=j,u_s=1\} = \sum_{s=1}^{N}\mathbbm{1}\{r_s=j,u'_s=1\}, \quad \forall j = 1,\ldots,J.
\]

\subsection{The Size of $\mathcal{U}_{\rm C}$} \label{app:B4}
\begin{proof}
We recall the definition of $\mathcal{U}_{\mathrm{C}}$. For $\mathbf{u}\in\{0,1\}^N$, define
\[
\overline{u}_j \;=\; \sum_{s=1}^{N}\mathbbm{1}\{r_s=j,\; u_s=1\},
\qquad
\overline{\mathbf{u}} \;=\; (\overline{u}_1,\ldots,\overline{u}_J)^{\intercal}.
\]
For the set of binary vectors $\{0,1\}^N$, we define the equivalence relation as $\mathbf{u}\sim\mathbf{u}'$ iff $\overline{\mathbf{u}}=\overline{\mathbf{u}'}$, and define the set of equivalence classes as $\mathcal{U}_{\mathrm{C}} \;=\; \{0,1\}^N/\!\sim$. Thus, to determine the size of $\mathcal{U}_{\mathrm{C}}$, it suffices to count the possible realizations of $\overline{\mathbf{u}}$ given the fixed vector $\mathbf{r}$ (equivalently, a fixed binary matrix $\trr$).
Let
\[
\mathcal{H}
\;=\;
\Bigl\{\mathbf{h}=(h_1,\ldots,h_J)^{\intercal}\,:\,
h_j\in\mathbb{Z}_{\ge 0},\; 0\le h_j\le N_{.j}\ \text{for } j=1,\ldots,J\Bigr\},
\]
and define the map
\[
f:\ \mathcal{U}_{\mathrm{C}}\longrightarrow \mathcal{H},
\qquad
f([\mathbf{u}]) \;=\; \trr^{\intercal}\mathbf{u},
\]
where $[\mathbf{u}]$ denotes the equivalence class of $\mathbf{u}$ and \(\trr\) is the \(N\times J\) binary representation of \(\mathbf{r}\) with entries \(\tilde r_{sj}:=\mathbbm{1}\{r_s=j\}\), so that
\[
(\trr^{\intercal}\mathbf{u})_j=\sum_{s=1}^N \tilde r_{sj}u_s.
\]
\emph{Well-definedness.} If $\mathbf{u}\sim\mathbf{u}'$, then $\overline{\mathbf{u}}=\overline{\mathbf{u}}'$, i.e., $\trr^{\intercal}\mathbf{u}=\trr^{\intercal}\mathbf{u}'$, so $f([\mathbf{u}])$ is independent of the representative.\\
\emph{Injectivity.} If $f([\mathbf{u}_1])=f([\mathbf{u}_2])$, then $\trr^{\intercal}\mathbf{u}_1=\trr^{\intercal}\mathbf{u}_2$, hence $\overline{\mathbf{u}}_1=\overline{\mathbf{u}}_2$ and $[\mathbf{u}_1]=[\mathbf{u}_2]$.\\
\emph{Image in $\mathcal{H}$}. For each $j$, $(\trr^{\intercal}\mathbf{u})_j=\sum_{s=1}^N \tilde r_{sj}u_s$ is a nonnegative integer bounded by the count $N_{.j}$ of indices with $r_s=j$, so $f([\mathbf{u}])\in\mathcal{H}$.\\
Since $f$ is well-defined and injective with $f(\mathcal{U}_{\mathrm{C}})\subseteq\mathcal{H}$, we have
\[
|\mathcal{U}_{\mathrm{C}}|
= |f(\mathcal{U}_{\mathrm{C}})|
\le |\mathcal{H}|.
\]
\noindent As we have established that the $|\mathcal{U}_{\rm{C}}| \leq |\mathcal{H}|$, we calculate: 
\[
|\mathcal{H}| = \prod_{j=1}^{J}(N_{.j}+1).
\]
Furthermore, assume the nontrivial case where we have at least two outcome levels, $J\ge2$, $N\ge2$. Then, since $\sum_{j=1}^J N_{.j}=N$, the AM–GM inequality gives us
\[
|\mathcal{U}_{\mathrm{C}}|\le |\mathcal H|
= \prod_{j=1}^J (N_{.j}+1)
\;\le\; \Big(\tfrac{N}{J}+1\Big)^{\!J}
\;\le\; N^{J}.
\]
\end{proof}

Sections~\ref{app:B1}--\ref{app:B4} characterize the candidate set $\mathcal{U}_{\rm C}$ that contains the maximizer $\mathbf{u}^+$ for general permutation-invariant tests and establish an upper bound on its size. Section~\ref{app:B5} then turns to the candidate set $\mathcal{U}_{\rm{o}}$ that contains the maximizer $\mathbf{u}^+$ for the ordinal test.

\subsection{Proof of Theorem~\mref{thm:score_test_maximizer_nonbinary}: The Candidate Set $\mathcal{U}_{\rm O}$ for Ordinal Tests} \label{app:B5}
We first review the definition of an arrangement-increasing function. 
\begin{defn}
A permutation-invariant function 
$T(\bm{X},\bm{Y})$ is arrangement-increasing (AI), also known as decreasing in transportation (DT), if
\[
T(\bm{X}, \prescript{}{ab}{\bm{Y}}) \;\geq\; T(\bm{X}, \bm{Y})
\quad\text{whenever}\quad
(X_a - X_b)\,(Y_a - Y_b) \;\leq\; 0,
\]
where $\prescript{}{ab}{\bm{Y}}$ denotes the vector obtained by exchanging the $a$-th and $b$-th entries of $\bm{Y}$. See \citet[Section 2.3, Definition 3]{rosenbaum2002observational} and \citet[Section 2]{hollander_proschan_sethuraman_1977}.
\end{defn}
We first show that
\[
T(\mathbf{Z},\rr) 
=  \sum_{i=1}^{I}\sum_{j=1}^{J} w_i v_j \sum_{s=1}^{N}\mathbbm{1}(Z_s = i, r_s = j)
\]
is AI under the ordinal test assumptions $w_1 \leq w_2 \leq \cdots \leq w_I$ and $v_1 \leq v_2 \leq \cdots \leq v_J$.

We therefore check that
\[
T\bigl(\prescript{}{ab}{\mathbf{Z}}, \rr\bigr) 
\;\ge\;
T(\mathbf{Z}, \rr)
\quad
\text{whenever}
\quad
(Z_a - Z_b)\,(r_a - r_b) \;\le\; 0.
\]
Let \(i_a = Z_a\), \(i_b = Z_b\), \(j_a = r_a\), and \(j_b = r_b\) be the realized treatment and outcome levels for subjects \(a\) and \(b\), respectively. Expanding the difference yields
\begin{align*}
T\bigl(\prescript{}{ab}{\mathbf{Z}}, \rr\bigr) 
  \;-\; T(\mathbf{Z}, \rr)
&=\;
w_{i_b}v_{j_a} \;+\; w_{i_a}v_{j_b}
\;-\; w_{i_a}v_{j_a} \;-\; w_{i_b}v_{j_b}
\\
&=\;
-\,(w_{i_a} - w_{i_b})\,(v_{j_a} - v_{j_b})
\\
&\ge 0,
\end{align*}
where the final inequality follows because, due to the ordinal test assumptions, 
\((w_{i_a} - w_{i_b})(v_{j_a} - v_{j_b})\) has the same sign as 
\((i_a - i_b)(j_a - j_b)\), and is therefore non-positive under the assumption \((Z_a - Z_b)(r_a - r_b) \le 0\).

Define the function:
\begin{align}
f_{\boldsymbol{\delta}} \colon \{1,\ldots,I\}^N \to \{\delta_1,\ldots,\delta_I\}^N,\quad
f_{\boldsymbol{\delta}}(\mathbf{z}) = (\delta_{z_1},\,\delta_{z_2},\,\dots,\,\delta_{z_N}).
\label{def:f_delta}
\end{align}
Then:
\[
\exp\!\bigl(\gamma \cdot (f_{\boldsymbol{\delta}}(\mathbf{z}))^\intercal \mathbf{u}\bigr)
\]
is AI in $\mathbf{z}$ and $\mathbf{u}$ whenever $\gamma > 0$ and 
$\delta_1 \leq \delta_2 \leq \cdots \leq \delta_I$.

\noindent Under the generic bias sensitivity model, the null distribution can be written as:
\begin{align}
\frac{\sum_{\mathbf{z}\in \mathcal{Z}(\mathbf{N}_{I.})}\mathbbm{1}\big\{T(\mathbf{z}, \rr)\geq \criticalt\big\}\exp\big\{\gamma (f_{\bm{\delta}}(\mathbf{z}))^\intercal \mathbf{u}\big\}}{\sum_{\mathbf{b} \in \mathcal{Z}(\mathbf{N}_{I.})}\exp\big\{\gamma \,(f_{\bm{\delta}}(\mathbf{b}))^\intercal \mathbf{u}\big\}}
\label{eq:significance_level_vector_form}
\end{align}
Since both $\exp(\gamma f_{\boldsymbol{\delta}}(\mathbf{z})^\intercal \mathbf{u})$ and 
$\mathbbm{1}\{T(\mathbf{z},\rr) \ge \criticalt\}$ are AI, 
\citet[Theorem~3.3]{hollander_proschan_sethuraman_1977} implies that 
\eqref{eq:significance_level_vector_form} is AI in $\mathbf{u}$ and $\rr$.
By Lemma~\ref{lem:corner_solution}, $\mathbf{u}^+ \in \{0,1\}^N$, and since 
\eqref{eq:significance_level_vector_form} is AI in $\mathbf{u}$ and $\rr$, a maximizer $\mathbf{u}^+$ must follow the same order as $\rr$.  
Thus, if $r_1 \leq r_2 \leq \cdots \leq r_N$, then
\[
\mathbf{u}^+ \in \mathcal{U}_{\rm{o}} 
\quad\text{where}\quad
\mathcal{U}_{\rm{o}} = \{\mathbf{u}\in \{0,1\}^N : u_1 \leq u_2 \leq \cdots \leq u_N\}.
\]

\subsection{Proof of Theorem~\mref{thm:score_test_maximizer_binary}: $\mathbf{u}^+$ for Sign-Score Tests.} \label{app:B6}
This proof proceeds by (i) defining a distributive lattice and (ii) establishing the premise of Holley's inequality. Together, these steps identify a maximizer of the p-value for sign-score tests under the sensitivity model~\eqref{eq:generic_bias_sensitivity_supp}.

\subsubsection*{Step 1. Defining the distributive lattice}
\label{subsec:def_lattice}
\noindent We first define a partially-ordered set of treatment assignments, similar to that in \citet[Theorem 1]{zhang2024sensitivity}. Let \(D^1\) and \(D^2\) be vectors of the treatment assigned to subjects with outcomes \(1\) and \(2\), respectively, arranged in ascending order. Define \(D = (D^1, D^2) =(D_1^1,\dots,D_{N_{.1}}^1,D_1^2,\dots,D_{N_{.2}}^2)\) and let:
\[
\Omega = \{D = (D^1, D^2) : D^1_1 \leq D^1_2 \leq \dots \leq D^1_{N_{.1}}; \; D^2_1 \leq D^2_2 \leq \dots \leq D^2_{N_{.2}}\},
\]
subject to the treatment margin constraints:
\[
|\{s : D^1_s = i\}| \leq N_{i.}, \quad |\{s : D^2_s = i\}| \leq N_{i.}, \quad \forall i,
\]
\[
|\{s : D^1_s = i\}| + |\{s : D^2_s = i\}| = N_{i.}, \quad \forall i.
\]

\noindent Note that \(\Omega\) has a smaller cardinality than \(\mathcal{Z}(\mathbf{N}_{I.})\): \(\mathcal{Z}(\mathbf{N}_{I.})\) distinguishes treatment assignments by permutations within subjects with outcomes 1 and 2, whereas \(\Omega\) retains only treatment assignments in ascending order.

\subsubsection*{Step 2. Establishing the lattice properties}
\label{subsec:lattice_properties}
\noindent Since \(D^1\) is uniquely determined by \(D^2\) because of its ascending order and the fixed treatment margin constraints, we define the partial order \(\preceq\) on \(\Omega\) as follows:
\[
D \preceq D^* \quad \text{if and only if} \quad D^2_s \leq D^{*2}_s \quad \forall  1 \leq s \leq N_{.2}.
\]
The join and meet of \(D, D^* \in \Omega\) are defined as:
\[
D \vee D^* = ((D \vee D^*)^1, (D \vee D^*)^2), \quad D \wedge D^* = ((D \wedge D^*)^1, (D \wedge D^*)^2),
\]
where the entries of the join and meet of $D$ and $D^*$ are defined by the entry-wise maximum and minimum:
\[
(D \vee D^*)^2_s = \max(D^2_s, D^{*2}_s), \quad (D \wedge D^*)^2_s = \min(D^2_s, D^{*2}_s), \quad \forall 1 \leq s  \leq N_{.2}.
\]

\noindent The conditions for a lattice are given in \citet[Chapter 1, Theorem 8]{Birkhoff1940}. It can be verified that \((\Omega, \vee, \wedge)\) satisfies:

\begin{itemize}
    \item \( D \vee D = D \), \( D \wedge D = D \)
    \item \( D \vee D^* = D^* \vee D \), \( D \wedge D^* = D^* \wedge D \)
    \item \( D \vee (D^* \vee D^{**}) = (D \vee D^*) \vee D^{**} \), \( D \wedge (D^* \wedge D^{**}) = (D \wedge D^*) \wedge D^{**} \)
    \item \( D \wedge (D \vee D^*) = D \), \quad \( D \vee (D \wedge D^*) = D \)
\end{itemize}

Thus, \((\Omega, \vee, \wedge)\) forms a lattice.

\subsubsection*{Step 3. Verifying distributivity}
\label{subsec:distributivity}
\noindent Next, we establish the distributivity of \((\Omega, \vee, \wedge)\). We verify that:
\[
D \vee (D^{*} \wedge D^{**}) = (D \vee D^{*}) \wedge (D \vee D^{**})
\]
for all \( D, D^*, D^{**} \in \Omega \). Based on our definitions, this is equivalent to showing that for all \( 1 \leq s \leq N_{.2} \),
\[
\max(D^2_s, \min(D^{*2}_s, D^{**2}_s)) = \min(\max(D^2_s, D^{*2}_s), \max(D^2_s, D^{**2}_s)).
\]
This verification follows from \citet[Page 60]{rosenbaum2002observational} and can be discussed through the following cases:

\begin{itemize}
    \item[(i)] If \( D^2_s \geq \min(D^{*2}_s, D^{**2}_s) \) and \( D^{*2}_s \leq D^2_s \leq D^{**2}_s \), both sides equal \( D^2_s \).
    \item[(ii)] If \( D^2_s \geq \min(D^{*2}_s, D^{**2}_s) \) and \( D^{**2}_s \leq D^2_s \leq D^{*2}_s \), both sides equal \( D^2_s \).
    \item[(iii)] If \( D^2_s < \min(D^{*2}_s, D^{**2}_s) \) and \( D^{*2}_s \leq D^{**2}_s \), both sides equal \( D^{*2}_s \).
    \item[(iv)] If \( D^2_s < \min(D^{*2}_s, D^{**2}_s) \) and \( D^{*2}_s > D^{**2}_s \), both sides equal \( D^{**2}_s \).
\end{itemize}
\noindent Thus, \((\Omega, \vee, \wedge)\) is distributive.

\subsubsection*{Step 4. Establishing isotonicity}
\label{subsec:isotonicity}
\noindent Now we recall the definition of an isotonic function.

\begin{defn}
A real-valued function \( f: \Omega \to \mathbb{R} \) on the distributive lattice \(\Omega\) is called isotonic if, for all \( d, d^* \in \Omega \), whenever \( d \preceq d^* \), it follows that \( f(d) \leq f(d^*) \).
\end{defn}

\par \noindent Rewriting the sign-score test:
\[
\sum_{i=1}^{I} w_i N_{i2} = \sum_{i=1}^{I}\sum_{s=1}^{N} w_i \mathbbm{1}(Z_s = i, r_s = 2).
\]
\par \noindent By our definition of \(\preceq\) on \(\Omega\), and since \(\rr\) is arranged in ascending order, the sign-score test is isotonic.

\subsubsection*{Step 5. Application of Holley's inequality}

\par \noindent Holley's inequality states that if \( f \) is an isotonic function on a finite distributive lattice \( \Omega \), and \( D \), \( D^* \) are two random variables on \( \Omega \) satisfying 
\begin{align}
\mathbb{P}(D = d \vee d^*) \mathbb{P}(D^* = d \wedge d^*) \geq \mathbb{P}(D = d) \mathbb{P}(D^* = d^*),
\label{eq:Holley_condition}
\end{align}
for all \( d, d^* \in \Omega \), then
\begin{align}
\mathbb{E}[f(D)] \geq \mathbb{E}[f(D^*)].
\label{eq:Holley_implication}
\end{align}

\noindent To proceed, we introduce the following notation. For any \( n \in \mathbb{Z}_{+} \), let \( \Pi_n \) denote the set of all \( n \)-dimensional vectors obtained by permuting the elements of \( \{1,2,\dots,n\} \). Thus, \( |\Pi_n| = n! \). 
We define a permutation operation for any \( \pi \in \Pi_n \) as:
\[
\pi(\bm{d}) = (d_{\pi_1}, d_{\pi_2}, \dots, d_{\pi_n}).
\]

\par \noindent Furthermore, we denote the sum of the entries in an \( n \)-dimensional vector \( \mathbf{d} \) as \( \operatorname{sum}(\mathbf{d}) \), that is:
\[
\operatorname{sum}(\mathbf{d}) = \sum_{s=1}^{n} d_s.
\]

\noindent We let \( D_{\bm{u}^+} \in \Omega \) denote the random variable with the treatment assignment distribution under model \eqref{eq:generic_bias_sensitivity_supp} corresponding to the proposed maximizer 
\begin{align}
\mathbf{u}^+ =  
(\underbrace{0, \dots, 0}_{N_{\cdot 1}}, \underbrace{1, \dots, 1}_{N_{\cdot 2}})^\intercal.
\label{eq:sign_score_proposed_maximizer}
\end{align}

\noindent Let \( D_{\bm{u}} \) be a random variable following the treatment assignment distribution under model \eqref{eq:generic_bias_sensitivity_supp} for any \( \mathbf{u} \in [0,1]^N \). We further denote the subvectors of \( \mathbf{u} \) corresponding to subjects with outcomes 1 and 2 as \( \mathbf{u}^1 \) and \( \mathbf{u}^2 \), respectively:
\[
\mathbf{u}^1 = (u_1, u_2, \dots, u_{N_{.1}})^\intercal \in [0,1]^{N_{.1}}, \quad 
\mathbf{u}^2 = (u_{N_{.1}+1}, \dots, u_N)^\intercal \in [0,1]^{N_{.2}}.
\]
\noindent We want to show that 
\begin{align}
\label{eq:Holley_conditioin_u_plus}
\mathbb{P}(D_{\bm{u}^+}=d\vee d^*) \mathbb{P}(D_{\bm{u}} = d \wedge d^*) \geq \mathbb{P}(D_{\bm{u}^+}=d) \mathbb{P}(D_{\bm{u}}=d^*),
\end{align}
for all $d, d^{*} \in \Omega$.  
\par \noindent We now explicitly write each probability term. First,  
\begin{align*}
\mathbb{P}(D_{\mathbf{u}^+} = d \vee d^*)
&= \Bigl\{\sum_{\bm{b} \in \mathcal{Z}(\mathbf{N}_{I.})}\exp\left\{\gamma \cdot \left(f_{\bm{\delta}}[\bm{b}]\right)^\intercal \mathbf{u}^+\right\}\Bigr\}^{-1}\cdot\; \Bigl\{\prod_{i=1}^{I} m_{\vee i}! (N_{i\cdot} - m_{\vee i})!\Bigr\}^{-1}\\
&\quad \cdot \sum_{\substack{\pi_1 \in \Pi_{N_{.1}} \\\pi_2 \in \Pi_{N_{.2}}}} 
\exp\left\{\gamma \cdot \left[
\left(f_{\bm{\delta}}\left[\pi_1\big((d \vee d^*)^1\big)\right]\right)^\intercal \mathbf{u}^{+1}
+
\left(f_{\bm{\delta}}\left[\pi_2\big((d \vee d^*)^2\big)\right]\right)^\intercal \mathbf{u}^{+2}
\right]\right\}\notag\\
& = \Big\{\sum_{\bm{b}\in \mathcal{Z}(\mathbf{N}_{I.})} \exp\left\{\gamma \cdot {\left(f_{\bm{\delta}}\left[\bm{b}\right]\right)}^\intercal \mathbf{u}^+\right\}\Big\}^{-1}
\cdot
\Big\{\prod_{i=1}^{I} m_{\vee i}! (N_{i\cdot}-m_{\vee i})!\Big\}^{-1} \notag \\
&\quad \cdot \sum_{\substack{\pi_1 \in \Pi_{N_{.1}} \\ \pi_2 \in \Pi_{N_{.2}}}} 
\exp\left\{\gamma \cdot \operatorname{sum}\left(f_{\bm{\delta}}\left[\left(d \vee d^*\right)^2\right]\right)\right\}. \notag 
\end{align*}

\noindent Here, \( f_{\bm{\delta}} \) is defined in \eqref{def:f_delta}. $\mathbf{u}^{+1}$ and $\mathbf{u}^{+2}$ denote the subvectors of $\mathbf{u}^+$ corresponding to outcome $1$ and $2$, respectively. \( m_{\vee i} \) denotes the number of times \( i \) appears in \( (d \vee d^*)^2 \) (The symbol $m$ abbreviates ``multiple''). That is:
\[
m_{\vee i} = \Big| \big\{ s \in \{1, \dots, N_{.2}\} : (d \vee d^*)^2_s = i \big\} \Big|, \quad i = 1,\dots,I.
\]

\noindent Since the treatment margins are fixed, the number of times \( i \) appears in \( (d \vee d^*)^1 \) is given by:
\[
N_{i\cdot} - m_{\vee i} = \Big| \big\{ s \in \{1, \dots, N_{.1}\} : (d \vee d^*)^1_s = i \big\} \Big|,
\]
for each \( i = 1, \dots, I \).

\noindent The first equality follows from model \eqref{eq:generic_bias_sensitivity_supp}, where we express the probability using $\pi_1$ and $\pi_2$. These permutations generate all possible treatment assignments in $\mathcal{Z }(\mathbf{N}_{I\cdot})$, whose subvectors corresponding to outcomes $1$ and $2$, after sorting in ascending order, 
coincide with $(d \vee d^*)^1$ and $(d \vee d^*)^2$, respectively.
 The division by \( \prod_{i=1}^{I} m_{\vee i}! \) accounts for identical assignments produced by multiple permutations \(\pi_2\), due to repeated elements in \( (d \vee d^*)^2 \). Specifically, for any \(\mathbf{z} \in \mathcal{Z}(\mathbf{N}_{I.})\),  
\[
\Big| \big\{ \pi_2 \in \Pi_{N_{.2}} : \pi_2((d \vee d^*)^2) = (z_{N_{.1}+1}, \dots, z_{N}) \big\} \Big| = \prod_{i=1}^{I} m_{\vee i}!.\] An analogous argument justifies the division by $\prod_{i=1}^{I}(N_{i\cdot}-m_{\vee i})!$.
The other probability terms are:
\begin{align*}
\mathbb{P}(D_{\mathbf{u}} = d \wedge d^*) 
&= \Big\{\sum_{\bm{b}\in \mathcal{Z}(\mathbf{N}_{I.})} \exp\left\{\gamma \cdot {\left(f_{\bm{\delta}}\left[\bm{b}\right]\right)}^\intercal \mathbf{u}\right\}\Big\}^{-1}
\cdot\; \Big\{\prod_{i=1}^{I} m_{\wedge i}! (N_{i\cdot}-m_{\wedge i})!\Big\}^{-1}\\
&\quad \cdot \sum_{\substack{\pi_1 \in \Pi_{N_{.1}} \\ \pi_2 \in \Pi_{N_{.2}}}} 
\exp\left\{\gamma \cdot \left[ {\left(f_{\bm{\delta}}\left[\pi_1 \left(\left(d \wedge d^*\right)^1\right)\right]\right)}^\intercal \mathbf{u}^{1} 
+ {\left(f_{\bm{\delta}}[\pi_2((d \wedge d^*)^2)]\right)}^\intercal \mathbf{u}^{2} \right] \right\};\notag\\[15pt]
\mathbb{P}(D_{\mathbf{u}^+} = d) 
&= \Big\{\sum_{\bm{b}\in \mathcal{Z}(\mathbf{N}_{I.})} \exp\left\{\gamma \cdot \; {\left(f_{\bm{\delta}}\left[\bm{b}\right]\right)}^\intercal \mathbf{u}^+\right\}\Big\}^{-1}
\cdot \Biggl\{\prod_{i=1}^{I} m_{i}! (N_{i\cdot}-m_{i})!\Biggr\}^{-1} \\
&\quad \cdot \sum_{\substack{\pi_1 \in \Pi_{N_{.1}} \\ \pi_2 \in \Pi_{N_{.2}}}} 
\exp\left\{\gamma \cdot \operatorname{sum}\left(f_{\bm{\delta}}\left[d^2\right]\right) \right\};\notag 
\\[15pt]
\mathbb{P}(D_{\mathbf{u}} = d^*) 
&= \Big\{\sum_{\bm{b}\in \mathcal{Z}(\mathbf{N}_{I.})} \exp\left\{\gamma \cdot {\left(f_{\bm{\delta}}\left[\bm{b}\right]\right)}^\intercal \mathbf{u}\right\}\Big\}^{-1}
\cdot \; \Big\{\prod_{i=1}^{I} m_{*i}! (N_{i\cdot}-m_{*i})!\Big\}^{-1} \\
&\quad \cdot \sum_{\substack{\pi_1 \in \Pi_{N_{.1}} \\ \pi_2 \in \Pi_{N_{.2}}}} 
\exp\left\{ \gamma \cdot \left[
\left(f_{\bm{\delta}}\left[\pi_1\left(d^{*1}\right)\right]\right)^\intercal \mathbf{u}^{1}
+\left(f_{\bm{\delta}}\left[\pi_2\left(d^{*2}\right)\right]\right)^\intercal \mathbf{u}^{2}
\right] \right\}. \notag
\end{align*}

\noindent Where \( m_{i} \), \( m_{*i} \), and \( m_{\wedge i} \) are defined similarly to \( m_{\vee i} \), for all \( i \in \{1, \ldots, I\} \):
\begin{equation*}
\begin{aligned}
m_{\wedge i} &= \Big| \big\{ s \in \{1, \dots, N_{.2}\} : (d \wedge d^*)^2_s = i \big\} \Big|, \\ 
m_{i} &= \Big| \big\{ s \in \{1, \dots, N_{.2}\} : d^2_s = i \big\} \Big|, \\ 
m_{*i} &= \Big| \big\{ s \in \{1, \dots, N_{.2}\} : d^{*2}_s = i \big\} \Big|.
\end{aligned}
\end{equation*}

\par \noindent To proceed, we prove the following two Lemmas. 

\begin{lem}
\label{lem:binomial_coefficient_inequality}
For all \( i \in \{1, \ldots, I\} \),
\begin{align*}
 m_{\wedge i}! (N_{i\cdot} - m_{\wedge i})!{m_{\vee i}!(N_{i\cdot} - m_{\vee i})!} \leq {m_i! (N_{i\cdot} - m_i)!}{m_{*i}!(N_{i\cdot}-m_{*i})!}.
\end{align*}
\end{lem}
\noindent We prove Lemma \ref{lem:binomial_coefficient_inequality} by establishing the following relationships.

\noindent \textbf{Relationship 1.} For all \( i \in \{1, \ldots, I\} \),
$m_{\wedge i} + m_{\vee i} = m_i + m_{*i}$.
\par \noindent Suppressing \( s \in \{1, \dots, N_{.2}\} \) in the following, we note:
\begin{align*}
m_{\vee i} &= |\{s : d_s^2 > d_s^{*2}, d_s^2 = i\}|
+ |\{s : d_s^2 = d_s^{*2} = i\}|
+ |\{s : d_s^2 < d_s^{*2}, d_s^{*2} = i\}|, \\
m_{\wedge i} &= |\{s : d_s^2 > d_s^{*2}, d_s^{*2} = i\}| 
+ |\{s : d_s^2 = d_s^{*2} = i\}|
+ |\{s : d_s^2 < d_s^{*2}, d_s^2 = i\}|,\\
m_i &= |\{s : d_s^2 > d_s^{*2}, d_s^2 = i\}|
+ |\{s : d_s^2 = d_s^{*2} = i\}|
+ |\{s : d_s^2 < d_s^{*2}, d_s^2 = i\}|, \\
m_{*i} &= |\{s : d_s^2 > d_s^{*2}, d_s^{*2} = i\}|
+ |\{s : d_s^2 = d_s^{*2} = i\}|
+ |\{s : d_s^2 < d_s^{*2}, d_s^{*2} = i\}|.
\end{align*}
Summing \( m_{\wedge i} \) and \( m_{\vee i} \), and subtracting \( m_i \) and \( m_{*i} \), we obtain:
\[
m_{\wedge i} + m_{\vee i} - m_i - m_{*i} = 0.
\]

\par \noindent \textbf{Relationship 2.} For all $i\in \{1,\ldots,I\}$, $\min(m_i, m_{*i}) \leq m_{\wedge i}, m_{\vee i} \leq \max(m_i, m_{*i}).$
\par \noindent \textbf{Claim 1. $\min(m_i, m_{*i})\leq m_{\vee i} \leq \max(m_i, m_{*i})$.}

\noindent Define the smallest index where \( i \) occurs in \( d^2 \) and \( d^{*2} \) with $o$ and $o^*$ (where $o$ stands for ``occurrence''):
\[
o = \min \{s : d^{2}_s = i\}, \quad
o^{*} = \min \{s : d^{*2}_s = i\}.
\]
Since $d^2$ and $d^{*2}$ are both arranged in ascending order, each vector can be divided into three consecutive regions: entries smaller than $i$, entries equal to $i$, and entries greater than $i$. Assume first that $m_i \leq m_{*i}$. We then consider the following cases, where in each range of indices we specify the inequalities satisfied by $d^2_s$ and $d^{*2}_s$ relative to $i$.
\par \noindent (i) If \( o = o^* \):
\begin{align*}
1 \leq s < o, \quad & d_{s}^2<i,\; d_{s}^{*2} < i; \\
o \leq s < o + m_{i}, \quad & d_{s}^2 = d_{s}^{*2} = i; \\
o + m_{i} \leq s < o + m_{*i}, \quad & d_{s}^2 > i, d_{s}^{*2} = i; \\
o + m_{*i} \leq s, \quad & d_{s}^2>i, \; d_{s}^{*2} > i.
\end{align*}
Since $(d \vee d^*)^2_s = \max(d^2_s,d^{*2}_s)$, it equals $i$ only for $o \leq s < o+m_i$, giving $m_{\vee i}=m_i$. 
\par \noindent (ii) If \( o < o^* \) and $o^{*} \leq o+m_i$:
\begin{align*}
1 \leq s < o, \quad & d_{s}^2<i,\; d_{s}^{*2} < i; \\
o \leq s < o^*, \quad & d_{s}^2 = i, d_{s}^{*2} < i; \\
o^* \leq s < o + m_{i}, \quad & d_{s}^2 = d_{s}^{*2} = i; \\
o + m_{i} \leq s < o^* + m_{*i}, \quad & d_{s}^2 > i, d_{s}^{*2} = i; \\
o^* + m_{*i} \leq s, \quad & d_{s}^2>i,\; d_{s}^{*2} > i.
\end{align*}
Since $(d \vee d^*)^2_s = \max(d^2_s,d^{*2}_s)$, it equals $i$ for $o \leq s < o^*$ and $o^*\leq s < o+m_i$, giving $m_{\vee i}=m_i$.
\par \noindent (iii) If $o < o^*$ and $o^* > o+m_i$:
\begin{align*}
1 \leq s < o, \quad & d_{s}^2<i,\; d_{s}^{*2} < i; \\
o \leq s < o+m_i, \quad & d_s^2 =i,\; d_s^{*2} <i;\\
o+m_i \leq s < o^*, \quad & d_s^2>i, \; d_s^{*2} = i;\\
o^{*}\leq s < o^* + m_{*i}, \quad & d_s^2 >i, \; d_s^{*2} = i; \\
o^{*}+m_{*i}\leq s, \quad & d_{s}^2 >i, \; d_{s}^{*2} > i.
\end{align*}
Similarly, $(d \vee d^*)^2_s$ equals $i$ for $o\leq s < o+m_i$, giving $m_{\vee i} = m_i$. 

\par \noindent (iv) If \( o > o^* \), we consider two cases.  
If \( o + m_i \geq o^* + m_{*i} \), then \( m_{\vee i} = m_{*i} \).  
If \( o + m_i < o^* + m_{*i} \), then \( m_{\vee i} = m_{i} + (o - o^*) \). The inequality \( \min(m_i, m_{*i}) \leq m_{\vee i} \leq \max(m_i, m_{*i}) \) follows from the conditions \( o > o^* \) and \( o + m_i < o^* + m_{*i} \). A symmetric argument shows that $\min(m_i, m_{*i}) \leq m_{\vee i} \leq \max(m_i,m_{*i})$ when $m_i>m_{*i}$ as well.

\par \noindent \textbf{Claim 2. $\min(m_i,m_{*i})\leq m_{\wedge i} \leq \max(m_i, m_{*i})$.}
\par \noindent 
Using Relationship 1 and the fact that $\max(x,y)+\min(x,y) = x+y$, Claim 1 gives:
\begin{align*}
m_{\wedge i}+m_{\vee i} - \max(m_i, m_{*i})
& \leq m_{\vee i} \leq m_{\wedge i}+m_{\vee i} - \min(m_i,m_{*i}).
\end{align*}
Rearranging terms gives
\[
\min(m_i,m_{*i}) \leq m_{\wedge i} \leq \max(m_i, m_{*i}).
\]

\par \noindent \textbf{Relationship 3.} The inequality of the product of the factorial terms

We argue that for $l, m, M, L \in \mathbb{Z}_{0+}$, and $m \leq l \leq  M \leq L$, we have:
\begin{align}
l!(L-l)!(M+m-l)!(L-(M+m-l))! \leq m!(L-m)!M!(L-M)!
\label{eq:factorial_inequality}
\end{align}
To see this, take the ratio of the left-hand side to the right-hand side, we have:
\begin{align*}
\frac{l!(L-l)!(M+m-l)!(L-M-m+l)!}{m!(L-m)!M!(L-M)!} 
= \prod_{k=1}^{l-m} \frac{m+k}{M+m-l+k} \cdot \frac{L-M+k}{L-l+k}.
\end{align*}
Since $l \leq M$, each factor in the product is at most one, and therefore the entire ratio is bounded above by one.

\par \noindent By Relationships~1 and~2, we may set 
$l = m_{\wedge i}$, $m = \min(m_i,m_{*i})$, 
$M = \max(m_i,m_{*i})$, and $L = N_{i\cdot}$, 
which gives $M+m-l = m_{\vee i}$ and satisfies the conditions in Relationship~3 that $m \leq l \leq M \leq L$. 
Applying Relationship~3 with these settings proves 
Lemma~\ref{lem:binomial_coefficient_inequality}.

\begin{lem}
\label{lem:exponential_inequality}
\noindent For each fixed $\pi_1 \in \Pi_{N_{.1}}$ and for each fixed $\pi_2 \in \Pi_{N_{.2}}$, we have:  
\begin{align*}
& \operatorname{sum}(f_{\bm{\delta}}[(d \vee d^*)^2]) + {(f_{\bm{\delta}}[\pi_1((d \wedge d^*)^1)])}^\intercal \mathbf{u}^1+{(f_{\bm{\delta}}[\pi_2((d \wedge d^*)^2)])}^\intercal \mathbf{u}^2 \\
& \quad \geq \operatorname{sum}(f_{\bm{\delta}}[d^2]) + {(f_{\bm{\delta}}[\pi_1(d^{*1})])}^\intercal \mathbf{u}^1 + {(f_{\bm{\delta}}[\pi_2(d^{*2})])}^\intercal \mathbf{u}^2.
\end{align*}
\end{lem}
Lemma \ref{lem:exponential_inequality} can be implied if we can show the following two inequalities: 
\begin{align}
{(f_{\bm{\delta}}[\pi_1((d \wedge d^*)^1)])}^\intercal \mathbf{u}^1 \geq (f_{\bm{\delta}}[\pi_1(d^{*1})])^\intercal \mathbf{u}^1
\label{eq:u_1_holley_step}
\end{align}
and 
\begin{equation}
\begin{aligned}
& \operatorname{sum}(f_{\bm{\delta}}[(d \vee d^*)^2]) + {f_{\bm{\delta}}[\pi_2((d \wedge d^*)^2)]}^\intercal \mathbf{u}^2\\
& \quad \geq \operatorname{sum}(f_{\bm{\delta}}(d^2))+ {(f_{\bm{\delta}}[\pi_2(d^{*2})])}^\intercal \mathbf{u}^2
\label{eq:sum_u_2_holley_step}
\end{aligned}
\end{equation}

\noindent To establish line \eqref{eq:u_1_holley_step}, we draw insights from \citet[Lemma 3]{zhang2024sensitivity}, which states that for any $d$, $d^* \in \Omega$, we have: 
\[
(d \wedge d^*)^1 \geq d^1, d^{*1} \text{ element-wise}.
\]
As a result, 
\[
\pi_{1}((d \wedge d^*)^1) \geq \pi_1(d^{*1}) \text{ element-wise} \rightarrow  f_{\bm{\delta}}[\pi_{1}((d \wedge d^*)^1)] \geq f_{\bm{\delta}}[\pi_{1}(d^{*1})] \text{ element-wise}.
\]
As $\mathbf{u}^1 \geq \bm{0}_{N_{.1}} \text{ element-wise}$, we obtain line \eqref{eq:u_1_holley_step}.
\par \noindent To establish line \eqref{eq:sum_u_2_holley_step}, we observe that for any \( x, y \in \mathbb{R} \), the following set equality holds:
\[
\{\max(x,y), \min(x,y)\} = \{x,y\}.
\]
Applying this to all \( s \in \{1,2,\dots, N_{\cdot2}\} \) and using the definitions of \( f_{\bm{\delta}} \), \(\vee\), and \(\wedge\), we obtain:
\[
(f_{\bm{\delta}}[(d \vee d^*)^2])_s + (f_{\bm{\delta}}[(d \wedge d^*)^2])_s = (f_{\bm{\delta}}[d^2])_s + (f_{\bm{\delta}}[d^{*2}])_s.
\]
Summing over all indices, we arrive at:
\begin{align}
    \operatorname{sum}(f_{\bm{\delta}}[(d \vee d^{*})^2]) + \operatorname{sum}(f_{\bm{\delta}}[(d \wedge d^*)^2]) = \operatorname{sum}(f_{\bm{\delta}}[d^2]) + \operatorname{sum}(f_{\bm{\delta}}[d^{*2}]).
    \label{eq:indicator_min_max_equality}
\end{align}

\par \noindent Since each element in \( \mathbf{u}^2 \) is less than or equal to \( 1 \), and by definition, for all $s$, \[( f_{\bm{\delta}}[\pi_2(d^{*2})] )_s - ( f_{\bm{\delta}}[\pi_2(d \wedge d^*)^2] )_s \geq 0,\]
we have:
\begin{align*}
    {(f_{\bm{\delta}}[\pi_2(d^{*2})])}^\intercal \mathbf{u}^2 - {(f_{\bm{\delta}}[\pi_2((d \wedge d^*)^2)])}^\intercal \mathbf{u}^2 
    &= {\left( f_{\bm{\delta}}[\pi_2(d^{*2})] - f_{\bm{\delta}}[\pi_2((d \wedge d^*)^2)] \right)}^\intercal \mathbf{u}^2 \\
    &\leq {\left( f_{\bm{\delta}}[\pi_2(d^{*2})] - f_{\bm{\delta}}[\pi_2((d \wedge d^*)^2)] \right)}^\intercal \bm{1}_{N_{\cdot 2}} \\
    &= \operatorname{sum}(f_{\bm{\delta}}[\pi_2(d^{*2})]) - \operatorname{sum}(f_{\bm{\delta}}[\pi_2((d \wedge d^{*})^2)]).
\end{align*}

Combine with result from line \eqref{eq:indicator_min_max_equality}, we have:
\begin{align*}
& {(f_{\bm{\delta}}[\pi_2(d^{*2})])}^\intercal \mathbf{u}^2-{(f_{\bm{\delta}}[\pi_2((d \wedge d^*)^2)])}^\intercal \mathbf{u}^2\\    
& \quad \quad \leq \operatorname{sum}(f_{\bm{\delta}}[(d \vee d^*)^2]) - \operatorname{sum}(f_{\bm{\delta}}[d^2]),
\end{align*}
which proves line \eqref{eq:sum_u_2_holley_step}.

\par \noindent The results from Lemma \ref{lem:binomial_coefficient_inequality} and Lemma \ref{lem:exponential_inequality} implies the premise in \eqref{eq:Holley_conditioin_u_plus}. Taking sign-score test as the isotonic function in \eqref{eq:Holley_implication}, we prove Theorem \mref{thm:score_test_maximizer_binary}.

\subsection{Remark on Simultaneous Ordering and Nonmonotone Sensitivity Parameters} \label{app:B_nonmonotone}
\change{Theorems~\mref{thm:score_test_maximizer_nonbinary} and~\mref{thm:score_test_maximizer_binary} assume that the treatment scores $w_1 \le \cdots \le w_I$ and sensitivity parameters $\delta_1 \le \cdots \le \delta_I$ are simultaneously nondecreasing across treatment levels. When treatment categories are nominal, this simultaneous monotonicity condition can often be satisfied by reordering the categories. For example, consider a binary outcome ($J=2$) and nominal treatment categories $(A, B, C)$. If an investigator assigns ordinal test scores $(0, 2, 1)$ to $(A, B, C)$ with sensitivity parameters $\bm{\delta} = (0, 1, 1)$, one can simply reorder the categories as $(A, C, B)$. Under this reordering, the scores become $(0, 1, 2)$ and $\bm{\delta} = (0, 1, 1)$, so both sequence vectors are simultaneously nondecreasing ($0 \le 1 \le 2$ and $0 \le 1 \le 1$), making Theorems~\mref{thm:score_test_maximizer_nonbinary} and~\mref{thm:score_test_maximizer_binary} directly applicable.

However, when a sensitivity pattern is inherently nonmonotone relative to a fixed score ordering, such as $\bm{\delta} = (0, 1, 0)$ for treatment scores $(0, 1, 2)$, no reordering of treatment categories can make both vectors simultaneously nondecreasing. In such cases, the arrangement-increasing property and the Holley's inequality used to shrink the candidate set in Theorems~\mref{thm:score_test_maximizer_nonbinary} and~\mref{thm:score_test_maximizer_binary} does not apply.

Fortunately, count-based test statistics include ordinal and sign-score tests as special cases, and Theorem~\mref{thm:set_size_of_U} does not require monotonicity of $\bm{\delta}$ or the scores. Therefore, when no compatible reordering exists to yield simultaneously nondecreasing scores and $\bm{\delta}$, Theorem~\mref{thm:set_size_of_U} applies directly to provide exact worst-case inference using the general candidate set $\mathcal{U}_{\rm C}$ of size at most $N^J$.}

\subsection{Integrating Multiple $\bm{\delta}$ Vectors (Statement and Proof)} \label{app:B8}

Suppose the investigator believes that $M$ vectors of sensitivity parameters
$\bm{\delta}^{1},\ldots, \boldsymbol{\delta}^{M}$ could plausibly describe the effects
of the unmeasured confounder under the sensitivity model~\eqref{eq:generic_bias_sensitivity_supp},
and wishes to integrate these $M$ vectors in a sensitivity analysis.
One natural way to do so is to compute the worst-case $p$-value across all
$M$ vectors of $\boldsymbol{\delta}$ and all possible $\mathbf{u}$.
For each $m = 1, \ldots, M$, let
$\boldsymbol{\delta}^{m} = (\delta^{m}_1, \ldots, \delta^{m}_I)$
denote the $m$th vector of sensitivity parameters, and let
\begin{equation} \label{eq:pvalue_at_u}
p_{\gamma,\boldsymbol{\delta}}(T_{\rm obs},\mathbf{u})
= \frac{\sum_{\mathbf{z} \in \mathcal{Z}(\mathbf{N}_{I\cdot})} \mathbbm{1}\{T(\mathbf{N}(\mathbf{z}, \rr)) \geq T_{\rm obs}\} \exp \left\{ \gamma \sum_{s=1}^{N} \sum_{i=1}^I \delta_i \mathbbm{1}\{z_s = i\} u_s \right\}}{\sum_{\mathbf{b} \in \mathcal{Z}(\mathbf{N}_{I\cdot})} \exp \left\{ \gamma \sum_{s=1}^{N} \sum_{i =1}^I \delta_i \mathbbm{1}\{b_s = i\} u_s \right\}}
\end{equation}
denote the exact, one-sided $p$-value under $\boldsymbol{\delta}$ at the unmeasured confounder
$\mathbf{u}$; that is, the null distribution $\alpha_{\boldsymbol{\delta}}(T,\rr,\mathbf{u})$ in
equation~\meqref{eq:worst_case_null} of the main text with the threshold set to the observed test
statistic $T_{\rm obs}$. By the definition of the worst-case $p$-value in
Section~\mref{sec:def_worst_case} of the main text, $\overline{p}_{\gamma,\boldsymbol{\delta}}(T_{\rm obs})
= \max_{\mathbf{u}\in[0,1]^N} p_{\gamma,\boldsymbol{\delta}}(T_{\rm obs},\mathbf{u})$.
The worst-case $p$-value across the $M$ vectors can then be formalized as
\begin{equation} \label{eq:integrate_delta_supp}
\max_{\mathbf{u} \in [0,1]^N,\; \boldsymbol{\delta} \in
\{\boldsymbol{\delta}^{1}, \dots, \boldsymbol{\delta}^{M}\}}
p_{\gamma,\boldsymbol{\delta}}(T_{\rm obs},\mathbf{u}).
\end{equation}
Corollary~\ref{thm:several_sensitivity_models} shows that the optimization problem in
\eqref{eq:integrate_delta_supp} can be solved by taking the maximum of the $M$ worst-case $p$-values.

\setcounter{corollary}{0}\renewcommand{\thecorollary}{\thesection.\arabic{corollary}}
\begin{corollary}[Worst-Case $P$-Value Across Multiple Vectors of $\bm{\delta}$]
\label{thm:several_sensitivity_models}
Let $\bm{\delta}^{1},\ldots,\bm{\delta}^{M}$ denote the $M$ vectors of $\bm{\delta}$ and for each
$\bm{\delta}^m$, let $\mathbf{u}_m = \operatorname{argmax}_{\mathbf{u}\in [0,1]^N}
p_{\gamma,\bm{\delta}^m}(T_{\rm obs},\mathbf{u})$
be the worst-case unmeasured confounder under $\bm{\delta}^m$, which can be found by the techniques
in Section~\mref{sec:Maximizer_Characterization_and_Bounds} of the main text, so that
$p_{\gamma,\bm{\delta}^m}(T_{\rm obs},\mathbf{u}_m) = \overline{p}_{\gamma,\bm{\delta}^m}(T_{\rm obs})$.
Then, the solution to equation \eqref{eq:integrate_delta_supp} is
\begin{equation} \label{eq:integrate_delta_result_supp}
\max_{\mathbf{u}\in [0,1]^N,\bm{\delta} \in \{\bm{\delta}^1,\dots,\bm{\delta}^M\} }
p_{\gamma,\bm{\delta}}(T_{\rm obs},\mathbf{u})
= \max\big(\overline{p}_{\gamma,\bm{\delta}^1}(T_{\rm obs}),\dots,\overline{p}_{\gamma,\bm{\delta}^M}(T_{\rm obs})\big).
\end{equation}
\end{corollary}
We can use Corollary~\ref{thm:several_sensitivity_models} to obtain the worst-case $p$-value across
$M$ vectors $\bm{\delta}$ by looking at the maximum $p$-value among the $M$ worst-case $p$-values
based on each $\bm{\delta}^{1}$,\ldots,$\bm{\delta}^M$.

We now prove Corollary~\ref{thm:several_sensitivity_models}. For simplicity, we present the proof
for the case with two sets of bias parameters, $\bm{\delta}^1$ and $\bm{\delta}^2$; the case with
more than two sets follows analogously.

We first show that
\begin{equation}
\max_{\mathbf{u}\in [0,1]^N,\; \bm{\delta}\in \{\bm{\delta}^1,\bm{\delta}^2\}}
p_{\gamma,\bm{\delta}}(T_{\rm obs},\mathbf{u})
\;\leq\;
\max\big(\overline{p}_{\gamma,\bm{\delta}^1}(T_{\rm obs}),\, \overline{p}_{\gamma,\bm{\delta}^2}(T_{\rm obs})\big),
\label{eq:upper_bounded_by_pairwise_max}
\end{equation}
where $\mathbf{u}_1 = \operatorname*{arg\,max}_{\mathbf{u}\in [0,1]^N} p_{\gamma,\bm{\delta}^1}(T_{\rm obs},\mathbf{u})$
so that $p_{\gamma,\bm{\delta}^1}(T_{\rm obs},\mathbf{u}_1) = \overline{p}_{\gamma,\bm{\delta}^1}(T_{\rm obs})$,
and $\mathbf{u}_2$ and $\overline{p}_{\gamma,\bm{\delta}^2}(T_{\rm obs})$ are defined analogously.

Pick an arbitrary $\mathbf{u}_0 \in [0,1]^N$.
By the definition of $\mathbf{u}_1$ and $\mathbf{u}_2$, we have
\begin{align*}
p_{\gamma,\bm{\delta}^1}(T_{\rm obs},\mathbf{u}_0)
&\leq p_{\gamma,\bm{\delta}^1}(T_{\rm obs},\mathbf{u}_1) = \overline{p}_{\gamma,\bm{\delta}^1}(T_{\rm obs})
 \leq \max\big(\overline{p}_{\gamma,\bm{\delta}^1}(T_{\rm obs}),\, \overline{p}_{\gamma,\bm{\delta}^2}(T_{\rm obs})\big),\\[4pt]
p_{\gamma,\bm{\delta}^2}(T_{\rm obs},\mathbf{u}_0)
&\leq p_{\gamma,\bm{\delta}^2}(T_{\rm obs},\mathbf{u}_2) = \overline{p}_{\gamma,\bm{\delta}^2}(T_{\rm obs})
 \leq \max\big(\overline{p}_{\gamma,\bm{\delta}^1}(T_{\rm obs}),\, \overline{p}_{\gamma,\bm{\delta}^2}(T_{\rm obs})\big).
\end{align*}
Combining these inequalities, we see that for all $\mathbf{u}_0 \in [0,1]^N$, both
$p_{\gamma,\bm{\delta}^1}(T_{\rm obs},\mathbf{u}_0)$ and $p_{\gamma,\bm{\delta}^2}(T_{\rm obs},\mathbf{u}_0)$
are bounded above by $\max\big(\overline{p}_{\gamma,\bm{\delta}^1}(T_{\rm obs}),\, \overline{p}_{\gamma,\bm{\delta}^2}(T_{\rm obs})\big)$.

On the other hand, $\max\big(\overline{p}_{\gamma,\bm{\delta}^1}(T_{\rm obs}), \overline{p}_{\gamma,\bm{\delta}^2}(T_{\rm obs})\big)$
equals either $\overline{p}_{\gamma,\bm{\delta}^1}(T_{\rm obs})$ or $\overline{p}_{\gamma,\bm{\delta}^2}(T_{\rm obs})$,
and each $\overline{p}_{\gamma,\bm{\delta}^j}(T_{\rm obs}) = p_{\gamma,\bm{\delta}^j}(T_{\rm obs},\mathbf{u}_j)$
is attained at $\mathbf{u}_j \in [0,1]^N$ by definition; Therefore,
\begin{equation}
\max_{\mathbf{u}\in [0,1]^N,\; \bm{\delta}\in \{\bm{\delta}^1,\bm{\delta}^2\}}
p_{\gamma,\bm{\delta}}(T_{\rm obs},\mathbf{u})
\;\geq\;
\max\big(\overline{p}_{\gamma,\bm{\delta}^1}(T_{\rm obs}),\, \overline{p}_{\gamma,\bm{\delta}^2}(T_{\rm obs})\big).
\label{eq:lower_bounded_by_pairwise_max}
\end{equation}
Combining~\eqref{eq:upper_bounded_by_pairwise_max} and~\eqref{eq:lower_bounded_by_pairwise_max} yields the desired equality.

\subsection{Proof of Corollary~\mref{cor:i2k}: Sharp Bound for the Stratified Sum Test} \label{app:B9}

This section derives Corollary~\mref{cor:i2k} in the main text, which establishes the exact worst-case null 
distribution of the stratified sum test statistic 
$g(\mathbf{T}) = g(T_1,\ldots,T_K)$ for an $I\times2\times K$ contingency table with binary outcomes. 
For each stratum $k=1,\ldots,K$, let $T_k = \sum_{i=1}^{I} w_{ki} N_{ki2}$ be a sign-score statistic with nondecreasing weights $w_{k1} \le \cdots \le w_{kI}$. We work under the generic bias sensitivity model, fixing the sensitivity parameters $\gamma>0$ and 
$\bm{\delta}$ with $\delta_1 \le \cdots \le \delta_I$.

Consider the sign-score statistic $T_k$ from the $k$-th stratum:
\begin{align}
T_{k} 
&= \sum_{i=1}^{I} w_{ki} N_{ki2} 
= \sum_{i=1}^{I} w_{ki} \,\mathbbm{1}\{Z_{ks}=i,\, r_{ks}=2\}, 
\quad 
w_{ki} \ge w_{k(i-1)} \;\; \forall i \in \{2,\ldots,I\}.
\end{align}
Theorem~\mref{thm:score_test_maximizer_binary} implies that, for any $\criticalt_k \in \mathbb{R}$, the upper bound on the one-sided $p$-value
\[
\mathbb{P}_{H_{k0}}\!\left(T_k \ge \criticalt_k \mid \mathcal{F}_{k},\, \mathcal{Z}(\mathbf{N}_{kI\cdot})\right)
\]
is attained at the maximizer
\begin{equation}
\mathbf{u}^{+}_{k} = \trr_{k 2},
\label{eq:each_table_maxizer_implicit}
\end{equation}
where $\trr_{k 2}$ corresponds to the second column of the binary matrix representation of the observed outcomes 
(see \eqref{eq:integer_dummy_mapping}). 
In other words, if we assume $r_{k1} \le r_{k2} \le \cdots \le r_{kN_k}$, so that the outcomes are arranged in ascending order, then
\begin{equation}
\mathbf{u}^+_k = (\underbrace{0, \dots, 0}_{N_{k\cdot 1}}, \underbrace{1, \dots, 1}_{N_{k\cdot 2}}),
\label{eq:each_table_maximizer_explicit}
\end{equation}
where $N_{k\cdot 1}$ and $N_{k\cdot 2}$ denote the outcome margins of the $k$-th table. 
Corollary \mref{cor:unique_corner_multivariate_hyper} (see Section \ref{app:C4} of the Supplementary Material for proof) says that the distribution of the cell counts
\begin{equation*}
\begin{aligned}
\bm{M}_{k} & = \Big(N_{k12},\ldots,N_{kI2}\Big) \\
& = \Big(\sum_{s=1}^{N_k} \mathbbm{1}\{Z_{ks}=1,r_{ks}=2\},\ldots,\sum_{s=1}^{N_k}\mathbbm{1}\{Z_{ks}=I,r_{ks}=2\}\Big)\\
& = \Big(\widetilde{\mathbf{Z}}_{k1}^\intercal \trr_{k2}, \dots,\widetilde{\mathbf{Z}}_{kI}^\intercal \trr_{k2}\Big)
\end{aligned}
\end{equation*} under the sensitivity model \eqref{eq:generic_bias_sensitivity_supp} with $\mathbf{u}^{+}_{k} = \trr_{k2}$ follows a multivariate extended hypergeometric distribution with parameters $\mathbf{N}_{kI\cdot}$, $N_{k\cdot 2}$, and odds $(\exp(\gamma \delta_1),\dots,\exp(\gamma\delta_{I}))$. Therefore, define $\overline{T}_{k}$ a random variable obtained with the linear transformation $(w_{k1}, \cdots, w_{kI})^\intercal\bm{M}_{k}$, then $\overline{T}_{k}$ is stochastically larger than $T_k$ under the generic bias sensitivity model fixing $\gamma$ and $\bm{\delta}$. In other words, for every \( k = 1,\dots,K \),
\begin{align*}
\max_{\mathbf{u}_k \in [0,1]^{N_k}}\mathbb{P}_{H_{k0}}(T_k \geq \criticalt_k \mid \mathcal{F}_{k}, {\mathcal{Z}}(\mathbf{N}_{kI\cdot})) 
= \mathbb{P}_{H_{k0}}(\overline{T}_{k} \geq \criticalt_k), \quad \forall \criticalt_k \in \mathbb{R}.
\label{eq:single_table_stochastic_ordering}
\end{align*}
Assuming independence of treatment assignments across strata and omitting the explicit dependence of the probability on $\mathbf{u}_1,\ldots,\mathbf{u}_K$ for notational simplicity, then
\begin{align*}
& \mathbb{P}_{H_{\text{joint}}}\!\left(T_1 \ge \criticalt_1, \ldots, T_K \ge \criticalt_K 
\mid \mathcal{F}_1, \ldots, \mathcal{F}_K, 
\mathcal{Z}(\mathbf{N}_{1I\cdot}), \ldots, \mathcal{Z}(\mathbf{N}_{KI\cdot})\right) \\
& \quad = \prod_{k=1}^{K} 
\mathbb{P}_{H_{k0}}\!\left(T_k \ge \criticalt_k 
\mid \mathcal{F}_k, \mathcal{Z}(\mathbf{N}_{kI\cdot})\right) \\
& \quad \leq \prod_{k=1}^{K} 
\mathbb{P}_{H_{k0}}\!\left(\overline{T}_k \ge \criticalt_k\right) \\
& \quad = 
\mathbb{P}_{H_{\text{joint}}}\!\left(\overline{T}_1 \ge \criticalt_1, \ldots, \overline{T}_K \ge \criticalt_K\right).
\end{align*}
Consequently, the random vector \( (\overline{T}_1,\dots, \overline{T}_K) \) is stochastically larger than the random vector $(T_1\allowbreak,\, T_2\allowbreak,\, \dots\allowbreak,\, T_K)$ under \( H_{\text{joint}} \), fixing the sensitivity parameter $\gamma>0$ and $\bm{\delta}$, evaluated at arbitrary $\mathbf{u}_k \in [0,1]^{N_k}$, $k=1,\ldots,K$.

By the typical property of stochastic ordering, for a monotone increasing function $g$:
\begin{equation}
\begin{aligned}
& \mathbb{P}_{H_{\text{joint}}}(
 g(T_1,\dots,T_K) \geq \criticalt
\mid   \mathcal{F}_{1},\dots,\mathcal{F}_{K},
{\mathcal{Z}}\bigl(\mathbf{N}_{1I\cdot}),\dots,
{\mathcal{Z}}\bigl(\mathbf{N}_{KI\cdot})) \\
& \quad = \mathbb{E}(\mathbbm{1}( g(T_1,\dots,T_K) \geq \criticalt)\mid \mathcal{F}_{1},\dots,\mathcal{F}_{K},
{\mathcal{Z}}\bigl(\mathbf{N}_{1I\cdot}),\dots,
{\mathcal{Z}}\bigl(\mathbf{N}_{KI\cdot}))  \\
& \quad \leq \mathbb{E}(\mathbbm{1}(g(\overline{T}_1,\dots, \overline{T}_{K})\geq \criticalt)) \\
& \quad = 
\mathbb{P}_{H_{\text{joint}}}(g(\overline{T}_1, \dots, \overline{T}_{K}) \geq \criticalt), \quad \forall \criticalt \in \mathbb{R}.
\label{eq:g_upper_bound}
\end{aligned}
\end{equation}

On the other hand, since each $\overline{T}_k$ is obtained from the worst-case 
$\mathbf{u}^+_k \in [0,1]^{N_k}$, which belongs to the maximization space, we have
\begin{equation}
\begin{aligned}
\max_{\substack{
\mathbf{u}_1 \in [0,1]^{N_1},\,\ldots,\,\mathbf{u}_{{K}} \in [0,1]^{N_{{K}}}
}} & \mathbb{P}_{H_{\mathrm{joint}}}\!\big(
g(\mathbf{T}) \ge c 
\mid \mathcal{F}_{1},\ldots,\mathcal{F}_{K},
\mathcal{Z}(\mathbf{N}_{1I\cdot}),\ldots,
\mathcal{Z}(\mathbf{N}_{KI\cdot})\big)\\
& \ge
\mathbb{P}_{H_{\mathrm{joint}}}\!\big(g(\overline{\mathbf{T}}) \ge c\big).
\label{eq:g_function_lower_bound}
\end{aligned}
\end{equation}

Combining \eqref{eq:g_upper_bound} and \eqref{eq:g_function_lower_bound}, we obtain 

\begin{equation*}
\begin{aligned}
\max_{\substack{
\mathbf{u}_1 \in [0,1]^{N_1},\,\ldots,\,\mathbf{u}_{{K}} \in [0,1]^{N_{{K}}}
}} & \mathbb{P}_{H_{\mathrm{joint}}}\!\big(
g(\mathbf{T}) \ge c 
\mid \mathcal{F}_{1},\ldots,\mathcal{F}_{K},
\mathcal{Z}(\mathbf{N}_{1I\cdot}),\ldots,
\mathcal{Z}(\mathbf{N}_{KI\cdot})\big)\\
& =
\mathbb{P}_{H_{\mathrm{joint}}}\!\big(g(\overline{\mathbf{T}}) \ge c\big).
\end{aligned}
\end{equation*}

\subsection{Sensitivity Analysis With Dichotomized Treatment Levels} \label{app:B10}
In this section, we study the implied sensitivity model from the generic bias sensitivity model after
the investigator collapses or only uses the extreme levels of treatment and conducts sensitivity analysis with the ``transformed''/dichotomized treatment variable. 
We recall that the generic bias sensitivity model discussed throughout the main text is defined by equations~\eqref{eq:generic_sensitivity_1} and~\eqref{eq:generic_sensitivity_2} (shown to be equivalent in Section~\ref{app:A} of the Supplementary Material):
\begin{equation}
\mathbb{P}(Z_s = i \mid \mathcal{F})
= 
\frac{\exp\{\xi_i(x_s) + \gamma\delta_i u_s\}}
{\sum_{i'=1}^{I} \exp\{\xi_{i'}(x_s) + \gamma \delta_{i'} u_s\}},
\quad
\gamma \ge 0,\;
\delta_i \in \{0,1\},\;
u_s \in [0,1].
\label{eq:generic_sensitivity_1}
\end{equation}
For any pair \(i,i' \in \{1,\dots,I\}\), \eqref{eq:generic_sensitivity_1} can be equivalently expressed as:
\begin{equation}
\log\!\left(
\frac{\mathbb{P}(Z_s = i' \mid \mathcal{F})}
{\mathbb{P}(Z_s = i \mid \mathcal{F})}
\right)
=
\xi_{i'}(x_s) - \xi_i(x_s) + \gamma (\delta_{i'} - \delta_i) u_s.
\label{eq:generic_sensitivity_2}
\end{equation}
\subsubsection*{Sensitivity Analysis Based on Collapsed Treatment Levels} 
This subsection shows that under the generic bias sensitivity model in \eqref{eq:generic_sensitivity_1}–\eqref{eq:generic_sensitivity_2}, the 
odds ratio obtained by collapsing treatment levels among two subjects \(s\) and \(s'\) with identical measured covariates \(x_s = x_{s'}\) 
is bounded above by \(\exp\{\gamma |u_s - u_{s'}|\}\), the odds-ratio bound implied by~\citet{rosenbaum1990sensitivity}. Therefore, if the original treatment assignment follows the generic bias sensitivity model parameterized by \(\gamma\) and \(\bm{\delta}\), a sensitivity analysis using dichotomized treatment variables (e.g., \citet{rosenbaum1990sensitivity}) continues to provide valid Type~I error control. Our result implies that while dichotomizing a multi-level treatment may reduce statistical power, it does not compromise size control.

To formalize this, we define sets of treatment levels corresponding to the investigator-defined dichotomization scheme. Let \(\mathcal{B}\) denote the set of treatment levels categorized as the binarized treatment and \(\mathcal{C}\) denote those categorized as the binarized control, such that
\begin{equation*}
\mathcal{B}, \mathcal{C} \subseteq \{1,\dots,I\}, 
\quad \mathcal{B} \cap \mathcal{C} = \emptyset, 
\quad \mathcal{B} \cup \mathcal{C} = \{1,2,\dots,I\}.
\end{equation*}
Importantly, we do not impose that all treatment levels within either $\mathcal{B}$ or $\mathcal{C}$ share the same bias indicator ($\delta_i = 0$ or $\delta_i = 1$), 
as investigators may collapse the treatment levels in a way that does not align with the underlying bias structure. 
Consider computing the collapsed odds ratio comparing two subjects $s$ and $s'$ with identical covariates $x_s = x_{s'} = x$. 
We define the following quantities:
\begin{align*}
b_1 &= \sum_{i \in \mathcal{B},\, \delta_i = 1} \exp\{\xi_i(x)\}, 
&\quad b_0 &= \sum_{i \in \mathcal{B},\, \delta_i = 0} \exp\{\xi_i(x)\}, \\
c_1 &= \sum_{i \in \mathcal{C},\, \delta_i = 1} \exp\{\xi_i(x)\}, 
&\quad c_0 &= \sum_{i \in \mathcal{C},\, \delta_i = 0} \exp\{\xi_i(x)\}.
\end{align*}
Therefore, we have:
\[
\frac{\mathbb{P}(Z_s \in \mathcal{B} \mid \mathcal{F}) \mathbb{P}(Z_{s'} \in \mathcal{C}\mid \mathcal{F})}
     {\mathbb{P}(Z_s \in \mathcal{C} \mid \mathcal{F}) \mathbb{P}(Z_{s'} \in \mathcal{B}\mid \mathcal{F})} = \frac{\left(b_1 \exp(\gamma u_s) + b_0\right)\left(c_1 \exp(\gamma u_{s'}) + c_0\right)}
         {\left(b_1 \exp(\gamma u_{s'}) + b_0\right)\left(c_1 \exp(\gamma u_{s}) + c_0\right)}. 
\]
Our goal is to show that the odds ratio between two subjects receiving either the binarized treatment or control is bounded as follows:
\begin{equation}
\begin{aligned}
\exp(-\gamma|u_s - u_{s'}|) & \leq 
\frac{\mathbb{P}(Z_s \in \mathcal{B}\mid \mathcal{F}) \mathbb{P}(Z_{s'} \in \mathcal{C}\mid \mathcal{F})}
     {\mathbb{P}(Z_s \in \mathcal{C}\mid \mathcal{F}) \mathbb{P}(Z_{s'} \in \mathcal{B}\mid \mathcal{F})} \\
& = \frac{\left(b_1 \exp(\gamma u_s) + b_0\right)\left(c_1 \exp(\gamma u_{s'}) + c_0\right)}
         {\left(b_1 \exp(\gamma u_{s'}) + b_0\right)\left(c_1 \exp(\gamma u_{s}) + c_0\right)} \\
& \leq \exp(\gamma |u_s - u_{s'}|) 
\end{aligned}
\label{eq:binarized_odds_ratio_bound}
\end{equation}
To do this, we first establish the following Lemma. 
\begin{lem}
\label{eq:odds_ratio_log_concave_inequality}
Let $a_0, a_1 \geq 0$ and $\gamma > 0$, and suppose that $a_0$ and $a_1$ are not both zero. 
Then, for any $x, y \in [0,1]$, we have
\begin{equation}
\exp(-\gamma |x - y|) 
\;\leq\; 
\frac{a_1 \exp(\gamma x) + a_0}{a_1 \exp(\gamma y) + a_0} 
\;\leq\; 
\exp(\gamma |x - y|).
\end{equation}
\end{lem}

\noindent\textit{Proof.}
If $x = y$, the inequality holds trivially. 
Suppose $x > y$. 
Since the logarithm function has a decreasing slope, we have
\begin{align*}
\frac{\log\!\left(a_1 \exp(\gamma x) + a_0\right) - \log\!\left(a_1 \exp(\gamma y) + a_0\right)}{x - y}
&\leq 
\frac{\log\!\left(a_1 \exp(\gamma x)\right) - \log\!\left(a_1 \exp(\gamma y)\right)}{x - y} \\
&= \frac{\gamma(x - y)}{x - y} = \gamma.
\end{align*}
Hence,
\[
\frac{a_1 \exp(\gamma x) + a_0}{a_1 \exp(\gamma y) + a_0}
\;\leq\; 
\exp\{\gamma (x - y)\}.
\]
For the lower bound, note that when $x > y$, the ratio 
$\frac{a_1 \exp(\gamma x) + a_0}{a_1 \exp(\gamma y) + a_0}$ 
is at least $1$. Hence, 
\[
\exp\{-\gamma |x - y|\} \leq 1 \leq 
\frac{a_1 \exp(\gamma x) + a_0}{a_1 \exp(\gamma y) + a_0}.
\]

Now consider $x < y$. 
Since $\exp\{\gamma |x - y|\} > 1$ while
\[
\frac{a_1 \exp(\gamma x) + a_0}{a_1 \exp(\gamma y) + a_0} \leq 1,
\]
we immediately obtain
\[
\frac{a_1 \exp(\gamma x) + a_0}{a_1 \exp(\gamma y) + a_0} 
\leq \exp\{\gamma |x - y|\}.
\]
To establish the lower bound, observe that
\begin{align*}
&\exp\{-\gamma |x - y|\} 
\leq \frac{a_1 \exp(\gamma x) + a_0}{a_1 \exp(\gamma y) + a_0} 
\\[3pt]
\Leftrightarrow\;& 
\gamma \geq 
\frac{\log\!\left(a_1 \exp(\gamma y) + a_0\right) - \log\!\left(a_1 \exp(\gamma x) + a_0\right)}{y - x},
\end{align*}
which again follows from the fact that the slope of the logarithm function decreases with its argument. 
\hfill$\square$

\medskip

We are now ready to prove inequality~\eqref{eq:binarized_odds_ratio_bound} by considering the following cases.

\par \noindent \textbf{Case I} \\
Assume the multinomial generic bias sensitivity model in \eqref{eq:generic_sensitivity_1} and \eqref{eq:generic_sensitivity_2} holds. 
If the collapsing is done correctly—so that the practitioner pools all treatment levels with $\delta_i = 1$ into $\mathcal{B}$—then $b_0 = c_1 = 0$.
\begin{align*}
\frac{\mathbb{P}(Z_s \in \mathcal{B} \mid \mathcal{F}) \mathbb{P}(Z_{s'} \in \mathcal{C}\mid \mathcal{F})}
     {\mathbb{P}(Z_s \in \mathcal{C}\mid \mathcal{F}) \mathbb{P}(Z_{s'} \in \mathcal{B}\mid \mathcal{F})} & = \frac{\left (b_1 \times \exp(\gamma u_s) + b_0\right)\times \left( c_1\exp(\gamma u_{s'}) + c_0\right)}{\left( b_1\times \exp(\gamma u_{s'}) + b_0 \right)\times \left(c_1\exp(\gamma u_{s}) + c_0\right)}\\
     & = \frac{b_1 c_0 \exp(\gamma u_s)}{b_1 c_0 \exp(\gamma u_{s'})} \\
     & = \exp(\gamma(u_s - u_{s'}))
\end{align*}
collapsed into the usual binary case. 
\par \noindent \textbf{Case II} \\
If the collapsing include some treatments with $\delta_i = 0$ or $1$ in $\mathcal{B}$, but only has $\delta_i =0$ in $\mathcal{C}$, then: 
\begin{align*}
\frac{\mathbb{P}(Z_s \in \mathcal{B}\mid \mathcal{F}) \mathbb{P}(Z_{s'} \in \mathcal{C}\mid \mathcal{F})}
     {\mathbb{P}(Z_s \in \mathcal{C}\mid \mathcal{F}) \mathbb{P}(Z_{s'} \in \mathcal{B}\mid \mathcal{F})} & = \frac{(b_1 \exp(\gamma u_s)+b_0) c_0}{\left( b_1\exp(\gamma u_{s'}) + b_0\right) c_0}\\
     & = \frac{b_1 \exp(\gamma u_s) + b_0}{b_1\exp(\gamma u_{s'})+ b_0}.
\end{align*}
Applying Lemma \ref{eq:odds_ratio_log_concave_inequality} gives us \eqref{eq:binarized_odds_ratio_bound}. The remaining partially aligned collapsing cases—such as when $\mathcal{B}$ contains treatment levels with $\delta_i \in \{0,1\}$ but $\mathcal{C}$ contains only $\delta_i = 1$, or vice versa—can be established analogously.

\par \noindent \textbf{Case III} \\
Suppose that $b_0$, $b_1$, $c_0$, and $c_1$ are all strictly positive. 
To establish our claim in \eqref{eq:binarized_odds_ratio_bound}, note first that the inequality in \eqref{eq:binarized_odds_ratio_bound} holds trivially when $u_s = u_{s'}$. 
Hence, it suffices to consider the two remaining cases: $u_s > u_{s'}$ and $u_s < u_{s'}$.

\smallskip
\noindent\textbf{Case 1:} $u_s > u_{s'}$. 
For the upper bound,
\begin{align*}
\frac{b_1 \exp(\gamma u_s) + b_0}{b_1 \exp(\gamma u_{s'}) + b_0} 
\frac{c_1 \exp(\gamma u_{s'}) + c_0}{c_1 \exp(\gamma u_s) + c_0}
&<
\frac{b_1 \exp(\gamma u_s) + b_0}{b_1 \exp(\gamma u_{s'}) + b_0} 
\;\leq\;
\exp\{\gamma |u_s - u_{s'}|\}.
\end{align*}
The first inequality follows from $u_s > u_{s'}$, and the second inequality applies Lemma~\ref{eq:odds_ratio_log_concave_inequality}. 

For the lower bound,
\[
\frac{b_1 \exp(\gamma u_s) + b_0}{b_1 \exp(\gamma u_{s'}) + b_0}  
\frac{c_1 \exp(\gamma u_{s'}) + c_0}{c_1 \exp(\gamma u_s) + c_0}
>
\frac{c_1 \exp(\gamma u_{s'}) + c_0}{c_1 \exp(\gamma u_s) + c_0}
\;\geq\;
\exp\{-\gamma |u_s - u_{s'}|\},
\]
where the first inequality follows from $u_s > u_{s'}$ and the second from Lemma~\ref{eq:odds_ratio_log_concave_inequality}.

\smallskip
\noindent\textbf{Case 2:} $u_s < u_{s'}$. 
The argument is analogous and yields the same bounds by symmetry.

\smallskip
\noindent Combining all three cases ($u_s = u_{s'}$, $u_s > u_{s'}$, and $u_s < u_{s'}$), we obtain the desired result in~\eqref{eq:binarized_odds_ratio_bound}.

\subsubsection*{Sensitivity Analysis Based on Taking Extreme Treatment Levels}
This subsection shows that under the generic bias sensitivity model in \eqref{eq:generic_sensitivity_1}–\eqref{eq:generic_sensitivity_2}, the 
odds ratio obtained based on the two extreme treatment levels among two subjects \(s\) and \(s'\) with identical measured covariates \(x_s = x_{s'}\) 
is bounded above by \(\exp\{\gamma |u_s - u_{s'}|\}\), the odds-ratio bound implied by~\citet{rosenbaum1990sensitivity}. 

Suppose the cross-cut table is formed by retaining two treatment levels $i$ and $i'$, 
which in the generic bias sensitivity model \eqref{eq:generic_sensitivity_1}–\eqref{eq:generic_sensitivity_2} 
are associated with bias parameters $\delta_i$ and $\delta_{i'}$, respectively. 
Consider two subjects $s$ and $s'$ with identical covariates, i.e., $x_s = x_{s'}$. 
The probability ratio of receiving treatment $i'$ versus $i$ between these two subjects is
\begin{align}
\frac{\mathbb{P}(Z_s = i' \mid \mathcal{F}) \, \mathbb{P}(Z_{s'} = i \mid \mathcal{F})}
     {\mathbb{P}(Z_s = i \mid \mathcal{F}) \, \mathbb{P}(Z_{s'} = i' \mid \mathcal{F})}
&= 
\frac{\exp\{\gamma \delta_{i'} u_s\} \exp\{\gamma \delta_i u_{s'}\}}
     {\exp\{\gamma \delta_i u_s\} \exp\{\gamma \delta_{i'} u_{s'}\}} \\
&= 
\exp\{\gamma (\delta_{i'} - \delta_i)(u_s - u_{s'})\}.
\label{eq:cross_cut_prob_ratio_generic_bias}
\end{align}

If $|\delta_{i'} - \delta_i| = 1$, then \eqref{eq:cross_cut_prob_ratio_generic_bias} gives
\begin{align}
\exp(-\gamma |u_s - u_{s'}|) 
\;\leq\;
\frac{\mathbb{P}(Z_s = i' \mid \mathcal{F}) \, \mathbb{P}(Z_{s'} = i \mid \mathcal{F})}
     {\mathbb{P}(Z_s = i \mid \mathcal{F}) \, \mathbb{P}(Z_{s'} = i' \mid \mathcal{F})}
\;\leq\;
\exp(\gamma |u_s - u_{s'}|),
\label{eq:generic_bias_implied_cross_cut_bound_1}
\end{align}
whereas if $|\delta_{i'} - \delta_i| = 0$, then
\begin{align}
\exp(-\gamma |u_s - u_{s'}|) 
\;\leq\; 
\frac{\mathbb{P}(Z_s = i' \mid \mathcal{F}) \, \mathbb{P}(Z_{s'} = i \mid \mathcal{F})}
     {\mathbb{P}(Z_s = i \mid \mathcal{F}) \, \mathbb{P}(Z_{s'} = i' \mid \mathcal{F})}
= 1 
\;\leq\; 
\exp(\gamma |u_s - u_{s'}|).
\label{eq:generic_bias_implied_cross_cut_bound_2}
\end{align}

In summary, \eqref{eq:generic_bias_implied_cross_cut_bound_1} and \eqref{eq:generic_bias_implied_cross_cut_bound_2} together imply that, 
under the generic bias model with sensitivity parameter $\gamma$ and bias vector $\bm{\delta} \in \{0,1\}^I$, 
conducting a binary sensitivity analysis on any two treatment levels in the cross-cut table remains valid.

\section{Proofs for Efficient Computation of the Exact Null Distribution} \label{app:C}

\change{This section proves Theorem~\mref{thm:recursive_probability}. We first re-express the worst-case $p$-value in terms of two kernel functions, a representation that holds for any $\bm{\delta} \in \mathbb{R}^I$ (Lemma~\ref{lem:kernel_general_delta}); Sections~\ref{app:C1}--\ref{app:C2} give closed-form formulas for the two kernels, and Section~\ref{app:C3} proves Theorem~\mref{thm:recursive_probability} by exploiting that $\bm{\delta}$ is binary in the sensitivity model~\eqref{eq:generic_bias_sensitivity_supp}.}
First, note that the derivations can be carried out using either the integer vector or the binary matrix representation of the treatment and outcome, 
with a one-to-one mapping between the two representations defined in Appendix~\ref{app:B}, equation~\eqref{eq:integer_dummy_mapping}. 
The choice of representation is solely a matter of convenience, depending on the form of the kernel function being analyzed. 
Second, once the outcome is fixed under Fisher’s sharp null hypothesis and a particular unmeasured confounder vector $\mathbf{u}$ is given, 
the only source of randomness is the permutation of treatments. 
Consequently, characterizing the probability amounts to counting the number of treatment assignments in the set $\mathcal{Z}(\mathbf{N}_{I\cdot})$ that satisfy the constraints imposed by $\mathbf{u}$ and the test statistic. 
Therefore, to facilitate discussion, we define the following two mathematical objects, which capture the cardinality of treatment permutations satisfying specific conditions related to $\mathbf{r}$ and $\mathbf{u}$.

\begin{itemize}
\item \(\mathbf{t} \in \mathbb{Z}_{\ge 0}^{I \times J}\): 
a nonnegative integer matrix whose \((i,j)\)-th entry is denoted by \(t_{ij}\), with row and column sums satisfying 
\(\mathbf{t}\mathbf{1}_J = \mathbf{N}_{I\cdot}\) and \((\mathbf{1}_I^\intercal \mathbf{t})^\intercal = \mathbf{N}_{\cdot J}\). 
Here, \(\mathbf{N}_{I\cdot}\) and \(\mathbf{N}_{\cdot J}\) denote the treatment and outcome margins defined in the main text. 
The notation \(\mathbf{t}\) is chosen to emphasize that we will later substitute specific test-defined contingency tables for \(\mathbf{t}\).

\item \(\mathbf{q} \in \mathbb{Z}_{\ge 0}^I\): 
a nonnegative integer vector whose \(i\)-th entry is \(q_i\), subject to the constraint 
\(\mathbf{1}_I^\intercal \mathbf{q} = \overline{u}\), where \(\overline{u} \in \mathbb{Z}_{\ge 0}\) is a given constant. 
We then define two kernel functions that re-express $\alpha(T,\mathbf{r},\mathbf{u})$ for a given $\mathbf{u}$:
\end{itemize}
\begin{align}
\operatorname{kernel}(\mathbf{q} \mid \mathbf{u}, \mathbf{N}_{I\cdot}) 
&= \left| \left\{ \mathbf{z} \in \mathcal{Z}(\mathbf{N}_{I\cdot}) 
\;\middle|\; \sum_{s=1}^N \mathbbm{1}(z_s = i,\ u_s = 1) = q_i,\ \forall i \right\} \right|,
\label{def:kernel_q_def} \\[1em]
\operatorname{kernel}(\mathbf{t}, \mathbf{q} \mid \mathbf{u}, \mathbf{N}_{I\cdot}, \mathbf{r}) 
&= \left| \left\{ \mathbf{z} \in \mathcal{Z}(\mathbf{N}_{I\cdot}) 
\;\middle|\; 
\begin{array}{l}
\sum_{s=1}^N \mathbbm{1}\{z_s = i,\ u_s = 1\} = q_i,\ \forall i, \\[2pt]
\sum_{s=1}^N \mathbbm{1}\{z_s = i,\ r_s = j\} = t_{ij},\ \forall i,j
\end{array}
\right\} \right|.
\label{def:kernel_t_q_def}
\end{align}

We suppress explicit indexing and write $\forall i$ and $\forall i, j$ in place of the more precise expressions $\forall i = 1,\ldots,I$ and $\forall j = 1,\ldots,J$, respectively.

The kernel functions in~\eqref{def:kernel_q_def} and~\eqref{def:kernel_t_q_def} count elements in ${\mathcal{Z}}(\mathbf{N}_{I\cdot})$ that satisfy constraints imposed by $\mathbf{q}$ alone, or by both $\mathbf{q}$ and $\mathbf{t}$, respectively. 

\begin{changeblock}
Recall from Section~\mref{sec:def_worst_case} of the main text that, for any $\bm{\delta}$ and a given $\mathbf{u}$, the $p$-value is given by equation \meqref{eq:raw_sens_test} of the main text,
\begin{equation*}
\alpha(T,\rr,\mathbf{u}) = \frac{\sum_{\mathbf{z} \in \mathcal{Z}(\mathbf{N}_{I\cdot})} \mathbbm{1}\{T(\mathbf{N}(\mathbf{z}, \rr)) \geq \criticalt\} \exp \left\{ \gamma \sum_{s=1}^{N} \sum_{i=1}^I \delta_i \mathbbm{1}\{z_s = i\} u_s \right\}}{\sum_{\mathbf{b} \in \mathcal{Z}(\mathbf{N}_{I\cdot})} \exp \left\{ \gamma \sum_{s=1}^{N} \sum_{i =1}^I \delta_i \mathbbm{1}\{b_s = i\} u_s \right\}},
\end{equation*}
a sum over all treatment assignments in $\mathcal{Z}(\mathbf{N}_{I\cdot})$. Many assignments contribute the same summand. For a binary $\mathbf{u}$, the exponent depends on $\mathbf{z}$ only through the vector $\mathbf{q}(\mathbf{z})$ whose $i$-th entry is $q_i(\mathbf{z}) = \sum_{s=1}^{N} \mathbbm{1}\{z_s = i, u_s = 1\}$, the number of subjects with $u_s = 1$ assigned to treatment level $i$, since $\sum_{s=1}^{N} \sum_{i=1}^{I} \delta_i \mathbbm{1}\{z_s = i\} u_s = \bm{\delta}^\intercal \mathbf{q}(\mathbf{z})$; and the indicator depends on $\mathbf{z}$ only through the table $\mathbf{N}(\mathbf{z}, \rr)$. Hence, instead of summing over every assignment, we can sum over the distinct values of $\mathbf{q}$ in the denominator, and of $(\mathbf{t}, \mathbf{q})$ in the numerator, and weight each value by the number of assignments that produce it. These numbers are exactly the kernels~\eqref{def:kernel_q_def} and~\eqref{def:kernel_t_q_def}, and the ranges of $\mathbf{q}$ and $\mathbf{t}$ are the sets $\Omega_{\mathbf{q}}$ and $\Omega_{\mathbf{t}}$ below. This gives the following re-expression of equation \meqref{eq:raw_sens_test} of the main text.
\begin{lem}[Kernel Computation of $\alpha(T,\rr,\mathbf{u})$ for General $\bm{\delta}$]
\label{lem:kernel_general_delta}
For a given $\mathbf{u} \in \{0,1\}^N$, let $\overline{u}=\sum_{s=1}^{N}\mathbbm{1}\{u_s=1\}$. For any $\bm{\delta} \in \mathbb{R}^I$, the exact null distribution of any count-based test statistic $T(\mathbf{N})$ can be rewritten as
\begin{align}
\alpha(T,\rr,\mathbf{u}) 
&= 
\dfrac{
\sum_{\mathbf{t} \in \Omega_{\mathbf{t}}} 
\sum_{\mathbf{q} \in \Omega_{\mathbf{q}}} 
\exp(\gamma \bm{\delta}^\intercal \mathbf{q}) 
\, \operatorname{kernel}(\mathbf{t}, \mathbf{q} \mid \mathbf{u}, \mathbf{N}_{I\cdot}, \rr)
}{
\sum_{\mathbf{q} \in \Omega_{\mathbf{q}}} 
\exp(\gamma \bm{\delta}^\intercal \mathbf{q}) 
\, \operatorname{kernel}(\mathbf{q} \mid \mathbf{u}, \mathbf{N}_{I\cdot})
},
\label{eq:kernel_general_delta}
\end{align}
where
\begin{align}
\Omega_{\mathbf{q}} 
&= 
\left\{
\mathbf{q} \in \mathbb{Z}_{\ge 0}^{I} \,\middle|\,
\sum_{i=1}^{I} q_i = \overline{u}, \;
\max(0, \overline{u} + N_{i\cdot} - N) \le q_i \le \min(\overline{u}, N_{i\cdot}), \;
\forall i
\right\}, \nonumber \\[0.4em]
\Omega_{\mathbf{t}} 
&= 
\left\{
\mathbf{t} \in \mathbb{Z}_{\ge 0}^{I \times J} \,\middle|\,
T(\mathbf{t}) \ge \criticalt, \;
\sum_{i=1}^{I} t_{ij} = N_{\cdot j}, \;
\sum_{j=1}^{J} t_{ij} = N_{i\cdot}, \;
\forall i,j
\right\}, \nonumber
\end{align}
and the kernel functions are defined in~\eqref{def:kernel_q_def} and~\eqref{def:kernel_t_q_def}.
\end{lem}
The following two lemmas, in Sections~\ref{app:C1} and~\ref{app:C2}, give closed-form formulas for the two kernels in~\eqref{eq:kernel_general_delta}, extending the results of \citet[Section~2]{pagano1981algorithm} to incorporate the unmeasured confounder~$\mathbf{u}$.
\end{changeblock}

\subsection{Derivation of the Formula for $\operatorname{kernel}(\mathbf{q} \mid \mathbf{u}, \mathbf{N}_{I\cdot})$} \label{app:C1}

We now establish the following lemma.
\begin{lem}[Closed-Form Formula of $\operatorname{kernel}(\mathbf{q} \mid \mathbf{u}, \mathbf{N}_{I\cdot})$]\mbox{}
\label{lem:kernel_q}
\noindent For a given \( \mathbf{u} \in \{0,1\}^N \), let 
\begin{align*}
\overline{u} = \sum_{s=1}^{N} u_s.
\label{eq:u_bar}
\end{align*}
Then, $\operatorname{kernel}(\mathbf{q} \mid \mathbf{u}, \mathbf{N}_{I\cdot})$ is given by
\begin{equation}
\operatorname{kernel}(\mathbf{q} \mid \mathbf{u}, \mathbf{N}_{I\cdot}) = 
\prod_{i=1}^{I-1} 
\binom{\overline{u} - A_i}{q_i} 
\binom{N - \overline{u} - B_i}{N_{i\cdot} - q_i},
\label{eq:kernel_q_formula}
\end{equation}
\noindent where  \(A_i = \sum_{e=1}^{i-1} q_e\), \(B_i = \sum_{e=1}^{i-1} (N_{e\cdot} - q_e)\), \(A_1 = 0\), and \(B_1 = 0\).
\end{lem}

The kernel computation follows an urn model. Among the \( N \) subjects, \( \overline{u} \) have \( u_s = 1 \), and the remaining \( N - \overline{u} \) have \( u_s = 0 \). The enumeration of treatment assignments satisfying the constraints imposed by \( \mathbf{q} \) is equivalent to assigning treatments based on \( \mathbf{u} \).

\noindent
\textbf{Treatment 1.} To ensure
\[
\sum_{s=1}^{N} \mathbbm{1}\{Z_s = 1, u_s = 1\}
= q_1,
\]
we select \( q_1 \) subjects with \( u_s = 1 \) from the \( \overline{u} \) available, which can be done in  
\[
\binom{\overline{u}}{q_1}
\]
ways. The remaining \( N_{1\cdot} - q_1 \) labels for treatment 1 must be assigned to subjects with \( u_s = 0 \), contributing  
\[
\binom{N - \overline{u}}{N_{1\cdot} - q_1}.
\]
Thus, the number of ways to assign treatment 1 is
\begin{equation}
\binom{\overline{u}}{q_1} 
\binom{N - \overline{u}}{N_{1\cdot} - q_1}. \label{eq:kernel_q_first_step}
\end{equation}

\noindent
\textbf{Treatment 2.} To ensure  
\[
\sum_{s=1}^{N}\mathbbm{1}\{Z_s=2,u_s=1\}= q_2,
\]
we select \( q_2 \) subjects with \( u_s = 1 \) from the \( \overline{u} - q_1 \) remaining after treatment 1, which can be done in  
\[
\binom{\overline{u} - q_1}{q_2}
\]
ways. Similarly, the remaining \( N_{2\cdot} - q_2 \) labels for treatment 2 must be assigned to subjects with \( u_s = 0 \), chosen from the \( (N - \overline{u}) - (N_{1\cdot} - q_1) \) available. This yields  
\[
\binom{(N - \overline{u}) - (N_{1\cdot} - q_1)}{N_{2\cdot} - q_2}.
\]

The process repeats for \( i = 3, 4, \dots, I-1 \). At each step, the number of remaining subjects with \( u_s = 1 \) decreases by  
\(\sum_{e=1}^{i-1} q_e\),  
and those with \( u_s = 0 \) decrease by  
\(\sum_{e=1}^{i-1} (N_{e\cdot} - q_e)\).  
Thus, for each \( i \in \{1,\dots,I-1\}\), the number of ways to make these assignments is  
\[
\binom{\overline{u} - A_i}{q_i}
\binom{(N - \overline{u}) - B_i}{N_{i\cdot} - q_i},
\]
where  
\( A_i = \sum_{e=1}^{i-1} q_e \) and  
\( B_i = \sum_{e=1}^{i-1} (N_{e\cdot} - q_e) \). 
Since both the treatment margins and \( \overline{u} \) are fixed, the last treatment group is fully determined once the first \( I-1 \) treatments are assigned. Multiplying over all \( i \) yields the claimed recursion formula.

\subsection{Derivation of the Formula for $\operatorname{kernel}(\mathbf{t},\mathbf{q} \mid \mathbf{u},\mathbf{N}_{I\cdot}, \mathbf{r})$} \label{app:C2}

We derive the expression for
\begin{equation}
\operatorname{kernel}(\mathbf{t}, \mathbf{q} \mid \mathbf{u}, \mathbf{N}_{I\cdot}, \mathbf{r})
=
\left|
\left\{
\mathbf{z} \in \mathcal{Z}(\mathbf{N}_{I\cdot})
\;\middle|\;
\begin{array}{l}
\displaystyle \sum_{s=1}^N \mathbbm{1}\{z_s = i,\, u_s = 1\} = q_i,\; \forall i, \\[4pt]
\displaystyle \sum_{s=1}^N \mathbbm{1}\{z_s = i,\, r_s = j\} = t_{ij},\; \forall i,j
\end{array}
\right\}
\right|.
\label{eq:integer_encoding_kernel_t_q}
\end{equation}
Specifically, we prove the following lemma.

\begin{lem}[Closed-Form Formula of $\operatorname{kernel}(\mathbf{t}, \mathbf{q} \mid \mathbf{u}, \mathbf{N}_{I\cdot}, \mathbf{r})$]
\label{lem:kernel_t_q}
Using the same notation and assumptions as in Lemma~\ref{lem:kernel_q}, define

\begin{equation*}
\overline{u}_j = \sum_{s=1}^{N} \mathbbm{1}\{r_s = j,\, u_s = 1\}, \quad j = 1,\ldots,J,
\quad \text{and let } \overline{u} = \sum_{j=1}^{J} \overline{u}_j.
\end{equation*}

Further, define the set
\begin{equation*}
\Omega_{\mathbf{n}} =
\Bigl\{
  \mathbf{n} = (n_{ij1})_{\substack{i = 1,\ldots,I-1 \\ j = 1,\ldots,J-1}} :
  \max\bigl(0,\, \overline{u}_j + t_{ij} - N_{\cdot j}\bigr)
  \le n_{ij1}
  \le \min\bigl(t_{ij},\, q_i,\, \overline{u}_j\bigr)
\Bigr\},
\end{equation*}
which implicitly depends on $\mathbf{t}$, $\mathbf{q}$, $\mathbf{u}$, and $\mathbf{r}$.

\noindent The kernel function can then be written as
\begin{equation*}
\operatorname{kernel}(\mathbf{t}, \mathbf{q} \mid \mathbf{u}, \mathbf{N}_{I\cdot}, \mathbf{r})
= \sum_{\mathbf{n} \in \Omega_{\mathbf{n}}} \mathcal{P}(\mathbf{n}),
\end{equation*}
where
\begin{equation}
\begin{aligned}
\mathcal{P}(\mathbf{n})
&= \prod_{i=1}^{I-1} \prod_{j=1}^{J-1}
  \binom{\overline{u}_j - C_{ij}}{n_{ij1}}
  \binom{N_{\cdot j} - \overline{u}_j - D_{ij}}{t_{ij} - n_{ij1}}  \\[4pt]
&\quad\times
  \prod_{i=1}^{I-1}
  \binom{\overline{u}_J - E_i}{\,q_i - \sum_{j=1}^{J-1} n_{ij1}\,}
  \binom{N_{\cdot J} - \overline{u}_J - F_i}{
    N_{i\cdot} - \sum_{j=1}^{J-1} t_{ij} - q_i + \sum_{j=1}^{J-1} n_{ij1}
  },
\end{aligned}
\label{eq:kernel_t_q_P_n}
\end{equation}
and
\begin{align*}
C_{ij} &= \sum_{e=1}^{i-1} n_{e j 1}, &
D_{ij} &= \sum_{e=1}^{i-1} (t_{e j} - n_{e j 1}), \\[2pt]
E_i &= \sum_{e=1}^{i-1} \Big(q_e - \sum_{j=1}^{J-1} n_{e j 1}\Big), &
F_i &= \sum_{e=1}^{i-1} \Big(N_{e\cdot} - \sum_{j=1}^{J-1} t_{e j}
        - \big(q_e - \sum_{j=1}^{J-1} n_{e j 1}\big)\Big).
\end{align*}
\end{lem}

\noindent
For notational convenience, we derive the above formula using the binary-matrix representation.
Specifically, we establish the equivalence of \eqref{eq:integer_encoding_kernel_t_q}:
\begin{equation*}
\operatorname{kernel}(\mathbf{t}, \mathbf{q} \mid \mathbf{u}, \mathbf{N}_{I\cdot}, \tilde{\mathbf{r}})
=
\Bigl|
\bigl\{
\tilde{\mathbf{z}} \in \widetilde{\mathcal{Z}}(\mathbf{N}_{I\cdot})
:\,
\tilde{\mathbf{z}}^\intercal\mathbf{u} = \mathbf{q},\;
\tilde{\mathbf{z}}^\intercal\tilde{\mathbf{r}} = \mathbf{t}
\bigr\}
\Bigr|,
\end{equation*}
where $\tilde{\mathbf{r}}$ is the binary matrix representation of $\mathbf{r}$, and each $\tilde{\mathbf{z}}\in\widetilde{\mathcal{Z}}(\mathbf{N}_{I\cdot})$ corresponds one-to-one to
$\mathbf{z}\in\mathcal{Z}(\mathbf{N}_{I\cdot})$ 
as defined in \eqref{eq:integer_dummy_mapping}–\eqref{eq:binary_matrix_permutation_treatment_set}.

\par The proof consists of two parts:  
(i) partitioning the space of $\widetilde{\mathcal{Z}}(\mathbf{N}_{I\cdot})$, and  
(ii) applying a combinatorial argument similar to that in Lemma~\ref{lem:kernel_q} to each partition.  
For notational convenience, we suppress the explicit dependence on $\tilde{\mathbf{z}} \in \widetilde{\mathcal{Z}}(\mathbf{N}_{I\cdot})$ throughout the proof.  
We also denote the index sets $E = \{1,\dots,I-1\}$ and $G = \{1,\dots,J-1\}$.

\subsubsection*{Review: the law of partitions}
\noindent Let $S$ be a set, and let $\{S_{1}, S_{2}, \dots, S_{n}\}$ be a partition of $S$. 
By definition, this means that $S_{c} \cap S_{c'} = \varnothing$ for any $c \neq c'$, and that $\bigcup_{c=1}^{n} S_{c} = S$. 
Recall that we use $\lvert \cdot \rvert$ to denote the cardinality of a set. 
Then, the law of partitions states that
\[
\lvert S \rvert 
\;=\;
\sum_{c=1}^{n} \lvert S_{c} \rvert.
\]

\subsubsection*{Forming partitions}
Fix $\mathbf{t}$, $\mathbf{q}$, $\mathbf{u}$, and $\trr$, and we define $\mathcal{A}$ and \(\mathcal{A}_{\mathbf{n}}\) for every 
\(\mathbf{n} \in \Omega_{\mathbf{n}}\) as:  
\begin{align}
\mathcal{A} & = 
\big\{\tilde{\mathbf{z}} : \tilde{\mathbf{z}}_i^\intercal \trr_j = t_{ij}, \, \tilde{\mathbf{z}}_i^\intercal \mathbf{u} = q_i, \forall i \in E, j \in G \big\}, 
\label{eq:definition_A}\\
\mathcal{A}_{\mathbf{n}} & = 
\big\{\tilde{\mathbf{z}}: \tilde{\mathbf{z}}_i^\intercal (\trr_j \circ \mathbf{u}) = n_{ij1}, \, \tilde{\mathbf{z}}_i^\intercal \trr_j = t_{ij}, \, \tilde{\mathbf{z}}_i^\intercal \mathbf{u} = q_i, \forall i \in E, \forall j \in G\big\}, \; \mathbf{n} \in \Omega_{\mathbf{n}}.
\label{eq:definition_A_n}
\end{align}
The $\circ$ denotes the Hadamard product. We aim to show that the sets \(\mathcal{A}_{\mathbf{n}}\), for all \(\mathbf{n} \in \Omega_{\mathbf{n}}\),  partition \(\mathcal{A}\).
\paragraph{$\bigcup_{\mathbf{n}\in \Omega_{\mathbf{n}}} \mathcal{A}_{\mathbf{n}}$ covers $\mathcal{A}$} 
\leavevmode \\[0.5em]
First note that for each \(i\) and \(j\),
\begin{align*}\tilde{\mathbf{z}}_i^\intercal 
    (\trr_j \circ \mathbf{u})
  \leq \min(\tilde{\mathbf{z}}_i^\intercal \trr_j, \tilde{\mathbf{z}}_i^\intercal \mathbf{u}, \trr_j^\intercal \mathbf{u}),
\end{align*}
\begin{align*}
0 \leq \tilde{\mathbf{z}}_i^\intercal \trr_j - \tilde{\mathbf{z}}_i^\intercal(\trr_j \circ \mathbf{u}) = \tilde{\mathbf{z}}_i^\intercal(\trr_j \circ(\bm{1}_{N}-\mathbf{u})) \leq \bm{1}_{N}^\intercal(\trr_j \circ (\bm{1}_{N} - \mathbf{u})).
\end{align*}
Therefore, given the constraints in $\mathcal{A}$, \(\tilde{\mathbf{z}}_i^\intercal \trr_j = t_{ij}\), \(\tilde{\mathbf{z}}_i^\intercal \mathbf{u} = q_i\) for all \(i\) and \(j\), and by definition $\overline{u}_j = \sum_{s=1}^{N}\mathbbm{1}\{r_s=j,u_s=1\} = \trr_j^\intercal \mathbf{u}$, 
\[
\max(0,\overline{u}_j+t_{ij}-N_{\cdot j}) \leq \tilde{\mathbf{z}}_i^\intercal(\trr_j \circ \mathbf{u}) \leq \min(t_{ij},q_i, \overline{u}_j).
\]

\par \noindent Since \(\Omega_{\mathbf{n}}\) includes all such possible configurations \(\mathbf{n}\), whose entries \(n_{ij1}\) range over the lower bound and upper bound imposed by $\tilde{\mathbf{z}}_i^\intercal \trr_i$, $\tilde{\mathbf{z}}_i^\intercal \mathbf{u}$, and $\trr_j^\intercal \mathbf{u}$, we obtain 
\begin{align*}\bigcup_{\mathbf{n} \in \Omega_{\mathbf{n}}} \mathcal{A}_{\mathbf{n}}=\mathcal{A}.
\end{align*}

\paragraph{ $\mathcal{A}_{\mathbf{n}}$ and $\mathcal{A}_{\mathbf{n}'}$ are disjoint.}
\leavevmode \\[0.5em]
Second, we aim to show that for $\mathbf{n},\mathbf{n}' \in \Omega_{\mathbf{n}}, \mathbf{n} \neq \mathbf{n}'$, 
\begin{align}
\mathcal{A}_{\mathbf{n}} \cap \mathcal{A}_{\mathbf{n}'} = \emptyset.
\label{eq:partition_intersection_empty}
\end{align}

\noindent We first consider a \(2 \times 2\) table, in which case the definition of $\mathcal{A}_{\mathbf{n}}$ becomes:
\[
\mathcal{A}_{\mathbf{n}} = \{\tilde{\mathbf{z}} : \tilde{\mathbf{z}}_1^\intercal (\trr_1 \circ \mathbf{u}) = n_{111},\; \tilde{\mathbf{z}}_1^\intercal \trr_1 = t_{11}, \;\tilde{\mathbf{z}}_1^\intercal \mathbf{u} = q_1\}.
\]
\textbf{Case I: If $\mathcal{A}_{\mathbf{n}} = \emptyset$.} In this case, \eqref{eq:partition_intersection_empty} gets satisfied trivially. \\  
\textbf{Case II: If $\Omega_{\mathbf{n}}$ contains only one element}. In this case, there is only one set of $\mathcal{A}_{\mathbf{n}}$, so \eqref{eq:partition_intersection_empty} gets satisfied.\\
\textbf{Case III: If $\Omega_{\mathbf{n}}$ contains at least two different elements.} Let $\mathbf{n}\neq \mathbf{n}', \mathbf{n},\mathbf{n}' \in \Omega_{\mathbf{n}}.$ Note that by \eqref{eq:definition_A_n}, this is to say $n_{111} \neq n'_{111}$ in the $2 \times 2$ table setting. Thus,
\begin{align*}
\mathcal{A}_{\mathbf{n}}
& = \Bigl\{\tilde{\mathbf{z}}:  \tilde{\mathbf{z}}_1^\intercal (\trr_1 \circ \mathbf{u}) = n_{111},  \quad \tilde{\mathbf{z}}_1^\intercal \trr_1 = t_{11}, \quad \tilde{\mathbf{z}}^\intercal \mathbf{u} = q_1\Bigr\}, \\
\mathcal{A}_{\mathbf{n}'}
& = \Bigl\{\tilde{\mathbf{z}} : \tilde{\mathbf{z}}_1^\intercal (\trr_1 \circ \mathbf{u}) = n'_{111}, \quad \tilde{\mathbf{z}}_1^\intercal \trr_1 = t_{11}, \quad \tilde{\mathbf{z}}^\intercal \mathbf{u} = q_1\Bigr\},
\end{align*}
each assumed to be nonempty. 

Suppose \(\tilde{\mathbf{z}} \in \mathcal{A}_{\mathbf{n}}\). By definition of \(\mathcal{A}_{\mathbf{n}}\), this is equivalent to  
\[
|s\in \{1,\dots,N\}:\tilde{z}_{s1}=1, \tilde{r}_{s1}=1, u_s=1| = n_{111}.
\]
In other words, the first column of $\tilde{\mathbf{z}}$ contains exactly $n_{111}$ entries equal to one at the indices
\[
\mathcal{I} := \{s \in \{1,\dots,N\} : \tilde{r}_{s1} = 1,\ u_s = 1\}.
\]
Since both $\trr$ and $\mathbf{u}$ are fixed, the index set $\mathcal{I}$ is fixed as well. Therefore, if $n_{111} \neq n'_{111}$, then $\tilde{\mathbf{z}}$ does not contain exactly $n'_{111}$ entries equal to one at the indices in $\mathcal{I}$, which implies $\tilde{\mathbf{z}} \notin \mathcal{A}_{\mathbf{n}'}$.

Consequently, for every \(\tilde{\mathbf{z}}\in \mathcal{A}_{\mathbf{n}}\), we have \(\tilde{\mathbf{z}}\notin \mathcal{A}_{\mathbf{n}'}\), implying that \(\mathcal{A}_{\mathbf{n}} \cap \mathcal{A}_{\mathbf{n}'} = \emptyset\) for \(\mathbf{n} \neq \mathbf{n}'\). To extend this proof to non \(2 \times 2\) table setting, note if \(\mathbf{n} \neq \mathbf{n}'\), there exists some \((i, j)\) such that \(n_{ij1} \neq n'_{ij1}\). The proof follows by applying the same argument to the \(i\)-th treatment column and \(j\)-th outcome column.

\paragraph{Establishing the formula}   
\begin{align}
& \bigl|\,
  \tilde{\mathbf{z}}
  : 
  \tilde{\mathbf{z}}^\intercal \trr= \mathbf{t},\,
  \tilde{\mathbf{z}}^\intercal \mathbf{u} = \mathbf{q}
\bigr| \notag \\
=
& \bigl|\,
  \tilde{\mathbf{z}}
  :
\tilde{\mathbf{z}}_i^\intercal \trr_j = t_{ij},\,
\tilde{\mathbf{z}}_i^\intercal  \mathbf{u} = q_i,
  \ \forall\, i \in E,\; j \in G
\bigr| \notag \\
=
&\sum_{\Omega_{\mathbf{n}}}
\bigl|\,
  \tilde{\mathbf{z}}
  :
\tilde{\mathbf{z}}_i^\intercal (\trr_j \circ \mathbf{u}) = n_{ij1},\,  \tilde{\mathbf{z}}_i^\intercal \trr_j = t_{ij},\,
\tilde{\mathbf{z}}_i^\intercal \mathbf{u} = q_i,\,
  \ \forall\, i \in E,\; j \in G
\bigr|.
\label{eq:compute_element_cardinality}
\end{align}
Where we obtain \eqref{eq:compute_element_cardinality} with the law of partitions. With some rewriting based on the constraints that $\mathbf{1}_{N} = \sum_{i=1}^{I}\tilde{\mathbf{z}}_i = \sum_{j=1}^{J}\trr_j$,  \eqref{eq:compute_element_cardinality} can be equivalently expressed as
\begin{equation*}
\sum_{\Omega_{\mathbf{n}}}
\Bigl|\,
\left\{
  \widetilde{\mathbf{z}} :
\begin{aligned}
&\tilde{\mathbf{z}}_i^\top 
  \bigl(\trr_j \circ \mathbf{u}\bigr) 
  = n_{ij1},
\;
\tilde{\mathbf{z}}_i^\top 
  \bigl(\trr_j \circ (\mathbf{1}_{N} - \mathbf{u})\bigr)
  = t_{ij} - n_{ij1},
&& \forall\, i \in E,\; j \in G,\\[6pt]
&\tilde{\mathbf{z}}_i^\top 
  \bigl(\trr_J \circ \mathbf{u}\bigr)
  = q_i - \sum_{j=1}^{J-1} n_{ij1},
&& \forall\, i \in E,\\[6pt]
&\tilde{\mathbf{z}}_i^\top\bigl(\trr_J \circ (\bm{1}_{N}- \mathbf{u})\bigr)
  = N_{i.} - \sum_{j=1}^{J-1} t_{ij}
    - q_i + \sum_{j=1}^{J-1} n_{ij1},
&& \forall\, i \in E
\end{aligned}
\right\}
\,\Bigr|.
\label{events_dof_sum}
\end{equation*}
This problem can be viewed as another urn model. Compared to Lemma \ref{lem:kernel_q}, each subject now has two key attributes: (a) their (fixed) outcome and (b) their unmeasured confounder value.  

We begin with subjects at \( r_s = 1 \) and \( u_s = 1 \). To satisfy \( \tilde{\mathbf{z}}_1^\intercal(\trr_1 \circ \mathbf{u}) = n_{111} \), we assign treatment level 1 to \( n_{111} \) out of \( \overline{u}_1 \) subjects, yielding the binomial coefficient
\begin{equation}
\binom{\overline{u}_1}{n_{111}}. \label{eq:kernel_t_q_first_binomial_u1}
\end{equation}
Next, to satisfy \( \tilde{\mathbf{z}}_2^\intercal(\trr_1 \circ \mathbf{u}) = n_{211} \), we assign treatment level 2 to \( n_{211} \) subjects from the remaining \( \overline{u}_1 - n_{111} \), yielding \[
\binom{\overline{u}_1 - n_{111}}{n_{211}}. \]
This process continues for additional treatment levels, with available subjects adjusted at each step, reflected in the term \( C_{ij} \).  

Next, for subjects at \( r_s = 1 \) and \( u_s = 0 \), to satisfy \( \tilde{\mathbf{z}}_1^\intercal (\trr_1 \circ (\mathbf{1}_N - \mathbf{u})) = t_{11} - n_{111} \), we assign treatment 1 to \( t_{11} - n_{111} \) subjects from the \( N_{.1} - \overline{u}_1 \) total, yielding
\begin{equation}
\binom{N_{.1} - \overline{u}_1}{t_{11} - n_{111}}. \label{eq:kernel_t_q_first_binomial_u0}
\end{equation}
Similarly, for \( \tilde{\mathbf{z}}_2^\intercal (\trr_1 \circ (\mathbf{1}_N - \mathbf{u})) = t_{21} - n_{211} \), we assign treatment 2 to \( t_{21} - n_{211} \) subjects from the remaining \( (N_{.1} - \overline{u}_1) - (t_{11} - n_{111}) \), yielding \[
\binom{(N_{.1} - \overline{u}_1) - (t_{11} - n_{111})}{t_{21} - n_{211}}. \] For more than two treatments, this process repeats, reflected in the term $D_{ij}$. By the same reasoning, the recursive terms \(E_i\) and \(F_i\) account for the remaining outcome level \(J\), reflecting the treatment assignments among subjects with \(u_s = 1\) and \(u_s = 0\), respectively.

We remark that, by the kernel expressions in Sections~\ref{app:C1}–\ref{app:C2}, only the counts 
$\overline{\mathbf{u}}$ are needed for computation. In particular, Lemma~\ref{lem:kernel_q} implies that the
denominator part of $\alpha$ depends only on
\[
\overline{u}\;:=\;\mathbf{1}_J^{\intercal}\overline{\mathbf{u}}
\;=\;\sum_{j=1}^J \overline{u}_j,
\]
the total number of ones in the unmeasured confounder vector, while Lemma~\ref{lem:kernel_t_q} implies that the numerator depends on the full vector
\[
\overline{\mathbf{u}}=(\overline{u}_1,\ldots,\overline{u}_J)^\intercal,
\qquad
\overline{u}_j:=\sum_{s=1}^{N}\mathbbm{1}\{u_s=1,\ r_s=j\}.
\]
From an algorithmic standpoint, it suffices to enumerate the feasible
$\overline{\mathbf{u}}$ consistent with the margins $\{N_{.j}\}_{j=1}^J$ (i.e.,
$0\le \overline{u}_j\le N_{.j}$ for each $j$). 

\begin{changeblock}
\subsection{Proof of Theorem~\mref{thm:recursive_probability}} \label{app:C3}

\noindent
Using the same notation as in Lemmas~\ref{lem:kernel_q} and~\ref{lem:kernel_t_q}, we restate the objects defined in Theorem~\mref{thm:recursive_probability}. Let $\bm{\delta} \in \{0,1\}^I$, $\mathcal{I}_1 = \{i : \delta_i = 1\}$, and $\overline{N} = \sum_{i \in \mathcal{I}_1} N_{i\cdot}$, the number of subjects assigned to a treatment level in $\mathcal{I}_1$. For a given $\mathbf{u} \in \{0,1\}^N$, $\overline{u}_j = \sum_{s=1}^{N}\mathbbm{1}\{r_s = j, u_s = 1\}$ and $\overline{u} = \sum_{j=1}^{J}\overline{u}_j$ are as in Lemma~\ref{lem:kernel_t_q}. For a table $\mathbf{t} \in \mathbb{Z}_{\ge 0}^{I \times J}$, $\overline{t}_j = \sum_{i \in \mathcal{I}_1} t_{ij}$ and $\overline{\mathbf{t}} = (\overline{t}_1, \ldots, \overline{t}_J)$. For an assignment $\mathbf{z} \in \mathcal{Z}(\mathbf{N}_{I\cdot})$, write $\mathbf{t}(\mathbf{z}) = \mathbf{N}(\mathbf{z},\rr)$ for the table it produces, $q_i(\mathbf{z}) = \sum_{s=1}^{N}\mathbbm{1}\{z_s = i, u_s = 1\}$ as in Lemma~\ref{lem:kernel_q}, and $\overline{q}(\mathbf{z}) = \sum_{i \in \mathcal{I}_1} q_i(\mathbf{z})$, the number of subjects with $u_s = 1$ assigned to a level in $\mathcal{I}_1$. The sets $\Omega_{\overline{q}}$, $\Omega_{\overline{\mathbf{n}}}$, and $\Omega_{\mathbf{t}}$ are as in Theorem~\mref{thm:recursive_probability}, and the multinomial constants, whose meaning is established at \eqref{eq:thm4_fibers} below, are
\begin{equation}
\kappa(\mathbf{t}) = \prod_{j=1}^{J} \frac{\overline{t}_j!\,(N_{\cdot j} - \overline{t}_j)!}{\prod_{i=1}^{I} t_{ij}!},
\qquad
\kappa_0 = \frac{\overline{N}!\,(N - \overline{N})!}{\prod_{i=1}^{I} N_{i\cdot}!}. \label{eq:kappa_def}
\end{equation}

\noindent
\textbf{Binary $\bm{\delta}$.} Since $\mathbf{u}$ is binary, $\sum_{s=1}^{N}\sum_{i=1}^{I}\delta_i \mathbbm{1}\{z_s = i\} u_s = \sum_{i \in \mathcal{I}_1} q_i(\mathbf{z}) = \overline{q}(\mathbf{z})$, so equation \meqref{eq:raw_sens_test} of the main text becomes
\begin{equation}
\alpha(T,\rr,\mathbf{u}) = \frac{\sum_{\mathbf{z} \in \mathcal{Z}(\mathbf{N}_{I\cdot})} \mathbbm{1}\{T(\mathbf{t}(\mathbf{z})) \ge \criticalt\}\, \exp(\gamma \overline{q}(\mathbf{z}))}{\sum_{\mathbf{z} \in \mathcal{Z}(\mathbf{N}_{I\cdot})} \exp(\gamma \overline{q}(\mathbf{z}))}.
\label{eq:alpha_grouped_by_z}
\end{equation}
The weight depends on $\mathbf{z}$ only through $\overline{q}(\mathbf{z})$, that is, only on whether each subject is assigned to a level in $\mathcal{I}_1$ or not.

\noindent
\textbf{Two treatment groups.} Consequently, Lemma~\ref{lem:kernel_general_delta} and the closed forms in Lemmas~\ref{lem:kernel_q} and~\ref{lem:kernel_t_q} only need to distinguish two treatment groups: $\mathcal{I}_1$, which we treat as the first treatment group, and $\{1,\ldots,I\} \setminus \mathcal{I}_1$, which we treat as the second. For $\mathbf{z} \in \mathcal{Z}(\mathbf{N}_{I\cdot})$, let $\overline{\mathbf{z}} \in \{1,2\}^N$ be the two-level assignment with $\overline{z}_s = 1$ if $z_s \in \mathcal{I}_1$ and $\overline{z}_s = 2$ otherwise. Then $\overline{\mathbf{z}}$ has exactly $\overline{N}$ entries equal to $1$ and $N - \overline{N}$ entries equal to $2$; we write $\overline{\mathbf{N}} = (\overline{N}, N - \overline{N})$ for these two-level treatment margins and, as for $\mathcal{Z}(\mathbf{N}_{I\cdot})$ in the main text, $\mathcal{Z}(\overline{\mathbf{N}})$ for the set of all two-level assignments of the $N$ subjects with margins $\overline{\mathbf{N}}$, so that $\overline{\mathbf{z}} \in \mathcal{Z}(\overline{\mathbf{N}})$. The quantities defined above for $I$-level assignments apply to two-level assignments with $I = 2$: $q_1(\overline{\mathbf{z}}) = \sum_{s=1}^{N} \mathbbm{1}\{\overline{z}_s = 1, u_s = 1\}$ is the number of subjects with $u_s = 1$ in the first group, so that $\overline{q}(\mathbf{z}) = q_1(\overline{\mathbf{z}})$ and $q_2(\overline{\mathbf{z}}) = \overline{u} - q_1(\overline{\mathbf{z}})$; $\mathbf{N}(\overline{\mathbf{z}}, \rr)$ is the $2 \times J$ table whose $(1,j)$ entry is $\sum_{s=1}^{N} \mathbbm{1}\{\overline{z}_s = 1, r_s = j\}$ and whose $(2,j)$ entry is $\sum_{s=1}^{N} \mathbbm{1}\{\overline{z}_s = 2, r_s = j\}$; and the kernels \eqref{def:kernel_q_def} and \eqref{def:kernel_t_q_def} with $\mathcal{Z}(\overline{\mathbf{N}})$ in place of $\mathcal{Z}(\mathbf{N}_{I\cdot})$ count two-level assignments, so that $\operatorname{kernel}((q_1, q_2) \mid \mathbf{u}, \overline{\mathbf{N}})$ is the number of $\overline{\mathbf{z}} \in \mathcal{Z}(\overline{\mathbf{N}})$ with $q_1(\overline{\mathbf{z}}) = q_1$, and $\operatorname{kernel}(\overline{\mathbf{t}}, (q_1, q_2) \mid \mathbf{u}, \overline{\mathbf{N}}, \rr)$ is the number of those with, in addition, $\mathbf{N}(\overline{\mathbf{z}}, \rr) = \overline{\mathbf{t}}$. Merging the $I$ treatment levels into the two groups turns the treatment margins $\mathbf{N}_{I\cdot}$ into $\overline{\mathbf{N}}$, the vector $\mathbf{q}$ into $(\overline{q}, \overline{u} - \overline{q})$, and each column $(t_{1j}, \ldots, t_{Ij})$ of $\mathbf{t}$ into $(\overline{t}_j, N_{\cdot j} - \overline{t}_j)$; we write $\overline{\mathbf{t}}(\mathbf{t})$ for the resulting $2 \times J$ table. Lemmas~\ref{lem:kernel_q} and~\ref{lem:kernel_t_q} then apply with $I = 2$ and $\mathbf{N}_{I\cdot} = \overline{\mathbf{N}}$, in which case the products over $i = 1, \ldots, I-1$ in \eqref{eq:kernel_q_formula} and in \eqref{eq:kernel_t_q_P_n} have the single factor $i = 1$, and the recursion terms vanish: $A_1 = B_1 = 0$ in Lemma~\ref{lem:kernel_q}, and $C_{1j} = D_{1j} = E_1 = F_1 = 0$ in Lemma~\ref{lem:kernel_t_q}. Thus \eqref{eq:kernel_q_formula} reduces to
\begin{equation}
\operatorname{kernel}((\overline{q}, \overline{u} - \overline{q}) \mid \mathbf{u}, \overline{\mathbf{N}}) = \binom{\overline{u}}{\overline{q}}\binom{N - \overline{u}}{\overline{N} - \overline{q}},
\label{eq:thm4_kernel_q_two_groups}
\end{equation}
the first-step factor \eqref{eq:kernel_q_first_step} with $\overline{q}$ in place of $q_1$ and $\overline{N}$ in place of $N_{1\cdot}$; and, writing $\overline{n}_j = n_{1j1}$ for $j \le J-1$ and $\overline{n}_J = \overline{q} - \sum_{j=1}^{J-1} \overline{n}_j$, the summand \eqref{eq:kernel_t_q_P_n} of Lemma~\ref{lem:kernel_t_q} reduces to
\begin{equation}
\mathcal{P}(\mathbf{n}) = \prod_{j=1}^{J} \binom{\overline{u}_j}{\overline{n}_j}\binom{N_{\cdot j} - \overline{u}_j}{\overline{t}_j - \overline{n}_j},
\label{eq:thm4_P_two_groups}
\end{equation}
the product over outcome levels of the first-step factors \eqref{eq:kernel_t_q_first_binomial_u1} and~\eqref{eq:kernel_t_q_first_binomial_u0} with $\overline{n}_j$ in place of $n_{1j1}$ and $\overline{t}_j$ in place of $t_{1j}$; for $j = J$ the two factors come from the last two factors of \eqref{eq:kernel_t_q_P_n} with $E_1 = F_1 = 0$, since $\overline{N} - \sum_{j=1}^{J-1} \overline{t}_j = \overline{t}_J$. These are the binomial coefficients in equation \meqref{eq:exact_distribution_kernel_form} of Theorem~\mref{thm:recursive_probability} in the main text. Merging, however, loses part of the combinatorics. Several treatment levels are collected in $\mathcal{I}_1$, and several in its complement, so Lemmas~\ref{lem:kernel_q} and~\ref{lem:kernel_t_q} applied with $I = 2$ count two-level assignments $\overline{\mathbf{z}}$, whereas \eqref{eq:alpha_grouped_by_z} sums over the $I$-level assignments $\mathbf{z}$. Each two-level assignment $\overline{\mathbf{z}}$ is the image of many $\mathbf{z}$: all the ways of distributing the subjects placed in the first group among the levels $i \in \mathcal{I}_1$, and the subjects placed in the second group among the levels $i \notin \mathcal{I}_1$, with the prescribed margins. To recover the sum over $\mathbf{z}$, each two-level count must therefore be multiplied back by the number of $\mathbf{z}$ that merge to the same $\overline{\mathbf{z}}$. These numbers are the cardinalities
\begin{equation}
\big|\{\mathbf{z} : \overline{\mathbf{z}} \text{ given}\}\big| = \kappa_0,
\qquad
\big|\{\mathbf{z} : \overline{\mathbf{z}} \text{ given},\ \mathbf{t}(\mathbf{z}) = \mathbf{t}\}\big| = \kappa(\mathbf{t}) \quad \text{whenever } \mathbf{N}(\overline{\mathbf{z}}, \rr) = \overline{\mathbf{t}}(\mathbf{t}),
\label{eq:thm4_fibers}
\end{equation}
The first count is obtained as follows. Fix $\overline{\mathbf{z}} \in \mathcal{Z}(\overline{\mathbf{N}})$. An assignment $\mathbf{z} \in \mathcal{Z}(\mathbf{N}_{I\cdot})$ merges to $\overline{\mathbf{z}}$ exactly when the $\overline{N}$ subjects with $\overline{z}_s = 1$ receive levels in $\mathcal{I}_1$ and the $N - \overline{N}$ subjects with $\overline{z}_s = 2$ receive levels outside $\mathcal{I}_1$, with $N_{i\cdot}$ subjects at each level $i$. The first group can be distributed among the levels $i \in \mathcal{I}_1$ with counts $N_{i\cdot}$ in the multinomial number $\overline{N}!/\prod_{i \in \mathcal{I}_1} N_{i\cdot}!$ of ways, and, independently, the second group among the levels $i \notin \mathcal{I}_1$ with counts $N_{i\cdot}$ in $(N - \overline{N})!/\prod_{i \notin \mathcal{I}_1} N_{i\cdot}!$ ways; the product is $\kappa_0$ in \eqref{eq:kappa_def}. The second count is the same reasoning once the table is fixed. If $\mathbf{t}(\mathbf{z}) = \mathbf{t}$ is required in addition, then within each outcome level $j$ the $\overline{t}_j$ subjects with $\overline{z}_s = 1$ and $r_s = j$ must be distributed among the levels $i \in \mathcal{I}_1$ with counts $t_{ij}$, in $\overline{t}_j!/\prod_{i \in \mathcal{I}_1} t_{ij}!$ ways, and the $N_{\cdot j} - \overline{t}_j$ subjects with $\overline{z}_s = 2$ and $r_s = j$ among the levels $i \notin \mathcal{I}_1$ with counts $t_{ij}$, in $(N_{\cdot j} - \overline{t}_j)!/\prod_{i \notin \mathcal{I}_1} t_{ij}!$ ways. The choices at different outcome levels are independent, and the row margins $\sum_{j=1}^{J} t_{ij} = N_{i\cdot}$ hold automatically because $\mathbf{t}$ has these margins, so the product over $j$ is $\kappa(\mathbf{t})$ in \eqref{eq:kappa_def}. Thus $\kappa_0$ counts the $I$-level assignments behind each two-level assignment, and $\kappa(\mathbf{t})$ counts those producing the table $\mathbf{t}$; both depend only on the margins and on $\mathbf{t}$, not on $\mathbf{u}$, which is why they sit outside the binomial coefficients in equation \meqref{eq:exact_distribution_kernel_form} of the main text.

\noindent
\textbf{Denominator.} Grouping the assignments by $\overline{q}(\mathbf{z})$, as in Lemma~\ref{lem:kernel_general_delta}, and then by $\overline{\mathbf{z}}$,
\begin{align}
\sum_{\mathbf{z} \in \mathcal{Z}(\mathbf{N}_{I\cdot})} \exp(\gamma \overline{q}(\mathbf{z}))
&= \sum_{\overline{q} \in \Omega_{\overline{q}}} \exp(\gamma \overline{q}) \, \big|\{\mathbf{z} : \overline{q}(\mathbf{z}) = \overline{q}\}\big| \nonumber \\
&= \sum_{\overline{q} \in \Omega_{\overline{q}}} \exp(\gamma \overline{q}) \, \kappa_0 \, \big|\{\overline{\mathbf{z}} \in \mathcal{Z}(\overline{\mathbf{N}}) : q_1(\overline{\mathbf{z}}) = \overline{q}\}\big|,
\label{eq:thm4_partition_q}
\end{align}
where $\overline{q}(\mathbf{z})$ ranges over $\Omega_{\overline{q}}$ because $\overline{q}(\mathbf{z}) \le \min(\overline{u}, \overline{N})$ and $\overline{N} - \overline{q}(\mathbf{z}) \le N - \overline{u}$, and the second equality is \eqref{eq:thm4_fibers}. The last cardinality is $\operatorname{kernel}((\overline{q}, \overline{u} - \overline{q}) \mid \mathbf{u}, \overline{\mathbf{N}})$ by definition~\eqref{def:kernel_q_def}, so \eqref{eq:thm4_kernel_q_two_groups} gives
\begin{equation}
\sum_{\mathbf{z} \in \mathcal{Z}(\mathbf{N}_{I\cdot})} \exp(\gamma \overline{q}(\mathbf{z}))
= \sum_{\overline{q} \in \Omega_{\overline{q}}} \exp(\gamma \overline{q})\, \kappa_0 \binom{\overline{u}}{\overline{q}}\binom{N - \overline{u}}{\overline{N} - \overline{q}}.
\label{eq:thm4_denominator}
\end{equation}

\noindent
\textbf{Numerator.} Grouping the assignments by $\mathbf{t}(\mathbf{z})$, then by $\overline{q}(\mathbf{z})$, as in Lemma~\ref{lem:kernel_general_delta}, and then by $\overline{\mathbf{z}}$,
\begin{equation}
\sum_{\mathbf{z} \in \mathcal{Z}(\mathbf{N}_{I\cdot})} \mathbbm{1}\{T(\mathbf{t}(\mathbf{z})) \ge \criticalt\}\, \exp(\gamma \overline{q}(\mathbf{z}))
= \sum_{\mathbf{t} \in \Omega_{\mathbf{t}}} \kappa(\mathbf{t}) \sum_{\overline{q} \in \Omega_{\overline{q}}} \exp(\gamma \overline{q}) \, \operatorname{kernel}\big(\overline{\mathbf{t}}(\mathbf{t}), (\overline{q}, \overline{u} - \overline{q}) \mid \mathbf{u}, \overline{\mathbf{N}}, \rr\big),
\label{eq:thm4_partition_t_q}
\end{equation}
by \eqref{eq:thm4_fibers} and definition~\eqref{def:kernel_t_q_def}. Lemma~\ref{lem:kernel_t_q} with $I = 2$ evaluates the kernel as $\sum_{\mathbf{n} \in \Omega_{\mathbf{n}}} \mathcal{P}(\mathbf{n})$ with $\mathcal{P}(\mathbf{n})$ as in \eqref{eq:thm4_P_two_groups}, and summing over $\overline{q} \in \Omega_{\overline{q}}$ and $\mathbf{n} \in \Omega_{\mathbf{n}}$ is the same as summing over $\overline{\mathbf{n}} \in \Omega_{\overline{\mathbf{n}}}$, with $\overline{n}_J = \overline{q} - \sum_{j=1}^{J-1} \overline{n}_j$. Hence
\begin{equation}
\sum_{\mathbf{z} \in \mathcal{Z}(\mathbf{N}_{I\cdot})} \mathbbm{1}\{T(\mathbf{t}(\mathbf{z})) \ge \criticalt\}\, \exp(\gamma \overline{q}(\mathbf{z}))
= \sum_{\mathbf{t} \in \Omega_{\mathbf{t}}} \sum_{\overline{\mathbf{n}} \in \Omega_{\overline{\mathbf{n}}}} \exp\Big(\gamma \sum_{j=1}^{J} \overline{n}_j\Big)\, \kappa(\mathbf{t}) \prod_{j=1}^{J} \binom{\overline{u}_j}{\overline{n}_j}\binom{N_{\cdot j} - \overline{u}_j}{\overline{t}_j - \overline{n}_j}.
\label{eq:thm4_numerator}
\end{equation}
Dividing \eqref{eq:thm4_numerator} by \eqref{eq:thm4_denominator} gives equation \meqref{eq:exact_distribution_kernel_form} of the main text, which proves Theorem~\mref{thm:recursive_probability}.

\paragraph{Equivalent forms of equation \meqref{eq:exact_distribution_kernel_form} of the main text.}
Starting from equation \meqref{eq:exact_distribution_kernel_form} of the main text, the numerator can be rearranged as follows. The inner sum over $\overline{\mathbf{n}}$ depends on $\mathbf{t}$ only through $\overline{\mathbf{t}} = (\overline{t}_1, \ldots, \overline{t}_J)$, and on $\mathbf{u}$ only through $(\overline{u}_1, \ldots, \overline{u}_J)$, so
\begin{align}
&\phantom{=} \sum_{\mathbf{t} \in \Omega_{\mathbf{t}}} \sum_{\overline{\mathbf{n}} \in \Omega_{\overline{\mathbf{n}}}} \exp\Big(\gamma \sum_{j=1}^{J} \overline{n}_j\Big)\, \kappa(\mathbf{t}) \prod_{j=1}^{J} \binom{\overline{u}_j}{\overline{n}_j}\binom{N_{\cdot j} - \overline{u}_j}{\overline{t}_j - \overline{n}_j} \nonumber \\
&= \sum_{\mathbf{t} \in \Omega_{\mathbf{t}}} \kappa(\mathbf{t}) \sum_{\overline{\mathbf{n}} \in \Omega_{\overline{\mathbf{n}}}} \exp\Big(\gamma \sum_{j=1}^{J} \overline{n}_j\Big) \prod_{j=1}^{J} \binom{\overline{u}_j}{\overline{n}_j}\binom{N_{\cdot j} - \overline{u}_j}{\overline{t}_j - \overline{n}_j} \label{eq:equiv_step1} \\[0.4em]
&= \sum_{\overline{\mathbf{t}}} \Bigg( \sum_{\substack{\mathbf{t} \in \Omega_{\mathbf{t}} \\ \sum_{i \in \mathcal{I}_1} t_{ij} = \overline{t}_j\ \forall j}} \kappa(\mathbf{t}) \Bigg) \nonumber \\
&\qquad \times \sum_{\overline{\mathbf{n}} \in \Omega_{\overline{\mathbf{n}}}} \exp\Big(\gamma \sum_{j=1}^{J} \overline{n}_j\Big) \prod_{j=1}^{J} \binom{\overline{u}_j}{\overline{n}_j}\binom{N_{\cdot j} - \overline{u}_j}{\overline{t}_j - \overline{n}_j}, \label{eq:equiv_step2}
\end{align}
where \eqref{eq:equiv_step1} moves $\kappa(\mathbf{t})$, which does not depend on $\overline{\mathbf{n}}$, outside the inner sum, and \eqref{eq:equiv_step2} groups the tables $\mathbf{t} \in \Omega_{\mathbf{t}}$ by the value of $\overline{\mathbf{t}}$, the outer sum running over the distinct values of $\overline{\mathbf{t}}$ attained on $\Omega_{\mathbf{t}}$; the inner sum is the same for all $\mathbf{t}$ with the same $\overline{\mathbf{t}}$, because both $\Omega_{\overline{\mathbf{n}}}$ and the summand depend on $\mathbf{t}$ only through $\overline{\mathbf{t}}$. Within the inner sum, the terms can be collected further by the value of $\overline{q} = \sum_{j=1}^{J} \overline{n}_j$, on which the exponential factor depends:
\begin{align}
\sum_{\overline{\mathbf{n}} \in \Omega_{\overline{\mathbf{n}}}} \exp\Big(\gamma \sum_{j=1}^{J} \overline{n}_j\Big) \prod_{j=1}^{J} \binom{\overline{u}_j}{\overline{n}_j}\binom{N_{\cdot j} - \overline{u}_j}{\overline{t}_j - \overline{n}_j}
&= \sum_{\overline{q} = 0}^{\min(\overline{u}, \overline{N})} \exp(\gamma \overline{q}) \sum_{\substack{\overline{\mathbf{n}} \in \Omega_{\overline{\mathbf{n}}} \\ \overline{n}_1 + \cdots + \overline{n}_J = \overline{q}}} \prod_{j=1}^{J} \binom{\overline{u}_j}{\overline{n}_j}\binom{N_{\cdot j} - \overline{u}_j}{\overline{t}_j - \overline{n}_j}, \label{eq:equiv_step3}
\end{align}
since $0 \le \sum_{j=1}^{J} \overline{n}_j \le \sum_{j=1}^{J} \min(\overline{u}_j, \overline{t}_j) \le \min(\overline{u}, \overline{N})$ for every $\overline{\mathbf{n}} \in \Omega_{\overline{\mathbf{n}}}$. That is, defining
\begin{align}
S(\overline{\mathbf{t}}, \mathbf{u}) &= \sum_{\overline{\mathbf{n}} \in \Omega_{\overline{\mathbf{n}}}} \exp\Big(\gamma \sum_{j=1}^{J} \overline{n}_j\Big) \prod_{j=1}^{J} \binom{\overline{u}_j}{\overline{n}_j}\binom{N_{\cdot j} - \overline{u}_j}{\overline{t}_j - \overline{n}_j} \nonumber \\
&= \sum_{\overline{q} = 0}^{\min(\overline{u}, \overline{N})} \exp(\gamma \overline{q}) \sum_{\substack{\overline{\mathbf{n}} \in \Omega_{\overline{\mathbf{n}}} \\ \overline{n}_1 + \cdots + \overline{n}_J = \overline{q}}} \prod_{j=1}^{J} \binom{\overline{u}_j}{\overline{n}_j}\binom{N_{\cdot j} - \overline{u}_j}{\overline{t}_j - \overline{n}_j}, \label{eq:S_def}\\[0.4em]
S_0(\mathbf{u}) &= \sum_{\overline{q} \in \Omega_{\overline{q}}} \exp(\gamma \overline{q}) \binom{\overline{u}}{\overline{q}}\binom{N - \overline{u}}{\overline{N} - \overline{q}}, \label{eq:S0_def}\\[0.4em]
\mathcal{W}(\overline{\mathbf{t}}) &= \sum_{\substack{\mathbf{t} \in \Omega_{\mathbf{t}} \\ \sum_{i \in \mathcal{I}_1} t_{ij} = \overline{t}_j\ \forall j}} \kappa(\mathbf{t}), \label{eq:W_def}
\end{align}
and noting that the denominator of equation \meqref{eq:exact_distribution_kernel_form} of the main text is $\kappa_0 S_0(\mathbf{u})$, we have
\begin{equation}
\alpha(T,\rr,\mathbf{u})
= \frac{\sum_{\mathbf{t} \in \Omega_{\mathbf{t}}} \kappa(\mathbf{t})\, S(\overline{\mathbf{t}}, \mathbf{u})}{\kappa_0\, S_0(\mathbf{u})}
= \frac{\sum_{\overline{\mathbf{t}}} \mathcal{W}(\overline{\mathbf{t}})\, S(\overline{\mathbf{t}}, \mathbf{u})}{\kappa_0\, S_0(\mathbf{u})},
\label{eq:alpha_W_form}
\end{equation}
where the first expression is \eqref{eq:equiv_step1} and the second is \eqref{eq:equiv_step2}. The function $\mathcal{W}$ depends on the observed margins and on the test statistic, but not on $\mathbf{u}$ or on $\gamma$.
\end{changeblock}

\subsection{Proof of Corollary \mref{cor:unique_corner_multivariate_hyper}} \label{app:C4}

Similar to the previous lemma, we use the binary matrix representation for notational convenience.

We recall that the probability mass function of an \(I\)-dimensional random vector \(\mathbf{t}\) following a multivariate extended hypergeometric distribution is given by \citep[Equations~(7)--(9)]{Fog2008}:
\[
\mathbb{P}(\mathbf{t} \mid n, \bm{m}, \bm{\omega}) = 
\frac{\prod_{i=1}^{I} \binom{m_i}{t_i} \omega_i^{t_i}}
     {\sum_{\bm{q} \in \mathcal{S}} \prod_{i=1}^{I} \binom{m_i}{q_i} \omega_i^{q_i}},
\]
where the probability mass function is parameterized by \(n\), \(\bm{m}\), and \(\bm{\omega}\),
and the corresponding distribution is defined over the support set
\[
\mathcal{S} = \Bigl\{ \bm{t} \in \mathbb{Z}_{0+}^I \,\Big|\, \sum_{i=1}^{I} t_i = n \Bigr\}.
\]

Assume that \(\mathbf{u}^+ = \tilde{\mathbf{r}}_j\) and that the sensitivity model in \eqref{eq:generic_bias_sensitivity_supp} is applied with \(\bm{\delta} = (\delta_1,\allowbreak \delta_2, \allowbreak \dots, \delta_I)\). 
Furthermore, note that the probability mass of an \(I \times 2\) contingency table is fully determined by the joint counts of a single column when the margins are fixed. 
The numerator of the probability mass function is then given by
\begin{align*}
& \sum_{\tilde{\mathbf{z}} \in \widetilde{\mathcal{Z}}(\mathbf{N}_{I\cdot})} 
    \mathbbm{1}\left\{
        \tilde{\mathbf{z}}_1^\intercal \tilde{\mathbf{r}}_j = t_1, \dots, 
        \tilde{\mathbf{z}}_I^\intercal \tilde{\mathbf{r}}_j = t_I
    \right\} 
    \exp\!\left\{\gamma \bm{\delta}^\intercal \tilde{\mathbf{z}}^\intercal \mathbf{u}^+\right\} \\
&= \sum_{\tilde{\mathbf{z}} \in \widetilde{\mathcal{Z}}(\mathbf{N}_{I\cdot})} 
    \mathbbm{1}\left\{
        \tilde{\mathbf{z}}_1^\intercal \tilde{\mathbf{r}}_j = t_1, \dots, 
        \tilde{\mathbf{z}}_I^\intercal \tilde{\mathbf{r}}_j = t_I
    \right\}
    \exp\!\left\{\gamma \bm{\delta}^\intercal \tilde{\mathbf{z}}^\intercal \tilde{\mathbf{r}}_j\right\} \\
&= \Bigl| 
    \bigl\{ \tilde{\mathbf{z}} \in \widetilde{\mathcal{Z}}(\mathbf{N}_{I\cdot}) 
    : \tilde{\mathbf{z}}_1^\intercal \tilde{\mathbf{r}}_j = t_1, \dots, 
      \tilde{\mathbf{z}}_I^\intercal \tilde{\mathbf{r}}_j = t_I \bigr\} 
  \Bigr| 
  \prod_{i=1}^I \exp\!\left\{\gamma \delta_i t_i\right\} \\
& \propto \prod_{i=1}^I \binom{N_{i\cdot}}{t_i} 
   \exp\!\left\{\gamma \delta_i t_i\right\},
   \qquad \sum_{i=1}^{I} t_i = N_{\cdot j}.
\end{align*}

The first equality uses the assumption that \(\mathbf{u}^+ = \tilde{\mathbf{r}}_j\); 
the second equality rewrites the summation as a cardinality; 
and note that when \(\tilde{\mathbf{z}}_i^\intercal \tilde{\mathbf{r}}_j = t_i\) for all \(i = 1,\dots,I\),
\[
\exp\!\left\{\gamma \bm{\delta}^\intercal (\tilde{\mathbf{z}}^\intercal \tilde{\mathbf{r}}_j)\right\} 
= \exp\!\left\{\gamma \sum_{i=1}^{I} \delta_i t_i\right\} 
= \prod_{i=1}^{I} \exp\!\left\{\gamma \delta_i t_i\right\}.
\]
The third equality follows since there are \(\binom{N_{i\cdot}}{t_i}\) possible treatment assignments for which \(\tilde{\mathbf{z}}_i^\intercal \tilde{\mathbf{r}}_j = t_i\) for all \(i\), under the constraint that 
\((\sum_{i=1}^{I}\tilde{\mathbf{z}}_i)^\intercal \tilde{\mathbf{r}}_j = \mathbf{1}_N^\intercal \tilde{\mathbf{r}}_j = N_{\cdot j}\).

The denominator of the probability is obtained by summing over all possible \((t_1,t_2,\dots,t_I)\), giving
\begin{align}
\mathbb{P}\bigl(
\tilde{\mathbf{Z}}_1^\intercal \tilde{\mathbf{r}}_j = t_1, \dots, 
\tilde{\mathbf{Z}}_I^\intercal \tilde{\mathbf{r}}_j = t_I
\mid \mathcal{F}, \mathbf{u}^+ = \tilde{\mathbf{r}}_j
\bigr)
= 
\frac{\prod_{i=1}^I \binom{N_{i\cdot}}{t_i} \exp\!\left\{\gamma \delta_i t_i\right\}}
     {\sum\limits_{q_1 + \dots + q_I = N_{\cdot j}}
      \Bigl(\prod_{i=1}^I \binom{N_{i\cdot}}{q_i} \exp\!\left\{\gamma \delta_i q_i\right\}\Bigr)}.
\label{multivariate_hyper_sensitivity}
\end{align}

This expression matches the general form of the multivariate extended hypergeometric distribution with parameters
\[
\bm{m} = \mathbf{N}_{I\cdot} = (N_{1\cdot}, N_{2\cdot}, \dots, N_{I\cdot}), 
\quad n = N_{\cdot j}, 
\quad \bm{\omega} = \exp(\gamma \bm{\delta}) = (\exp(\gamma \delta_1), \exp(\gamma \delta_2), \dots, \exp(\gamma \delta_I)).
\]

Therefore, if \(\mathbf{u}^+ = \tilde{\mathbf{r}}_2\) as given in Theorem~\mref{thm:score_test_maximizer_binary}, 
the cell counts \((N_{12}, \ldots, N_{I2})\) from an \(I\times 2\) contingency table 
under the generic bias sensitivity model follow a multivariate extended hypergeometric distribution 
with parameters \(\mathbf{N}_{I\cdot}\), \(N_{\cdot 2}\), and odds \((\exp(\gamma \delta_1),\ldots,\exp(\gamma \delta_I))\).

\section{Proofs for Exact Moments of Ordinal Tests} \label{app:D}
\label{sec:Appendix_exact_moments}

Similarly to Appendices~\ref{app:B} and~\ref{app:C}, the derivations can be performed using either an integer representation or a binary matrix representation. For notational convenience, we adopt the binary matrix representation in these derivations, while the more intuitive integer representation is maintained in the main text.

We further note that all derivations in this appendix assume that Fisher’s sharp null hypothesis $H_0$ holds and that the treatment assignment follows the generic bias sensitivity model, so the resulting distribution depends on the fixed unmeasured confounder vector $\mathbf{u}$. For notational simplicity, we suppress this dependence in what follows. In the main text, however, we retain the subscript $H_0$ to emphasize that the expectations and probabilities are taken under the sharp null hypothesis, in contrast to the favorable situation discussed in the power analysis of Section~\mref{sec:power_main} of the main text.

We state here the exact first and second moments of cell counts in an \( I \times J \) contingency table under Fisher's null hypothesis in equation~\meqref{eq:null_sharp} of the main text; the remainder of this appendix proves the result. These moments can be used to construct a normal approximation of ordinal and sign-score tests and complement our results on characterizing the exact worst-case null distribution.
\setcounter{thm}{0}\renewcommand{\thethm}{\thesection.\arabic{thm}}
\begin{thm}[Exact First and Second Moments of $N_{ij}$]\mbox{}
\label{prop:first_moment_table}
For every $\mathbf{u}\in \{0,1\}^N$,
let $\overline{u}_j = \sum_{s=1}^{N}\mathbbm{1}\{r_s=j,u_s=1\}$ and $\overline{u}=\sum_{s=1}^{N}\mathbbm{1}\{u_s=1\}$. For $k \in \{0,1\}$, let $\rho_{j,k} = \sum_{s=1}^N \mathbbm{1}\{r_s=j, u_s = k\} / \sum_{s=1}^N \mathbbm{1}\{u_s = k\}$ and $w_{j,k} = \frac{1}{n_k-1}\sum_{s:\,u_s=k}\big(\mathbbm{1}\{r_s=j\} - \rho_{j,k}\big)^2$, where $n_k = \sum_{s=1}^N \mathbbm{1}\{u_s=k\}$. Under equation~\meqref{eq:null_sharp} of the main text and given $\mathbf{u}$, the mean and the variance of the cell counts $N_{ij}$ are
\begin{equation*}
\begin{aligned}
\mathbb{E}_{H_0}\big[N_{ij} \big]
&= \frac{\overline{u}_j}{\overline{u}}\, \mathbb{E}_{H_0} \big[ Q_i \big] + \frac{N_{\cdot j} - \overline{u}_j}{N - \overline{u}} \biggl( N_{i\cdot} - \mathbb{E}_{H_0} \big[ Q_i\big] \biggr), \\
\text{Var}_{H_0}\big[N_{ij} \big]
&= (w_{j,1} - w_{j,0})\, \mathbb{E}_{H_0} \bigl[ Q_i \bigr]  - \Big( \frac{w_{j,1}}{\overline{u}} + \frac{w_{j,0}}{N - \overline{u}} \Big) \Big( \mathbb{E}_{H_0}^2\big[ Q_i\big] +\text{Var}_{H_0}\big[Q_i\big] \Big)  \\
&\; + \frac{N_{i\cdot} \bigl( N - \overline{u} - N_{i\cdot} + 2\, \mathbb{E}_{H_0} \big[ Q_i\big] \bigr) w_{j,0}}{N - \overline{u}} + \text{Var}_{H_0} \big[ Q_i \big]
\cdot \bigl( \rho_{j,1} - \rho_{j,0} \bigr)^2, \\
Q_{i} &= \sum_{s=1}^{N} \mathbbm{1}\{Z_s=i, u_s=1\}.
\end{aligned}
\end{equation*}
The moments of $Q_i$'s can be computed from the exact, joint distribution of $Q_i$, which is
\begin{equation*}
\begin{aligned}
\mathbb{P} \Big( (Q_1,\ldots,Q_I) = \mathbf{q} \;\Big|\; \mathbf{u},\ \mathbf{Z} \in \mathcal{Z}(\mathbf{N}_{I\cdot}) \Big)
= \frac{\exp(\gamma \bm{\delta}^\intercal \mathbf{q}) \cdot \operatorname{kernel}(\mathbf{q} \mid \mathbf{N}_{I\cdot}, \mathbf{u})}{\sum_{\mathbf{q} \in \Omega_{\mathbf{q}}} \exp(\gamma \bm{\delta}^\intercal \mathbf{q}) \cdot \operatorname{kernel}(\mathbf{q} \mid \mathbf{N}_{I\cdot}, \mathbf{u})}.
\end{aligned}
\end{equation*}
The formulas for the covariances of $N_{ij}$ are in Section~\ref{app:D3}.
\end{thm}
An immediate consequence of Theorem~\ref{prop:first_moment_table} is that we can derive the exact moments for ordinal tests. For example, the exact, first moment of the ordinal test under the null in equation~\meqref{eq:null_sharp} of the main text is
$\mathbb{E}_{H_0} \left[\sum_{i=1}^{I} \sum_{j=1}^J w_i v_j N_{ij} \right] = \sum_{i=1}^{I} \sum_{j=1}^J w_i v_j \mathbb{E}_{H_0}[N_{ij}].$
Computing $\mathbb{E}_{H_0}[N_{ij}]$ above only requires computing $\mathbb{E}_{H_0}[Q_{i}]$, which can be computed using the joint distribution specified in Theorem~\ref{prop:first_moment_table}. Importantly, the joint distribution depends on only one kernel function $\operatorname{kernel}(\mathbf{q} \mid \mathbf{N}_{I\cdot}, \mathbf{u})$, which is easier to compute than the two sums in equation~\meqref{eq:exact_distribution_kernel_form} of the main text.

\subsection{Proof of the First Moment} \label{app:D1}

Before deriving the first moment of the cell counts, we first present the distribution of the random vector
\begin{align*}
(Q_1,\ldots, Q_I) 
&= \bigl(\sum_{s=1}^{N}\mathbbm{1}\{Z_s=1,u_s=1\},\ldots,\sum_{s=1}^{N}\mathbbm{1}\{Z_s=I,u_s=1\}\bigr)\\
&= (\widetilde{\mathbf{Z}}_1^\intercal \mathbf{u},\ldots, \widetilde{\mathbf{Z}}_I^\intercal \mathbf{u}).
\end{align*}
Similar to the exact computation of $\alpha(T,\rr,\mathbf{u})$, to achieve higher computational efficiency, 
we express the distribution of this vector in terms of the kernels. 

\par \noindent \textbf{Distribution of $(Q_1,\ldots,Q_I)$}\\  
Consider a fixed vector \(\mathbf{u} \in \{0,1\}^N\) with \(\sum_{s=1}^N\mathbbm{1}\{u_s=1\} = \overline{u}\). 
Under the generic bias sensitivity model in \eqref{eq:generic_bias_sensitivity_supp}, 
\begin{equation*}
\label{eq:distribution_Zu}
\begin{aligned}
\mathbb{P} \Big( (Q_1,\ldots,Q_I) = \mathbf{q} \;\Big|\; \mathbf{u},\ \mathbf{Z} \in \mathcal{Z}(\mathbf{N}_{I\cdot}) \Big)
= \frac{\exp(\gamma \bm{\delta}^\intercal \mathbf{q}) \cdot \operatorname{kernel}(\mathbf{q} \mid \mathbf{N}_{I\cdot}, \mathbf{u})}
       {\sum_{\mathbf{q} \in \Omega_{\mathbf{q}}} \exp(\gamma \bm{\delta}^\intercal \mathbf{q}) \cdot \operatorname{kernel}(\mathbf{q} \mid \mathbf{N}_{I\cdot}, \mathbf{u})}.
\end{aligned}
\end{equation*}

The support \(\Omega_{\mathbf{q}}\) is given by
\[
\Omega_{\mathbf{q}} = \left\{ \mathbf{q} \in \mathbb{Z}_{\geq 0}^{I} \;\middle|\; 
\max(0, N_{i\cdot} + \overline{u} - N) \leq q_i \leq \min(N_{i\cdot}, \overline{u}),\ \forall i,\ 
\mathbf{1}^\intercal \mathbf{q} = \overline{u} \right\}.
\]

Here we start to derive the first moment of the cell counts in terms of the moments involving $Q_i$'s. The proof of the first moment adapts the approach of \citet[Proposition~20, Chapter~4]{rosenbaum2002observational} 
to derive the expectation of each cell count. We consider one binary column vector 
\(\widetilde{\mathbf{Z}}_i\) and one binary column vector \(\tilde{\mathbf{r}}_j\) at a time.

By the law of total expectation,
\[
\mathbb{E}\left[\widetilde{\mathbf{Z}}_i^\intercal \tilde{\mathbf{r}}_j\right] 
= \mathbb{E}\left[\mathbb{E}\left[\widetilde{\mathbf{Z}}_i^\intercal \tilde{\mathbf{r}}_j 
  \mid \widetilde{\mathbf{Z}}_i^\intercal \mathbf{u}\right]\right].
\]
We recall that the exact conditional distribution of $\mathbf{Z}$ under the generic bias sensitivity model is
\begin{equation}
\mathbb{P}\!\left(\mathbf{Z} = \mathbf{z} 
     \mid \mathcal{F}, \mathcal{Z}(\mathbf{N}_{I\cdot})\right)
= 
\frac{\exp \!\left\{ \gamma \sum_{s=1}^{N} \sum_{i =1}^{I} \delta_i \mathbbm{1}(z_s = i) u_s \right\}}
     {\sum_{\mathbf{b} \in \mathcal{Z}(\mathbf{N}_{I\cdot})} 
      \exp \!\left\{ \gamma \sum_{s=1}^{N} \sum_{i=1}^{I} \delta_i \mathbbm{1}(b_s = i) u_s \right\}}.
\label{eq:conditional_distribution_treatment_with_sensitivity_supp}
\end{equation}

Given \(Q_i = q_i\), 
the column vector \(\widetilde{\mathbf{Z}}_i\) follows a uniform distribution over the subset 
of binary column vectors that satisfy \(Q_i = \widetilde{\mathbf{Z}}_i^\intercal \mathbf{u} = q_i\), 
where \(\widetilde{\mathbf{Z}}_i\) denotes the \(i\)-th column of 
\(\widetilde{\mathbf{Z}} \in \widetilde{\mathcal{Z}}(\mathbf{N}_{I\cdot})\). 
In other words, for all 
\(\tilde{\mathbf{z}}, \tilde{\mathbf{w}} \in \widetilde{\mathcal{Z}}(\mathbf{N}_{I\cdot})\) such that 
\(\tilde{\mathbf{z}}_i^\intercal \mathbf{u} = \tilde{\mathbf{w}}_i^\intercal \mathbf{u} = q_i\),
\[
\mathbb{P}\bigl(\widetilde{\mathbf{Z}}_i = \tilde{\mathbf{z}}_i \mid Q_i = q_i\bigr)
=
\mathbb{P}\bigl(\widetilde{\mathbf{Z}}_i = \tilde{\mathbf{w}}_i \mid Q_i = q_i\bigr).
\]
\par\noindent Consequently,
\begin{equation}
\begin{aligned}
\mathbb{E}\!\left[\widetilde{\mathbf{Z}}_i^\intercal \tilde{\mathbf{r}}_j \mid Q_i=q_i\right]
&= \sum_{s=1}^{N} \tilde{r}_{sj}\,\mathbb{E}\!\left[\widetilde{Z}_{si}\mid Q_i=q_i\right] \\
& = \sum_{s:u_s=1}\tilde{r}_{sj}\,\mathbb{E}\!\left[\widetilde{Z}_{si}\mid Q_i=q_i\right] 
   +\sum_{s:u_s=0}\tilde{r}_{sj}\,\mathbb{E}\!\left[\widetilde{Z}_{si}\mid Q_i=q_i\right] \\
&= \sum_{s:u_s=1} \tilde{r}_{sj}\,\frac{q_i}{\overline{u}}
   \;+\; \sum_{s:u_s=0} \tilde{r}_{sj}\,\frac{N_{i\cdot}-q_i}{N-\overline{u}} \\
&= q_i\,\rho_{j,1} + (N_{i\cdot}-q_i)\,\rho_{j,0},
\end{aligned}
\label{eq:conditional_expectation_cell_count_on_u}
\end{equation}
where 
\begin{align}
\rho_{j,1} 
= \frac{1}{\overline{u}}\sum_{s: u_s=1} \tilde{r}_{sj}, 
\quad
\rho_{j,0} 
= \frac{1}{N - \overline{u}}\sum_{s: u_s=0} \tilde{r}_{sj},
\label{def:q_outcome_score}
\end{align}
are the averages of the \( j \)-th outcome scores among subjects with \( u_s = 1 \) and \( u_s = 0 \), respectively. 
The third equality follows from the uniformity: to satisfy $\widetilde{\mathbf{Z}}_i^\intercal \mathbf{u} = q_i$, 
we must choose $q_i$ subjects with $u_s=1$ and $N_{i\cdot}-q_i$ subjects with $u_s=0$ to receive treatment $i$, 
and every such selection is equally likely under the uniform randomization. In other words, once we condition on \(Q_i = q_i\), 
\(\widetilde{\mathbf{Z}}_i^\intercal \tilde{\mathbf{r}}_j\) can be viewed as
the sum of \(q_i\) of the \(j\)-th outcome scores randomly selected from those with \(u_s=1\), 
plus the sum of \((N_{i\cdot}-q_i)\) \(j\)-th outcome scores randomly selected from those with \(u_s=0\).

Taking the expectation over \(Q_i\) and using the definition 
\(\overline{u}_j = \tilde{\mathbf{r}}_j^\intercal \mathbf{u}\), we obtain
\[
\mathbb{E}\left[\widetilde{\mathbf{Z}}_i^\intercal \tilde{\mathbf{r}}_j\right]
= \frac{\overline{u}_j}{\overline{u}} \,\mathbb{E}\left[Q_i\right]
+ \frac{N_{.j} - \overline{u}_j}{N - \overline{u}}
  \left(N_{i\cdot} - \mathbb{E}\left[Q_i\right]\right).
\]

\subsection{Proof of the Second Moment} \label{app:D2}
This proof derives the variance of each cell count.
Let \( Q_i = \widetilde{\mathbf{Z}}_i^\intercal \mathbf{u} \), as defined previously.  
Applying the law of total variance:  
\begin{align}
\operatorname{Var}\left[\widetilde{\mathbf{Z}}_i^\intercal \trr_j\right] = \mathbb{E}\left[\operatorname{Var}\left(\widetilde{\mathbf{Z}}_i^\intercal \trr_j \mid Q_i\right)\right] + \operatorname{Var}\left[\mathbb{E}\left(\widetilde{\mathbf{Z}}_i^\intercal \trr_j \mid Q_i\right)\right].
\label{eq:the_law_of_total_variance}
\end{align}

First, consider \(\operatorname{Var}\left[\widetilde{\mathbf{Z}}_i^\intercal \trr_j \mid Q_i = q_i\right]\), which represents the variance of a sum consisting of \( q_i \) of the \( j \)-th outcome scores randomly selected from the \( \overline{u} \) subjects with \( u_s = 1 \), plus \( N_{i.} - q_i \) of the \( j \)-th outcome scores randomly selected from the \( N - \overline{u} \) subjects with \( u_s = 0 \).

Recall that for a finite population \( A = (a_1, \dots, a_N) \), 
the variance of the sum of \( n \) elements sampled without replacement is given by
\[
\operatorname{Var}\!\left[\sum_{s=1}^N \mathbbm{1}\{s \in \mathcal{S}\}\, a_s \right] 
= \frac{n(N-n)}{N} \operatorname{Var}[A],
\]
where \(\mathcal{S} \subset \{1,\dots,N\}\) denotes a random subset of size \(n\), and
\[
\operatorname{Var}[A] 
= \frac{1}{N-1} \sum_{s=1}^N (a_s - \overline{a})^2, 
\quad \text{with} \quad 
\overline{a} = \frac{1}{N} \sum_{s=1}^N a_s.
\]

\par \noindent Applying this result to our setting gives:
\begin{align}
\operatorname{Var}\left[\widetilde{\mathbf{Z}}_i^\intercal \trr_j \mid Q_i=q_i\right] = \frac{q_i(\overline{u}-q_i)}{\overline{u}} w_{j,1} + \frac{(N_{i.}-q_i)(N-\overline{u}-N_{i.}+q_i)}{N-\overline{u}} w_{j,0}.
\label{eq:conditional_variance}
\end{align}
where 
\begin{align}
w_{j,1} = \frac{1}{\overline{u} - 1} \sum_{s:u_s=1} (\tilde{r}_{sj} - \rho_{j,1})^2,  
\quad \text{and} \quad  
w_{j,0} = \frac{1}{N - \overline{u} - 1} \sum_{s:u_s=0} (\tilde{r}_{sj} - \rho_{j,0})^2.
\end{align}

\par \noindent Substituting the results from \eqref{eq:conditional_expectation_cell_count_on_u} and \eqref{eq:conditional_variance} into \eqref{eq:the_law_of_total_variance}, we obtain:
\begin{align*}
\operatorname{Var}\left[\widetilde{\mathbf{Z}}_i^\intercal \trr_j\right] 
& = \mathbb{E}\left[\frac{Q_i(\overline{u}-Q_i)}{\overline{u}} w_{j,1} + \frac{(N_{i.}-Q_i)(N-\overline{u}-N_{i.}+Q_i)}{N-\overline{u}} w_{j,0} \right] \\
&\quad + \operatorname{Var}\left[Q_i \rho_{j,1} + (N_{i.}-Q_i) \rho_{j,0}\right]\\
&= (w_{j,1}-w_{j,0}) \mathbb{E}\left[Q_i\right]- \left(\operatorname{Var}\left[Q_i\right] 
+ \mathbb{E}^2\left[Q_i\right]\right)
\Big(\frac{w_{j,1}}{\overline{u}} + \frac{w_{j,0}}{N-\overline{u}}\Big) \\
& \quad + \frac{N_{i.}\left(N-\overline{u}-N_{i.}+2\mathbb{E}[Q_i]\right)w_{j,0}}{N-\overline{u}}+ \operatorname{Var}[Q_i]\left(\rho_{j,1}-\rho_{j,0}\right)^2. 
\end{align*}

\subsection{Proof of the Covariance between Cell Counts} \label{app:D3}
\noindent We here provide formulas for the covariance between cell counts under the generic bias sensitivity model in \eqref{eq:generic_bias_sensitivity_supp} at a given $\mathbf{u}\in \{0,1\}^N$. Recall that $N_{ij} = \sum_{s=1}^{N}\mathbbm{1}\{Z_s=i,r_s=j\} = \widetilde{\mathbf{Z}}_i^\intercal \trr_j$.
\par \noindent \textbf{Case 1-1: different outcome levels $j \neq j'$, same treatment level, $1<\overline{u}<N-1$} 
\begin{equation*}
\begin{aligned}
\text{Cov}\big[N_{ij}, 
N_{ij'} \big]
& = \frac{{\rho}_{j,1} {\rho}_{j',1}}{\overline{u} - 1} 
\Bigl(\mathbb{E}\big[Q_i^2 \bigr] 
- \overline{u}\cdot \mathbb{E}\bigl[Q_i \bigr]\Bigr) \\  
&\quad + \frac{\rho_{j,0} \rho_{j',0}}{N - \overline{u} - 1} 
N_{i.}(N_{i.} - N + \overline{u}) \\  
&\quad + \frac{\rho_{j,0} \rho_{j',0}}{N - \overline{u} - 1} 
\Bigl(\mathbb{E}\bigl[Q_i^2\bigr] 
- (2 N_{i.} - N + \overline{u}) 
\mathbb{E}\bigl[Q_i \bigr]\Bigr) \\  
&\quad + \Bigl(\rho_{j,1} \rho_{j',1} 
+ \rho_{j,0} \rho_{j',0} 
- \rho_{j,1} \rho_{j',0} 
- \rho_{j',1} \rho_{j,0}\Bigr) 
\text{Var}\bigl[Q_i\bigr].
\end{aligned}
\label{eq:first_pair_covariance}
\end{equation*}
\par \noindent \textbf{Case 1-2: different outcome levels $j \neq j'$, same treatment level, $\overline{u}=N-1$} \mbox{}
\begin{equation*}
\begin{aligned}
\text{Cov}\big[N_{ij}, 
N_{ij'} \big]& = \frac{\rho_{j,1} \rho_{j',1}}{(\overline{u} - 1)} 
\Bigl(\mathbb{E}\left[Q_i^2 \right] 
- \overline{u} \cdot \mathbb{E}\left[Q_i\right]\Bigr)\\  
&\quad + \Bigl(\rho_{j,1} \rho_{j',1} 
+ \rho_{j,0} \rho_{j',0} 
- \rho_{j,1} \rho_{j',0} 
- \rho_{j',1} \rho_{j,0}\Bigr) 
\text{Var}\left[Q_i\right]. 
\end{aligned}
\end{equation*}
\par \noindent \textbf{Case 1-3: different outcome levels $j \neq j'$, same treatment level, $\overline{u}=1$} \mbox{}
\begin{equation*}
\begin{aligned}
\text{Cov}\big[N_{ij}, 
N_{ij'} \big]& = \frac{\rho_{j,0} \rho_{j',0}}{(N - \overline{u} - 1)} 
N_{i.}\left(N_{i.} - N + \overline{u}\right) \\  
&\quad + \frac{\rho_{j,0} \rho_{j',0}}{(N - \overline{u} - 1)} 
\Bigl(\mathbb{E}\left[Q_i^2\right] 
- (2 N_{i.} - N + \overline{u}) 
\mathbb{E}\left[Q_i\right]\Bigr) \\  
&\quad + \Bigl(\rho_{j,1} \rho_{j',1} 
+ \rho_{j,0} \rho_{j',0} 
- \rho_{j,1} \rho_{j',0} 
- \rho_{j',1} \rho_{j,0}\Bigr) 
\text{Var}\left[Q_i\right].
\end{aligned}
\end{equation*}
\par \noindent \textbf{Case 1-4: different outcome levels $j \neq j'$, same treatment level, $\overline{u}=0$ or $N$} \mbox{}
\begin{equation*}
\text{Cov}\big[N_{ij}, 
N_{ij'}\big] = \frac{-N_{\cdot j}N_{\cdot j'}N_{i\cdot}(N-N_{i\cdot})}{N^2(N-1)}.
\end{equation*}

\noindent \textbf{Case 2-1: same outcome level, different treatment levels $i \neq i'$, $1< \overline{u}<N-1$}  
\begin{equation*}
\begin{aligned}
\text{Cov}\big[N_{ij}, 
N_{i'j} \big] & = \Big(\rho_{j,1}^2 + \rho_{j,0}^2 - 2 \rho_{j,1} \rho_{j,0}\Big) 
\text{Cov}\big[Q_i,
Q_{i'}\big]  \\  
&\quad - \frac{\overline{u}_j(\overline{u}-\overline{u}_j)}{\overline{u}^2(\overline{u}-1)}
\mathbb{E}\big[Q_i \cdot 
Q_{i'}\big]  \\   
&\quad+\frac{(N_{.j} - \overline{u}_j) (N_{.j}-\overline{u}_j-N+\overline{u})}{(N - \overline{u})^2 (N - \overline{u} - 1)} \Big(\mathbb{E}\big[Q_i\cdot Q_{i'} \big] + N_{i.}N_{i'.}\Big) \\
&\quad-\frac{(N_{.j} - \overline{u}_j) (N_{.j}-\overline{u}_j-N+\overline{u})}{(N - \overline{u})^2 (N - \overline{u} - 1)} \Big(N_{i.}\mathbb{E}\big[Q_{i'}\big] +N_{i'.}\mathbb{E}\big[Q_{i}\big]\Big) 
\end{aligned}
\end{equation*}
\noindent \textbf{Case 2-2: same outcome level, different treatment levels $i \neq i'$, $\overline{u} = N-1$}  \mbox{}
\begin{equation*}
\begin{aligned}
\text{Cov}\big[N_{ij}, 
N_{i'j} \big] & =  
\left(\rho_{j,1}^2 + \rho_{j,0}^2 - 2 \rho_{j,1} \rho_{j,0}\right) 
\text{Cov}\left[Q_i, 
Q_{i'}\right]  \\  
&\quad - \frac{\overline{u}_j(\overline{u}-\overline{u}_j)}{\overline{u}^2(\overline{u}-1)}
\mathbb{E}\left[Q_i \cdot 
Q_{i'}\right].
\end{aligned}
\end{equation*}
\par \noindent \textbf{Case 2-3: same outcome level, different treatment levels $i \neq i'$, $\overline{u}=1$} \mbox{}
\begin{equation*}
\begin{aligned}
\text{Cov}\big[N_{ij}, 
N_{i'j} \big] & =  
\Big(\rho_{j,1}^2 + \rho_{j,0}^2 - 2 \rho_{j,1} \rho_{j,0}\Big) 
\text{Cov}\left[Q_i, 
Q_{i'}\right]  \\  
&\quad+\frac{(N_{.j} - \overline{u}_j) (N_{.j}-\overline{u}_j-N+\overline{u})}{(N - \overline{u})^2 (N - \overline{u} - 1)} \Big(\mathbb{E}\left[Q_i\cdot Q_{i'}\right] + N_{i.}N_{i'.}\Big) \\
&\quad-\frac{(N_{.j} - \overline{u}_j) (N_{.j}-\overline{u}_j-N+\overline{u})}{(N - \overline{u})^2 (N - \overline{u} - 1)} \Big(N_{i.}\mathbb{E}\left[Q_{i'}\right] +N_{i'.}\mathbb{E}\left[Q_i\right]\Big). 
\end{aligned}
\end{equation*}
\par \noindent \textbf{Case 2-4: same outcome level, different treatment levels $i \neq i'$, $\overline{u} = 0$ or $N$}
\begin{equation*}
\text{Cov}\big[N_{ij}, 
N_{i'j} \big] = \frac{-N_{\cdot j}(N-N_{\cdot j})N_{i\cdot}N_{i'\cdot}}{N^2(N-1)}.
\end{equation*}

\noindent \textbf{Case 3-1: different outcome level $j \neq j'$, different treatment levels $i \neq i'$, $1<\overline{u}<N-1$}  
\begin{equation*}
\begin{aligned}
\text{Cov}\big[N_{ij}, N_{i'j'}\big]
 & =  
\Big(\rho_{j,1} \rho_{j',1} + \rho_{j,0} \rho_{j',0} - \rho_{j,1} \rho_{j',0} - \rho_{j',1} \rho_{j,0}\Big) \text{Cov}\big[Q_i, Q_{i'} \big] \\
& \quad +\frac{\rho_{j,1} \rho_{j',1}}{(\overline{u} - 1)} \mathbb{E}\big[Q_i \cdot Q_{i'}\big]  \\
&\quad + \frac{\rho_{j,0} \rho_{j',0}}{(N - \overline{u} - 1)}\Big(\mathbb{E}
\big[Q_i \cdot Q_{i'}\big]+N_{i.}N_{i'.}\Big) \\
& \quad - \frac{\rho_{j,0} \rho_{j',0}}{(N - \overline{u} - 1)}\Big(N_{i.} \mathbb{E}\big[Q_{i'}\big]+ N_{i'.} \mathbb{E}\big[Q_i\big]\Big)  
\end{aligned}
\end{equation*}
\par \noindent \textbf{Case 3-2: different outcome level $j \neq j'$, different treatment levels $i \neq i'$, $\overline{u} = N-1$}\mbox{}
\begin{equation*}
\begin{aligned}
\text{Cov}\big[N_{ij}, N_{i'j'} \big]& =  
\Big(\rho_{j,1} \rho_{j',1} + \rho_{j,0} \rho_{j',0} - \rho_{j,1} \rho_{j',0} - \rho_{j',1} \rho_{j,0}\Big) \text{Cov}\left[Q_i, Q_{i'}\right] \\
& \quad +\frac{\rho_{j,1} \rho_{j',1}}{(\overline{u} - 1)} \mathbb{E}\left[Q_i \cdot Q_{i'}\right].
\end{aligned}
\end{equation*}
\par \noindent \textbf{Case 3-3: different outcome level $j \neq j'$, different treatment levels $i \neq i'$, $\overline{u}=1$}
\begin{equation*}
\begin{aligned}
\text{Cov}\big[N_{ij}, N_{i'j'} \big] & =  
\Big(\rho_{j,1} \rho_{j',1} + \rho_{j,0} \rho_{j',0} - \rho_{j,1} \rho_{j',0} - \rho_{j',1} \rho_{j,0}\Big) \text{Cov}\left[Q_i, Q_{i'}\right] \\
&\quad + \frac{\rho_{j,0} \rho_{j',0}}{(N - \overline{u} - 1)}\Big(\mathbb{E}
\big(Q_i \cdot Q_{i'}\big)+N_{i.}N_{i'.}\Big) \\
& \quad - \frac{\rho_{j,0} \rho_{j',0}}{(N - \overline{u} - 1)}\Big(N_{i.} \mathbb{E}\left[Q_{i'} \right]+ N_{i'.} \mathbb{E}\left[Q_{i}\right]\Big). 
\end{aligned}
\end{equation*}
\par \noindent \textbf{Case 3-4: different outcome level $j \neq j'$, different treatment levels $i \neq i'$, $\overline{u} = 0$ or $N$} 
\begin{equation*}
\text{Cov}_{H_0}\big[N_{ij}, N_{i'j'} \big]
= \frac{N_{\cdot j}N_{\cdot j'}N_{i\cdot}N_{i'\cdot}}{N^2(N-1)}.
\end{equation*}

\bigskip
\bigskip

We now derive the covariance for cell counts that share the same treatment 
but differ in outcome levels. Let \(\widetilde{\mathbf{Z}}_i^\intercal \mathbf{u} = Q_i\). Using the law of total covariance, 
\begin{align*}
\text{Cov}\left[\widetilde{\mathbf{Z}}_i^\intercal \trr_j,\widetilde{\mathbf{Z}}_i^\intercal \trr_{j'}\right] 
&= \mathbb{E}\left[\text{Cov}\left(\widetilde{\mathbf{Z}}_i^\intercal \trr_j, \widetilde{\mathbf{Z}}_i^\intercal \trr_{j'} \mid Q_i\right)\right] + \text{Cov}\left[\mathbb{E}\left(\widetilde{\mathbf{Z}}_i^\intercal \trr_j \mid Q_i\right), \mathbb{E}\left(\widetilde{\mathbf{Z}}_i^\intercal \trr_{j'} \mid Q_i\right)\right]
\end{align*}
\noindent Compute the second term: 
\begin{align*}
&\text{Cov}\left[\mathbb{E}\left(\widetilde{\mathbf{Z}}_i^\intercal \trr_j \mid Q_i\right), \mathbb{E}\left(\widetilde{\mathbf{Z}}_i^\intercal \trr_{j'} \mid Q_i\right)\right] \notag\\
&= \text{Cov}\Big[\rho_{j,1}Q_i + \rho_{j,0}(N_{i.}-Q_i), \;\rho_{j',1}Q_i + \rho_{j',0}(N_{i.}-Q_i)\Big] \notag\\
& = \left(\rho_{j,1}\rho_{j',1} + \rho_{j,0}\rho_{j',0} - \rho_{j,0}\rho_{j',1} - \rho_{j',0}\rho_{j,1}\right) \text{Var}[Q_i].
\end{align*}

\noindent
Here, $\rho_{j,1}$ and $\rho_{j,0}$ are as defined in \eqref{def:q_outcome_score}. 
The first equality follows from \eqref{eq:conditional_expectation_cell_count_on_u}. 
The second equality applies the linearity of covariance and identifies 
$\mathrm{Cov}\left[Q_i, Q_i\right] = \mathrm{Var}[Q_i]$.

To compute $\mathbb{E}\left[\mathrm{Cov}\left(\widetilde{\mathbf{Z}}_i^\intercal \trr_j, \widetilde{\mathbf{Z}}_i^\intercal \trr_j \mid Q_i \right)\right]$, we define
\begin{align*}
\mathbbm{1}_{si} = 
\begin{cases}
1, & \text{if the $s$-th subject receives treatment $i$},\\
0, & \text{otherwise}.
\end{cases}
\label{eq:indicator_function_s_subject_treatment_i}
\end{align*}
for $s = 1,\dots,N$. 
Using this indicator function to rewrite the cell count, we obtain
\begin{align*}
\widetilde{\mathbf{Z}}_i^\intercal \trr_j 
& = \sum_{s : u_s=1} \mathbbm{1}_{si}\,\tilde{r}_{sj} 
   +\sum_{s: u_s=0} \mathbbm{1}_{si}\,\tilde{r}_{sj}; \\
\widetilde{\mathbf{Z}}_i^\intercal \trr_{j'} & =\sum_{s:u_s=1}\mathbbm{1}_{si}\tilde{r}_{sj'} + \sum_{s:u_s=0}\mathbbm{1}_{si}\tilde{r}_{sj'}.
\end{align*}

Recall when conditioning on $Q_i = q_i$, 
$\widetilde{\mathbf{Z}}_i^\intercal \trr_{j}$ is as 
drawing $q_i$ $j$-th outcome scores from individuals with $u_s=1$ 
and $(N_{i.} - q_i)$ scores from those with $u_s=0$.  
\noindent We now compute the covariance between \(\mathbbm{1}_{ai}\) and \(\mathbbm{1}_{bi}\), $1 \leq a,b \leq N$, $u_a = u_b=1$, $a\neq b$, Suppressing explicit conditioning on \(Q_i= q_i\) in the following computations:  
\begin{align}
\text{Cov}\left[\mathbbm{1}_{ai},\mathbbm{1}_{bi}\right] &= \mathbb{P}(\mathbbm{1}_{ai}=1)\mathbb{P}(\mathbbm{1}_{bi}=1 \mid \mathbbm{1}_{ai}=1)-\mathbb{P}(\mathbbm{1}_{ai}=1)\mathbb{P}(\mathbbm{1}_{bi}=1) \notag\\
&= \frac{q_i}{\overline{u}}\frac{q_i-1}{(\overline{u}-1)} - \frac{q_i}{\overline{u}}\frac{q_i}{\overline{u}}.
\end{align}
\noindent The first equality follows from the definition of covariance and Bayes' rule. The second equality follows because, among the \(\overline{u}\) subjects with \(u_s = 1\), the probability that a subject receives treatment \(i\), given that \(Q_i= q_i\), is $q_i/\overline{u}$. After assigning one subject, the probability for the next becomes $(q_i-1)/(\overline{u}-1)$. 

For the covariance between \(\mathbbm{1}_{ai}\) and \(\mathbbm{1}_{bi}\), where \(1 \leq a,b \leq N\), \(u_a=1\), \(u_b=0\), and conditioning on \(Q_i= q_i\), we have  
\begin{align*}
\text{Cov}\left[\mathbbm{1}_{ai},\mathbbm{1}_{bi}\right] = 0.
\end{align*}
This holds because, given \(\widetilde{\mathbf{Z}}_i^\intercal \mathbf{u} = q_i\), the treatment assignments for subjects with \( u_s = 1 \) and those with \( u_s = 0 \) behave as if they were independent. Organize all scenarios as follows.  \begin{align}
    \text{Cov}\left[\mathbbm{1}_{ai},\mathbbm{1}_{bi} \mid Q_i= q_i\right] = 
    \begin{cases}
        \displaystyle \frac{q_i(q_i-\overline{u})}{\overline{u}^2(\overline{u}-1)}, 
        & \text{if } u_a = u_b = 1, \\[5pt]
        0, 
        & \text{if } u_a = 1, u_b = 0, \\[5pt]
        0, 
        & \text{if } u_a = 0, u_b = 1, \\[5pt]
        \displaystyle \frac{(N_{i \cdot} - q_i)(N_{i \cdot }-q_i-N+\overline{u})}{(N-\overline{u})^2(N-\overline{u}-1)}, 
        & \text{if } u_a = u_b = 0.
    \end{cases}
\label{eq:first_case_pair_covariance}
\end{align}

\noindent Expanding the conditional covariance, we have:  
\begin{align}
&\text{Cov}\left[\widetilde{\mathbf{Z}}_i^\intercal \trr_j, \widetilde{\mathbf{Z}}_i^\intercal \trr_{j'} \mid Q_i=q_i\right] \notag\\
&= \sum_{a:u_a=1}\sum_{b:u_b=1}\tilde{r}_{aj}\tilde{r}_{bj'} \text{Cov}\left[\mathbbm{1}_{ai},\mathbbm{1}_{bi}\right] + \sum_{a:u_a=1}\sum_{b:u_b=0}\tilde{r}_{aj}\tilde{r}_{bj'} \text{Cov}\left[\mathbbm{1}_{ai},\mathbbm{1}_{bi}\right] \notag\\
&\quad + \sum_{a:u_a=0}\sum_{b:u_b=1}\tilde{r}_{aj}\tilde{r}_{bj'} \text{Cov}\left[\mathbbm{1}_{ai},\mathbbm{1}_{bi}\right] + \sum_{a:u_a=0}\sum_{b:u_b=0}\tilde{r}_{aj}\tilde{r}_{bj'} \text{Cov}\left[\mathbbm{1}_{ai},\mathbbm{1}_{bi}\right] \notag\\
&= \sum_{a:u_a=1}\sum_{b:u_b=1}\tilde{r}_{aj}\tilde{r}_{bj'} \text{Cov}\left[\mathbbm{1}_{ai},\mathbbm{1}_{bi}\right] + \sum_{a:u_a=0}\sum_{b:u_b=0}\tilde{r}_{aj}\tilde{r}_{bj'} \text{Cov}\left[\mathbbm{1}_{ai},\mathbbm{1}_{bi}\right] \notag\\
&= \overline{u}_j\overline{u}_{j'}\left(\frac{q_i(q_i-\overline{u})}{\overline{u}^2(\overline{u}-1)}\right) + (N_{.j}-\overline{u}_{j})(N_{.j'}-\overline{u}_{j'})\left(\frac{(N_{i \cdot} - q_i)(N_{i \cdot }-q_i-N+\overline{u})}{(N-\overline{u})^2(N-\overline{u}-1)}\right).
\end{align}

\noindent The first equality follows from the linearity of covariance. The second equality holds by \eqref{eq:first_case_pair_covariance}. The third equality follows by counting occurrences of \(\tilde{r}_{aj}\) and \(\tilde{r}_{bj'}\) being both ones among individuals with \(u_s = 1\) and \(u_s = 0\), respectively. By definition, these quantities are  
\( \overline{u}_j \) and \(\overline{u}_{j'} \) for individuals with \( u_s = 1 \), and \( N_{\cdot j} - \overline{u}_j \) and \( N_{\cdot j'} - \overline{u}_{j'} \) for those with \( u_s = 0 \).

\noindent After algebraic simplifications and expressing terms using the definitions of \(\rho_{j,1}\) and \(\rho_{j,0}\) in \eqref{eq:conditional_expectation_cell_count_on_u}, we get:  
\begin{align*}
& \mathbb{E} \left[ \text{Cov} \left( \widetilde{\mathbf{Z}}_i^\intercal \trr_j,  \widetilde{\mathbf{Z}}_i^\intercal \trr_{j'} \mid Q_i \right) \right] \\[5pt]
    &= \frac{\rho_{j,1} \rho_{j',1}}{(\overline{u}-1)}
    \left( \mathbb{E} \left[ Q_i^2 \right] - \overline{u} \cdot  \mathbb{E} \left[ Q_i \right] \right) \\[8pt]
    &\quad + \frac{\rho_{j,0} \rho_{j',0}}{(N - \overline{u} - 1)}
    \Big( N_{i.} (N_{i.} - N + \overline{u}) - (2N_{i.} - N + \overline{u}) \mathbb{E} \left[ Q_i\right] 
    + \mathbb{E} \left[Q_i^2 \right] \Big).
\end{align*}
Combining this with the term $\text{Cov}\left[ \mathbb{E} ( \widetilde{\mathbf{Z}}_i^\intercal \trr_{j} \mid Q_i), 
\mathbb{E}( \widetilde{\mathbf{Z}}_i^\intercal \trr_{j'} \mid Q_i)\right]$ yields the desired result.  

\vspace{5 mm}
\par The proof to case 2 is similar. Define $Q_i=\widetilde{\mathbf{Z}}_i^\intercal \mathbf{u}$ and $Q_{i'}=\widetilde{\mathbf{Z}}_{i'}^\intercal \mathbf{u}$. 
\begin{align}
\text{Cov}\left[\mathbb{E}(\widetilde{\mathbf{Z}}_i^\intercal \trr_j\mid Q_i,Q_{i'}),\mathbb{E}(\widetilde{\mathbf{Z}}_{i'}^\intercal \trr_{j}\mid Q_i,Q_{i'})\right] & = \text{Cov}\left[(\mathbb{E}(\widetilde{\mathbf{Z}}_i^\intercal \trr_j\mid Q_i),\mathbb{E}(\widetilde{\mathbf{Z}}_{i'}^\intercal \trr_{j}\mid Q_{i'})\right] \notag\\ & = \left(\rho_{j,1}^2 + \rho_{j,0}^2-2\left(\rho_{j,1}\rho_{j,0}\right)\right)\text{Cov}\left[Q_i, Q_{i'}\right].
\label{eq:second_case_cov_conditional_expectation}
\end{align}
The first equality is due to $\widetilde{\mathbf{Z}}_i^\intercal \mathbf{u} = Q_i$ being sufficient for $\widetilde{\mathbf{Z}}_i$ under the sensitivity model. Thus, $\mathbb{P}(\widetilde{\mathbf{Z}}_i \mid Q_i, Q_{i'}) = \mathbb{P}(\widetilde{\mathbf{Z}}_i \mid Q_{i})$. Similarly for $\widetilde{\mathbf{Z}}_{i'}^\intercal \mathbf{u} = Q_{i'}$ and 
$\widetilde{\mathbf{Z}}_{i'}$. 
\par \noindent For the conditional covariance:
\begin{align}
& \text{Cov}\left[\widetilde{\mathbf{Z}}_i^\intercal \trr_j, \widetilde{\mathbf{Z}}_{i'}^\intercal \trr_{j}\mid Q_i=q_i,Q_{i'}=q_{i'}\right] \notag \\
& = \sum_{a:u_a=1}\sum_{b:u_b=1}\tilde{r}_{aj}\tilde{r}_{bj}\text{Cov}\left[{\mathbbm{1}_{ai},\mathbbm{1}_{bi'}}\right] + \sum_{a:u_a=0}\sum_{b:u_b=0}\tilde{r}_{aj}\tilde{r}_{bj}\text{Cov}\left[\mathbbm{1}_{ai}, \mathbbm{1}_{bi'}\right] \notag \\
& = \sum_{a:u_a=1}\tilde{r}_{aj}\text{Cov}\left[\mathbbm{1}_{ai},\mathbbm{1}_{ai'}\right] + \sum_{a:u_a=1, b:u_b=1, a\neq b}\tilde{r}_{aj}\tilde{r}_{bj}\text{Cov}\left[\mathbbm{1}_{ai},\mathbbm{1}_{bi'}\right] \notag \\
& \quad + \sum_{a:u_a=0}\tilde{r}_{aj}\text{Cov}\left[\mathbbm{1}_{ai},\mathbbm{1}_{ai'}\right] + \sum_{a:u_a=0, b:u_b=0, a\neq b}\tilde{r}_{aj}\tilde{r}_{bj}\text{Cov}\left[\mathbbm{1}_{ai},\mathbbm{1}_{bi'}\right] \notag \\
    & = \overline{u}_j \left(-\frac{q_i q_{i'}}{\overline{u}^2} \right) 
    + \overline{u}_j (\overline{u}_j - 1) \left(\frac{q_i q_{i'}}{\overline{u}^2 (\overline{u} - 1)} \right) 
    + (N_{\cdot j} - \overline{u}_j) \left(-\frac{(N_{i \cdot} - q_i)(N_{i' \cdot} - q_{i'})}{(N - \overline{u})^2} \right) \notag \\
    & \quad + (N_{\cdot j} - \overline{u}_j)(N_{\cdot j} - \overline{u}_j - 1) 
    \left(\frac{(N_{i \cdot} - q_i)(N_{i' \cdot} - q_{i'})}{(N - \overline{u})^2 (N - \overline{u} - 1)} \right).
\label{eq:second_pair_conditional_cov}
\end{align}

\par \noindent Where we apply, for $a \neq b$ and $i \neq i'$:
\begin{align*}
\text{Cov}\left[\mathbbm{1}_{ai},\mathbbm{1}_{bi'} \mid Q_i=q_i, Q_{i'}=q_{i'}\right] &= 
    \begin{cases}
        \displaystyle \frac{q_i q_{i'}}{\overline{u}^2 (\overline{u}-1)}, 
        & \text{if } u_a = u_b = 1, \\[5pt]
        0, 
        & \text{if } u_a = 1, u_b = 0, \\[5pt]
        0, 
        & \text{if } u_a = 0, u_b = 1, \\[5pt]
        \displaystyle \frac{(N_{i \cdot}-q_i)(N_{i' \cdot }-q_{i'})}{(N-\overline{u})^2(N-\overline{u}-1)}, 
        & \text{if } u_a = u_b = 0.
    \end{cases}\\
    \text{Cov}\left[\mathbbm{1}_{ai},\mathbbm{1}_{ai'} \mid Q_i=q_i, Q_{i'}=q_{i'}\right] &= 
    \begin{cases}
        \displaystyle \frac{-q_i q_{i'}}{\overline{u}^2}, & \text{if } u_a = 1, \\[5pt]
        \displaystyle \frac{-(N_{i \cdot}-q_i)(N_{i' \cdot }-q_{i'})}{(N-\overline{u})^2}, 
        & \text{if } u_a = 0.
    \end{cases}
\end{align*}
Combining \eqref{eq:second_case_cov_conditional_expectation} and \eqref{eq:second_pair_conditional_cov} with the law of total covariance gives the result. The proof to case 3 is the same thus omitted.  

\section{Further Discussion: On Whether $\mathbf{u}^+ \in \{0,1\}^N$} \label{app:E}
\label{sec:Appendix_whether_corner}
\par \noindent From our earlier derivation in Lemma~\ref{lem:corner_solution} in Section~\ref{app:B2} of the Supplementary Material, we showed that under the generic bias sensitivity model, the maximizer $\mathbf{u}^+$ can always be taken to be a binary vector, i.e., $\mathbf{u}^+ \in \{0,1\}^N$. This greatly reduces the computational burden. However, under more general sensitivity models, the maximizer $\mathbf{u}^+$ may not lie in $\{0,1\}^N$.

\subsection{A Counterexample} \label{app:E1}
Consider a case with sample size $6$. Let the treatments be $1,1,2,2,3,4$. In other words, the treatment levels are $i \in \{1,2,3,4\}$ with treatment margins $\mathbf{N}_{4\cdot} = (2,2,1,1)$, and let the outcomes be $\mathbf{r} = (1.4, 1.4, 2.1, 2.1, 3.5, 4.7)$. Suppose treatment assignment follows the general sensitivity model
\[
P(Z_s = i \mid \mathcal{F})
=\frac{\exp\{\xi_i(x_s) + \gamma\,\phi(i)\,u_s\}}
{\sum_{i'=1}^I \exp\{\xi_{i'}(x_s) + \gamma\,\phi(i')\,u_s\}},
\quad \gamma=2,\; u_s \in [0,1],
\]
and here we specify the dose function as $\phi(i)=i$. (See Section~\mref{sec:sens_model_ibyj} for further discussion of different choices of $\phi(i)$.) 
Consider the ordinal test
\[
T = \sum_{s=1}^{6} Z_s r_s \;=\; \sum_{i=1}^{4}\sum_{j=1}^{4} w_i v_j N_{ij},
\]
where $(w_1,w_2,w_3,w_4)=(1,2,3,4)$ and $(v_1,v_2,v_3,v_4)=(1.4,2.1,3.5,4.7)$. Take the critical value $\criticalt = 40$. The corresponding p-value is
\[
\frac{\sum_{\mathbf{z}\in \mathcal{Z}(\mathbf{N}_{4\cdot})} \mathbbm{1}\!\left\{\sum_{s=1}^{6} z_s r_s \geq 40\right\}\exp(\gamma \cdot \mathbf{z}^\intercal \mathbf{u})}{\sum_{\mathbf{b}\in \mathcal{Z}(\mathbf{N}_{4\cdot})}\exp(\gamma \cdot \mathbf{b}^\intercal \mathbf{u})}.
\]
\par \noindent In this case, the maximizer is
\[
\mathbf{u}^+ = (0.0,\, 0.0,\, 0.4674,\, 0.4674,\, 0.9073,\, 1.0),
\]
which is not binary and therefore does not belong to $\{0,1\}^N$.

\subsection{Limitation of the Proof of Lemma \ref{lem:corner_solution}} \label{app:E2}
The following demonstrates an attempt to generalize the proof in \citet[Proposition 2]{rosenbaum1990sensitivity}. 
Recall that the p-value under the general sensitivity model at $\mathbf{u}$ is
\[
\alpha(T,\rr,\mathbf{u}) 
= \frac{\sum_{\mathbf{z}\in \mathcal{Z}(\mathbf{N}_{I\cdot})} 
\mathbbm{1}\{T(\mathbf{z},\rr)\geq \criticalt\}
\exp(\gamma \cdot \phi(\mathbf{z})^\intercal \mathbf{u})}
{\sum_{\mathbf{z}\in \mathcal{Z}(\mathbf{N}_{I\cdot})}
\exp(\gamma \cdot \phi(\mathbf{z})^\intercal \mathbf{u})}.
\]
where
\[
\mathcal{Z}(\mathbf{N}_{I\cdot}) = \big\{\mathbf{z} \in \{1,2,\dots,I\}^N : 
|\{s \in \{1,\dots,N\} : z_s=i\}| = N_{i\cdot}, \, i=1,\dots,I \big\}
\]
denotes the set of $N$-dimensional vectors satisfying the treatment margin constraints.  
Fix an index $s \in \{1,\dots,N\}$, and define
\[
\Omega_{s}^i = \{\mathbf{z} \in \mathcal{Z}(\mathbf{N}_{I\cdot}) : z_s = i\}, 
\quad i = 1,\dots,I.
\]
For an arbitrary function $\phi: \{1,2,\dots,I\} \to \mathbb{R}$, we extend $\phi$ component-wise to vectors:
\[
\phi(\mathbf{z}) = [\phi(z_1), \phi(z_2), \dots, \phi(z_N)].
\]
Fix an arbitrary $\mathbf{u}\in [0,1]^N$ and define
\begin{align*}
A_{s}^i &= \sum_{\mathbf{z}\in \Omega_{s}^i} 
\mathbbm{1}\{T(\mathbf{z}, \rr)\geq \criticalt\}
\exp(\gamma \cdot \phi(\mathbf{z})^\intercal \mathbf{u}),\\
D_{s}^i &= \sum_{\mathbf{z}\in \Omega_{s}^i}
\exp(\gamma \cdot \phi(\mathbf{z})^\intercal \mathbf{u}).
\end{align*}
With these definitions, $\alpha(T,\rr,\mathbf{u})$ takes the form
\[
\alpha(T,\rr,\mathbf{u}) = 
\frac{\sum_{i=1}^{I} A_{s}^i}{\sum_{i=1}^{I} D_{s}^i}.
\]

\par \noindent In \citet[Proposition 2]{rosenbaum1990sensitivity}, $\mathbf{u}^+ \in \{0,1\}^N$ is established by showing that the first-order derivative of $\alpha(T,\rr, \mathbf{u}+\Delta \mathbf{e}_s)$ with respect to $\Delta\in \mathbb{R}$ has a sign independent of $\Delta$, where
\[
\alpha(T,\rr, \mathbf{u} + \Delta \mathbf{e}_s) = \frac{\sum_{i=1}^{I}A_s^i \exp(\gamma \Delta \phi(i))}{\sum_{i=1}^{I}D_{s}^i \exp(\gamma \Delta \phi(i))},
\]
and $\mathbf{e}_s$ is an $N$-dimensional vector with one in the $s$-th entry and zeros elsewhere. Assuming $\gamma >0$, one can show that the first-order partial derivative
$\frac{\partial \alpha(T,\rr, \mathbf{u}+\Delta \mathbf{e}_s)}{\partial \Delta}$ has its sign depending on
\begin{align}
\sum_{i,i' \in \{1,2,\dots,I\}, \;i > i'} \exp(\gamma \Delta(\phi(i)+\phi(i')))(\phi(i)-\phi(i')) \big(A_s^iD_{s}^{i'}-A_{s}^{i'}D_{s}^{i}\big).
\label{eq:sign_difference}
\end{align}
In the binary treatment case, \eqref{eq:sign_difference} reduces to
\begin{align}
\exp(\gamma \Delta (\phi(1)+\phi(2)))(\phi(2)-\phi(1))(A_s^2D_s^{1}-A_s^{1}D_s^2).
\label{eq:binary_treatment_derivative_sign}
\end{align}
In a dichotomized sensitivity model, for example, take $\phi(i) = 1$ if $i \geq k$ and $\phi(i)=0$ if $i < k$, for a fixed $1 \leq k\leq I-1$, \eqref{eq:sign_difference} becomes
\begin{align}
\exp(\gamma \Delta)\Big(\sum_{i=k}^{I}\sum_{i'=1}^{k-1}A_s^{i}D_{s}^{i'} - \sum_{i'=1}^{k-1}\sum_{i=k}^{I}A_s^{i'}D_s^{i}\Big).
\label{eq:dichotomized_partial_derivative}
\end{align}
In both \eqref{eq:binary_treatment_derivative_sign} and \eqref{eq:dichotomized_partial_derivative}, the sign is independent of $\Delta$. However, note that $A_s^i$ and $D_s^i$ vary with different choices of $\mathbf{u}\in [0,1]^N$. Thus, the choice of $\phi$ that ensures the sign of \eqref{eq:sign_difference} remains independent of $\Delta$, while allowing $A_s^i$ and $D_s^i$ to vary, is limited.

\section{Simulation Results} \label{app:F}
\label{sec:Appendix_simulations}
\noindent Section~\ref{app:F} provides additional details on various simulations. 
Section~\ref{app:F1} illustrates the computational savings of obtaining exact $p$-values using Theorem~\mref{thm:recursive_probability}.
Section~\ref{app:F2} shows that the table-based sampling method on the space of contingency tables $\Omega_{\mathbf{t}}$ converges to the true $p$-value with fewer iterations than the random sampling method on the space of treatment assignments $\mathcal{Z}(\mathbf{N}_{I\cdot})$. 
Section~\ref{app:F3} reports the $p$-values of the various tests discussed in Section~\mref{sec:power_main} of the main text, applied to the ECLS-K data analyzed in Section~\mref{sec:data_analysis_main} of the main text. Sections~\ref{app:F4} and onward present further simulations on the statistical power and size control of the tests.

\subsection{Advantages of Theorem~\mref{thm:recursive_probability} in Exact Computation} \label{app:F1}
For exact $p$-value computation, the key advantage of Theorem~\mref{thm:recursive_probability} is that it avoids iterating over all $\mathbf{z} \in \mathcal{Z}(\mathbf{N}_{I\cdot})$, instead performing a search over the table space $\Omega_{\mathbf{t}}$. When computing exact $p$-values based on permutations of treatment assignments, we use the \texttt{multicool} library \citep{multicool} to generate permuted treatment vectors one at a time. The \texttt{multicool} library implements the loopless algorithm described in \citet{Williams2009Loopless} and is written in C++ for efficiency. All reported timings were obtained on the Posit Cloud (8 CPUs, 4 GB RAM).

\par Table~\ref{tb:exact_computation_time_permutation_and_kernel} presents the results of the speed comparison. The first three rows correspond to an observed table with $(N_{11}, N_{12}, N_{13}, N_{21}, N_{22}, N_{23}, N_{31}, N_{32}, N_{33}) =$ $(2, 3, 0, 0, 1, 4, 0, 1, 4)$, yielding treatment margins $\mathbf{N}_{3\cdot} = (5, 5, 5)$ and outcome margins $\mathbf{N}_{\cdot 3} = (2, 5, 8)$. The $p$-value is based on the test statistic $T(\mathbf{N}) = \sum_{i=1}^{3} \sum_{j=1}^{3} w_i v_j N_{ij}$, where $(w_1, w_2, w_3) = (0, 1, 2)$ and $(v_1, v_2, v_3) = (0, 1, 2)$. Unlike the main text, where we specify a corner $\mathbf{u}$ directly, Theorem~\mref{thm:set_size_of_U} shows that it suffices to search over $\mathcal{U}_{\rm C}$. Furthermore, there is a one-to-one mapping between $\rr$ and $\trr$ as in Figure~\ref{fig:integer_dummy_mapping} in Appendix~\ref{app:B}; therefore, it is enough to specify a corner of $\mathbf{u}$ using the quantifier $\trr^\intercal \mathbf{u}$. For a cleaner presentation, we adopt this quantifier in Table~\ref{tb:exact_computation_time_permutation_and_kernel}. The first three rows of the table report results for three corner vectors $\mathbf{u}$ such that $\trr^\intercal \mathbf{u} = (0, 0, 3), (0, 0, 7), (0, 5, 8)$. The $p$-values obtained with \change{equation~\meqref{eq:raw_sens_test} of the main text} require, on average, \change{$19.647$, $19.303$, and $19.439$} seconds, respectively. In contrast, the same $p$-values can be obtained with Theorem~\mref{thm:recursive_probability} in just \change{$0.00050$, $0.00048$, and $0.00050$} seconds.

\par Moreover, as sample size increases, the runtime of \change{equation~\meqref{eq:raw_sens_test} of the main text} grows much faster than that of Theorem~\mref{thm:recursive_probability}. For instance, in the seventh row of Table~\ref{tb:exact_computation_time_permutation_and_kernel}, the $p$-value for a table with treatment margins $\mathbf{N}_{3\cdot} = (6, 6, 6)$ is computed in \change{$450.974$} seconds using \change{equation~\meqref{eq:raw_sens_test} of the main text}; this represents about a $23$-fold increase compared to a table with $\mathbf{N}_{3\cdot} = (5, 5, 5)$. In contrast, Theorem~\mref{thm:recursive_probability} computes the same value in just \change{$0.00060$} seconds, with runtime increasing by only about \change{$1.2$}-fold, which is far more modest. In this example, Theorem~\mref{thm:recursive_probability} achieves a speedup factor of approximately \change{$450.974 / 0.00060 \approx 750{,}000$}.

\begin{table}[ht]
\centering
\small
\begin{tabular}{@{}C{22mm} c d{2.2} c c@{}}
\toprule
\thead{Observed\\Table} & $\trr^\intercal \mathbf{u}$ 
& \multicolumn{1}{c}{\(p\)-value}
& \multicolumn{2}{c}{Computation Time (in seconds)} \\ 
\cmidrule(lr){4-5}
& & & \multicolumn{1}{c}{Theorem~\mref{thm:recursive_probability}} & \multicolumn{1}{c}{\change{\makecell{Equation~\meqref{eq:raw_sens_test}\\ of the main text}}} \\
\midrule
\multirow{3}{*}{{%
  $\begin{bmatrix}
    2 & 3 & 0 \\
    0 & 1 & 4 \\
    0 & 1 & 4
  \end{bmatrix}$%
}} & $(0,0,3)$  & 0.01 & \change{$0.00050 \pm 0.00009$} & \change{$19.647 \pm 0.411$} \\  
   & $(0,0,7)$  & 0.03 & \change{$0.00048 \pm 0.00004$} & \change{$19.303 \pm 0.263$} \\ 
   & $(0,5,8)$  & 0.02 & \change{$0.00050 \pm 0.00003$} & \change{$19.439 \pm 0.266$} \\
\midrule
\multirow{3}{*}{{%
  $\begin{bmatrix}
    2 & 2 & 1 \\
    1 & 2 & 2 \\
    1 & 2 & 2
  \end{bmatrix}$%
}} & $(0,0,2)$  & 0.36 & \change{$0.00066 \pm 0.00004$} & \change{$19.157 \pm 0.210$} \\  
   & $(0,3,5)$  & 0.52 & \change{$0.00067 \pm 0.00005$} & \change{$19.293 \pm 0.255$} \\ 
   & $(0,6,5)$  & 0.53 & \change{$0.00066 \pm 0.00003$} & \change{$19.478 \pm 0.359$} \\
\midrule
\multirow{3}{*}{{%
  $\begin{bmatrix}
    3 & 2 & 1 \\
    0 & 2 & 4 \\
    0 & 1 & 5
  \end{bmatrix}$%
}} & $(0,0,5)$  & 0.01 & \change{$0.00060 \pm 0.00005$} & \change{$450.974 \pm 3.373$} \\     
   & $(0,0,10)$ & 0.04 & \change{$0.00059 \pm 0.00003$} & \change{$456.565 \pm 7.369$} \\
   & $(0,4,10)$ & 0.03 & \change{$0.00060 \pm 0.00003$} & \change{$460.657 \pm 4.571$} \\
  \midrule
\multirow{3}{*}{{%
  $\begin{bmatrix}
    3 & 3 & 0 \\
    1 & 2 & 3 \\
    2 & 3 & 1
  \end{bmatrix}$%
}} & $(0,0,4)$  & 0.49 & \change{$0.00083 \pm 0.00008$} & \change{$456.546 \pm 5.646$} \\   
   & $(0,3,4)$  & 0.52 & \change{$0.00083 \pm 0.00003$} & \change{$457.841 \pm 2.968$} \\
   & $(0,8,4)$  & 0.58 & \change{$0.00082 \pm 0.00005$} & \change{$448.732 \pm 6.384$} \\
\midrule
\multirow{2}{*}{{%
  $\begin{bmatrix}
    4 & 6 & 0 \\
    1 & 3 & 6
  \end{bmatrix}$%
}} & $(4,9,6)$  & 0.01 & \change{$0.00044 \pm 0.00009$} & \change{$4.796 \pm 0.081$} \\
   & $(0,1,6)$  & 0.05 & \change{$0.00040 \pm 0.00002$} & \change{$4.783 \pm 0.038$} \\
\midrule
\multirow{2}{*}{{%
  $\begin{bmatrix}
    2 & 4 & 4 \\
    2 & 2 & 6
  \end{bmatrix}$%
}} & $(3,6,10)$ & 0.46 & \change{$0.00045 \pm 0.00004$} & \change{$4.919 \pm 0.112$} \\ 
     & $(0,0,7)$  & 0.67 & \change{$0.00044 \pm 0.00003$} & \change{$5.006 \pm 0.194$} \\ 
\bottomrule
\end{tabular}
\caption{A computational comparison between two approaches to evaluating the exact $p$-value: one based on Theorem~\mref{thm:recursive_probability} and the other on \change{equation~\meqref{eq:raw_sens_test} of the main text, which iterates over all treatment assignments in $\mathcal{Z}(\mathbf{N}_{I\cdot})$}. Each block corresponds to a fixed observed table, with rows representing different quantifiers $\trr^\intercal \mathbf{u}$. \change{The mean $\pm$ standard deviation of the computation time, measured in seconds, is evaluated over $10$ runs on Posit Cloud with $8$ CPUs and $4$~GB RAM.}}
\label{tb:exact_computation_time_permutation_and_kernel}
\end{table}

\FloatBarrier
\subsection{Advantages of Theorem~\mref{thm:recursive_probability} in Sampling-Based Estimation} \label{app:F2}
In addition to exact computation, one may approximate the probability using sampling-based methods. Similar to the exact setting in Section~\ref{app:F1}, sampling-based estimators achieve better performance when operating in the space of contingency tables $\Omega_{\bm t}$ rather than in the treatment-assignment space $\mathcal{Z}(\mathbf{N}_{I\cdot})$.

To formalize the sampling-based estimators for comparison, we first introduce the proposal distribution used in sequential importance sampling. Let $h(\bm t)$ denote the probability of sampling table $\bm t$ under the SIS-G algorithm \citep{Eisinger2017}, and\change{, following Theorem~\mref{thm:recursive_probability} in its equivalent form \eqref{eq:alpha_W_form},} define
\begin{changeblock}
\begin{equation*}
l(\bm t)
\;=\;
\kappa(\bm t)\, S(\overline{\bm t}, \mathbf{u}),
\end{equation*}
\end{changeblock}
\change{with $\kappa(\bm t)$ as in \eqref{eq:kappa_def} and $S(\overline{\bm t}, \mathbf{u})$ as in \eqref{eq:S_def},} as the numerator contribution from $\bm t$.

The importance-sampling estimator is:
\begin{align*}
\hat{\alpha}_{\mathrm{SIS}}(T,\rr,\mathbf{u})
&= \frac{1}{M\,\mathcal{C}(\mathbf{u})}\sum_{m=1}^{M}
\mathbbm{1}\!\bigl\{T(\bm t_m) \ge \criticalt\bigr\}\,
\frac{l(\bm t_m)}{h(\bm t_m)},
\end{align*}
where $\{\bm t_1,\dots,\bm t_M\}$ are tables drawn from the proposal $h(\bm t)$, and the normalizing constant is\change{, with $S_0(\mathbf{u})$ as in \eqref{eq:S0_def},}
\begin{changeblock}
\[
\mathcal{C}(\mathbf{u}) 
\;=\; 
\kappa_0\, S_0(\mathbf{u}).
\]
\end{changeblock}

A self-normalized version of the estimator is:
\begin{align*}
\hat{\alpha}_{\mathrm{snSIS}}(T,\rr,\mathbf{u})
&= \frac{\displaystyle
\sum_{m=1}^{M}
\mathbbm{1}\!\bigl\{T(\bm t_m)\ge \criticalt\bigr\}\,\dfrac{l(\bm t_m)}{h(\bm t_m)}
}{
\displaystyle \sum_{m=1}^{M}\dfrac{l(\bm t_m)}{h(\bm t_m)}
}.
\end{align*}

Finally, an alternative estimator based on sampling the treatment directly from $\mathcal{Z}(\mathbf{N}_{I\cdot})$ is:
\begin{equation*}
\hat{\alpha}_{\mathrm{PermTreat}}(T,\rr,\mathbf{u})
= \frac{\displaystyle \sum_{m=1}^{M} \mathbbm{1}\!\bigl\{T(\mathbf{z}_{m},\rr) \ge \criticalt\bigr\}\,
\exp\!\Big\{\gamma \sum_{s=1}^{N}\sum_{i=1}^{I} \delta_i\, \mathbbm{1}\{z_{m,s}=i\}\, u_s\Big\}}
{\displaystyle \sum_{m=1}^{M} \exp\!\Big\{\gamma \sum_{s=1}^{N}\sum_{i=1}^{I} \delta_i\, \mathbbm{1}\{z_{m,s}=i\}\, u_s\Big\}},
\end{equation*}
where each $\mathbf{z}_m$ is drawn uniformly from $\mathcal{Z}(\mathbf{N}_{I\cdot})$, and $z_{m,s}$ denotes the $s$th entry of $\mathbf{z}_m$.

\begin{changeblock}
The following presents the progression of $\hat{\alpha}_{\text{SIS}}$, 
$\hat{\alpha}_{\text{snSIS}}$, and $\hat{\alpha}_{\text{PermTreat}}$ 
over $10{,}000$ Monte Carlo runs estimating the $p$-value for a given 
observed table and a given $\mathbf{u}$, together with the exact $p$-value 
computed via Theorem~\mref{thm:recursive_probability}. 
We consider $3$ by $4$ tables and use the test statistic 
\[
T(\mathbf{N}) = \sum_{i=1}^{3}\sum_{j=1}^{4} w_i v_j N_{ij}
\]
with $(w_1,w_2,w_3) = (1,2,3)$ and $(v_1,v_2,v_3,v_4) = (1,2,3,4)$ as in 
Section~\ref{app:G1}, and $\bm{\delta} = (0,1,1)$. The sample sizes range 
from $N = 100$ to $N = 168$, so that the exact $p$-value remains computable 
and the precision of each estimator can be assessed against it, while the computation of the exact $p$-value takes from about one second to about $43$ seconds. Timings were recorded under the same environment as in 
Section~\ref{app:F1} on Posit Cloud (8 CPUs, 4~GB RAM).
\end{changeblock}   

\begin{changeblock}
In general, a desirable property of a sampling-based estimator is that it 
stabilizes near the true value with relatively few iterations and remains 
stable thereafter. Figures~\ref{fig:plot_convergence_S1}--\ref{fig:plot_convergence_P2} 
show the progression of the three estimators. The vertical axis limits differ across panels, and our focus is on their comparative convergence behavior in various settings. 

Figures~\ref{fig:plot_convergence_S1} and \ref{fig:plot_convergence_S2} 
consider the larger tables ($N = 150$ and $N = 168$), for which the computation of the exact p-value takes $20$ to $43$ seconds, whereas each 
sampling-based estimator finishes within about three seconds; all three 
estimators approach the exact $p$-value. Figures~\ref{fig:plot_convergence_P1} 
and \ref{fig:plot_convergence_P2} consider smaller tables ($N = 100$) with a 
stronger association, for which tables in $\Omega_{\mathbf{t}}$ are rare 
under uniform sampling of treatment assignments; $\hat{\alpha}_{\text{PermTreat}}$ 
then remains far below the exact $p$-value after $10{,}000$ runs, because it 
must explore the much larger space of $\mathcal{Z}(\mathbf{N}_{I\cdot})$ and 
only a handful of its draws fall in $\Omega_{\mathbf{t}}$. Across all figures, 
$\hat{\alpha}_{\text{snSIS}}$ performs best: it exhibits lower variance than 
$\hat{\alpha}_{\text{SIS}}$ and settles close to the true $p$-value. For this 
reason, we suggest using table-sampling methods for sampling-based estimation when the exact p-value computation is infeasible.

The computation times reported in the captions should not be read as 
contradicting Section~\mref{sec:exact_worst_null} of the main text. All three 
estimators use the same number of Monte Carlo runs, so 
$\hat{\alpha}_{\text{PermTreat}}$, whose draws are cheap, reports the shortest 
time; it nevertheless searches the much larger space 
$\mathcal{Z}(\mathbf{N}_{I\cdot})$, and exhausting that space is infeasible 
compared with the exact computation of 
Theorem~\mref{thm:recursive_probability}, as 
Table~\mref{tb:main_text_exact_computation_time_permutation_and_kernel} of the 
main text and Table~\ref{tb:exact_computation_time_permutation_and_kernel} show.
\end{changeblock}

\begin{figure}[ht]
    \centering
    \includegraphics[width=0.9\textwidth]{plot_convergence_experiment_S1.png}
    \caption{\change{Performance of $\hat{\alpha}_{\text{SIS}}$, $\hat{\alpha}_{\text{snSIS}}$, 
and $\hat{\alpha}_{\text{PermTreat}}$ over $10{,}000$ runs based on an 
observed $3$ by $4$ table with $N = 150$ and rows 
$(N_{11},N_{12},N_{13},N_{14}) = (18,13,10,10)$, 
$(N_{21},N_{22},N_{23},N_{24}) = (12,12,12,12)$, and 
$(N_{31},N_{32},N_{33},N_{34}) = (9,11,14,17)$, given $\mathbf{u}$ with 
$\trr^{\intercal}\mathbf{u} = (10,10,30,30)$ and $\gamma = 0.5$ 
$(\Gamma = 1.649)$. The black dashed line indicates the exact $p$-value. 
The exact $p$-value is $0.042$, requiring $19.82$ seconds to compute, 
whereas $\hat{\alpha}_{\text{SIS}}$, $\hat{\alpha}_{\text{snSIS}}$, and 
$\hat{\alpha}_{\text{PermTreat}}$ yield estimates of $0.079$, $0.043$, and 
$0.056$ within $2.00$, $2.15$, and $0.45$ seconds, respectively.}}
    \label{fig:plot_convergence_S1}
\end{figure}

\begin{figure}[ht]
    \centering
    \includegraphics[width=0.9\textwidth]{plot_convergence_experiment_S2.png}
    \caption{\change{Performance of $\hat{\alpha}_{\text{SIS}}$, $\hat{\alpha}_{\text{snSIS}}$, 
and $\hat{\alpha}_{\text{PermTreat}}$ over $10{,}000$ runs based on an 
observed $3$ by $4$ table with $N = 168$ and rows 
$(N_{11},N_{12},N_{13},N_{14}) = (18,15,13,10)$, 
$(N_{21},N_{22},N_{23},N_{24}) = (14,14,14,14)$, and 
$(N_{31},N_{32},N_{33},N_{34}) = (10,13,15,18)$, given $\mathbf{u}$ with 
$\trr^{\intercal}\mathbf{u} = (11,11,32,32)$ and $\gamma = 0.5$ 
$(\Gamma = 1.649)$. The black dashed line indicates the exact $p$-value. 
The exact $p$-value is $0.058$, requiring $43.29$ seconds to compute, 
whereas $\hat{\alpha}_{\text{SIS}}$, $\hat{\alpha}_{\text{snSIS}}$, and 
$\hat{\alpha}_{\text{PermTreat}}$ yield estimates of $0.072$, $0.055$, and 
$0.061$ within $2.07$, $2.55$, and $0.42$ seconds, respectively.}}
    \label{fig:plot_convergence_S2}
\end{figure}

\begin{figure}[ht]
    \centering
    \includegraphics[width=0.9\textwidth]{plot_convergence_experiment_P1.png}
    \caption{\change{Performance of $\hat{\alpha}_{\text{SIS}}$, $\hat{\alpha}_{\text{snSIS}}$, 
and $\hat{\alpha}_{\text{PermTreat}}$ over $10{,}000$ runs based on an 
observed $3$ by $4$ table with $N = 100$ and rows 
$(N_{11},N_{12},N_{13},N_{14}) = (13,10,7,4)$, 
$(N_{21},N_{22},N_{23},N_{24}) = (8,8,8,8)$, and 
$(N_{31},N_{32},N_{33},N_{34}) = (5,6,9,14)$, given $\mathbf{u}$ with 
$\trr^{\intercal}\mathbf{u} = (3,6,16,22)$ and $\gamma = 1.000$ 
$(\Gamma = 2.718)$. The black dashed line indicates the exact $p$-value. 
The exact $p$-value is $0.021$, requiring $1.64$ seconds to compute, 
whereas $\hat{\alpha}_{\text{SIS}}$, $\hat{\alpha}_{\text{snSIS}}$, and 
$\hat{\alpha}_{\text{PermTreat}}$ yield estimates of $0.037$, $0.016$, and 
$0.003$ within $0.97$, $1.24$, and $0.28$ seconds, respectively.}}
    \label{fig:plot_convergence_P1}
\end{figure}

\begin{figure}[ht]
    \centering
    \includegraphics[width=0.9\textwidth]{plot_convergence_experiment_P2.png}
    \caption{\change{Performance of $\hat{\alpha}_{\text{SIS}}$, $\hat{\alpha}_{\text{snSIS}}$, 
and $\hat{\alpha}_{\text{PermTreat}}$ over $10{,}000$ runs based on an 
observed $3$ by $4$ table with $N = 100$ and rows 
$(N_{11},N_{12},N_{13},N_{14}) = (14,10,6,4)$, 
$(N_{21},N_{22},N_{23},N_{24}) = (8,8,8,8)$, and 
$(N_{31},N_{32},N_{33},N_{34}) = (4,6,10,14)$, given $\mathbf{u}$ with 
$\trr^{\intercal}\mathbf{u} = (0,8,18,24)$ and $\gamma = 1.000$ 
$(\Gamma = 2.718)$. The black dashed line indicates the exact $p$-value. 
The exact $p$-value is $0.014$, requiring $1.04$ seconds to compute, 
whereas $\hat{\alpha}_{\text{SIS}}$, $\hat{\alpha}_{\text{snSIS}}$, and 
$\hat{\alpha}_{\text{PermTreat}}$ yield estimates of $0.012$, $0.013$, and 
$0.0001$ within $0.80$, $1.16$, and $0.28$ seconds, respectively.}}
    \label{fig:plot_convergence_P2}
\end{figure}
\FloatBarrier

\subsection{Additional Discussion for the Data Analysis} \label{app:F3}
This section first introduces how the treatment and outcome weights can be 
determined and then reports additional results for the data analysis. We 
consider two approaches. The first set of weights was pre-specified, 
independent of the data. The weights $(w_1,w_2,w_3)=(0.00,0.25,1.50)$ 
encode the belief that center-based care has a much stronger effect than the 
other care types, while the outcome weights $(v_1,v_2,v_3)=(0.00,1.00,1.50)$ 
reflect that “relative size” and “ordinality and sequence” both involve 
advanced pattern recognition and are conceptually closer to each other than 
to “number and shape.”

The second set of weights was estimated using a profile likelihood method 
solved with the Gurobi software \citep{gurobi}, based on alternating 
maximization over the treatment weights and the outcome weights. For 
identifiability, we fix $w_1=0$ and $v_1=0$ and optimize over $(w_2,w_3)$ 
and $(v_2,v_3)$ in turn. Let $N_{kij}$ denote the observed count in stratum 
$k \in \{1,2\}$, treatment level $i \in \{1,2,3\}$, and outcome level 
$j \in \{1,2,3\}$. Define
\[
\mathbf{w}=(w_1,w_2,w_3)^{\intercal},\qquad 
\mathbf{v}=(v_1,v_2,v_3)^{\intercal},
\]
and collect nuisance parameters as
\[
\bm{\lambda}=\big(\lambda_1,\lambda_2,\ 
\{\lambda^Z_{ki}\}_{k=1,2;\,i=1,2,3},\ 
\{\lambda^r_{kj}\}_{k=1,2;\,j=1,2,3}\big),
\]
where $\lambda_k$ are stratum intercepts, $\lambda^Z_{ki}$ are treatment 
main effects (by stratum), and $\lambda^r_{kj}$ are outcome main effects 
(by stratum). The Poisson log-likelihood is
\begin{equation}
\ell(\bm{\lambda},\mathbf{w},\mathbf{v})
=\sum_{k=1}^{2}\sum_{i=1}^{3}\sum_{j=1}^{3}
\Big\{N_{kij}\log\!\big(\mathbb{E}[N_{kij}]\big)
- \mathbb{E}[N_{kij}]\Big\},
\label{eq:profile_ll}
\end{equation}
with linear predictor
\begin{equation}
\log\!\big(\mathbb{E}[N_{kij}]\big)
=\lambda_k+\lambda^Z_{ki}+\lambda^r_{kj}+w_i v_j,
\qquad k=1,2,\; i,j=1,2,3.
\label{eq:profile_lp}
\end{equation}

Maximization alternates between the following two subproblems, re-estimating 
the nuisance parameters $\bm{\lambda}$ in each step:
\begin{align*}
\text{Step A:}&\quad
\max_{\bm{\lambda},\,w_2,w_3}\ \ell(\bm{\lambda},\mathbf{w},\mathbf{v})
\quad \text{with $(v_2,v_3)$ fixed},\\[4pt]
\text{Step B:}&\quad
\max_{\bm{\lambda},\,v_2,v_3}\ \ell(\bm{\lambda},\mathbf{w},\mathbf{v})
\quad \text{with $(w_2,w_3)$ fixed}.
\end{align*}
We note that joint maximization over both the treatment weights $\mathbf{w}$ 
and the outcome weights $\mathbf{v}$ may not be feasible because the 
log-likelihood is not concave when both sets of weights are free; see 
\citet[Section~10.5.2]{agresti2012categorical}.

The following results from Table~\ref{tb:additional_p_value_data_analysis_appendix} 
show that the collapsed Fisher’s exact test $2\times 2$ (V1) and the cross-cut 
test are unable to detect a treatment effect in the data analysis, while the 
$2\times 2$ (V2) test detects one only for girls and only at $\Gamma = 1$. In contrast, the 
$3\times 2$ (V2) test is more robust to unmeasured confounding and 
provides evidence of positive treatment effects of attending pre-K 
center-based care or relative care on children’s math achievement at 
kindergarten entry. Similar to Case~I in Section~\mref{sec:power_main} of the main text, binarizing the outcome but not the treatment can still yield a 
powerful test. When the outcome is binary but the treatment has multiple 
levels, Theorem~\mref{thm:score_test_maximizer_binary} and Corollary~\mref{cor:unique_corner_multivariate_hyper} can be applied to reduce the 
computational burden of obtaining the exact $p$-value.

\begin{table}[ht]
\centering
\begin{tabular}{llcccccccc}
\toprule
\multirow{2}{*}{Test} & \multirow{2}{*}{} & \multicolumn{8}{c}{$\Gamma$ values} \\
\cmidrule(lr){3-10}
& & 1.0 & 1.5 & 2.0 & 2.5 & 3.0 & 3.5 & 4.0 & 4.5 \\
\midrule
\multirow{2}{*}{$2\times 2$ (V1)} 
& Girls &  0.283&  0.417&  0.511&  0.581&  0.633&  0.674&  0.707&  0.734\\
& Boys  &  0.466&  0.645&  0.751&  0.817&  0.859&  0.889&  0.910& 0.926\\
\midrule
\multirow{2}{*}{$2\times 2$ (V2)}
& Girls &  0.011&  0.057&  0.139&    0.240&           0.343&  0.438&  0.524&  0.596\\
& Boys  &  0.602&  0.859&  0.950&    0.981&           0.993&  0.997&  0.998&  0.999\\
\midrule
\multirow{2}{*}{Cross-Cut}
& Girls &  0.146&  0.247&  0.332&    0.402&           0.459&  0.508&  0.548&  0.583\\
& Boys  &  0.309&  0.481&  0.602&    0.687&           0.749&  0.794&  0.828&  0.854\\
\midrule
\multirow{2}{*}{$3\times 2$ (V1)}
& Girls &  0.287&  0.372&  0.427&    0.466&           0.495&  0.517&  0.534&  0.548\\
& Boys  &  0.188&  0.259&  0.304&    0.335&           0.356&  0.372&  0.384&  0.393\\
\midrule
\multirow{2}{*}{$3\times 2$ (V2)}
& Girls &  0.004&  0.012&  0.023&    0.036&           0.050&  0.064&  0.077&  0.089\\
& Boys  &  0.013&  0.034&  0.062&    0.092&           0.121 &  0.149&  0.173&  0.196\\
\bottomrule
\end{tabular}
\caption{The p-values for the data analysis using different tests. The treatment weights $(w_1,w_2,w_3) = (0,0.25,1.5)$ for the $3\times 2$ (V1), and $3\times 2$ (V2) tests.}
\label{tb:additional_p_value_data_analysis_appendix}
\end{table}

\FloatBarrier

\subsection{Power Analysis: Simulation Steps} \label{app:F4}
This section provides additional details on the simulation procedure used in the power analysis. In particular, we describe how to generate outcomes for contingency tables under fixed treatment margins using the log–linear model introduced in Section~\mref{sec:power_main} of the main text, and how to compute the empirical power of a test. 

Recall that in Section~\mref{sec:power_main}, the data-generating process corresponds to a setting with a nonzero treatment effect but no unmeasured confounding. Specifically, the expected cell counts are given by the log–linear model
\begin{equation*}
\log\!\big(\mathbb{E}[N_{ij}]\big)
\;=\;
\lambda + \lambda_i^{Z} + \lambda_j^{r} + \beta w^*_i v^*_j,
\quad
w^*_1 \leq w^*_2 \leq \cdots \leq w^*_I, 
\quad
v^*_1 \leq v^*_2 \leq \cdots \leq v^*_J.
\end{equation*}
After computing $\mathbb{E}[N_{ij}]$ for all $i=1,2,3$ and $j=1,2,3$, the simulation proceeds as follows.

\par \noindent \textbf{Simulation Steps:}
\begin{itemize}[leftmargin=*]
\item \textbf{Transform each mean \(\mathbb{E}[N_{ij}]\) into a conditional probability.}  
From the log–linear model, the probability of outcome \(j\) given treatment \(i\) is
\[
\mathbb{P}(r_s = j \mid Z_s = i)
\;=\;
\frac{\mathbb{E}[N_{ij}]}{\sum_{j'=1}^3 \mathbb{E}[N_{ij'}]},
\]
which specifies the outcome distribution among subjects receiving treatment \(i\).

\item \textbf{Generate a contingency table with fixed treatment margins.}  
Let \(N_{i\cdot}\) denote the row margin for treatment level \(i\). For each treatment \(i\), we sample \(N_{i\cdot}\) outcomes independently from a categorical distribution with probabilities \(\mathbb{P}(r_s=j \mid Z_s=i)\), \(j=1,2,3\). In the $3 \times 3$ setting, this yields three rows corresponding to \(i=1,2,3\).

\item \textbf{Compute the worst-case p-value at fixed sensitivity parameters \(\Gamma\) and \(\bm{\delta}\).}  
For each simulated table, compute the worst-case p-value under the sensitivity model with specified \(\Gamma\) and \(\bm{\delta}\). This requires evaluating the p-value under every candidate unmeasured confounder consistent with the test—for example, each \(\mathbf{u} \in \mathcal{U}_{\mathrm{o}}\) for the ordinal test (Theorem~\mref{thm:score_test_maximizer_nonbinary}), or the single \(\mathbf{u}^+\) for the sign-score test (Theorem~\mref{thm:score_test_maximizer_binary}). The maximum of these values is taken as the final worst-case p-value.

\item \textbf{Reject or fail to reject the null hypothesis.}  
Reject the null if the worst-case p-value is less than or equal to $0.05$. Repeating this procedure $1{,}000$ times yields an empirical power estimate, defined as the proportion of rejections across the $1{,}000$ simulated datasets.
\end{itemize}

\subsection{Power of Ordinal Tests: Collapsed, Non-Collapsed, and Cross-Cut Tables} \label{app:F5}

We fix $\beta = 1$ and record the bias setup $\bm{\delta}$, the row effects $\lambda_i^{Z}$, the column effects $\lambda_j^{r}$, and the treatment margins $\mathbf{N}_{3\cdot}$ in the table captions. This section examines the power of sensitivity analysis under different data-generating weights $(w_i^*, v_j^*)$, across sample sizes and treatment margins, to evaluate which tests generally perform best. We observe that the $3 \times 3$ (Opt) test usually achieves the highest power across sample sizes, treatment margins, and data-generating processes. The cross-cut test performs well only when the treatment or outcome weights exhibit a clear progression. Collapsed tests can also retain power if levels with similar weights are combined.

For example, in Table~\ref{tb:sparse_table_6}, the data-generating weights are $(w^*_1, w^*_2, w^*_3) = (0, 1.5, 1.5)$ and $(v^*_1, v^*_2, v^*_3) = (0, 0, 1.5)$, with $\bm{\delta} = (0, 1, 1)$. By definition of these tests introduced in Section~\mref{sec:power_main} of the main text, the $3 \times 3$ (Opt), $3 \times 2$ (V1), and $2 \times 2$ (V1) tests yield the same value of the test statistic for any realized table. Moreover, because the collapsed treatment levels share the same $\delta_i$ value, the sensitivity model is unchanged, and these three tests therefore have identical power in this setting.

By contrast, Table~\ref{tb:sparse_table_12} uses the same data-generating weights but a different bias parameter, $\bm{\delta} = (0, 0, 1)$. By definition of these tests in Section~\mref{sec:power_main}, the $3 \times 3$ (Opt), $3 \times 2$ (V1), and $2 \times 2$ (V1) tests still yield the same value of the test statistic, since $w^*_2 = w^*_3$. However, collapsing the second and third treatment levels now merges $\delta_2 = 0$ with $\delta_3 = 1$, thereby altering the sensitivity model. As a result, the $3 \times 2$ (V1) test retains the same power as the $3 \times 3$ (Opt) test, while the $2 \times 2$ (V1) test exhibits lower power when $\Gamma > 1$. This example highlights an important principle: collapsing levels preserves power only when the collapsing is consistent with both the test design (i.e., the structure of $w_i^*$ and $v_j^*$) and the bias specification (i.e., $\bm{\delta}$). See Section~\ref{app:B10} of the Supplementary Material for a more detailed explanation of this effect.

Tables \ref{tb:weighted_four_60}, \ref{tb:weighted_four_v2_60}, \ref{tb:weighted_four_v8_60}, and \ref{tb:weighted_four_v10_60} present the numerical values underlying the power curves in Section~\mref{sec:power_main} of the main text with $N = 60$. With the same data-generating weights $w_i^*$ and $v_j^*$, Tables \ref{tb:weighted_four_120}, \ref{tb:weighted_four_v2_120}, \ref{tb:weighted_four_v8_120}, and \ref{tb:weighted_four_v10_120} report the results for $N = 120$. Consistent with the main text, the $3 \times 3$ (Opt) test performs strongly across all cases: in Cases~I, III, and IV, it has the highest or essentially tied-highest power across all $\gamma$; in Case~II, it has the highest power for all $\gamma$ when $N = 60$ and for $\gamma \leq 1.5$ when $N = 120$, becoming the second-best thereafter. The cross-cut test performs well in Cases~I and II, but shows relatively low power in Cases~III and IV. Binarized outcome options may perform well when some outcome levels exhibit similar trends, as in the $3 \times 2$ (V2) test in Case~I and the $3 \times 2$ (V1) test in Cases~III and IV. Theorem~\mref{thm:score_test_maximizer_binary} and Corollary~\mref{cor:unique_corner_multivariate_hyper} apply to these multi-level treatment and binary outcome tests.

For completeness, Tables \ref{tb:weighted_four_sparse_3} and \ref{tb:weighted_four_sparse_4} present the power analysis when the treatment margins are unbalanced, while Table \ref{tb:sparse_table_4} reports results for sparse tables, i.e., when some cell counts are likely to be zero. Finally, Section~\ref{app:F6} considers the power of tests when different weights are attached to non-collapsed tables.

\begin{table}[H]
    \centering
    \small

    \renewcommand{\arraystretch}{1.2} 
    \begin{tabular}{c c | c c c c c c}
        \toprule
        $\gamma$ & $\Gamma = e^\gamma$ & 3×3 (Opt) & 3×2 (V1) & 3×2 (V2) & 2×2 (V1) & 2×2 (V2) & Cross-Cut \\
        \midrule
        0.00  & 1.00  & 0.995 & 0.703 & 0.994 & 0.600 & 0.983 & 0.943\\
        0.25  & 1.28  & 0.988 & 0.549 & 0.983 & 0.439 & 0.953 & 0.908\\
        0.50  & 1.65  & 0.960 & 0.375 & 0.948 & 0.278 & 0.879 & 0.828\\
        0.75  & 2.12  & 0.908 & 0.244 & 0.887 & 0.125 & 0.797 & 0.734\\
        1.00  & 2.72  & 0.829 & 0.133 & 0.795 & 0.074 & 0.694 & 0.609\\
        1.25  & 3.49  & 0.720 & 0.081 & 0.669 & 0.036 & 0.572 & 0.478\\
        1.50  & 4.48  & 0.593 & 0.046 & 0.487 & 0.013 & 0.356 & 0.346\\
        1.75  & 5.75  & 0.421 & 0.024 & 0.330 & 0.006 & 0.206 & 0.235\\
        2.00  & 7.39  & 0.288 & 0.012 & 0.224 & 0.002 & 0.124 & 0.124\\
        2.15  & 8.58  & 0.205 & 0.011 & 0.126 & 0.001 & 0.068 & 0.077\\
        2.30  & 9.97  & 0.130 & 0.007 & 0.068 & 0.000 & 0.051 & 0.045\\
        \bottomrule
    \end{tabular}
    \caption{$(w^*_1,w^*_2,w^*_3)=(0,1.7,2.45)$; $(v^*_1,v^*_2,v^*_3)=(0,1.25,1.4)$; $(\lambda_1^Z, \lambda_2^Z, \lambda_3^Z) = (1,0,0)$; $(\lambda_1^r, \lambda_2^r, \lambda_3^r) = (1,0.2,0)$; $\bm{\delta} = (0,1,1)$; $\mathbf{N}_{3\cdot}=(20,20,20)$. Case I in Section \mref{sec:power_main} of the main text.}
    \label{tb:weighted_four_60}
\end{table}

\begin{table}[H]
    \centering
    \small 
    \renewcommand{\arraystretch}{1.2} 
    \begin{tabular}{c c | c c c c c c }
        \toprule
        $\gamma$ & $\Gamma = e^\gamma$ & 3×3 (Opt) & 3×2 (V1) & 3×2 (V2) & 2×2 (V1) & 2×2 (V2) & Cross-Cut \\
        \midrule
        0.00  & 1.00  & 0.996 & 0.914 & 0.995& 0.823 & 0.965 & 0.989\\
        0.25  & 1.28  & 0.993 & 0.856 & 0.980& 0.697 & 0.917 & 0.972\\
        0.50  & 1.65  & 0.980 & 0.669 & 0.944& 0.500 & 0.817 & 0.939\\
        0.75  & 2.12  & 0.949 & 0.524 & 0.880& 0.325 & 0.700 & 0.888\\
        1.00  & 2.72  & 0.885 & 0.398 & 0.769& 0.193 & 0.555 & 0.804\\
        1.25  & 3.49  & 0.787 & 0.272 & 0.631& 0.074 & 0.428 & 0.690\\
        1.50  & 4.48  & 0.664 & 0.183 & 0.435& 0.035 & 0.220 & 0.556\\
        1.75  & 5.75  & 0.528 & 0.120 & 0.295& 0.018 & 0.115 & 0.431\\
        2.00  & 7.39  & 0.364 & 0.068 & 0.144& 0.010 & 0.067 & 0.290\\
        2.15  & 8.58  & 0.266 & 0.049 & 0.097& 0.004 & 0.036 & 0.187\\
        2.30  & 9.97  & 0.197 & 0.035 & 0.065& 0.003 & 0.028 & 0.125\\
        \bottomrule
    \end{tabular}
\caption{$(w^*_1,w^*_2,w^*_3)=(0,1,2)$; $(v^*_1,v^*_2,v^*_3)=(0,1.5,2)$; ($\lambda_1^Z$, $\lambda_2^Z$, $\lambda_3^Z$) = $(1,0,0)$; ($\lambda_1^r$, $\lambda_2^r$, $\lambda_3^r$) = $(1,0.2,0)$; $\bm{\delta} = (0,1,1)$; $\mathbf{N}_{3\cdot} = (20,20,20)$. Case II in Section \mref{sec:power_main} of the main text.}
\label{tb:weighted_four_v2_60}
\end{table}

\begin{table}[H]
    \centering
    \small 
    \renewcommand{\arraystretch}{1.2} 
    \begin{tabular}{c c | c c c c c c }
        \toprule
        $\gamma$ & $\Gamma = e^\gamma$ & 3×3 (Opt) & 3×2 (V1) & 3×2 (V2) & 2×2 (V1) & 2×2 (V2) & Cross-Cut \\
        \midrule
        0.00  & 1.00  &  0.963&  0.962&  0.680&  0.548&  0.253& 0.718\\
        0.25  & 1.28  &  0.943&  0.937&  0.573&  0.377&  0.142& 0.607\\
        0.50  & 1.65  &  0.897&  0.896&  0.467&  0.170&  0.070& 0.461\\
        0.75  & 2.12  &  0.834&  0.828&  0.347&  0.086&  0.027& 0.353\\
        1.00  & 2.72  &  0.767&  0.740&  0.260&  0.031&  0.015& 0.225\\
        1.25  & 3.49  &  0.683&  0.666&  0.170&  0.007&  0.001& 0.131\\
        1.50  & 4.48  &  0.597&  0.572&  0.102&  0.005&  0.001& 0.074\\
        1.75  & 5.75  &  0.507&  0.506&  0.059&  0.001&  0.000& 0.033\\
        2.00  & 7.39  &  0.429&  0.429&  0.020&  0.001&  0.000& 0.015\\
        2.15  & 8.58  &  0.377&  0.369&  0.014&  0.000&  0.000& 0.005\\
        2.30  & 9.97  &  0.342&  0.340&  0.012&  0.000&  0.000& 0.004\\
        \bottomrule
    \end{tabular}
\caption{$(w^*_1,w^*_2,w^*_3) = (0,0.2,1.5)$; $(v^*_1,v^*_2,v^*_3) = (0,0.2,1.5)$; $(\lambda_1^Z, \lambda_2^Z, \lambda_3^Z) = (0,0,0)$; $(\lambda_1^r,\lambda_2^r,\lambda_3^r) = (0,0,0)$; $\bm{\delta} = (0,1,1)$; $\mathbf{N}_{3\cdot} = (20,20,20)$. Case III in Section \mref{sec:power_main} of the main text.}
\label{tb:weighted_four_v8_60}
\end{table}

\begin{table}[H]
    \centering
    \small 
    \renewcommand{\arraystretch}{1.2} 
    \begin{tabular}{c c | c c c c c c }
        \toprule
        $\gamma$ & $\Gamma = e^\gamma$ & 3×3 (Opt) & 3×2 (V1) & 3×2 (V2) & 2×2 (V1) & 2×2 (V2) & Cross-Cut \\
        \midrule
        0.00  & 1.00  &  0.980&  0.976     &  0.743      &  0.657&  0.338& 0.793   \\
        0.25  & 1.28  &  0.957&  0.962     &  0.651     &  0.485 &  0.203& 0.685 \\
        0.50  & 1.65  &  0.930&  0.921     &  0.532     &  0.261 &  0.105& 0.564  \\
        0.75  & 2.12  &  0.860&  0.858     &  0.413    &   0.138 &  0.047& 0.444 \\
        1.00  & 2.72  &  0.788&  0.766     &  0.292    &  0.050  &  0.022& 0.311 \\
        1.25  & 3.49  &  0.707&  0.693     &  0.194     &  0.017 &  0.006& 0.198 \\
        1.50  & 4.48  &  0.597&  0.594     &  0.118    &  0.006&  0.003& 0.115 \\
        1.75  & 5.75  &  0.502&  0.497     &  0.053    &  0.003&  0.000& 0.055 \\
        2.00  & 7.39  &  0.408&  0.398     &  0.020    &  0.001&  0.000& 0.022 \\
        2.15  & 8.58  &  0.359&  0.350     &  0.013     &  0.000&  0.000& 0.009 \\
        2.30  & 9.97  &  0.315&  0.314     &  0.006     &  0.000&  0.000& 0.004 \\
        \bottomrule
    \end{tabular}
\caption{$(w^*_1,w^*_2,w^*_3) = (0,0.3,1.6)$; $(v^*_1,v^*_2,v^*_3) = (0,0.3,1.6)$; $(\lambda_1^Z, \lambda_2^Z, \lambda_3^Z) = (0,0,0)$; $(\lambda_1^r,\lambda_2^r,\lambda_3^r) = (0,0,0)$; $\bm{\delta} = (0,1,1)$; $\mathbf{N}_{3\cdot} = (20,20,20)$. Case IV in Section \mref{sec:power_main} of the main text.}
\label{tb:weighted_four_v10_60}
\end{table}



\begin{table}[H]
    \centering
    \small

    \renewcommand{\arraystretch}{1.2} 
    \begin{tabular}{c c | c c c c c c}
        \toprule
        $\gamma$ & $\Gamma = e^\gamma$ & 3×3 (Opt) & 3×2 (V1) & 3×2 (V2) & 2×2 (V1) & 2×2 (V2) & Cross-Cut \\
        \midrule
        0.00  & 1.00  &       1.000& 0.938 & 1.000 & 0.916 & 1.000 & 0.999\\
        0.25  & 1.28  &       1.000& 0.807 & 1.000 & 0.735 & 0.999 & 0.997\\
        0.50  & 1.65  &       0.999& 0.617 & 0.999 & 0.509 & 0.998 & 0.992\\
        0.75  & 2.12  &       0.995& 0.403 & 0.994 & 0.275 & 0.993 & 0.979\\
        1.00  & 2.72  &       0.989& 0.217 & 0.987 & 0.116 & 0.957 & 0.946\\
        1.25  & 3.49  &       0.961& 0.092 & 0.952 & 0.036 & 0.881 & 0.880\\
        1.50  & 4.48  &       0.871& 0.021 & 0.825 & 0.011 & 0.694 & 0.758\\
        1.75  & 5.75  &       0.685& 0.010 & 0.643 & 0.005 & 0.481 & 0.603\\
        2.00  & 7.39  &       0.484& 0.004 & 0.413 & 0.003 & 0.274 & 0.433\\
        2.15  & 8.58  &       0.358& 0.002 & 0.314 & 0.000 & 0.171 & 0.337\\
        2.30  & 9.97  &       0.234& 0.002 & 0.184 & 0.000 & 0.102 & 0.247\\
        \bottomrule
    \end{tabular}
    \caption{$(w^*_1,w^*_2,w^*_3)=(0,1.7,2.45)$; $(v^*_1,v^*_2,v^*_3)=(0,1.25,1.4)$; $(\lambda_1^Z, \lambda_2^Z, \lambda_3^Z) = (1,0,0)$; $(\lambda_1^r, \lambda_2^r, \lambda_3^r) = (1,0.2,0)$; $\bm{\delta} = (0,1,1)$; $\mathbf{N}_{3\cdot}=(40,40,40)$. Case I in Section \mref{sec:power_main} of the main text.}
    \label{tb:weighted_four_120}
\end{table}

\begin{table}[H]
    \centering
    \small 
    \renewcommand{\arraystretch}{1.2} 
    \begin{tabular}{c c | c c c c c c }
        \toprule
        $\gamma$ & $\Gamma = e^\gamma$ & 3×3 (Opt) & 3×2 (V1) & 3×2 (V2) & 2×2 (V1) & 2×2 (V2) & Cross-Cut \\
        \midrule
        0.00  & 1.00  &       1.000   &       0.999&      1.000&       0.991&       0.999&  1.000    \\
        0.25  & 1.28  &       1.000   &       0.977&      1.000&       0.946&       0.998&  1.000   \\
        0.50  & 1.65  &       1.000   &       0.937&      0.998&       0.808&       0.989&  1.000    \\
        0.75  & 2.12  &       0.997   &       0.833&      0.995&       0.602&       0.963&  0.996    \\
        1.00  & 2.72  &       0.993   &       0.677&      0.982&       0.343&       0.874&  0.990    \\
        1.25  & 3.49  &       0.978   &       0.507&      0.930&       0.163&       0.733&  0.973    \\
        1.50  & 4.48  &       0.936   &       0.333&      0.817&       0.059&       0.487&  0.936    \\
        1.75  & 5.75  &       0.819  &        0.185&      0.615&       0.022&       0.279&  0.839    \\
        2.00  & 7.39  &       0.651   &       0.092&      0.388&       0.007&       0.127&  0.711    \\
        2.15  & 8.58  &       0.540   &       0.066&      0.249&       0.004&       0.067&  0.610    \\
        2.30  & 9.97  &       0.407  &        0.047&      0.179&       0.003&       0.044&  0.511    \\
        \bottomrule
    \end{tabular}
\caption{$(w^*_1,w^*_2,w^*_3)=(0,1,2)$; $(v^*_1,v^*_2,v^*_3)=(0,1.5,2)$; ($\lambda_1^Z$, $\lambda_2^Z$, $\lambda_3^Z$) = $(1,0,0)$; ($\lambda_1^r$, $\lambda_2^r$, $\lambda_3^r$) = $(1,0.2,0)$; $\bm{\delta} = (0,1,1)$; $\mathbf{N}_{3\cdot} = (40,40,40)$. Case II in Section \mref{sec:power_main} of the main text.}
\label{tb:weighted_four_v2_120}
\end{table}

\begin{table}[H]
    \centering
    \small

    \renewcommand{\arraystretch}{1.2} 
    \begin{tabular}{c c | c c c c c c}
        \toprule
        $\gamma$ & $\Gamma = e^\gamma$ & 3×3 (Opt) & 3×2 (V1) & 3×2 (V2) & 2×2 (V1) & 2×2 (V2) & Cross-Cut \\
        \midrule
        0.00  & 1.00  &  1.000& 1.000& 0.924& 0.836& 0.459& 0.961\\
        0.25  & 1.28  &  0.999& 0.999& 0.870& 0.622& 0.240& 0.907\\
        0.50  & 1.65  &  0.993& 0.993& 0.773& 0.360& 0.103& 0.827\\
        0.75  & 2.12  &  0.978& 0.978& 0.653& 0.157& 0.019& 0.696\\
        1.00  & 2.72  &  0.955& 0.952& 0.493& 0.046& 0.004& 0.523\\
        1.25  & 3.49  &  0.913& 0.910& 0.329& 0.015& 0.001& 0.343\\
        1.50  & 4.48  &  0.851& 0.841& 0.192& 0.006& 0.000& 0.211\\
        1.75  & 5.75  &  0.764& 0.750& 0.098& 0.003& 0.000& 0.091\\
        2.00  & 7.39  &  0.639& 0.630& 0.048& 0.001& 0.000& 0.042\\
        2.15  & 8.58  &  0.576& 0.562& 0.028& 0.000& 0.000& 0.022\\
        2.30  & 9.97  &  0.513& 0.495& 0.010& 0.000& 0.000& 0.015\\
        \bottomrule
    \end{tabular}
    \caption{$(w^*_1,w^*_2,w^*_3)=(0,0.2,1.5)$; $(v^*_1,v^*_2,v^*_3)=(0,0.2,1.5)$; $(\lambda_1^Z, \lambda_2^Z, \lambda_3^Z) = (0,0,0)$; $(\lambda_1^r, \lambda_2^r, \lambda_3^r) = (0,0,0)$; $\bm{\delta} = (0,1,1)$; $\mathbf{N}_{3\cdot}=(40,40,40)$. Case III in Section \mref{sec:power_main} of the main text.}
    \label{tb:weighted_four_v8_120}
\end{table}

\begin{table}[H]
    \centering
    \small

    \renewcommand{\arraystretch}{1.2} 
    \begin{tabular}{c c | c c c c c c}
        \toprule
        $\gamma$ & $\Gamma = e^\gamma$ & 3×3 (Opt) & 3×2 (V1) & 3×2 (V2) & 2×2 (V1) & 2×2 (V2) & Cross-Cut \\
        \midrule
        0.00  & 1.00  &       1.000   & 1.000& 0.957& 0.915& 0.577& 0.985\\
        0.25  & 1.28  &       1.000   & 0.999& 0.915& 0.752& 0.356& 0.955\\
        0.50  & 1.65  &       0.998   & 0.996& 0.835& 0.508& 0.156& 0.896\\
        0.75  & 2.12  &       0.990   & 0.984& 0.716& 0.272& 0.063& 0.807\\
        1.00  & 2.72  &       0.971   & 0.963& 0.555& 0.089& 0.010& 0.664\\
        1.25  & 3.49  &       0.931   & 0.928& 0.368& 0.022& 0.003& 0.487\\
        1.50  & 4.48  &       0.870   & 0.852& 0.224& 0.008& 0.000& 0.317\\
        1.75  & 5.75  &       0.772   & 0.743& 0.102& 0.004& 0.000& 0.177\\
        2.00  & 7.39  &       0.637   & 0.620& 0.034& 0.002& 0.000& 0.084\\
        2.15  & 8.58  &       0.558   & 0.536& 0.017& 0.001& 0.000& 0.051\\
        2.30  & 9.97  &       0.481   & 0.460& 0.010& 0.000& 0.000& 0.032\\
        \bottomrule
    \end{tabular}
    \caption{$(w^*_1,w^*_2,w^*_3)=(0,0.3,1.6)$; $(v^*_1,v^*_2,v^*_3)=(0,0.3,1.6)$; $(\lambda_1^Z, \lambda_2^Z, \lambda_3^Z) = (0,0,0)$; $(\lambda_1^r, \lambda_2^r, \lambda_3^r) = (0,0,0)$; $\bm{\delta} = (0,1,1)$; $\mathbf{N}_{3\cdot}=(40,40,40)$. Case IV in Section \mref{sec:power_main} of the main text.}
    \label{tb:weighted_four_v10_120}
\end{table}

\begin{table}[H]
    \centering
    \small 
    \renewcommand{\arraystretch}{1.2} 
    \begin{tabular}{c c | c c c c c c}
        \toprule
        $\gamma$ & $\Gamma = e^\gamma$ & 3×3 (Opt) & 3×2 (V1) & 3×2 (V2) & 2×2 (V1) & 2×2 (V2) & Cross-Cut \\
        \midrule
        0.00  & 1.00  &  0.816     &  0.791     &  0.405     & 0.233      &  0.115   & 0.426 \\
        0.25  & 1.28  &  0.723     &  0.712     &  0.287     & 0.141      &  0.055   & 0.290 \\
        0.50  & 1.65  &  0.622     &  0.607     &  0.208     & 0.061      &  0.020   & 0.191 \\
        0.75  & 2.12  &  0.535     &  0.515     &  0.134     & 0.023      &  0.004   & 0.108 \\
        1.00  & 2.72  &  0.452     &  0.429     &  0.083     & 0.004      &  0.002   & 0.055 \\
        1.25  & 3.49  &  0.369     &  0.355     &  0.048     & 0.001      &  0.000   & 0.026 \\
        1.50  & 4.48  &  0.284     &  0.271     &  0.022     & 0.001      &  0.000   & 0.013 \\
        1.75  & 5.75  &  0.245     &  0.232     &  0.014     & 0.001      &  0.000   & 0.006 \\
        2.00  & 7.39  &  0.192     &  0.186     &  0.007     & 0.001      &  0.000   & 0.002 \\
        2.15  & 8.58  &  0.156     &  0.153     &  0.004     & 0.001      &  0.000   & 0.002\\
        2.30  & 9.97  &  0.139     &  0.137     &  0.004     & 0.000      &  0.000   & 0.001 \\
        \bottomrule
    \end{tabular}
\caption{$(w^*_1,w^*_2,w^*_3) = (0.01,0.015,1.2)$; $(v^*_1,v^*_2,v^*_3) = (0.01,0.015,1.2)$; $(\lambda_1^Z, \lambda_2^Z, \lambda_3^Z) = (0,0,0)$; $(\lambda_1^r,\lambda_2^r,\lambda_3^r) = (0,0,0)$; $\bm{\delta} = (0,1,1)$; $\mathbf{N}_{3\cdot} = (20,20,20)$.}
\label{tb:sparse_table_7}
\end{table}

\begin{table}[H]
    \centering
    \small 
    \renewcommand{\arraystretch}{1.2} 
     \begin{tabular}{c c | c c c c c c }
        \toprule
        $\gamma$ & $\Gamma = e^\gamma$ & 3×3 (Opt) & 3×2 (V1) & 3×2 (V2) & 2×2 (V1) & 2×2 (V2) & Cross-Cut \\
        \midrule
        0.00  & 1.00  &  0.809     &  0.809     &  0.424     &   0.747     &  0.324 & 0.441   \\
        0.25  & 1.28  &  0.664     &  0.664     &  0.259     &   0.605     &  0.200 & 0.309   \\
        0.50  & 1.65  &  0.468     &  0.468     &  0.155     &   0.362     &  0.093 & 0.196  \\
        0.75  & 2.12  &  0.285     &  0.285     &  0.070     &   0.215     &  0.045 & 0.106  \\
        1.00  & 2.72  &  0.149     &  0.149     &  0.032     &   0.093     &  0.014 & 0.054  \\
        1.25  & 3.49  &  0.065     &  0.065     &  0.007     &   0.035     &  0.004 & 0.026 \\
        1.50  & 4.48  &  0.027     &  0.027     &  0.002     &   0.013     &  0.001 & 0.009  \\
        1.75  & 5.75  &  0.009     &  0.009     &  0.001     &   0.005     &  0.000 & 0.006  \\
        2.00  & 7.39  &  0.005     &  0.005     &  0.000     &   0.004     &  0.000 & 0.002   \\
        2.15  & 8.58  &  0.003     &  0.003     &  0.000     &   0.001     &  0.000 & 0.002   \\
        2.30  & 9.97  &  0.002     &  0.002     &  0.000     &   0.001     &  0.000 & 0.002   \\
        \bottomrule
    \end{tabular}
\caption{$(w^*_1,w^*_2,w^*_3) = (0,1.1,1.2)$; $(v^*_1,v^*_2,v^*_3) = (0,0,1.2)$; $(\lambda_1^Z, \lambda_2^Z, \lambda_3^Z) = (0,0,0)$; $(\lambda_1^r,\lambda_2^r,\lambda_3^r) = (0,0,0)$; $\bm{\delta} = (0,1,1)$; $\mathbf{N}_{3\cdot} = (20,20,20)$.}
\label{tb:sparse_table_8}
\end{table}

\begin{table}[H]
    \centering
    \small 
    \renewcommand{\arraystretch}{1.2} 
    \begin{tabular}{c c | c c c c c c}
        \toprule
        $\gamma$ & $\Gamma = e^\gamma$ & 3×3 (Opt) & 3×2 (V1) & 3×2 (V2) & 2×2 (V1) & 2×2 (V2) & Cross-Cut \\
        \midrule
        0.00  & 1.00  &  0.978&  0.543&  0.962&  0.407&  0.903 & 0.915\\
        0.25  & 1.28  &  0.949&  0.410&  0.944&  0.262&  0.869 & 0.880\\
        0.50  & 1.65  &  0.922&  0.300&  0.912&  0.145&  0.811 & 0.822\\
        0.75  & 2.12  &  0.870&  0.198&  0.846&  0.084&  0.707 & 0.743\\
        1.00  & 2.72  &  0.801&  0.134&  0.764&  0.040&  0.604 & 0.634\\
        1.25  & 3.49  &  0.713&  0.097&  0.673&  0.006&  0.447 & 0.496\\
        1.50  & 4.48  &  0.592&  0.070&  0.537&  0.000&  0.357 & 0.361\\
        1.75  & 5.75  &  0.463&  0.049&  0.406&  0.000&  0.211 & 0.270\\
        2.00  & 7.39  &  0.318&  0.038&  0.279&  0.000&  0.115 & 0.189\\
        2.15  & 8.58  &  0.247&  0.033&  0.206&  0.000&  0.086 & 0.149\\
        2.30  & 9.97  &  0.184&  0.031&  0.142&  0.000&  0.063 & 0.110\\
        \bottomrule
    \end{tabular}
\caption{$(w^*_1,w^*_2,w^*_3)=(0,1.7 ,2.45)$; $(v^*_1,v^*_2,v^*_3)=(0,1.25,1.4)$; ($\lambda_1^Z$, $\lambda_2^Z$, $\lambda_3^Z$) = $(1,0,0)$; $(\lambda_1^r$, $\lambda_2^r$, $\lambda_3^r)$ = $(1,0.2,0)$; $\bm{\delta} = (0,1,1)$; $\mathbf{N}_{3\cdot} =(10,40,40).$}
\label{tb:weighted_four_sparse_1}
\end{table}

\begin{table}[H]
    \centering
    \small 
    \renewcommand{\arraystretch}{1.2} 
    \begin{tabular}{c c | c c c c c c}
        \toprule
        $\gamma$ & $\Gamma = e^\gamma$ & 3×3 (Opt) & 3×2 (V1) & 3×2 (V2) & 2×2 (V1) & 2×2 (V2) & Cross-Cut \\
        \midrule
        0.00  & 1.00  &  0.851&  0.766&  0.751&  0.417&  0.425 & 0.717\\
        0.25  & 1.28  &  0.772&  0.661&  0.649&  0.252&  0.302 & 0.604\\
        0.50  & 1.65  &  0.670&  0.556&  0.519&  0.142&  0.149 & 0.476\\
        0.75  & 2.12  &  0.568&  0.447&  0.404&  0.061&  0.066 & 0.370\\
        1.00  & 2.72  &  0.460&  0.356&  0.279&  0.034&  0.026 & 0.229\\
        1.25  & 3.49  &  0.362&  0.293&  0.177&  0.012&  0.003 & 0.145\\
        1.50  & 4.48  &  0.262&  0.216&  0.103&  0.007&  0.001 & 0.072\\
        1.75  & 5.75  &  0.184&  0.168&  0.059&  0.003&  0.000 & 0.038\\
        2.00  & 7.39  &  0.126&  0.147&  0.027&  0.000&  0.000 & 0.021\\
        2.15  & 8.58  &  0.095&  0.133&  0.018&  0.000&  0.000 & 0.011\\
        2.30  & 9.97  &  0.072&  0.116&  0.017&  0.000&  0.000 & 0.009\\
        \bottomrule
    \end{tabular}
\caption{$(w^*_1,w^*_2,w^*_3)=(0,0.3 ,1)$; $(v^*_1,v^*_2,v^*_3)=(0,1 ,2)$; ($\lambda_1^Z$, $\lambda_2^Z$, $\lambda_3^Z$) = $(1,0,0)$; $(\lambda_1^r$, $\lambda_2^r$, $\lambda_3^r)$ = $(1,0.2,0)$; $\bm{\delta} = (0,1,1)$; $\mathbf{N}_{3 \cdot} = (20,20,20).$ }
\label{tb:weighted_four_v3}
\end{table}

\begin{table}[H]
    \centering
    \small 
    \renewcommand{\arraystretch}{1.2} 
    \begin{tabular}{c c | c c c c c c }
        \toprule
        $\gamma$ & $\Gamma = e^\gamma$ & 3×3 (Opt) & 3×2 (V1) & 3×2 (V2) & 2×2 (V1) & 2×2 (V2) & Cross-Cut \\
        \midrule
        0.00  & 1.00  &  0.444&  0.197&  0.409&  0.136&  0.333 & 0.183\\
        0.25  & 1.28  &  0.275&  0.102&  0.256&  0.058&  0.214 & 0.104\\
        0.50  & 1.65  &  0.156&  0.050&  0.146&  0.025&  0.097 & 0.056\\
        0.75  & 2.12  &  0.062&  0.024&  0.054&  0.008&  0.030 & 0.034\\
        1.00  & 2.72  &  0.024&  0.012&  0.023&  0.005&  0.014 & 0.014\\
        1.25  & 3.49  &  0.011&  0.008&  0.011&  0.000&  0.007 & 0.007\\
        1.50  & 4.48  &  0.005&  0.003&  0.004&  0.000&  0.001 & 0.003\\
        1.75  & 5.75  &  0.001&  0.001&  0.001&  0.000&  0.001 & 0.001\\
        2.00  & 7.39  &  0.001&  0.001&  0.000&  0.000&  0.000 & 0.001\\
        2.15  & 8.58  &  0.000&  0.001&  0.000&  0.000&  0.000 & 0.001\\
        2.30  & 9.97  &  0.000&  0.001&  0.000&  0.000&  0.000 & 0.001\\
        \bottomrule
    \end{tabular}
\caption{$(w^*_1,w^*_2,w^*_3)=(0,0.8 ,0.9)$; $(v^*_1,v^*_2,v^*_3)=(0,0.9 ,1)$; ($\lambda_1^Z$, $\lambda_2^Z$, $\lambda_3^Z$) = $(1,0,0)$; $(\lambda_1^r$, $\lambda_2^r$, $\lambda_3^r)$ = $(1,0.2,0)$; $\bm{\delta} = (0,1,1)$; $\mathbf{N}_{3 \cdot} = (20,20,20)$.}
\label{tb:weighted_four_v5}
\end{table}

\begin{table}[H]
    \centering
    \small
    \renewcommand{\arraystretch}{1.2} 
    \begin{tabular}{c c | c c c c c c}
        \toprule
        $\gamma$ & $\Gamma = e^\gamma$ & 3×3 (Opt) & 3×2 (V1) & 3×2 (V2) & 2×2 (V1) & 2×2 (V2) & Cross-Cut \\
        \midrule
        0.00  & 1.00 &  0.991 & 0.847 &  0.980 &0.629  &  0.867 & 0.966\\
        0.25  & 1.28 &  0.987 & 0.772 &  0.966 &0.471 &   0.807 & 0.945\\
        0.50  & 1.65 &  0.976&  0.695 &  0.937 &0.309 &   0.729 & 0.908\\
        0.75  & 2.12 &  0.960&  0.591 &  0.891 &0.182 &   0.593 & 0.858 \\
        1.00  & 2.72 &  0.919&  0.506 &  0.826 &0.107 &   0.470 & 0.807 \\
        1.25  & 3.49 &  0.872&  0.404 &  0.748 &0.054 &   0.320 & 0.721 \\
        1.50  & 4.48 &  0.809&  0.336 &  0.671 &0.011 &   0.224 & 0.598 \\
        1.75  & 5.75 &  0.705&  0.282 &  0.491 &0.001 &    0.125& 0.477 \\
        2.00  & 7.39 &  0.587&  0.246 &  0.394 &0.000&    0.073 & 0.356\\
        2.15  & 8.58 &  0.519&  0.231 &  0.284 &0.000&    0.055 & 0.281\\
        2.30  & 9.97 &  0.445 &  0.216 &  0.267 &0.000&    0.033& 0.226\\
        \bottomrule
    \end{tabular}
    \caption{$(w^*_1,w^*_2,w^*_3)=(0,1 ,2)$; $(v^*_1,v^*_2,v^*_3)=(0,1.5,2)$; ($\lambda_1^Z$, $\lambda_2^Z$, $\lambda_3^Z$) = $(1,0,0)$; $(\lambda_1^r$, $\lambda_2^r$, $\lambda_3^r)$ = $(1,0.2,0)$; $\bm{\delta} = (0,1,1)$; $\mathbf{N}_{3\cdot} =(10,40,40).$}
    \label{tb:weighted_four_sparse_3}
\end{table}

\begin{table}[H]
    \centering
    \small
    \renewcommand{\arraystretch}{1.2} 
    \begin{tabular}{c c | c c c c c c}
        \toprule
        $\gamma$ & $\Gamma = e^\gamma$ & 3×3 (Opt) & 3×2 (V1) & 3×2 (V2) & 2×2 (V1) & 2×2 (V2) & Cross-Cut \\
        \midrule
        0.00  & 1.00 &  0.998  & 0.920 & 0.994  & 0.902  & 0.989  & 0.949\\
        0.25  & 1.28 &  0.992 &  0.829 & 0.980  & 0.728 &  0.958  & 0.912\\
        0.50  & 1.65 &  0.974 &  0.669 &  0.951 & 0.537  &  0.904 & 0.854\\
        0.75  & 2.12 &  0.931 &  0.496 &  0.863 &  0.337 &  0.751 & 0.753\\
        1.00  & 2.72 &  0.829 &  0.314 &  0.702 & 0.170  &  0.580 & 0.607\\
        1.25  & 3.49 &  0.669 &  0.173 &  0.505 & 0.059 &  0.335  & 0.442\\
        1.50  & 4.48 &  0.464 &  0.080 &  0.257 & 0.020&  0.159   & 0.293\\
        1.75  & 5.75 &  0.262 &  0.031 &  0.111&  0.007 &  0.066  &  0.142\\
        2.00  & 7.39 &  0.140 &  0.012 &  0.054 & 0.002  &  0.025 & 0.057\\
        2.15  & 8.58 &  0.091 &  0.007 &  0.022 & 0.001 &  0.010  & 0.029\\
        2.30  & 9.97 &  0.057 &  0.006 &  0.012 & 0.001 &  0.004 & 0.013\\
        \bottomrule
    \end{tabular}
    \caption{$(w^*_1,w^*_2,w^*_3)=(0,1 ,2)$; $(v^*_1,v^*_2,v^*_3)=(0,1.5,2)$; ($\lambda_1^Z$, $\lambda_2^Z$, $\lambda_3^Z$) = $(1,0,0)$; $(\lambda_1^r$, $\lambda_2^r$, $\lambda_3^r)$ = $(1,0.2,0)$; $\bm{\delta} = (0,1,1)$; $\mathbf{N}_{3\cdot} =(40,40,10).$}
    \label{tb:weighted_four_sparse_4}
\end{table}

\begin{table}[H]
    \centering
    \small 
    \renewcommand{\arraystretch}{1.2} 
    \begin{tabular}{c c | c c c c c c }
        \toprule
        $\gamma$ & $\Gamma = e^\gamma$ & 3×3 (Opt) & 3×2 (V1) & 3×2 (V2) & 2×2 (V1) & 2×2 (V2) & Cross-Cut \\
        \midrule
        0.00  & 1.00  &  0.973     &  0.973     &  0.651     &  0.973     &  0.651  & 0.722 \\
        0.25  & 1.28  &  0.941     &  0.941     &  0.501     &  0.941     &  0.501  & 0.614\\
        0.50  & 1.65  &  0.882     &  0.882     &  0.329     &  0.882     &  0.329  & 0.473 \\
        0.75  & 2.12  &  0.745     &  0.745     &  0.241     &  0.745     &  0.241  & 0.358 \\
        1.00  & 2.72  &  0.569     &  0.569     &  0.135     &  0.569     &  0.135  & 0.233 \\
        1.25  & 3.49  &  0.412     &  0.412     &  0.082     &  0.412     &  0.082  & 0.136 \\
        1.50  & 4.48  &  0.219     &  0.219     &  0.034     &  0.219     &  0.034  & 0.074 \\
        1.75  & 5.75  &  0.120     &  0.120     &  0.010     &  0.120     &  0.010  & 0.038 \\
        2.00  & 7.39  &  0.054     &  0.054     &  0.005     &  0.054     &  0.005  & 0.018 \\
        2.15  & 8.58  &  0.028     &  0.028     &  0.002     &  0.028     &  0.002  & 0.007 \\
        2.30  & 9.97  &  0.020     &  0.020     &  0.001     &  0.020     &  0.001  & 0.004 \\
        \bottomrule
    \end{tabular}
\caption{$(w^*_1,w^*_2,w^*_3) = (0,1.5,1.5)$; $(v^*_1,v^*_2,v^*_3) = (0,0,1.5)$; $(\lambda_1^Z, \lambda_2^Z, \lambda_3^Z) = (0,0,0)$; $(\lambda_1^r,\lambda_2^r,\lambda_3^r) = (0,0,0)$; $\bm{\delta} = (0,1,1)$; $\mathbf{N}_{3\cdot} = (20,20,20)$.}
\label{tb:sparse_table_6}
\end{table}

\begin{table}[H]
    \centering
    \small 
    \renewcommand{\arraystretch}{1.2} 
    \begin{tabular}{c c | c c c c c c }
        \toprule
        $\gamma$ & $\Gamma = e^\gamma$ & 3×3 (Opt) & 3×2 (V1) & 3×2 (V2) & 2×2 (V1) & 2×2 (V2) & Cross-Cut \\
        \midrule
        0.00  & 1.00  &  0.973     &  0.973     &  0.651     &  0.973     &  0.651  & 0.722   \\
        0.25  & 1.28  &  0.959     &  0.959     &  0.608     &  0.941     &  0.501  & 0.614  \\
        0.50  & 1.65  &  0.949     &  0.949     &  0.501     &  0.882     &  0.329  & 0.473  \\
        0.75  & 2.12  &  0.921     &  0.921     &  0.469     &  0.745     &  0.241  & 0.358 \\
        1.00  & 2.72  &  0.902     &  0.902     &  0.468     &  0.569     &  0.135  & 0.233 \\
        1.25  & 3.49  &  0.858     &  0.858     &  0.338     &  0.412     &  0.082  & 0.136 \\
        1.50  & 4.48  &  0.817     &  0.817     &  0.325     &  0.219     &  0.034  & 0.074 \\
        1.75  & 5.75  &  0.792     &  0.792     &  0.323     &  0.120     &  0.010  & 0.038 \\
        2.00  & 7.39  &  0.774     &  0.774     &  0.323     &  0.054     &  0.005  & 0.018 \\
        2.15  & 8.58  &  0.752     &  0.752     &  0.305     &  0.028     &  0.002  & 0.007 \\
        2.30  & 9.97  &  0.688     &  0.688     &  0.274     &  0.020     &  0.001  & 0.004 \\
        \bottomrule
    \end{tabular}
\caption{$(w^*_1,w^*_2,w^*_3) = (0,1.5,1.5)$; $(v^*_1,v^*_2,v^*_3) = (0,0,1.5)$; $(\lambda_1^Z, \lambda_2^Z, \lambda_3^Z) = (0,0,0)$; $(\lambda_1^r,\lambda_2^r,\lambda_3^r) = (0,0,0)$; $\bm{\delta} = (0,0,1)$; $\mathbf{N}_{3\cdot} = (20,20,20)$. }
\label{tb:sparse_table_12}
\end{table}

\FloatBarrier

\subsection{Power of Ordinal Tests: Non-Collapsed Table Comparisons} \label{app:F6}
Here we examine the influence of attaching different weights $w_i$ and $v_j$ to the ordinal test 
\[
T = \sum_{i=1}^{3}\sum_{j=1}^{3} w_i v_j N_{ij}.
\]
We compare the performance of five test statistics. The $3 \times 3$ (Opt) is included in each favorable situation, taking $w_i = w_i^*$ and $v_j = v_j^*$ as in the data-generating process. We further compare it with: 
\begin{itemize}
    \item Type 2: $(w_1,w_2,w_3) = (0,1,2);\; (v_1,v_2,v_3) = (0,1,2)$.
    \item Type 3: $(w_1,w_2,w_3) = (0,1,3);\; (v_1,v_2,v_3) = (0,1,1)$.
    \item Type 4: $(w_1,w_2,w_3) = (0,1,2);\; (v_1,v_2,v_3) = (0,1,3)$.
    \item Type 5: $(w_1,w_2,w_3) = (0,2,3);\; (v_1,v_2,v_3) = (0,2,3)$.
\end{itemize}

Simulations show that ordinal tests with the oracle weights $w_i^*$ and $v_j^*$ are usually a decent choice for maintaining high power; moreover, even when the assigned scores are not optimal, the ordinal test usually performs comparably to or better than the collapsed or cross-cut test.
\begin{table}[H]  
    \centering
    \small 
    \renewcommand{\arraystretch}{1.2} 
    \begin{tabular}{c c | c c c c c}
        \toprule
        $\gamma$ & $\Gamma = e^\gamma$ & 3×3 (Opt) & Type 2& Type 3& Type 4 & Type 5\\
        \midrule
        0.00  & 1.00  &  0.995&  0.958&  0.984&  0.907  & 0.989\\
        0.25  & 1.28  &  0.988&  0.920&  0.969&  0.836  & 0.973\\
        0.50  & 1.65  &  0.960&  0.867&  0.933&  0.734  & 0.948\\
        0.75  & 2.12  &  0.908&  0.782&  0.837&  0.613& 0.883\\
        1.00  & 2.72  &  0.829&  0.680&  0.755&  0.501& 0.806\\
        1.25  & 3.49  &  0.720&  0.566&  0.653&  0.370& 0.706\\
        1.50  & 4.48  &  0.593&  0.452&  0.527&  0.281& 0.583\\
        1.75  & 5.75  &  0.421&  0.326&  0.361&  0.208& 0.418\\
        2.00  & 7.39  &  0.288&  0.227&  0.215&  0.146& 0.278\\
        2.15  & 8.58  &  0.205&  0.178&  0.166&  0.120& 0.205\\
        2.30  & 9.97  &  0.130&  0.136&  0.085&  0.088& 0.145\\
        \bottomrule
    \end{tabular}
    \caption{$(w^*_1,w^*_2,w^*_3)=(0,1.7 ,2.45)$; $(v^*_1,v^*_2,v^*_3)=(0,1.25,1.4)$; ($\lambda_1^Z$, $\lambda_2^Z$, $\lambda_3^Z$) = $(1,0,0)$; $(\lambda_1^r$, $\lambda_2^r$, $\lambda_3^r)$ = $(1,0.2,0)$; $\bm{\delta} = (0,1,1)$; $\mathbf{N}_{3\cdot} = (20,20,20)$. Case I in Section \mref{sec:power_main} of the main text.}
    \label{tb:weighted_four_type}
\end{table}

\begin{table}[H]  
    \centering
    \small 
    \renewcommand{\arraystretch}{1.2} 
    \begin{tabular}{c c | c c c c c}
        \toprule
        $\gamma$ & $\Gamma = e^\gamma$ & 3×3 (Opt) & Type 2& Type 3& Type 4 & Type 5\\
        \midrule
        0.00  & 1.00  &  0.996&  0.994&  0.993&   0.983 & 0.996 \\
        0.25  & 1.28  &  0.993&  0.976&  0.982&   0.959& 0.989\\
        0.50  & 1.65  &  0.980&  0.955&  0.961&   0.914 & 0.970\\
        0.75  & 2.12  &  0.949&  0.909&  0.884&   0.849 & 0.930\\
        1.00  & 2.72  &  0.885&  0.846&  0.805&   0.743 & 0.851\\
        1.25  & 3.49  &  0.787&  0.738&  0.692&   0.625 & 0.745\\
        1.50  & 4.48  &  0.664&  0.631&  0.568&   0.511 & 0.620\\
        1.75  & 5.75  &  0.528&  0.516&  0.382&   0.398 & 0.464\\
        2.00  & 7.39  &  0.364&  0.381&  0.222&   0.306 & 0.312\\
        2.15  & 8.58  &  0.266&  0.308&  0.175&   0.252 & 0.224\\
        2.30  & 9.97  &  0.197&  0.254&  0.091&   0.212 & 0.152\\
        \bottomrule
    \end{tabular}
    \caption{$(w^*_1,w^*_2,w^*_3)=(0,1 ,2)$; $(v^*_1,v^*_2,v^*_3)=(0,1.5 ,2)$; $(\lambda_1^Z$, $\lambda_2^Z$, $\lambda_3^Z$) = $(1,0,0)$; $(\lambda_1^r$, $\lambda_2^r$, $\lambda_3^r)$ = $(1,0.2,0)$; $\bm{\delta} = (0,1,1)$; $\mathbf{N}_{3\cdot} = (20,20,20)$. Case II in Section \mref{sec:power_main} of the main text.}
    \label{tb:weighted_four_v2}
\end{table}

\begin{table}[H]  
    \centering
    \small 
    \renewcommand{\arraystretch}{1.2} 
    \begin{tabular}{c c | c c c c c}
        \toprule
        $\gamma$ & $\Gamma = e^\gamma$ & 3×3 (Opt) & Type 2 & Type 3 & Type 4 & Type 5  \\
        \midrule
        0.00  & 1.00  &  0.851&  0.822&  0.732&  0.820 & 0.756\\
        0.25  & 1.28  &  0.772&  0.695&  0.609&  0.700 & 0.611\\
        0.50  & 1.65  &  0.670&  0.572&  0.494&  0.583 & 0.466\\
        0.75  & 2.12  &  0.568&  0.447&  0.350&  0.481 & 0.321\\
        1.00  & 2.72  &  0.460&  0.323&  0.235&  0.363 & 0.181\\
        1.25  & 3.49  &  0.362&  0.205&  0.138&  0.261 & 0.095\\
        1.50  & 4.48  &  0.262&  0.127&  0.075&  0.166 & 0.042\\
        1.75  & 5.75  &  0.184&  0.068&  0.035&  0.118 & 0.017\\
        2.00  & 7.39  &  0.126&  0.033&  0.017&  0.074 & 0.009\\
        2.15  & 8.58  &  0.095&  0.026&  0.015&  0.050& 0.004\\
        2.30  & 9.97  &  0.072&  0.016&  0.008&  0.032& 0.001\\
        \bottomrule
    \end{tabular}
    \caption{$(w^*_1,w^*_2,w^*_3)=(0,0.3,1)$; $(v^*_1,v^*_2,v^*_3)=(0,1,2)$; ($\lambda_1^Z$, $\lambda_2^Z$, $\lambda_3^Z$) = $(1,0,0)$; $(\lambda_1^r$, $\lambda_2^r$, $\lambda_3^r)$ = $(1,0.2,0)$; $\bm{\delta} = (0,1,1)$;  $\mathbf{N}_{3\cdot} = (20,20,20)$. 
 }
 \label{tb:weighted_four_v3_type}
\end{table}

\begin{table}[H]  
    \centering
    \small 
    \renewcommand{\arraystretch}{1.2} 
    \begin{tabular}{c c | c c c c c}
        \toprule
        $\gamma$ & $\Gamma = e^\gamma$ & 3×3 (Opt) & Type 2 & Type 3 & Type 4 & Type 5\\
        \midrule
        0.00  & 1.00  &  0.639&  0.592&  0.546&  0.562 & 0.632 \\
        0.25  & 1.28  &  0.495&  0.467&  0.413&  0.435 & 0.483\\
        0.50  & 1.65  &  0.362&  0.341&  0.288&  0.319 & 0.351\\
        0.75  & 2.12  &  0.221&  0.213&  0.166&  0.212 & 0.216\\
        1.00  & 2.72  &  0.111&  0.132&  0.088&  0.139 & 0.110\\
        1.25  & 3.49  &  0.052&  0.075&  0.048&  0.086 & 0.053\\
        1.50  & 4.48  &  0.023&  0.038&  0.024&  0.052 & 0.021\\
        1.75  & 5.75  &  0.010&  0.016&  0.008&  0.025 & 0.010\\
        2.00  & 7.39  &  0.003&  0.005&  0.006&  0.015 & 0.003\\
        2.15  & 8.58  &  0.002&  0.003&  0.005&  0.010 & 0.001\\
        2.30  & 9.97  &  0.000&  0.002&  0.002&  0.005 & 0.000\\
        \bottomrule
    \end{tabular}
    \caption{$(w^*_1,w^*_2,w^*_3)=(0,1 ,1.5)$; $(v^*_1,v^*_2,v^*_3)=(0,0.7 ,1)$; ($\lambda_1^Z$, $\lambda_2^Z$, $\lambda_3^Z$) = $(1,0,0)$; $(\lambda_1^r$, $\lambda_2^r$, $\lambda_3^r)$ = $(1,0.2,0)$; $\bm{\delta} = (0,1,1)$;  $\mathbf{N}_{3\cdot} = (20,20,20)$.}
    \label{tb:weighted_four_v4_type}
\end{table}

\subsection{Power of Ordinal Tests under Sparsity: Collapsed, Non-Collapsed, and Cross-Cut Tables} \label{app:F7}
This section compares the performance of tests when the table is sparse (i.e., when some cells are zero). To quantify sparsity, we record the proportion of times each of the nine cells is zero across values of $\Gamma$ and iterations, under each favorable situation. We then examine whether the proposed optimal test remains the most powerful under sparse tables. 

The percentage of zeros occurring in each cell is calculated as follows. We consider $11$ different values of $\Gamma$, and for each $\Gamma$, we generate $1{,}000$ tables. The percentage of zeros in a given cell is the number of times that cell equals $0$, divided by the total number of iterations, $11 \times 1{,}000 = 11{,}000$.

\begin{table}[H]
    \centering
    \small 
    \renewcommand{\arraystretch}{1.2} 
    \begin{tabular}{c c | c c c c c c}
        \toprule
        $\gamma$ & $\Gamma = e^\gamma$ & 3×3 (Opt) & 3×2 (V1) & 3×2 (V2) & 2×2 (V1) & 2×2 (V2) & Cross-Cut \\
        \midrule
        0.00  & 1.00  & 0.990 & 0.975 & 0.880 & 0.925 & 0.791 & 0.949\\
        0.25  & 1.28  & 0.971 & 0.949 & 0.808 & 0.855 & 0.667 & 0.912\\
        0.50  & 1.65  & 0.936 & 0.874 & 0.716 & 0.719 & 0.516 & 0.854\\
        0.75  & 2.12  & 0.880 & 0.759 & 0.575 & 0.547 & 0.425 & 0.753\\
        1.00  & 2.72  & 0.786 & 0.614 & 0.448 & 0.325 & 0.312 & 0.607\\
        1.25  & 3.49  & 0.641 & 0.437 & 0.318 & 0.168 & 0.213 & 0.442\\
        1.50  & 4.48  & 0.488 & 0.272 & 0.142 & 0.077 & 0.102 & 0.293\\
        1.75  & 5.75  & 0.320 & 0.161 & 0.078 & 0.040 & 0.054 & 0.142\\
        2.00  & 7.39  & 0.209 & 0.066 & 0.035 & 0.018 & 0.027 & 0.057\\
        2.15  & 8.58  & 0.145 & 0.056 & 0.017 & 0.011 & 0.011 & 0.029\\
        2.30  & 9.97  & 0.099 & 0.025 & 0.003 & 0.005 & 0.002 & 0.013\\
        \bottomrule
    \end{tabular}
\caption{$\mathbf{N}_{3 \cdot} = (20,20,20)$. For the data-generating process, $(w^*_1,w^*_2,w^*_3) = (0,1,2)$; $(v^*_1,v^*_2,v^*_3) = (0,1,2)$; $(\lambda_1^{Z}, \lambda_2^Z, \lambda_3^Z) = (0,0,0)$; $(\lambda_1^r,\lambda_2^r, \lambda_3^r)=(0,0,0)$; $\bm{\delta} = (0,1,1)$. The overall proportion across $\Gamma$ and iterations that the cells $((N_{11}, N_{12}, N_{13}, N_{21}, N_{22}, N_{23}, N_{31}, N_{32}, N_{33})$ being zero are $(0\%, 0\%, 0\%, 14\%, 1\%, 0\%, 73\%, 7\%, 0\%)$. In other words, this data-generating process produces a high frequency of zeros in $N_{31}$.} 
\label{tb:sparse_table_1}
\end{table}

\begin{table}[H]
    \centering
    \small 
    \renewcommand{\arraystretch}{1.2} 
    \begin{tabular}{c c | c c c c c c}
        \toprule
        $\gamma$ & $\Gamma = e^\gamma$ & 3×3 (Opt) & 3×2 (V1) & 3×2 (V2) & 2×2 (V1) & 2×2 (V2) & Cross-Cut \\
        \midrule
        0.00  & 1.00  &  0.954&  0.936&  0.271&  0.330&  0.131 & 0.289\\
        0.25  & 1.28  &  0.934&  0.899&  0.174&  0.193&  0.057 & 0.174\\
        0.50  & 1.65  &  0.890&  0.860&  0.081&  0.088&  0.021 & 0.121\\
        0.75  & 2.12  &  0.842&  0.806&  0.038&  0.039&  0.011 & 0.057\\
        1.00  & 2.72  &  0.762&  0.730&  0.012&  0.014&  0.007 & 0.017\\
        1.25  & 3.49  &  0.682&  0.663&  0.006&  0.008&  0.002 & 0.004\\
        1.50  & 4.48  &  0.560&  0.522&  0.002&  0.000&  0.000 & 0.000\\
        1.75  & 5.75  &  0.374&  0.356&  0.000&  0.000&  0.000 & 0.000\\
        2.00  & 7.39  &  0.273&  0.265&  0.000&  0.000&  0.000 & 0.000\\
        2.15  & 8.58  &  0.204&  0.194&  0.000&  0.000&  0.000 & 0.000\\
        2.30  & 9.97  &  0.200&  0.194&  0.000&  0.000&  0.000 & 0.000\\
        \bottomrule
    \end{tabular}
\caption{$\mathbf{N}_{3 \cdot} = (20,20,20)$. For the data-generating process, $(w^*_1,w^*_2,w^*_3) = (1,1,2)$; $(v^*_1,v^*_2,v^*_3) = (0,0.5,3.5)$; $(\lambda_1^{Z}, \lambda_2^Z, \lambda_3^Z) = (0,0,1)$; $(\lambda_1^r,\lambda_2^r, \lambda_3^r)=(2,2,0)$; $\bm{\delta} = (0,1,1)$. The overall proportion across $\Gamma$ and iterations that the cells $(N_{11}, N_{12}, N_{13}, N_{21}, N_{22}, N_{23}, N_{31}, N_{32}, N_{33})$ being zero are $(5\%, 0\%, 0\%, 6\%, 0\%, 0\%, 87\%, 69\%, 0\%)$. In other words, this data-generating process produces a high frequency of zeros in $N_{31}$ and $N_{32}$.} 
\label{tb:sparse_table_4}
\end{table}
We note that even in the presence of sparsity, the aforementioned principle still holds: the $3 \times 3$ (Opt) test performs the best.

\subsection{Size of Tests: Exact vs. Normal Approximation of P-Values} \label{app:F8}
This paper not only develops the exact null distribution, but also derives the exact first and second moments of ordinal tests, which can be used to construct a normal approximation to the worst-case p-value. For example, for ordinal tests, we may consider each $\mathbf{u} \in \mathcal{U}_{\rm O}$, where
\[
\mathcal{U}_{\rm O} = \{(0,\ldots,0), (0,\ldots,0,1), \ldots, (1,\ldots,1)\}.
\]  
At each candidate $\mathbf{u}$, instead of computing the exact upper bound on the p-value via Theorem~\mref{thm:recursive_probability} in Section~\mref{sec:exact_worst_null}, we approximate it by  
\[
1 - \Phi\!\left(\frac{T_{\text{obs}} - \mathbb{E}_{\mathbf{u}}[T]}{\mathrm{sd}_{\mathbf{u}}(T)}\right),
\]  
where $\mathbb{E}_{\mathbf{u}}[T]$ and $\mathrm{sd}_{\mathbf{u}}(T)$ are the exact mean and standard deviation of the test statistic under $\mathbf{u}$, computed from the moment formulas in Section~\ref{app:D} of the Supplementary Material. The normally-approximated worst-case p-value is then the maximum of this expression over all $\mathbf{u} \in \mathcal{U}_{\rm O}$.

For the sign-score test with binary outcomes, a worst-case confounder $\mathbf{u}^+$ is given by
\[
\mathbf{u}^+ = (\underbrace{0,\ldots,0}_{N_{\cdot 1}}, \underbrace{1,\ldots,1}_{N_{\cdot 2}}),
\]
so the worst-case normal-approximated p-value  is
\[
1 - \Phi\!\left(\frac{T_{\text{obs}} - \mathbb{E}_{\mathbf{u}^+}[T]}{\mathrm{sd}_{\mathbf{u}^+}(T)}\right).
\]

The normal approximation offers computational advantages over the exact method by avoiding the sum over $\Omega_{\mathbf{t}}$ in Theorem~\mref{thm:recursive_probability}. The following simulations illustrate the differences between the two methods in terms of type I error rate control.

We conduct a simulation study using the exact worst-case unmeasured confounder identified in Theorem~\mref{thm:score_test_maximizer_binary}. Specifically, we generate data 1,000 times under the generic bias sensitivity model at the worst-case $\mathbf{u}^+$, using $\bm{\delta} = (0,0,1)$ and varying values of $\gamma$. Importantly, Corollary~\mref{cor:unique_corner_multivariate_hyper} implies that the data can be generated from a multivariate extended hypergeometric distribution; see \citet{Fog2008} and the \texttt{R} package \texttt{BiasedUrn} for details. For each iteration, we focus on a sign-score test with treatment weights $(w_1, w_2, w_3) = (0,1,2)$ and outcome weights $(v_1, v_2) = (0,1)$. We then compare the type I error rate based on the exact worst-case p-value and the normal-approximated worst-case p-value.

\begin{figure}[ht]
    \centering
    \includegraphics[width=0.9\textwidth]{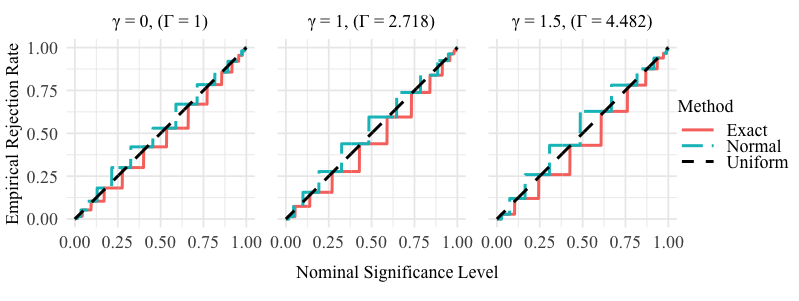}
    \caption{Type I error rate of the ordinal test on a $3\times 2$ table based on exact and normal-approximated worst-case p-values. Treatment margins are $\mathbf{N}_{3\cdot} = (60,10,20)$ and outcome margins are $\mathbf{N}_{\cdot 2} = (15,75)$. $\gamma \in \{0,1,1.5\}$.}
    \label{fig:size_control_exact_and_normal_scenario_1}
\end{figure}

Figure~\ref{fig:size_control_exact_and_normal_scenario_1} shows the empirical type I error rates when comparing exact and normal-approximated worst-case p-values. The rejection rates are computed as 
\[
\frac{1}{1000}\sum_{m=1}^{1000}\mathbbm{1}\{p \leq \alpha\}, 
\]
for $\alpha \in [0,1]$, where $p$ is either the exact or the normal-approximated worst-case p-value. The black 45-degree line represents the ideal case of uniform $p$-values under the null, and deviations above this line indicate anti-conservative behavior. Figure~\ref{fig:size_control_exact_and_normal_scenario_1} demonstrates that the normal approximation leads to inflated type I error rates, and that this inflation tends to worsen as the sensitivity parameter $\gamma$ increases. 

\begin{figure}[ht]
    \centering
    \includegraphics[width=0.9\textwidth]{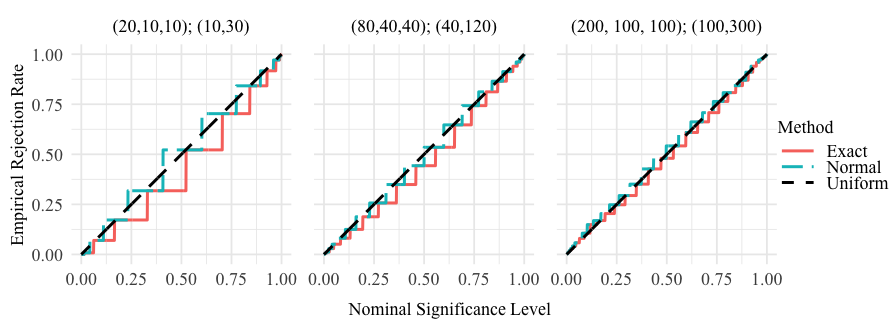}
    \caption{Type I error rate of the ordinal test on a $3\times 2$ table based on exact and normal-approximated worst-case p-values when margins increase proportionally. Treatment margins $\mathbf{N}_{3\cdot}$ grow from $(20,10,10)$ to $(80,40,40)$ and $(200,100,100)$, while outcome margins $\mathbf{N}_{\cdot 2}$ grow from $(10,30)$ to $(40,120)$ and $(100,300)$. $\gamma =1$.}
    \label{fig:size_control_exact_and_normal_scenario_4}
\end{figure}

\begin{figure}[ht]
    \centering
    \includegraphics[width=1.0\textwidth]{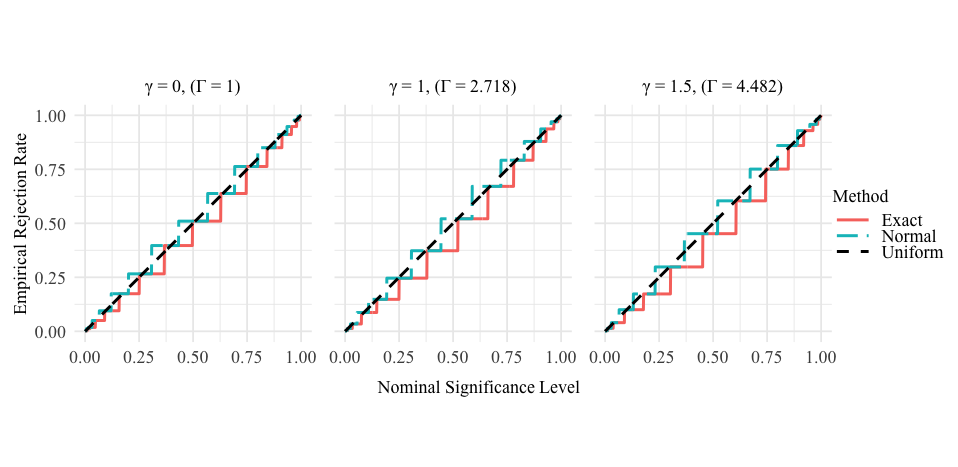}
    \caption{Type I error rate of the ordinal test on a $3\times 2$ table based on exact and normal-approximated worst-case p-values. Treatment margins are $\mathbf{N}_{3\cdot} = (30,30,40)$ and outcome margins are $\mathbf{N}_{\cdot 2} = (15,85)$. $\gamma \in \{0,1,1.5\}$.}
    \label{fig:size_control_exact_and_normal_scenario_2}
\end{figure}

\begin{figure}[ht]
    \centering
    \includegraphics[width=1.0\textwidth]{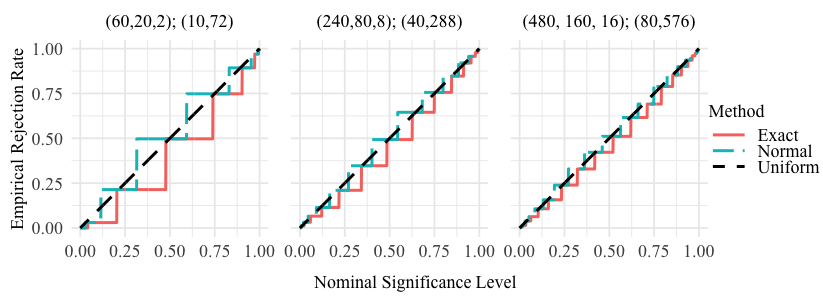}
    \caption{Type I error rate of the ordinal test on a $3\times 2$ table based on exact and normal-approximated worst-case p-values.  Treatment margins $\mathbf{N}_{3\cdot}$ grow from $(60,20,2)$ to $(240,80,8)$ and $(480,160,16)$, while outcome margins $\mathbf{N}_{\cdot 2}$ grow from $(10,72)$ to $(40,288)$ and $(80,576)$. $\gamma = 1$.}
    \label{fig:size_control_exact_and_normal_scenario_3}
\end{figure}

Larger sample sizes may mitigate this anti-conservativeness. Figure~\ref{fig:size_control_exact_and_normal_scenario_4} shows that, when $\gamma = 1$ ($\Gamma = 2.718$), the type I error rate curve from the normal approximation converges toward the 45-degree line as both treatment and outcome margins increase. Additional margin settings are considered in Figures~\ref{fig:size_control_exact_and_normal_scenario_2} and~\ref{fig:size_control_exact_and_normal_scenario_3}.

\FloatBarrier

\subsection{Size of Tests under an Average Null Effect} \label{app:F9}
We now examine empirical rejection rates under the setting $\beta = 0$ in the linear-by-linear association model in Section \mref{sec:power_main} of the main text. Importantly, $\beta = 0$ does not correspond to Fisher's sharp null hypothesis, which is the basis for our exact worst-case p-value calculation. The key distinction is that, under Fisher's sharp null, the outcome margins remain fixed across Monte Carlo iterations, whereas under $\beta = 0$, the outcome margins vary. In other words, $\beta = 0$ implies only that there is no average interaction between treatment and outcome levels.  

These simulations illustrate the connection between Fisher's sharp null hypothesis and the notion of no average treatment effect, thereby providing a more flexible framework for evaluating test size. We do not report the empirical size of tests under Fisher’s sharp null, since in that setting size control is already guaranteed by the exactness of the procedures.  

In what follows, we continue to report the weights $w_i^*$ and $v_j^*$ used to construct the ordinal tests. Although these weights are not part of the data-generating process under $\beta = 0$, listing them clarifies which test statistic is applied. For instance, if we set $(w^*_1,w^*_2,w^*_3)=(0,1.7,2.45)$ and $(v^*_1,v^*_2,v^*_3)=(0,1.25,1.4)$, then the $3\times 3$ (Opt) test statistic is  
\[
T(\mathbf{N}) = \sum_{i=1}^3 \sum_{j=1}^3 w^*_i v^*_j N_{ij}.
\]
Thus, Table~\ref{tb:size_test_1} has essentially the same data-generating process as Table~\ref{tb:size_test_4} (because $\beta=0$), but we assess size control across different ordinal tests—collapsed and non-collapsed—defined with respect to these $w_i^*$’s and $v_j^*$’s. As in the main text, we reject the null hypothesis if the worst-case the p-value is less than or equal to $\alpha=0.05$.  

Finally, consistent with the main text, the $3\times 3$ (Opt) and the two $3\times 2$ tests use $\bm{\delta} = (0,1,1)$, whereas the two Fisher's exact tests use $\bm{\delta} = (0,1)$ in the sensitivity analysis. If size is properly controlled, we expect the rejection rate of all tests under consideration to be no larger than $\alpha = 0.05$. One can see that, even though we form our rejection rule using the distribution under Fisher's sharp null, the rejection rate remains roughly at $\alpha = 0.05$.

\begin{table}[H]
    \centering
    \small
    \renewcommand{\arraystretch}{1.2} 
    \begin{tabular}{c c | c c c c c c}
        \toprule
        $\gamma$ & $\Gamma = e^\gamma$ & 3×3 (Opt) & 3×2 (V1) & 3×2 (V2) & 2×2 (V1) & 2×2 (V2) & Cross-Cut\\
        \midrule
        0.00  & 1.00  & 0.059 & 0.039 & 0.059 & 0.021 & 0.003 & 0.016\\
        0.25  & 1.28  & 0.023 & 0.024 & 0.021 & 0.009 & 0.009 & 0.008\\
        0.50  & 1.65  & 0.008 & 0.012 & 0.008 & 0.004 & 0.005 & 0.003\\
        0.75  & 2.12  & 0.001 & 0.006 & 0.001 & 0.002 & 0.001 & 0.002\\
        1.00  & 2.72  & 0.001 & 0.002 & 0.001 & 0.000 & 0.001 & 0.000\\
        1.25  & 3.49  & 0.000 & 0.001 & 0.001 & 0.000 & 0.000 & 0.000\\
        1.50  & 4.48  & 0.000 & 0.001 & 0.000 & 0.000 & 0.000 & 0.000\\
        1.75  & 5.75  & 0.000 & 0.000 & 0.000 & 0.000 & 0.000 & 0.000\\
        2.00  & 7.39  & 0.000 & 0.000 & 0.000 & 0.000 & 0.000 & 0.000\\
        2.15  & 8.58  & 0.000 & 0.000 & 0.000 & 0.000 & 0.000 & 0.000\\
        2.30  & 9.97  & 0.000 & 0.000 & 0.000 & 0.000 & 0.000 & 0.000\\
        \bottomrule
    \end{tabular}
    \caption{$\beta=0$;  \((w_1^*,w_2^*,w_3^*)=(0,1.7,2.45)\); \((v_1^*,v_2^*,v_3^*) = (0,1.25,1.4)\); $\mathbf{N}_{3\cdot} = (20, 20, 20)$; $(\lambda_{1}^Z, \lambda_2^Z, \lambda_{3}^Z) = (1,0,0)$; $(\lambda_{1}^{r}, \lambda_2^{r}, \lambda_3^r) = (1,0.2,0)$.} 
    \label{tb:size_test_1}
\end{table}

\begin{table}[H]
    \centering
    \small

    \renewcommand{\arraystretch}{1.2} 
    \begin{tabular}{c c | c c c c c c}
        \toprule
        $\gamma$ & $\Gamma = e^\gamma$ & 3×3 (Opt) & 3×2 (V1) & 3×2 (V2) & 2×2 (V1) & 2×2 (V2) & Cross-Cut\\
        \midrule
        0.00  & 1.00  & 0.051 & 0.041 & 0.046 & 0.007 & 0.023 & 0.009\\
        0.25  & 1.28  & 0.026 & 0.025 & 0.024 & 0.002 & 0.011 & 0.003\\
        0.50  & 1.65  & 0.012 & 0.014 & 0.012 & 0.000 & 0.005 & 0.001\\
        0.75  & 2.12  & 0.007 & 0.009 & 0.007 & 0.000 & 0.001 & 0.000\\
        1.00  & 2.72  & 0.005 & 0.005 & 0.005 & 0.000 & 0.000 & 0.000\\
        1.25  & 3.49  & 0.004 & 0.004 & 0.003 & 0.000 & 0.000 & 0.000\\
        1.50  & 4.48  & 0.003 & 0.004 & 0.002 & 0.000 & 0.000 & 0.000\\
        1.75  & 5.75  & 0.003 & 0.003 & 0.000 & 0.000 & 0.000 & 0.000\\
        2.00  & 7.39  & 0.000 & 0.003 & 0.000 & 0.000 & 0.000 & 0.000\\
        2.15  & 8.58  & 0.000 & 0.003 & 0.000 & 0.000 & 0.000 & 0.000\\
        2.30  & 9.97  & 0.000 & 0.003 & 0.000 & 0.000 & 0.000 & 0.000\\
        \bottomrule
    \end{tabular}
    \caption{$\beta=0$; \((w_1^*,w_2^*,w_3^*)=(0,1.7,2.45)\); \((v_1^*,v_2^*,v_3^*) = (0,1.25,1.4)\); $\mathbf{N}_{3\cdot} = (10, 10, 40)$; $(\lambda_{1}^Z, \lambda_2^Z, \lambda_{3}^Z) = (1,0,0)$; $(\lambda_{1}^{r}, \lambda_2^{r}, \lambda_3^r) = (1,0.2,0)$.}
    \label{tb:size_test_2}
\end{table}

\begin{table}[H]
    \centering
    \small

    \renewcommand{\arraystretch}{1.2} 
    \begin{tabular}{c c | c c c c c c}
        \toprule
        $\gamma$ & $\Gamma = e^\gamma$ & 3×3 (Opt) & 3×2 (V1) & 3×2 (V2) & 2×2 (V1) & 2×2 (V2) & Cross-Cut\\
        \midrule
        0.00  & 1.00  & 0.051 & 0.038 & 0.047 & 0.020 & 0.032 &0.019\\
        0.25  & 1.28  & 0.025 & 0.012 & 0.022 & 0.007 & 0.018 & 0.007\\
        0.50  & 1.65  & 0.010 & 0.005 & 0.007 & 0.000 & 0.002 & 0.003\\
        0.75  & 2.12  & 0.002 & 0.001 & 0.001 & 0.000 & 0.000 & 0.001\\
        1.00  & 2.72  & 0.000 & 0.000 & 0.000 & 0.000 & 0.000 & 0.000\\
        1.25  & 3.49  & 0.000 & 0.000 & 0.000 & 0.000 & 0.000 & 0.000\\
        1.50  & 4.48  & 0.000 & 0.000 & 0.000 & 0.000 & 0.000 & 0.000\\
        1.75  & 5.75  & 0.000 & 0.000 & 0.000 & 0.000 & 0.000 & 0.000\\
        2.00  & 7.39  & 0.000 & 0.000 & 0.000 & 0.000 & 0.000 & 0.000\\
        2.15  & 8.58  & 0.000 & 0.000 & 0.000 & 0.000 & 0.000 & 0.000\\
        2.30  & 9.97  & 0.000 & 0.000 & 0.000 & 0.000 & 0.000 & 0.000\\
        \bottomrule
    \end{tabular}
    \caption{$\beta=0$; \((w_1^*,w_2^*,w_3^*)=(0,1.7,2.45)\); \((v_1^*,v_2^*,v_3^*) = (0,1.25,1.4)\); $\mathbf{N}_{3\cdot} = (40, 10, 10)$; $(\lambda_{1}^Z, \lambda_2^Z, \lambda_{3}^Z) = (1,0,0)$; $(\lambda_{1}^{r}, \lambda_2^{r}, \lambda_3^r) = (1,0.2,0)$.} 
    \label{tb:size_test_3}
\end{table}

\begin{table}[H]
    \centering
    \small
    \renewcommand{\arraystretch}{1.2} 
    \begin{tabular}{c c | c c c c c c}
        \toprule
        $\gamma$ & $\Gamma = e^\gamma$ & 3×3 (Opt) & 3×2 (V1) & 3×2 (V2) & 2×2 (V1) & 2×2 (V2) & Cross-Cut  \\
        \midrule
        0.00  & 1.00  & 0.050 & 0.029 & 0.048 & 0.021 & 0.030 & 0.016\\
        0.25  & 1.28  & 0.024 & 0.016 & 0.025 & 0.009 & 0.009 & 0.008\\
        0.50  & 1.65  & 0.009 & 0.009 & 0.007 & 0.004 & 0.005 & 0.003\\
        0.75  & 2.12  & 0.001 & 0.003 & 0.001 & 0.002 & 0.001 & 0.002\\
        1.00  & 2.72  & 0.001 & 0.002 & 0.001 & 0.000 & 0.001 & 0.000\\
        1.25  & 3.49  & 0.001 & 0.001 & 0.001 & 0.000 & 0.000 & 0.000\\
        1.50  & 4.48  & 0.000 & 0.001 & 0.000 & 0.000 & 0.000 & 0.000\\
        1.75  & 5.75  & 0.000 & 0.000 & 0.000 & 0.000 & 0.000 & 0.000\\
        2.00  & 7.39  & 0.000 & 0.000 & 0.000 & 0.000 & 0.000 & 0.000\\
        2.15  & 8.58  & 0.000 & 0.000 & 0.000 & 0.000 & 0.000 & 0.000\\
        2.30  & 9.97  & 0.000 & 0.000 & 0.000 & 0.000 & 0.000 & 0.000\\
        \bottomrule
    \end{tabular}
    \caption{$\beta=0$; \((w_1^*,w_2^*,w_3^*)=(0,1,2)\); \((v_1^*,v_2^*,v_3^*) = (0,1.5,2)\); $\mathbf{N}_{3\cdot} = (20, 20, 20)$; $(\lambda_{1}^Z, \lambda_2^Z, \lambda_{3}^Z) = (1,0,0)$; $(\lambda_{1}^{r}, \lambda_2^{r}, \lambda_3^r) = (1,0.2,0)$.}  
    \label{tb:size_test_4}
\end{table}

\begin{table}[H]
    \centering
    \small
    \renewcommand{\arraystretch}{1.2} 
    \begin{tabular}{c c | c c c c c c }
        \toprule
        $\gamma$ & $\Gamma = e^\gamma$ & 3×3 (Opt) & 3×2 (V1) & 3×2 (V2) & 2×2 (V1) & 2×2 (V2) & Cross-Cut\\
        \midrule
        0.00  & 1.00  & 0.049 & 0.034 & 0.038 & 0.007 & 0.023 & 0.009 \\
        0.25  & 1.28  & 0.026 & 0.023 & 0.018 & 0.002 & 0.011 & 0.003\\
        0.50  & 1.65  & 0.010 & 0.013 & 0.011 & 0.000 & 0.005 & 0.001\\
        0.75  & 2.12  & 0.006 & 0.009 & 0.005 & 0.000 & 0.001 & 0.000\\
        1.00  & 2.72  & 0.005 & 0.005 & 0.005 & 0.000 & 0.000 & 0.000\\
        1.25  & 3.49  & 0.005 & 0.004 & 0.004 & 0.000 & 0.000 & 0.000\\
        1.50  & 4.48  & 0.004 & 0.004 & 0.003 & 0.000 & 0.000 & 0.000\\
        1.75  & 5.75  & 0.003 & 0.003 & 0.002 & 0.000 & 0.000 & 0.000\\
        2.00  & 7.39  & 0.001 & 0.003 & 0.001 & 0.000 & 0.000 & 0.000\\
        2.15  & 8.58  & 0.001 & 0.003 & 0.000 & 0.000 & 0.000 & 0.000\\
        2.30  & 9.97  & 0.000 & 0.003 & 0.000 & 0.000 & 0.000 & 0.000\\
        \bottomrule
    \end{tabular}
    \caption{$\beta=0$; \((w_1^*, w_2^*,w_3^*)=(0,1,2)\); \((v_1^*,v_2^*,v_3^*) = (0,1.5,2)\); $\mathbf{N}_{3\cdot} = (10, 10, 40)$; $(\lambda_{1}^Z, \lambda_2^Z, \lambda_{3}^Z) = (1,0,0)$; $(\lambda_{1}^{r}, \lambda_2^{r}, \lambda_3^r) = (1,0.2,0)$.}  
    \label{tb:size_test_5}
\end{table}

\begin{table}[H]
    \centering
    \small
    \renewcommand{\arraystretch}{1.2} 
    \begin{tabular}{c c | c c c c c c}
        \toprule
        $\gamma$ & $\Gamma = e^\gamma$ & 3×3 (Opt) & 3×2 (V1) & 3×2 (V2) & 2×2 (V1) & 2×2 (V2) & Cross-Cut \\
        \midrule
        0.00  & 1.00  & 0.051 & 0.030 & 0.045 & 0.020 & 0.032 & 0.019\\
        0.25  & 1.28  & 0.017 & 0.009 & 0.014 & 0.007 & 0.018 & 0.007\\
        0.50  & 1.65  & 0.007 & 0.004 & 0.007 & 0.000 & 0.002 & 0.003\\
        0.75  & 2.12  & 0.003 & 0.001 & 0.002 & 0.000 & 0.000 & 0.001\\
        1.00  & 2.72  & 0.001 & 0.000 & 0.000 & 0.000 & 0.000 & 0.000\\
        1.25  & 3.49  & 0.000 & 0.000 & 0.000 & 0.000 & 0.000 & 0.000\\
        1.50  & 4.48  & 0.000 & 0.000 & 0.000 & 0.000 & 0.000 & 0.000\\
        1.75  & 5.75  & 0.000 & 0.000 & 0.000 & 0.000 & 0.000 & 0.000\\
        2.00  & 7.39  & 0.000 & 0.000 & 0.000 & 0.000 & 0.000 & 0.000\\
        2.15  & 8.58  & 0.000 & 0.000 & 0.000 & 0.000 & 0.000 & 0.000\\
        2.30  & 9.97  & 0.000 & 0.000 & 0.000 & 0.000 & 0.000 & 0.000\\
        \bottomrule
    \end{tabular}
    \caption{$\beta=0$; \((w_1^*,w_2^*,w_3^*)=(0,1,2)\); \((v_1^*,v_2^*,v_3^*) = (0,1.5,2)\); $\mathbf{N}_{3\cdot} = (40, 10, 10)$; $(\lambda_{1}^Z, \lambda_2^Z, \lambda_{3}^Z) = (1,0,0)$; $(\lambda_{1}^{r}, \lambda_2^{r}, \lambda_3^r) = (1,0.2,0)$.}  
    \label{tb:size_test_6}
\end{table}


\newpage
\begin{changeblock}
\section{Computational Feasibility of the Exact Worst-Case $p$-Value} \label{app:G}
This section reports the running times and memory consumption of exact worst-case $p$-values as a function of sample size $N$ and table dimension $I \times J$. Section~\ref{app:G1} considers ordinal and sign-score test statistics, for which Theorems~\mref{thm:score_test_maximizer_nonbinary} and~\mref{thm:score_test_maximizer_binary} reduce the candidate set of unmeasured confounders to $N+1$ elements in $\mathcal{U}_{\rm O}$ or to a single candidate $\mathbf{u}^+$. Section~\ref{app:G2} examines the computational feasibility of the chi-square test, a non-ordinal, count-based test, to which Theorem~\mref{thm:set_size_of_U} applies.  \change{Section~\ref{app:G3} compares the running time of the exact $p$-value under Lemma~\ref{lem:kernel_general_delta} and under Theorem~\mref{thm:recursive_probability}.}

All simulations in Section~\ref{app:G} leverage Theorem~\mref{thm:recursive_probability} to reduce computational burden, replacing iteration over treatment assignment vectors $\mathbf{z} \in \mathcal{Z}(\mathbf{N}_{I\cdot})$ with enumeration over tables $\mathbf{t}$ in the rejection region $\Omega_{\mathbf{t}}$ and over the vectors $\overline{\mathbf{n}} \in \Omega_{\overline{\mathbf{n}}}$, which record, for each outcome level, how many subjects with $u_s = 1$ are assigned to a treatment level in $\mathcal{I}_1$ (see Section~\mref{sec:sens_ibyj} for details). Importantly, searching through tables in the rejection region $\Omega_{\mathbf{t}}$ is also required in exact calculations of cumulative distribution functions or $p$-values for discrete variables in randomized controlled trial (RCT) settings \citep{pagano1981algorithm, ChenEtAl2005, Eisinger2017}. Because this step contributes significantly to the computational burden and forms a direct link to exact RCT inference, we also examine the cardinality of these sets in the following cases. 

Conversely, because $\Omega_{\overline{\mathbf{n}}}$ lacks a counterpart in exact RCT $p$-value computations, we do not report its cardinality. We note, however, that $|\Omega_{\overline{\mathbf{n}}}|$ is at most $\prod_{j=1}^{J}(N_{\cdot j}+1)$, polynomial in $N$ of degree $J$, since each coordinate $\overline{n}_j$ ranges over at most $N_{\cdot j}+1$ values.

For each dimension we fix one observed table for each of the sample sizes $N = 20$, $50$, $100$, $120$, $150$, $180$, and $200$. The tables exhibit a positive association between treatment level and outcome level with margins that grow proportionally with $N$, and every table has an exact $p$-value below $0.05$ at $\gamma = 0$, so that a sensitivity analysis is meaningful; the tables themselves are displayed in the results.

\subsection{Ordinal and Sign-Score Test Statistics} \label{app:G1}

\paragraph{Setup.}
We evaluate the ordinal test statistic $T(\mathbf{N}) = \sum_{i=1}^{I}\sum_{j=1}^{J} w_i v_j N_{ij}$ using category-level scores $w_i = i$ and $v_j = j$, under sensitivity parameters $\bm{\delta} = (0, 1, \ldots, 1)$. The worst-case $p$-value is computed at $\gamma = 1$, with the baseline $p$-value at $\gamma = 0$ reported for reference. Notably, for any $\gamma > 0$, as long as the scores $w_i$ and $v_j$ and the sensitivity parameters $\delta_i$ are nondecreasing, the unmeasured confounder candidate set ($\mathcal{U}_{\rm O}$ for ordinal tests or the single candidate $\mathbf{u}^+$ for sign-score tests) is invariant to the specific choice of scores or sensitivity vector $\bm{\delta}$, and the rejection region $\Omega_{\mathbf{t}}$ is invariant to $\gamma$ and $\bm{\delta}$ (though not to the scores). Consequently, the running times in this section are indicative of ordinal or sign-score analyses with the same table dimension and sample size.

\paragraph{Understanding the outputs.}
In Tables~\ref{tb:feasibility_ordinal_3x3}--\ref{tb:feasibility_ordinal_5x2}, rows and columns represent treatment and outcome levels, respectively, for the observed contingency tables. The two primary drivers of computational cost are presented side by side: $|\mathcal{U}|$, the number of candidate unmeasured confounders evaluated ($N+1$ candidates in $\mathcal{U}_{\rm O}$ under Theorem~\mref{thm:score_test_maximizer_nonbinary}, or a single candidate under Theorem~\mref{thm:score_test_maximizer_binary}), and $|\Omega_{\mathbf{t}}|$, the number of tables $\mathbf{t}$ sharing the observed margins that satisfy the rejection criterion $T(\mathbf{t}) \ge T_{\rm obs}$. We further report the \emph{peak memory increase} for each upper bound on the $p$-value computation, defined as the difference between the baseline memory before computation and the maximum memory consumed during execution. Each table reports both the baseline $p$-value assuming no unmeasured confounding ($\gamma = 0$) and the worst-case $p$-value upper bound ($\gamma = 1$). The final column records the total execution time required to obtain the upper bound under $\gamma = 1$ for the specified $\bm{\delta}$ setting. Algorithmically, computing the worst-case $p$-value requires enumerating all candidates $\mathbf{u} \in \mathcal{U}_{\rm O}$ (or the single candidate $\mathbf{u}^+$) and, for each candidate $\mathbf{u}$, summing probability masses across all tables $\mathbf{t} \in \Omega_{\mathbf{t}}$. Consequently, the overall computational complexity scales as $O(|\mathcal{U}| \times |\Omega_{\mathbf{t}}| \times |\Omega_{\overline{\mathbf{n}}}|)$.

To contextualize these efficiency gains, consider what each group of results contributes. The worst-case $p$-value is the maximum of equation \meqref{eq:raw_sens_test} of the main text over $\mathbf{u} \in [0,1]^N$. Without Theorems~\mref{thm:set_size_of_U}--\mref{thm:score_test_maximizer_binary}, this is a maximization over a continuum, which is a non-convex global optimization problem (Section~\mref{sec:def_worst_case} of the main text); there is no finite set of candidates to enumerate, and the objective has no finite worst-case cost to report. Theorems~\mref{thm:set_size_of_U}--\mref{thm:score_test_maximizer_binary} replace the continuum by a finite candidate set $\mathcal{U}$, with $|\mathcal{U}| \le N^J$, $|\mathcal{U}| = N + 1$, or $|\mathcal{U}| = 1$ depending on the test statistic. Without Theorem~\mref{thm:recursive_probability}, the total cost is then $|\mathcal{U}|$ evaluations of equation \meqref{eq:raw_sens_test} of the main text, each a sum over the $|\mathcal{Z}(\mathbf{N}_{I\cdot})|$ treatment assignments. Theorem~\mref{thm:recursive_probability} replaces each of these sums by equation \meqref{eq:exact_distribution_kernel_form} of the main text, whose number of terms is governed by $|\Omega_{\mathbf{t}}|$ rather than by $|\mathcal{Z}(\mathbf{N}_{I\cdot})|$. Theorem~\mref{thm:recursive_probability} cannot stand alone: it presupposes a binary $\mathbf{u} \in \{0,1\}^N$, which Theorems~\mref{thm:set_size_of_U}--\mref{thm:score_test_maximizer_binary} show to be sufficient, and it does not by itself bound the number of candidates. The total cost of the worst-case $p$-value is therefore the product of the two factors reported side by side in the tables, $|\mathcal{U}|$ and the cost of one evaluation of equation \meqref{eq:exact_distribution_kernel_form} of the main text, the latter governed by $|\Omega_{\mathbf{t}}|$.

As noted at the beginning of Section~\ref{app:G}, the computational burden of aggregating probability masses across tables in the rejection region $\Omega_{\mathbf{t}}$ is not unique to sensitivity analysis, but is a standard challenge in exact RCT $p$-value evaluation. Prior literature addresses this burden either by reformulating probability expressions for exact computation \citep{pagano1981algorithm} (similar in spirit to our Theorem~\mref{thm:recursive_probability}, which rewrites the sum over treatment assignments as a sum over tables) or by employing sampling methods \citep{ChenEtAl2005, Eisinger2017} (see Section~\ref{app:F2} for our adaptation to sensitivity analysis). Beyond these algorithmic strategies for mitigating the complexity of $\Omega_{\mathbf{t}}$, we highlight a key practical insight: a larger rejection region $\Omega_{\mathbf{t}}$ yields a larger baseline $p$-value under $\gamma = 0$, and the worst-case upper bound under unmeasured confounding ($\gamma > 0$) is necessarily even higher. Consequently, if the baseline $p$-value under $\gamma = 0$ already exceeds the significance level, the worst-case bound will automatically exceed it as well, offering no additional decision-making value. Omitting the upper-bound calculation in such cases is therefore both practical and eliminates unnecessary computation over $\Omega_{\mathbf{t}}$.

\paragraph{Summary}
Tables~\ref{tb:feasibility_ordinal_3x3}--\ref{tb:feasibility_ordinal_5x2} show the computational burden of obtaining the upper bound on the $p$-value for $3\times 3$, $3 \times 4$, $4 \times 2$, and $5\times 2$ contingency tables across sample sizes $N = 20, 50, 100, 120, 150, 180, 200$. Our experiments were conducted in a Posit Cloud environment configured with $4$~GB RAM and $8$~CPUs. The computational burden scales as $O(|\mathcal{U}| \times |\Omega_{\mathbf{t}}|\times |\Omega_{\overline{\mathbf{n}}}|)$, as suggested by Theorems~\mref{thm:set_size_of_U}--\mref{thm:recursive_probability} in the main text.

First, regarding computational feasibility, all ordinal tests with binary outcomes, the so-called sign-score tests, in Tables~\ref{tb:feasibility_ordinal_4x2} and \ref{tb:feasibility_ordinal_5x2} obtain the upper bound on the $p$-value within $0.5$ seconds. Tables~\ref{tb:feasibility_ordinal_3x3} and \ref{tb:feasibility_ordinal_3x4} further detail the computational burden for ordinal tests with nonbinary outcomes. Taking Table~\ref{tb:feasibility_ordinal_3x4} as an example, when $N=20$, the computation time is within $0.023$ seconds, and when $N = 200$, it remains within $100.1$ seconds. Consequently, for practitioners implementing an ordinal test, the computational burden is quite light.

Second, examining the sources of computational savings, we note that ordinal tests with binary outcomes incur the lightest computational burden and display the mildest growth relative to sample size. In Table~\ref{tb:feasibility_ordinal_5x2}, for instance, the computational time grows from $0.019$ to $0.038$ and $0.129$ seconds at $N=20, 150$, and $200$, respectively. This demonstrates the efficiency enabled by Theorem~\mref{thm:score_test_maximizer_binary}, which reduces the unmeasured confounder search space to a single candidate, $\mathbf{u}^+$. In contrast, ordinal tests with nonbinary outcomes exhibit higher growth in computation time as $N$ increases. In Table~\ref{tb:feasibility_ordinal_3x4}, the computation time to obtain the upper bound on the $p$-value is $0.023$ seconds at $N=20$, whereas at $N=100$ and $N=200$, it increases to $1.384$ and $100.099$ seconds, respectively. This growth in execution time stems from increases in $|\Omega_{\mathbf{t}}|$ and $|\Omega_{\overline{\mathbf{n}}}|$, as well as the expansion of the unmeasured confounder search space $\mathcal{U}_{\rm O}$, where $|\mathcal{U}_{\rm O}| = N+1$ according to Theorem~\mref{thm:score_test_maximizer_nonbinary}.

Finally, while computational time grows with sample size $N$, these experiments highlight the critical importance of considering computational burden in exact sensitivity analysis. We refer readers to Section~\mref{sec:exact_worst_null} and Table~\mref{tb:main_text_exact_computation_time_permutation_and_kernel} in the main text for details on the substantial computational savings provided by Theorem~\mref{thm:recursive_probability}.
\end{changeblock}

\begin{changetable}[htbp]
\centering
\scriptsize\renewcommand{\arraystretch}{1.1}\setlength{\tabcolsep}{4pt}
\resizebox{0.95\textwidth}{!}{%
\begin{tabular}{@{}c C{21mm} r r c c r@{}}
\toprule
$N$ & \thead{Observed\\table} & \thead{Candidates\\$|\mathcal{U}|$} & $|\Omega_{\mathbf{t}}|$ & \thead{$p$-value\\$\gamma=0$} & \thead{Worst-case\\$p$-value\\$\gamma=1$} & \thead{Total time\\(seconds)} \\
\midrule
20 & $\begin{bmatrix} 6 & 2 & 0 \\ 2 & 1 & 1 \\ 0 & 2 & 6 \end{bmatrix}$ & 21 & 14 & 0.0003 & 0.0043 & 0.018 \\
\addlinespace[5pt]
50 & $\begin{bmatrix} 10 & 5 & 3 \\ 5 & 5 & 5 \\ 2 & 5 & 10 \end{bmatrix}$ & 51 & 1,832 & 0.0015 & 0.0446 & 0.004 \\
\addlinespace[5pt]
100 & $\begin{bmatrix} 17 & 10 & 7 \\ 11 & 11 & 10 \\ 7 & 10 & 17 \end{bmatrix}$ & 101 & 40,555 & 0.0021 & 0.1373 & 0.055 \\
\addlinespace[5pt]
120 & $\begin{bmatrix} 20 & 13 & 8 \\ 13 & 13 & 13 \\ 8 & 13 & 19 \end{bmatrix}$ & 121 & 82,633 & 0.0011 & 0.1267 & 0.115 \\
\addlinespace[5pt]
150 & $\begin{bmatrix} 24 & 16 & 11 \\ 16 & 16 & 16 \\ 11 & 16 & 24 \end{bmatrix}$ & 151 & 217,504 & 0.0011 & 0.1756 & 0.259 \\
\addlinespace[5pt]
180 & $\begin{bmatrix} 28 & 19 & 14 \\ 20 & 19 & 19 \\ 14 & 19 & 28 \end{bmatrix}$ & 181 & 481,744 & 0.0013 & 0.2518 & 0.531 \\
\addlinespace[5pt]
200 & $\begin{bmatrix} 30 & 22 & 15 \\ 22 & 22 & 22 \\ 15 & 22 & 30 \end{bmatrix}$ & 201 & 731,521 & 0.0009 & 0.2481 & 0.752 \\
\bottomrule
\end{tabular}}
\caption{Ordinal test, $3 \times 3$ tables, $w = v = (1,2,3)$, $\bm{\delta} = (0,1,1)$; worst case over the $N+1$ candidates of $\mathcal{U}_{\rm O}$ (Theorem~\mref{thm:score_test_maximizer_nonbinary}). $p$-values at $\gamma = 0$ are all below $0.003$; worst-case $p$-values at $\gamma = 1$ range from $0.004$ ($N=20$) to $0.25$ ($N=200$). Total time grows from $0.02$ to $0.75$ seconds; peak memory increase below 1~MB in every run.}
\label{tb:feasibility_ordinal_3x3}
\end{changetable}

\begin{changetable}[htbp]
\centering
\scriptsize\renewcommand{\arraystretch}{1.1}\setlength{\tabcolsep}{4pt}
\resizebox{0.95\textwidth}{!}{%
\begin{tabular}{@{}c C{27mm} r r c c r@{}}
\toprule
$N$ & \thead{Observed\\table} & \thead{Candidates\\$|\mathcal{U}|$} & $|\Omega_{\mathbf{t}}|$ & \thead{$p$-value\\$\gamma=0$} & \thead{Worst-case\\$p$-value\\$\gamma=1$} & \thead{Total time\\(seconds)} \\
\midrule
20 & $\begin{bmatrix} 5 & 2 & 1 & 0 \\ 1 & 1 & 1 & 1 \\ 0 & 1 & 2 & 5 \end{bmatrix}$ & 21 & 52 & 0.0004 & 0.0064 & 0.023 \\
\addlinespace[5pt]
50 & $\begin{bmatrix} 9 & 5 & 3 & 1 \\ 4 & 3 & 4 & 3 \\ 1 & 3 & 5 & 9 \end{bmatrix}$ & 51 & 15,392 & 0.0001 & 0.0061 & 0.040 \\
\addlinespace[5pt]
100 & $\begin{bmatrix} 14 & 10 & 6 & 4 \\ 8 & 8 & 8 & 8 \\ 4 & 6 & 10 & 14 \end{bmatrix}$ & 101 & 1,904,311 & 0.0001 & 0.0334 & 1.384 \\
\addlinespace[5pt]
120 & $\begin{bmatrix} 17 & 11 & 8 & 5 \\ 9 & 9 & 10 & 10 \\ 5 & 8 & 11 & 17 \end{bmatrix}$ & 121 & 5,910,478 & 0.0001 & 0.0361 & 4.051 \\
\addlinespace[5pt]
150 & $\begin{bmatrix} 20 & 14 & 10 & 7 \\ 12 & 12 & 12 & 12 \\ 7 & 10 & 14 & 20 \end{bmatrix}$ & 151 & 25,940,796 & 0.0001 & 0.0620 & 15.098 \\
\addlinespace[5pt]
180 & $\begin{bmatrix} 23 & 17 & 12 & 9 \\ 14 & 14 & 15 & 15 \\ 9 & 12 & 17 & 23 \end{bmatrix}$ & 181 & 84,359,549 & 0.0001 & 0.0919 & 48.001 \\
\addlinespace[5pt]
200 & $\begin{bmatrix} 24 & 19 & 14 & 11 \\ 16 & 16 & 16 & 16 \\ 11 & 14 & 19 & 24 \end{bmatrix}$ & 201 & 191,626,478 & 0.0004 & 0.2200 & 100.099 \\
\bottomrule
\end{tabular}}
\caption{Ordinal test, $3 \times 4$ tables, $w = (1,2,3)$, $v = (1,2,3,4)$, $\bm{\delta} = (0,1,1)$; worst case over the $N+1$ candidates of $\mathcal{U}_{\rm O}$ (Theorem~\mref{thm:score_test_maximizer_nonbinary}). $p$-values at $\gamma = 0$ are all below $0.0005$; worst-case $p$-values at $\gamma = 1$ range from $0.006$ to $0.22$. Total time grows from $0.02$ to $100$ seconds, dominated by the enumeration of $\Omega_{\mathbf{t}}$ ($1.9 \times 10^{8}$ tables at $N = 200$); peak memory increase below 1~MB in every run.}
\label{tb:feasibility_ordinal_3x4}
\end{changetable}

\begin{changetable}[htbp]
\centering
\scriptsize\renewcommand{\arraystretch}{1.1}\setlength{\tabcolsep}{4pt}
\resizebox{0.95\textwidth}{!}{%
\begin{tabular}{@{}c C{21mm} r r c c r@{}}
\toprule
$N$ & \thead{Observed\\table} & \thead{Candidates\\$|\mathcal{U}|$} & $|\Omega_{\mathbf{t}}|$ & \thead{$p$-value\\$\gamma=0$} & \thead{Worst-case\\$p$-value\\$\gamma=1$} & \thead{Total time\\(seconds)} \\
\midrule
20 & $\begin{bmatrix} 5 & 1 \\ 3 & 1 \\ 1 & 3 \\ 1 & 5 \end{bmatrix}$ & 1 & 15 & 0.0080 & 0.0520 & 0.020 \\
\addlinespace[5pt]
50 & $\begin{bmatrix} 10 & 3 \\ 7 & 5 \\ 5 & 7 \\ 3 & 10 \end{bmatrix}$ & 1 & 245 & 0.0029 & 0.0535 & 0.001 \\
\addlinespace[5pt]
100 & $\begin{bmatrix} 18 & 8 \\ 14 & 10 \\ 10 & 14 \\ 8 & 18 \end{bmatrix}$ & 1 & 2,534 & 0.0018 & 0.1010 & 0.001 \\
\addlinespace[5pt]
120 & $\begin{bmatrix} 21 & 10 \\ 16 & 13 \\ 13 & 16 \\ 10 & 21 \end{bmatrix}$ & 1 & 4,821 & 0.0023 & 0.1518 & 0.002 \\
\addlinespace[5pt]
150 & $\begin{bmatrix} 26 & 13 \\ 20 & 16 \\ 16 & 20 \\ 13 & 26 \end{bmatrix}$ & 1 & 9,607 & 0.0012 & 0.1569 & 0.004 \\
\addlinespace[5pt]
180 & $\begin{bmatrix} 30 & 16 \\ 24 & 20 \\ 20 & 24 \\ 16 & 30 \end{bmatrix}$ & 1 & 17,628 & 0.0014 & 0.2167 & 0.009 \\
\addlinespace[5pt]
200 & $\begin{bmatrix} 33 & 18 \\ 27 & 22 \\ 22 & 27 \\ 18 & 33 \end{bmatrix}$ & 1 & 24,306 & 0.0010 & 0.2210 & 0.008 \\
\bottomrule
\end{tabular}}
\caption{Sign-score test, $4 \times 2$ tables, $w = (1,2,3,4)$, $v = (1,2)$, $\bm{\delta} = (0,1,1,1)$; the worst case $\mathbf{u}^+$ is known in closed form (Theorem~\mref{thm:score_test_maximizer_binary}), so the candidate set has one element. $p$-values at $\gamma = 0$ are all below $0.009$; worst-case $p$-values at $\gamma = 1$ range from $0.052$ to $0.22$. Every run completes within $0.02$ seconds; peak memory increase below 1~MB.}
\label{tb:feasibility_ordinal_4x2}
\end{changetable}

\begin{changetable}[htbp]
\centering
\tiny\renewcommand{\arraystretch}{1.1}\setlength{\tabcolsep}{4pt}
\resizebox{0.95\textwidth}{!}{%
\begin{tabular}{@{}c C{21mm} r r c c r@{}}
\toprule
$N$ & \thead{Observed\\table} & \thead{Candidates\\$|\mathcal{U}|$} & $|\Omega_{\mathbf{t}}|$ & \thead{$p$-value\\$\gamma=0$} & \thead{Worst-case\\$p$-value\\$\gamma=1$} & \thead{Total time\\(seconds)} \\
\midrule
20 & $\begin{bmatrix} 5 & 0 \\ 3 & 1 \\ 1 & 1 \\ 1 & 3 \\ 0 & 5 \end{bmatrix}$ & 1 & 6 & 0.0003 & 0.0020 & 0.019 \\
\addlinespace[4pt]
50 & $\begin{bmatrix} 9 & 2 \\ 6 & 3 \\ 5 & 5 \\ 3 & 6 \\ 2 & 9 \end{bmatrix}$ & 1 & 692 & 0.0006 & 0.0109 & 0.001 \\
\addlinespace[4pt]
100 & $\begin{bmatrix} 15 & 6 \\ 12 & 8 \\ 9 & 9 \\ 8 & 12 \\ 6 & 15 \end{bmatrix}$ & 1 & 19,787 & 0.0014 & 0.0510 & 0.008 \\
\addlinespace[4pt]
120 & $\begin{bmatrix} 17 & 8 \\ 14 & 9 \\ 12 & 12 \\ 9 & 14 \\ 8 & 17 \end{bmatrix}$ & 1 & 46,952 & 0.0020 & 0.0843 & 0.017 \\
\addlinespace[4pt]
150 & $\begin{bmatrix} 21 & 10 \\ 18 & 12 \\ 14 & 14 \\ 12 & 18 \\ 10 & 21 \end{bmatrix}$ & 1 & 112,085 & 0.0008 & 0.0690 & 0.038 \\
\addlinespace[4pt]
180 & $\begin{bmatrix} 25 & 12 \\ 21 & 15 \\ 17 & 17 \\ 15 & 21 \\ 12 & 25 \end{bmatrix}$ & 1 & 237,430 & 0.0005 & 0.0696 & 0.079 \\
\addlinespace[5pt]
200 & $\begin{bmatrix} 27 & 14 \\ 23 & 17 \\ 19 & 19 \\ 17 & 23 \\ 14 & 27 \end{bmatrix}$ & 1 & 397,695 & 0.0009 & 0.1144 & 0.129 \\
\bottomrule
\end{tabular}}
\caption{Sign-score test, $5 \times 2$ tables, $w = (1,\ldots,5)$, $v = (1,2)$, $\bm{\delta} = (0,1,1,1,1)$; the worst case $\mathbf{u}^+$ is known in closed form (Theorem~\mref{thm:score_test_maximizer_binary}). Peak memory increase below 1~MB.}
\label{tb:feasibility_ordinal_5x2}
\end{changetable}
\newpage
\begin{changeblock}
\subsection{General Count-Based Test Statistics} \label{app:G2}
 
\paragraph{Setup}
The test statistic is the chi-square statistic $T(\mathbf{N}) = \sum_{i=1}^{I}\sum_{j=1}^{J} (N_{ij} - E_{ij})^2 / E_{ij}$ with $E_{ij} = N_{i\cdot} N_{\cdot j} / N$, and the sensitivity parameters are $\bm{\delta} = (0, 1, \ldots, 1)$ as in Section~\ref{app:G1}. The chi-square statistic is count-based but not arrangement-increasing, so Theorems~\mref{thm:score_test_maximizer_nonbinary} and~\mref{thm:score_test_maximizer_binary} do not apply and the worst case must be searched over the full candidate set of Theorem~\mref{thm:set_size_of_U}. In particular, Theorem~\mref{thm:set_size_of_U} states that for any count-based statistic the worst-case unmeasured confounder can be found among the equivalence classes indexed by $\overline{\mathbf{u}} = \trr^\intercal\mathbf{u} = (\overline{u}_1, \ldots, \overline{u}_J)$, $0 \le \overline{u}_j \le N_{\cdot j}$, so that
\[
|\mathcal{U}_{\rm C}| = \prod_{j=1}^{J} (N_{\cdot j} + 1) \le N^J.
\]
Similar as before, if a test is count-based, then the unmeasured confounder candidate set $\mathcal{U}_{\rm C}$ and the rejection region $\Omega_{\mathbf{t}}$ are invariant under different $\bm{\delta}$ or any $\gamma>0$, so the running times are representative of any value of $\gamma$ or $\bm{\delta}$.
 
\paragraph{Understanding the outputs.}
Tables~\ref{tb:feasibility_chisq_3x3}--\ref{tb:feasibility_chisq_5x2} have the same layout as those of Section~\ref{app:G1}, with the candidate column now reporting $|\mathcal{U}_{\rm C}| = \prod_j (N_{\cdot j}+1)$.
 
\paragraph{Summary}
Tables~\ref{tb:feasibility_chisq_3x3}--\ref{tb:feasibility_chisq_5x2} present the computation times (in seconds) for chi-square tests across $3\times 3$, $3\times 4$, $4\times 2$, and $5\times 2$ contingency tables at $N=20, 50, 100, 120, 150, 180, 200$. Regarding computational feasibility, when the outcome is binary ($J=2$), computing the upper bound on the $p$-value for $N \le 200$ took less than 5 seconds across all examined cases. However, as the outcome level $J$ increases, the computational burden grows substantially. For example, in Table~\ref{tb:feasibility_chisq_3x3} ($J=3$), the computation takes $0.021$~seconds when $N=20$, but increases to $97.1$~seconds for $N=100$ and $7{,}918.9$~seconds for $N=200$.

As in Section~\ref{app:G1}, computation time scales up with both table dimension and sample size $N$. In particular, for a general count-based test, computation time grows as $O(|\mathcal{U}_{\rm C}| \times |\Omega_{\mathbf{t}}| \times |\Omega_{\overline{\mathbf{n}}}|)$, where $|\mathcal{U}_{\rm C}| = \prod_{j=1}^{J}(N_{\cdot j}+1)$ is polynomial in the outcome level $J$ (Theorem~\mref{thm:set_size_of_U}). In Table~\ref{tb:feasibility_chisq_3x4}, computing the upper bound on the $p$-value for $N=100$ already takes more than $7{,}421$~seconds. As a rule of thumb, we recommend a general count-based test only if (i) the outcome is binary ($J=2$), or (ii) $J=3$ and $N \le 200$. Contingency tables with higher dimensions or larger sample sizes may lack feasibility for exact sensitivity analysis with a general count-based test. When this computational burden poses constraints, readers may refer to Section~\mref{sec:power_analysis_size_control} and \citet{agresti2012categorical}, which suggest that ordinal tests might be more powerful when the treatment has a monotone effect on the outcome; alternatively, readers can adopt the sampling method introduced in Section~\ref{app:F2} if the upper bound on the $p$-value based on a count-based test is deemed necessary.
\end{changeblock}

\begin{changetable}[htbp]
\centering
\scriptsize\renewcommand{\arraystretch}{1.1}\setlength{\tabcolsep}{4pt}
\resizebox{0.95\textwidth}{!}{%
\begin{tabular}{@{}c C{21mm} r r c c r@{}}
\toprule
$N$ & \thead{Observed\\table} & \thead{Candidates\\$|\mathcal{U}_{\rm C}|$} & $|\Omega_{\mathbf{t}}|$ & \thead{$p$-value\\$\gamma=0$} & \thead{Worst-case\\$p$-value\\$\gamma=1$} & \thead{Total time\\(seconds)} \\
\midrule
20 & $\begin{bmatrix} 7 & 1 & 0 \\ 2 & 1 & 1 \\ 0 & 1 & 7 \end{bmatrix}$ & 360 & 52 & 0.0030 & 0.0151 & 0.021 \\
\addlinespace[5pt]
50 & $\begin{bmatrix} 12 & 5 & 2 \\ 5 & 4 & 4 \\ 2 & 4 & 12 \end{bmatrix}$ & 5,320 & 7,674 & 0.0037 & 0.0481 & 1.676 \\
\addlinespace[5pt]
100 & $\begin{bmatrix} 20 & 10 & 5 \\ 10 & 10 & 10 \\ 5 & 10 & 20 \end{bmatrix}$ & 40,176 & 152,407 & 0.0008 & 0.0641 & 97.1 \\
\addlinespace[5pt]
120 & $\begin{bmatrix} 23 & 12 & 7 \\ 12 & 12 & 12 \\ 7 & 12 & 23 \end{bmatrix}$ & 68,413 & 322,861 & 0.0013 & 0.1126 & 329.9 \\
\addlinespace[5pt]
150 & $\begin{bmatrix} 28 & 15 & 9 \\ 16 & 15 & 15 \\ 9 & 15 & 28 \end{bmatrix}$ & 131,652 & 782,592 & 0.0004 & 0.1083 & 1243.7 \\
\addlinespace[5pt]
180 & $\begin{bmatrix} 32 & 19 & 11 \\ 19 & 19 & 19 \\ 11 & 18 & 32 \end{bmatrix}$ & 226,233 & 1,659,002 & 0.0003 & 0.1352 & 3990.5 \\
\addlinespace[5pt]
200 & $\begin{bmatrix} 35 & 21 & 13 \\ 21 & 21 & 21 \\ 13 & 21 & 34 \end{bmatrix}$ & 309,120 & 2,548,996 & 0.0005 & 0.2134 &  7918.9\\
\bottomrule
\end{tabular}}
\caption{Chi-square test, $3 \times 3$ tables, $\bm{\delta} = (0,1,1)$; worst case over the full candidate set $\mathcal{U}_{\rm C}$ of Theorem~\mref{thm:set_size_of_U}, $|\mathcal{U}_{\rm C}| = \prod_j (N_{\cdot j}+1)$. $p$-values at $\gamma = 0$ are all below $0.004$; worst-case $p$-values at $\gamma = 1$ range from $0.015$ ($N=20$) to 0.2134 ($N=200$). Total time grows from $0.02$ seconds at $N = 20$ to $330$ seconds at $N = 120$ and 7918.9 at $N = 200$; peak memory increases are near zero for $N\leq 150$ and are 29 MB and 42 MB for $N = 180$ and $N = 200$, respectively.}
\label{tb:feasibility_chisq_3x3}
\end{changetable}

\begin{changetable}[htbp]
\centering
\scriptsize\renewcommand{\arraystretch}{1.1}\setlength{\tabcolsep}{4pt}
\resizebox{0.95\textwidth}{!}{%
\begin{tabular}{@{}c C{27mm} r r c c r@{}}
\toprule
$N$ & \thead{Observed\\table} & \thead{Candidates\\$|\mathcal{U}_{\rm C}|$} & $|\Omega_{\mathbf{t}}|$ & \thead{$p$-value\\$\gamma=0$} & \thead{Worst-case\\$p$-value\\$\gamma=1$} & \thead{Total time\\(seconds)} \\
\midrule
20 & $\begin{bmatrix} 7 & 1 & 0 & 0 \\ 1 & 1 & 1 & 1 \\ 0 & 0 & 1 & 7 \end{bmatrix}$ & 729 & 97 & 0.0008 & 0.0051 & 0.031 \\
\addlinespace[5pt]
50 & $\begin{bmatrix} 11 & 5 & 2 & 1 \\ 3 & 3 & 3 & 3 \\ 1 & 2 & 5 & 11 \end{bmatrix}$ & 30,976 & 149,168 & 0.0015 & 0.0166 & 62.538 \\
\addlinespace[5pt]
100 & $\begin{bmatrix} 18 & 10 & 5 & 3 \\ 7 & 7 & 7 & 7 \\ 3 & 5 & 10 & 18 \end{bmatrix}$ & 444,889 & 12,007,077 & 0.0002 & 0.0159 & 7421.433 \\
\addlinespace[5pt]
120 & $\begin{bmatrix} 20 & 11 & 7 & 4 \\ 9 & 9 & 9 & 9 \\ 4 & 7 & 11 & 20 \end{bmatrix}$ & 906,304 & 39,588,845 & 0.0004 & 0.0399 &  55887.464\\
\addlinespace[5pt]
150 & $\begin{bmatrix} 24 & 15 & 9 & 5 \\ 11 & 11 & 11 & 11 \\ 5 & 9 & 15 & 24 \end{bmatrix}$ & 2,178,576 & X & X & X & X \\
\addlinespace[5pt]
180 & $\begin{bmatrix} 27 & 17 & 11 & 7 \\ 14 & 14 & 14 & 14 \\ 7 & 11 & 17 & 27 \end{bmatrix}$ & 4,439,449 & X & X & X & X \\
\addlinespace[5pt]
200 & $\begin{bmatrix} 29 & 19 & 13 & 8 \\ 15 & 16 & 16 & 15 \\ 8 & 13 & 19 & 29 \end{bmatrix}$ & 6,744,409 & X & X & X & X \\
\bottomrule
\end{tabular}}
\caption{Chi-square test, $3 \times 4$ tables, $\bm{\delta} = (0,1,1)$; worst case over $\mathcal{U}_{\rm C}$ (Theorem~\mref{thm:set_size_of_U}). Total time grows from 0.031 seconds at $N = 20$ to 55887.464 seconds at $N = 120$; entries marked X were not computed due to anticipated long computation time; peak memory increases are $\approx 0$ MB at $N=20$, and $1$ MB, $38$ MB, and $45$ MB at $N=50, 100, 120$, respectively.}
\label{tb:feasibility_chisq_3x4}
\end{changetable}

\begin{changetable}[htbp]
\centering
\scriptsize\renewcommand{\arraystretch}{1.1}\setlength{\tabcolsep}{4pt}
\resizebox{0.95\textwidth}{!}{%
\begin{tabular}{@{}c C{21mm} r r c c r@{}}
\toprule
$N$ & \thead{Observed\\table} & \thead{Candidates\\$|\mathcal{U}_{\rm C}|$} & $|\Omega_{\mathbf{t}}|$ & \thead{$p$-value\\$\gamma=0$} & \thead{Worst-case\\$p$-value\\$\gamma=1$} & \thead{Total time\\(seconds)} \\
\midrule
20 & $\begin{bmatrix} 6 & 0 \\ 3 & 1 \\ 1 & 3 \\ 0 & 6 \end{bmatrix}$ & 121 & 12 & 0.0007 & 0.0032 & 0.015 \\
\addlinespace[5pt]
50 & $\begin{bmatrix} 12 & 2 \\ 7 & 4 \\ 4 & 7 \\ 2 & 12 \end{bmatrix}$ & 676 & 854 & 0.0007 & 0.0115 & 0.025 \\
\addlinespace[5pt]
100 & $\begin{bmatrix} 21 & 6 \\ 14 & 9 \\ 9 & 14 \\ 6 & 21 \end{bmatrix}$ & 2,601 & 8,898 & 0.0002 & 0.0253 & 0.253 \\
\addlinespace[5pt]
120 & $\begin{bmatrix} 23 & 8 \\ 17 & 12 \\ 12 & 17 \\ 8 & 23 \end{bmatrix}$ & 3,721 & 17,038 & 0.0009 & 0.0760 & 0.508 \\
\addlinespace[5pt]
150 & $\begin{bmatrix} 28 & 11 \\ 21 & 15 \\ 15 & 21 \\ 11 & 28 \end{bmatrix}$ & 5,776 & 33,726 & 0.0006 & 0.1002 & 1.303 \\
\addlinespace[5pt]
180 & $\begin{bmatrix} 33 & 14 \\ 25 & 18 \\ 18 & 25 \\ 14 & 33 \end{bmatrix}$ & 8,281 & 58,742 & 0.0005 & 0.1386 & 2.804 \\
\addlinespace[5pt]
200 & $\begin{bmatrix} 36 & 16 \\ 27 & 21 \\ 21 & 27 \\ 16 & 36 \end{bmatrix}$ & 10,201 & 81,634 & 0.0007 & 0.1969 & 4.515 \\
\bottomrule
\end{tabular}}
\caption{Chi-square test, $4 \times 2$ tables, $\bm{\delta} = (0,1,1,1)$; worst case over $\mathcal{U}_{\rm C}$ (Theorem~\mref{thm:set_size_of_U}), $|\mathcal{U}_{\rm C}| = (N_{\cdot 1}+1)(N_{\cdot 2}+1)$. $p$-values at $\gamma = 0$ are all below $0.001$; worst-case $p$-values at $\gamma = 1$ range from $0.003$ ($N=20$) to $0.20$ ($N=200$). Total time grows from $0.02$ to $4.5$ seconds; peak memory increase below 1~MB in every run.}
\label{tb:feasibility_chisq_4x2}
\end{changetable}

\begin{changetable}[htbp]
\centering
\scriptsize\renewcommand{\arraystretch}{1.1}\setlength{\tabcolsep}{4pt}
\resizebox{0.95\textwidth}{!}{%
\begin{tabular}{@{}c C{21mm} r r c c r@{}}
\toprule
$N$ & \thead{Observed\\table} & \thead{Candidates\\$|\mathcal{U}_{\rm C}|$} & $|\Omega_{\mathbf{t}}|$ & \thead{$p$-value\\$\gamma=0$} & \thead{Worst-case\\$p$-value\\$\gamma=1$} & \thead{Total time\\(seconds)} \\
\midrule
20 & $\begin{bmatrix} 5 & 0 \\ 3 & 1 \\ 1 & 1 \\ 1 & 3 \\ 0 & 5 \end{bmatrix}$ & 121 & 84 & 0.0107 & 0.0251 & 0.015 \\
\addlinespace[4pt]
50 & $\begin{bmatrix} 10 & 2 \\ 6 & 3 \\ 4 & 4 \\ 3 & 6 \\ 2 & 10 \end{bmatrix}$ & 676 & 6,000 & 0.0102 & 0.0585 & 0.022 \\
\addlinespace[4pt]
100 & $\begin{bmatrix} 17 & 5 \\ 12 & 7 \\ 9 & 9 \\ 7 & 12 \\ 5 & 17 \end{bmatrix}$ & 2,601 & 100,826 & 0.0029 & 0.0642 & 0.253 \\
\addlinespace[4pt]
120 & $\begin{bmatrix} 20 & 6 \\ 15 & 8 \\ 11 & 11 \\ 8 & 15 \\ 6 & 20 \end{bmatrix}$ & 3,721 & 201,398 & 0.0005 & 0.0298 & 0.407 \\
\addlinespace[4pt]
150 & $\begin{bmatrix} 24 & 8 \\ 18 & 11 \\ 14 & 14 \\ 11 & 18 \\ 8 & 24 \end{bmatrix}$ & 5,776 & 501,984 & 0.0005 & 0.0513 & 1.092 \\
\addlinespace[4pt]
180 & $\begin{bmatrix} 28 & 10 \\ 22 & 13 \\ 17 & 17 \\ 13 & 22 \\ 10 & 28 \end{bmatrix}$ & 8,281 & 1,031,776 & 0.0002 & 0.0425 & 2.182\\
\addlinespace[4pt]
200 & $\begin{bmatrix} 30 & 12 \\ 24 & 15 \\ 19 & 19 \\ 15 & 24 \\ 12 & 30 \end{bmatrix}$ & 10,201 & 1,599,652 & 0.0005 & 0.0951 & 3.284 \\
\bottomrule
\end{tabular}}
\caption{Chi-square test, $5 \times 2$ tables, $\bm{\delta} = (0,1,1,1,1)$; worst case over $\mathcal{U}_{\rm C}$. peak memory increase at all $N \leq 200$ cases are less than $1$ MB.}
\label{tb:feasibility_chisq_5x2}
\end{changetable}

\pagebreak

\subsection{Computational Gain of Lemma~\ref{lem:kernel_general_delta} and Theorem~\mref{thm:recursive_probability}} \label{app:G3}
\begin{changeblock}
\paragraph{Setup.}
Both Lemma~\ref{lem:kernel_general_delta} and Theorem~\mref{thm:recursive_probability} reformulate the null distribution in \meqref{eq:raw_sens_test} of the main text, which iterates over every treatment assignment in $\mathcal{Z}(\mathbf{N}_{I\cdot})$, and both represent this paper's effort toward computational saving. Table~\mref{tb:main_text_exact_computation_time_permutation_and_kernel} of the main text and Table~\ref{tb:exact_computation_time_permutation_and_kernel} compare Theorem~\mref{thm:recursive_probability} with \meqref{eq:raw_sens_test}; here we compare \eqref{eq:kernel_general_delta} of Lemma~\ref{lem:kernel_general_delta}, which sums over $\Omega_{\mathbf{t}} \times \Omega_{\mathbf{q}}$, with \meqref{eq:exact_distribution_kernel_form} of Theorem~\mref{thm:recursive_probability}, evaluated in the form \eqref{eq:alpha_W_form}. The observed tables are a $3 \times 3$ table with $N = 86$ (rows $(12,18,5)$, $(6,12,6)$, $(6,6,15)$) and the two subgroup tables of Section~\mref{sec:data_analysis}, under the ordinal statistic with the stated scores and under the chi-square statistic. The two representations agree on every $p$-value to seven significant digits; Table~\ref{tb:legacy_vs_current} reports the running time of the exact $p$-value at one $\mathbf{u}$.

\paragraph{Conclusion.}
The source of the computational saving is the same throughout: whenever a margin constraint reduces the space iterated over in the probability expression, a computational gain follows. Lemma~\ref{lem:kernel_general_delta} imposes the margin constraints that define $\Omega_{\mathbf{t}}$ and $\Omega_{\mathbf{q}}$ in \eqref{eq:kernel_general_delta}, together with those on $\Omega_{\mathbf{n}}$ in Lemma~\ref{lem:kernel_t_q}; these sets track every treatment level. Theorem~\mref{thm:recursive_probability} uses the fact that $\bm{\delta} \in \{0,1\}^I$ to merge the treatment levels into the two groups $\mathcal{I}_1$ and its complement, so that $\Omega_{\mathbf{q}}$ and $\Omega_{\mathbf{n}}$ are replaced by the scalar range $\Omega_{\overline{q}}$ and the $J$-dimensional set $\Omega_{\overline{\mathbf{n}}}$ in \meqref{eq:exact_distribution_kernel_form}. As stated in Section~\mref{sec:Maximizer_Characterization_and_Bounds} of the main text, both results apply once Theorems~\mref{thm:set_size_of_U}--\mref{thm:score_test_maximizer_binary} have located a maximizer among the binary vertices, so they provide an additional saving, from the reduction of the space of $\mathbf{z}$, on top of the reduction of the space of $\mathbf{u}$.
\end{changeblock}

\begin{changetable}[htbp]
\centering
\scriptsize\renewcommand{\arraystretch}{1.1}\setlength{\tabcolsep}{5pt}
\begin{tabular}{@{}l c r r r@{}}
\toprule
Table & $N$ & Lemma~\ref{lem:kernel_general_delta} (s) & Theorem~\mref{thm:recursive_probability} (s) & Ratio \\
\midrule
$3 \times 3$, $N = 86$, ordinal $(0,1,2.5) \times (0,1,2)$, $\gamma = 1$      & 86  & 31.32  & 0.136 & 230 \\
$3 \times 3$, $N = 86$, chi-square, $\gamma = 1$                              & 86  & 109.34 & 1.309 & 84 \\
\addlinespace[3pt]
Black girls, ordinal $(0,1,2.5) \times (0,1,2)$, $\gamma = 0.5$              & 94  & 10.37 & 0.021 & 494 \\
Black boys, ordinal $(0,1,2.5) \times (0,1,2)$, $\gamma = 0.5$               & 112 & 26.76 & 0.061 & 439 \\
Black girls, prior scores, $\Gamma = 2$                                      & 94  & 10.84 & 0.024 & 452 \\
Black boys, prior scores, $\Gamma = 2$                                       & 112 & 27.66 & 0.061 & 453 \\
Black girls, profile scores, $\Gamma = 2$                                    & 94  & 10.22 & 0.022 & 465 \\
Black boys, profile scores, $\Gamma = 2$                                     & 112 & 28.74 & 0.058 & 496 \\
\bottomrule
\end{tabular}
\caption{Running time, in seconds, of the exact $p$-value at one $\mathbf{u}$ under \eqref{eq:kernel_general_delta} and under \meqref{eq:exact_distribution_kernel_form}, on Posit Cloud (8 CPUs, 4~GB RAM). Ratio: the former divided by the latter.}
\label{tb:legacy_vs_current}
\end{changetable}

\titleformat{\section}[block]{\normalfont\Large\bfseries}{}{0pt}{}
\bibliography{references}

\end{document}

%% file: macros.tex
\providecommand{\thesisversion}{0}

\if0\thesisversion
  \usepackage{xr}
  \newcommand{\sref}[1]{\ref{supp-#1}}
  \newcommand{\seqref}[1]{\eqref{supp-#1}}
  \newcommand{\mref}[1]{\ref{main-#1}}
  \newcommand{\meqref}[1]{\eqref{main-#1}}
\else
  \newcommand{\sref}[1]{\ref{#1}}
  \newcommand{\seqref}[1]{\eqref{#1}}
  \newcommand{\mref}[1]{\ref{#1}}
  \newcommand{\meqref}[1]{\eqref{#1}}
\fi

\newtheorem{thm}{Theorem}         
\newtheorem{defn}{Definition}      
\newtheorem{lem}{Lemma}            
\newtheorem{proposition}{Proposition} 
\newtheorem{corollary}{Corollary}

\newcommand{\criticalt}{c} 
\newcommand{\rr}{\mathbf{r}}
\newcommand{\trr}{\tilde{\mathbf{r}}}

\newcommand{\trZ}{\widetilde{\mathbf{Z}}}

